\documentclass[floats,floatfix,amssymb,prd,twocolumn,superscriptaddress,nofootinbib,nolongbibliography,reprint]{revtex4-1}

\usepackage{amsmath}
\usepackage{amsfonts}
\usepackage{amssymb}
\usepackage{mathrsfs}
\usepackage{graphicx}
\usepackage{bm}
\usepackage{color}
\usepackage{commath}
\allowdisplaybreaks
\usepackage{makecell}

\usepackage{color}
\definecolor{linkcolor}{rgb}{0.1,0.3,0.6}
\usepackage[unicode, colorlinks=true, linkcolor=linkcolor, citecolor=linkcolor, filecolor=linkcolor, urlcolor=linkcolor, linktocpage, breaklinks]{hyperref}

\def\F{{\cal F}}

\def\l{{\ell}}
\def\lm{{\ell m}}
\def\lmn{{\ell m n}}
\def\ha{{\hat{a}}}
\def\TEOB{\texttt{TEOBResumS}}
\def\SEOB{\texttt{SEOBNR}}
\def\Dali{\texttt{ \TEOB{-Dalí}}}
\def\RWZ{{\texttt{RWZhyp}}}
\def\Teuk{{\texttt{Teukode}}}

\def\tA22{{t_{A_{22}}^{\rm peak}}}
\def\tAlm{{t_{A_\lm }^{\rm peak}}}
\def\td2omg0{{ t_{\dot{\omega}_{22}}^{\rm max}}}
\def\tLSO{{t_{\rm LSO}}}

\def\d2Amx{{\ddot{A}_{\rm peak}}}

\def\tLR{{t_{\rm LR}}}
\def\tNQC{{t_\lm^{\rm NQC}}}

\def\t{{\tau}}

\def\tsep{{t_{s}}}
\def\esep{{e_{s}}}
\def\xisep{{\xi_{s}}}
\def\tplunge{{t_{\ddot{r}=0}}}

\def\to{{t_0}}
\def\hS{{{}^S h}}
\def\barh{{\bar{h}}}
\def\hNQC{{\hat{h}_\lm^{\rm NQC}}}

\def\HKerr{{\hat{H}^{\rm eq}_{\rm Kerr}}}
\def\Ylm{{{}_{-2}Y_\lm}}

\newcommand\be{\begin{equation}}
\newcommand\ee{\end{equation}}

\def\hline{{\rule{\linewidth}{0.2pt}}\\}

\DeclareMathOperator{\sech}{sech}

\newcommand{\orcid}[1]{\href{https://orcid.org/#1}{
\includegraphics[width=10pt]{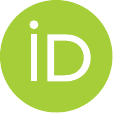}
}}

\usepackage[nolist,nohyperlinks]{acronym}

\newacro{bbh}[BBH]{binary black hole}
\newacro{bh}[BH]{black hole}
\newacroplural{bh}[BHs]{black holes}
\newacro{emri}[EMRI]{extreme mass ratio inspiral}
\newacroplural{emri}[EMRIs]{extreme mass ratio inspirals}
\newacro{eob}[EOB]{effective-one-body}
\newacro{eos}[EoS]{equation of state}
\newacro{gsf}[GSF]{Gravitational Self Force}
\newacro{gw}[GW]{gravitational-wave}
\newacroplural{gw}[GWs]{gravitational waves}
\newacro{hm}[HM]{Higher mode}
\newacroplural{hm}[HMs]{higher modes}
\newacro{imr}[IMR]{inspiral-merger-ringdown}
\newacro{lr}[LR]{light-right}
\newacro{lso}[LSO]{last stable orbit}
\newacro{lvk}[LVK]{LIGO-Virgo-KAGRA}
\newacro{nqc}[NQCs]{next-to-quasicircular corrections}
\newacro{nr}[NR]{numerical relativity}
\newacro{pn}[PN]{post-Newtonian}
\newacro{pm}[PM]{post-Minkowskian}
\newacro{qnm}[QNM]{quasi-normal mode}
\newacro{qc}[QC]{quasi-circular}
\newacro{rwz}[RWZ]{Regge-Wheeler-Zerilli}

\allowdisplaybreaks

\begin{document}

\title{Ringdown modeling for effective-one-body waveforms in the test-mass limit for eccentric equatorial orbits around a Kerr black hole}

\author{Simone \surname{Albanesi}\orcid{0000-0001-7345-4415}}
\affiliation{Theoretisch-Physikalisches Institut, Friedrich-Schiller-Universit{\"a}t Jena, 07743, Jena, Germany}
\affiliation{INFN sezione di Torino, Torino, 10125, Italy}
\author{Sebastiano \surname{Bernuzzi}\orcid{0000-0002-2334-0935}}
\affiliation{Theoretisch-Physikalisches Institut, Friedrich-Schiller-Universit{\"a}t Jena, 07743, Jena, Germany}
\author{Alessandro \surname{Nagar}\orcid{0000-0001-7998-2673}}
\affiliation{INFN sezione di Torino, Torino, 10125, Italy}
\affiliation{Institut des Hautes Etudes Scientifiques, 35 Route de Chartres, Bures-sur-Yvette, 91440, France}
\begin{abstract}
We study the plunge and merger of a non-spinning particle falling into a Kerr black hole
following an eccentric planar inspiral. The dynamics is driven by an effective-one-body radiation
reaction, and the corresponding numerical inspiral-merger-ringdown waveforms are obtained by solving
the Teukolsky equation with the 2+1 time-domain code \Teuk{}.
We then analyze in detail the plunge and merger phases, modeling the merger-ringdown waveform
using closed-form ans\"atze.
Crucially, our modeling starts from a point closely related to the light-ring crossing, rather than
from the amplitude peaks. This choice allows us to neglect the impact of the relativistic anomaly 
at the separatrix-crossing, and to extend the modeling to high spins and high eccentricities.
We model all the multipoles with $m\geq 1$ up to $\ell=4$, as well as the $(2,0)$, $(5,5)$, $(5,4)$,
and $(5,3)$ modes, including spherical-spheroidal mode-mixing and the beating between co-rotating and
counter-rotating quasi-normal modes.
The post-merger waveform model is then employed to complete an effective-one-body inspiral-plunge
waveform, thus providing a complete description.
Our model, built using elliptic-like configurations for the merger-ringdown phase, naturally extends to
dynamical capture scenarios without any further modification.
Finally, we provide insights into the extension of this framework to generic mass ratios, arguing 
that a time closely related to the inflection point of the (2,2) waveform frequency could be used 
as anchoring point for the ringdown modeling.
\end{abstract}

\date{\today}

\maketitle

% reset all acronyms
\acresetall

%=======================================================================================================
\section{Introduction}
%=======================================================================================================
\label{sec:introduction}
The direct detection of \acp{gw} by the \ac{lvk} collaboration has inaugurated an era of precision 
studies of compact-binary coalescences, with the observed signals predominantly originating from binary black hole mergers~\cite{LIGOScientific:2016aoc,LIGOScientific:2018mvr,LIGOScientific:2020ibl,LIGOScientific:2021djp,LIGOScientific:2025slb}. 
Extracting reliable source parameters from these observations requires waveform models that are both 
highly accurate and computationally efficient. Since the Einstein equations do not admit closed-form solutions describing 
the full coalescence of two compact objects, all state-of-the-art semi-analytical waveform models necessarily rely 
on numerical information to complement analytic approximations in the strong-field regime. An accurate description 
of the \ac{gw} source is needed to fully achieve the scientific goals of current 
detectors~\cite{VIRGO:2014yos,LIGOScientific:2014pky,Aso:2013eba} 
and future ones, such as the Einstein Telescope~\cite{Punturo:2010zz,ET:2025xjr}, Cosmic Explorer~\cite{Reitze:2019iox}
and LISA~\cite{LISA:2017pwj}.

Among the various modeling strategies developed to date, the \ac{eob} framework has emerged as one of the most successful 
approaches for \acp{gw} emitted by compact
binaries~\cite{Buonanno:1998gg,Buonanno:2000ef,Damour:2000we,Damour:2001tu,Buonanno:2005xu,Damour:2015isa}. 
By recasting the two-body problem into the motion of an effective particle in a deformed black hole spacetime, the \ac{eob} 
formalism provides a unified description of compact binary evolutions. While its analytical 
formulation captures the inspiral and plunge dynamics, the approach can be extended to merger and ringdown
by including numerical information, allowing the model to reproduce the late stages of the coalescence with remarkable 
fidelity~\cite{Buonanno:2006ui,Damour:2007xr}. 
Historically,  \ac{eob} models have been developed for describing quasi-circular black hole
binaries, and have been mostly developed in two model families: 
\TEOB{}~\cite{Damour:2014sva,Nagar:2015xqa,Nagar:2018zoe,Nagar:2018gnk,Nagar:2019wds,
Nagar:2018plt,Rettegno:2019tzh,
Riemenschneider:2021ppj,Nagar:2022icd,Damour:2025uka}
and \SEOB{}~\cite{Buonanno:2007pf,Buonanno:2009qa,Pan:2011gk,
Pan:2009wj,Taracchini:2012ig,Taracchini:2013rva,Bohe:2016gbl,
Cotesta:2018fcv,Ossokine:2020kjp,Khalil:2023kep,Pompili:2023tna,Buonanno:2024byg}.
In recent years many efforts have been devoted to the development of \ac{eob} models to describe
eccentric black hole binaries, considering both hyperbolic~\cite{Damour:2014afa,Hopper:2022rwo,Buonanno:2024vkx} and elliptic-like 
orbits~\cite{Hinderer:2017jcs,Chiaramello:2020ehz,Nagar:2020xsk,Nagar:2021gss,Ramos-Buades:2021adz,Nagar:2021xnh,
Nagar:2022fep,Andrade:2023trh,Nagar:2024oyk,Gamboa:2024imd,Gamboa:2024hli}, 
and the transition between the two regimes~\cite{Albanesi:2024xus}.
Similarly, \ac{eob} models have been extended to describe precessing dynamics~\cite{Pan:2003qt,Akcay:2020qrj,Gamba:2021ydi,Ramos-Buades:2023ehm,Estelles:2025zah}, eventually 
combining these effects with non-circular motion~\cite{Gamba:2024cvy,Albanesi:2025txj,Gamba:2025qfg,Chandra:2025jfc,Chiaramello:2025bhi}.

A crucial aspect of any \ac{eob} model is the enhancement and completion through the inclusion of numerical
information, which can be extracted from \ac{nr} simulations~\cite{Pretorius:2005gq,Campanelli:2005dd,Baker:2005vv}.
Current models usually rely on simulations of the 
SXS catalog~\cite{Scheel:2014ina,Blackman:2015pia,Lovelace:2014twa,Boyle:2019kee,Scheel:2025jct,SXS:catalog}. 
Notably, \ac{eob} models also benefits from numerical data obtained in perturbation theory. 
Indeed, a distinctive strength of the \ac{eob} approach lies in its smooth connection between the comparable-mass 
regime and the test-mass limit. The deformation of the effective metric depends indeed on the symmetric mass ratio 
$\nu=\mu/M$, where $\mu=m_1 m_2/(m_1+m_2)$ is the reduced mass and $M=m_1+m_2$ is the total mass.
This property has motivated extensive investigations of the extreme-mass-ratio limit as a controlled theoretical laboratory 
in which to explore and validate modeling prescriptions relevant also for comparable-mass binaries (see, e.g., Refs.~\cite{Nagar:2006xv,Damour:2007xr,Bernuzzi:2010ty,Bernuzzi:2010xj,Yunes:2010zj,Barausse:2011kb,Taracchini:2014zpa,
Albanesi:2021rby,Albanesi:2023bgi,Nagni:2025cdw,Nishimura:2026nse}). 
Beyond its role as a testing ground, this limit is of direct astrophysical interest, as \acp{emri} are among 
the key targets of the future space-based detector LISA~\cite{Babak:2017tow,Berry:2019wgg}.
Accurately modeling \acp{emri} requires a precise description of the particle's dynamics, 
also including the self-force effects of the particle's own gravitational field. 
Significant progress in this direction has been achieved within the \ac{gsf} program~\cite{Pound:2015tma,Barack:2018yvs,vandeMeent:2017bcc,VanDeMeent:2018cgn,Warburton:2021kwk,Kuchler:2024esj,Kuchler:2025hwx,Honet:2025dho}.
These results have recently been also employed into an \ac{eob} framework and compared with
\ac{gsf}-evolved quasi-circular 
inspirals~\cite{Albertini:2022rfe,Albertini:2022dmc,vandeMeent:2023ols,Albertini:2023aol,Albertini:2024rrs,Albertini:2024agg,Leather:2025nhu}. 
In the present work, however, we adopt a complementary perspective and focus on the \ac{eob}-based construction of 
\ac{gw} in the test-mass limit, without directly relying on \ac{gsf} evolution. 
Specifically, we study the gravitational radiation emitted by a non-spinning test particle undergoing an eccentric inspiral 
followed by a plunge into a Kerr black hole. Extending earlier analyses restricted to Schwarzschild spacetime~\cite{Albanesi:2023bgi}, 
we investigate how black hole spin and orbital eccentricity interplay in shaping the late-time dynamics and the resulting waveform. 
Several previous works studied plunging geodesics in Kerr~\cite{Levin:2008yp,Mummery:2023hlo,Li:2023bgn},
also considering off-plane dynamics~\cite{Dyson:2023fws} and including the spin of the particle~\cite{Piovano:2026wpz}.
In this paper, instead, we study non-conservative equatorial dynamics of non-spinning test masses 
by including a radiation reaction force in our equations of motion. However, it should be noted that for long 
and accurate evolutions, one should also consider physical effects that are neglected here, such as 
self-correction in the conservative dynamics, horizon absorption, and higher-order corrections~\cite{Albertini:2023aol}.
Despite these caveats, our setup allows us to study the plunge, merger and ringdown of these systems. 

The numerical waveforms used to build our ringdown model are obtained by solving the Teukolsky equation in the time domain, 
yielding the Weyl scalar $\Psi_4$ at infinity, from which the \ac{gw} multipoles are reconstructed. While for Schwarzschild
we used the \RWZ{} time-domain code~\cite{Bernuzzi:2010ty,Bernuzzi:2012ku}, for numerical waveforms in Kerr we use
\Teuk{}~\cite{Harms:2014dqa}.
We then adopt and extend a phenomenological ringdown model inspired by earlier works~\cite{Damour:2014yha,DelPozzo:2016kmd}, 
which allows for a more accurate representation of the waveform across the transition from plunge to ringdown. We construct and analyze the 
dominant $(2,2)$ mode, as well as higher multipoles with $m\geq1$ up to $\ell=4$, including also the $(5,5)$, $(5,4)$, 
and $(5,3)$ modes. We also model the (2,0) mode following the prescription introduced in Ref.~\cite{Albanesi:2024fts}. 
Spherical-spheroidal mode-mixing and beating between co-rotating 
and counter-rotating \acp{qnm} are also modeled. We also briefly discuss 
post-merger tails~\cite{Price:1971fb,Leaver:1986gd,Andersson:1996cm}, but we do not 
include them in our modelization.
The key novelty of our approach is the use of an anchoring point (or starting time) 
for the ringdown modeling that is strictly related to 
the \ac{lr} crossing, rather than to the amplitude peaks, as typically done in the past literature. 
As we will show, this choice not only improves the 
modeling for prograde quasi-circular systems with high spins, but also allows us to neglect, in the eccentric case, 
the value of the relativistic anomaly at the transition between stable and unstable orbits, $\xisep$. Indeed, while 
the location of the amplitude peaks can strongly depend on $\xisep$ for highly rotating systems, the waveform after the \ac{lr} crossing 
does not, thus showing that a significant portion of the waveform after the (2,2) amplitude peak is strongly source-driven.
The ringdown modeling is then used to complete the inspiral-plunge \ac{eob} waveform, thus obtaining a full \ac{imr} waveform.

The paper is organized as follows. In Sec.~\ref{sec:eccentric_systems} we discuss the particle’s dynamics, focusing in 
particular on the transition from inspiral to plunge, and briefly describe how the numerical Teukolsky waveforms are 
obtained. We also discuss the relevance of the relativistic anomaly at the separatrix crossing and how it can be neglected. 
In Sec.~\ref{sec:postmrg_waveform} we highlight the \ac{qnm} structure of the ringdown waveform and describe in detail our 
phenomenological modeling, providing a closed-form description of this signal across the parameter space. In Sec.~\ref{sec:eob} 
we discuss how the model obtained here is used to complete the \ac{eob} waveform, and in Sec.~\ref{sec:comparisons} we compare 
our full \ac{imr} \ac{eob} waveform with the numerical results obtained along the same dynamics, both for elliptic-like systems 
and for dynamical captures. We conclude in Sec.~\ref{sec:conclusions}, where we also discuss how the present work can provide 
useful insights for the comparable-mass case.

The rescaled phase-space variables that we use in this work are related to the physical ones by $t=T/(GM)$, 
$r=R/(GM)$, $p_{r}=P_{R}/\mu$ and $p_\varphi=P_\varphi/(\mu GM)$. We
will use geometric units $G=c=1$.

%=======================================================================================================
\section{From the eccentric inspiral to plunge, merger and ringdown}
\label{sec:eccentric_systems}
%=======================================================================================================
We start by describing how we obtain bound dynamics in the equatorial plane of a Kerr black hole and how
they are employed to obtained the corresponding waveform by solving numerically
the Teukolsky equation~\cite{Teukolsky:1973ha}. The numerical set up is similar to the one considered in 
Ref.~\cite{Albanesi:2021rby}, but we recall it for completeness. 
We then focus on the transition from inspiral to plunge, and the subsequent merger and ringdown. 
The logic outlined here is similar to Sec.~II of Ref.~\cite{Albanesi:2023bgi}, but while we were there considering
only the Schwarzschild case, here we consider the more general Kerr spacetime.  

%-------------------------------------------------------------------------------------------------------
\subsection{Equatorial bound orbits in Kerr spacetime}
\label{sec:kerr_dynamics}
%-------------------------------------------------------------------------------------------------------
We consider a non-spinning test-particle of mass $\mu$ which orbits a Kerr black hole (BH) with
dimensionless spin $\ha\equiv J_{\rm BH}/M^2$, where $J_{\rm BH}$ is the
angular momentum of the BH and $M\gg \mu$ its mass. If not specified otherwise,
we always consider $\nu\equiv \mu/M=10^{-3}$. Note that we consider $\ha\in (-1,1)$, with negative
values corresponding to $J_{\rm BH}$ anti-aligned with the orbital angular momentum.
In the equatorial case, the Hamiltonian which 
describes the conservative contributions can be written as a sum of 
orbital contribution with all the spin-spin contributions and a spin-orbit term~\cite{Damour:2014sva},
\be
\HKerr =  \sqrt{A \del{r} \del{1 + \frac{p_{\varphi}^2}{r_{c}^2}}+p_{r_*}^2} + \frac{2\ha p_\varphi}{r r_c^2},
\ee
where $r$ is the radius, $p_\varphi$ the orbital angular momentum, and $p_{r_*}$ is the conjugate 
momentum of the tortoise coordinate $r_*$, defined as $p_{r_*} = \sqrt{A/B}\;p_r$, being $p_r$ 
the radial momentum. The metric functions $A(r)$ and $B(r)$ are
\begin{align}
A(r) & = \frac{1+2 u_c}{1+2 u}\left(1- 2 u_c\right) \ , \\
B(r) & = \frac{1}{1-2 u + \ha^2 u^2},
\end{align}
where $u=1/r$, $u_c=1/r_c$ and $r_c$ is the centrifugal 
radius~\cite{Damour:2014sva}, which reads
\be
r_c^2=r^2+\ha^2+2\frac{\ha^2}{r}.
\ee

To model the loss of energy and angular momentum of the system due to the emission of \acp{gw}, 
we introduce a \ac{pn} radiation-reaction force $\F = \left(\F_r, \F_\varphi \right)$~\cite{Bini:2012ji}.
The prescription adopted here is that of Refs.~\cite{Chiaramello:2020ehz,Albanesi:2021rby}, in which the circular contribution 
is strengthened through appropriate resummation techniques~\cite{Damour:2008gu,Nagar:2016ayt,Messina:2018ghh}, while the non-circular corrections 
are expressed in terms of explicit time derivatives of the radius, $r$, and the orbital frequency, $\Omega$.
As anticipated, in this work we focus on the late stages of the evolution, in particular on 
the merger and ringdown phases. Therefore, we do not discuss in detail the accuracy of this 
prescription, which has nonetheless been extensively analyzed in previous studies~\cite{Albanesi:2021rby}.
A similar prescription is also employed in the \ac{eob} model \Dali{}~\cite{Nagar:2024oyk,Albanesi:2025txj}.
However, we stress that to evolve long and accurate astrophysical \acp{emri}, this prescription has to be improved, as
detailed for example in Ref.~\cite{Albertini:2023aol}. In particular, one should include 
$\nu$-corrections in the Hamiltonian~\cite{Nagar:2022fep} and higher-multipole
corrections and horizon absorption effects in the radiative part. 
This latter phenomenon could be particularly relevant for
extreme-mass-ratio systems and eccentric 
dynamics~\cite{Hughes:2001jr,Isoyama:2017tbp,Datta:2023wsn,Datta:2024vll}.
Analytical results for generic planar orbits are now available, and will 
be therefore considered in future work to evolve accurate 
inspirals~\cite{Chiaramello:2024unv,Gamba:horizonII}.
Finally, for highly eccentric prograde orbits around fast rotating black holes, it is possible 
to observe burst of radiation linked to \ac{qnm} excitation, known as 
{\it wiggles}~\cite{Kojima:10.1143,Rifat:2019fkt,Thornburg:2019ukt,Albanesi:2021rby}. However,
this effect is also not taken into account by our PN-based radiation reaction. 

With the caveats outlined above, the equation of motion
of a test-mass in the equatorial plane of a Kerr black hole can be written 
as~\cite{Damour:2014sva,Harms:2016ctx}
\begin{subequations}
\label{eq:eom}
\begin{align}
\dot{r} &=\left(\frac{A}{B}\right)^{1 / 2} \frac{\partial \HKerr }{\partial p_{r_{*}}}  , \\ 
\dot{\varphi} &=\frac{\partial \HKerr}{\partial p_{\varphi}} \equiv \Omega  , \label{eq:freq} \\ 
\dot{p}_{r_*} &=\left(\frac{A}{B}\right)^{1 / 2} \left( \hat{\F}_r -  \frac{\partial \HKerr}{\partial r} \right)  , \\
\dot{p}_\varphi &=\hat{\F}_{\varphi},
\end{align}
\end{subequations}
where the hat on the components of the radiation reaction denotes 
a $\nu$-normalization, $\hat{\F}_{r,\varphi} \equiv \F_{r,\varphi} /\nu $.
Note that Eqs.~\eqref{eq:eom} can be easily solved with standard ODE solvers.
The only technicality which is worth noting, is that the time-derivatives
of $r$ and $\Omega$ in $\F_\varphi$ are obtained with an analytical iterative procedure,
as detailed in Appendix~A of Ref.~\cite{Nagar:2024oyk}.

Since we consider planar orbits, each dynamics is fully characterized by two constants of motion. 
Note that these quantities actually slowly evolve due to the presence of the dissipative
radiation reaction force $\F$. A natural choice of constants of motion would be 
the energy $\hat{E}$, given by the on-shell value of the Hamiltonian, and the
orbital angular momentum $p_\varphi$. However, by providing numerical values of these two
quantities it is not intuitive to understand which kind of orbit we are considering.
It is therefore useful to define the semilatus rectum $p$ and the eccentricity $e$ as 
\be
\label{eq:ep_definition}
e  = \frac{r_+ - r_-}{r_+ + r_-}, \quad \quad p  = \frac{2 r_+ r_-}{r_+ + r_-} ,
\ee
where $r_\pm $ are the two radial turning points.
Using these two quantities, the radial motion can be thus parametrized as 
\be
r\left(\xi \right) =  \frac{p}{1-e \cos{\xi}},
\ee
where $\xi$ is the relativistic anomaly. With this 
choice\footnote{Other works consider $r\left(\xi \right) = p/(1+e \cos{\xi})$, such
that $\xi=0$ corresponds to the periastron. See, e.g., Ref.~\cite{Faggioli:2025hff}.}, 
$\xi=0$ correspond to the apastron, $r_+ = p/(1-e)$, while $\xi=\pi$ corresponds
to the periastron, $r_-=p/(1+e)$. 
Stable orbits are allowed for $p\geq p_s$, where 
$p_s$ is known as separatrix and, in the equatorial case, can be 
found as a root of the following polynomial~\cite{OShaughnessy:2002tbu,Stein:2019buj}
\begin{align}
& p_s^2 (p_s - 6 - 2 e)^2 + \ha^4 (e-3)^2 (e+1)^2  \\ 
& - 2 \ha^2 (1+e) p_s \left[14 + 2 e^2 + p_s (3 - e)\right] = 0 . \nonumber
\end{align} 
Note that in the Schwarzschild case, we simply have $p_s=6+2e$. 
If $p<p_s$, the orbit is no longer stable and the particle is doomed to plunge, even
if dissipative effects are not taken into account. Indeed, in this last case the periastron is 
no longer defined, and therefore $e$ and $p$ are no longer defined either.

The radial turning points $r_\pm$ can be computed 
at any time if $p>p_s$ by considering the radial effective potential
\be
\label{eq:energy}
V(r;p_\varphi) = \HKerr (r, p_\varphi, p_{r_*}=0).
\ee
Indeed, for any pair $(\hat{E}, p_\varphi)$, $r_\pm$ are the radii which satisfy the two equations
$\hat{E}=V(r_\pm,p_\varphi)$. Therefore, as long as the motion is bound (i.e, as long
the periastron is defined), we have a clear map between $(\hat{E}, p_\varphi)$ 
and $(e,p)$. However, while $(\hat{E}, p_\varphi)$ are defined along
the whole evolution, $(e,p)$ cease to be defined at the separatrix crossing
$p=p_s$, which inevitably occurs when dissipative effects are taken into account. 
Through this paper, we will denote the eccentricity and the relativistic anomaly
at the separatrix-crossing time as $\esep$ and $\xisep$. 

Note that in terms of the potential $V(r)$, the separatrix-crossing occurs
when the energy $\hat{E}$ is equal to the local maximum $V_{\rm max}$, which corresponds 
to unstable circular orbits. Therefore, in the special case in which the separatrix crossing 
occurs when the particle is close to the periastron, a long-lived circular whirl is observed 
before the onset of the plunge. However, the condition $\hat{E}=V_{\rm max}$ can, in principle, 
be satisfied at any point along the orbit, and the occurrence of such a whirl phase is therefore 
not guaranteed. We will return to this point in Sec.~\ref{subsec:plunge}.

We finally note that care must be taken when using eccentricity in General Relativity, 
since an eccentricity directly tied to the dynamics cannot be defined in a gauge-invariant manner. 
Nevertheless, recent studies have introduced definitions of eccentricity based on the waveform, 
thereby providing a notion linked to an observable quantity~\cite{Ramos-Buades:2019uvh,Ramos-Buades:2022lgf,Shaikh:2023ypz,Shaikh:2025tae}.
Despite the drawbacks discussed above, a dynamical eccentricity can still be employed internally within 
a waveform model to characterize certain late-time features of the signal. 
We will discuss this aspect in more detail in Sec.~\ref{subsec:global}, also highlighting the associated caveats.

%-------------------------------------------------------------------------------------------------------
\subsection{Numerical waveforms}
\label{subsec:num_waves}
%-------------------------------------------------------------------------------------------------------
%
\begin{figure*}[t]
	\begin{center}
	\includegraphics[width=0.320\textwidth]{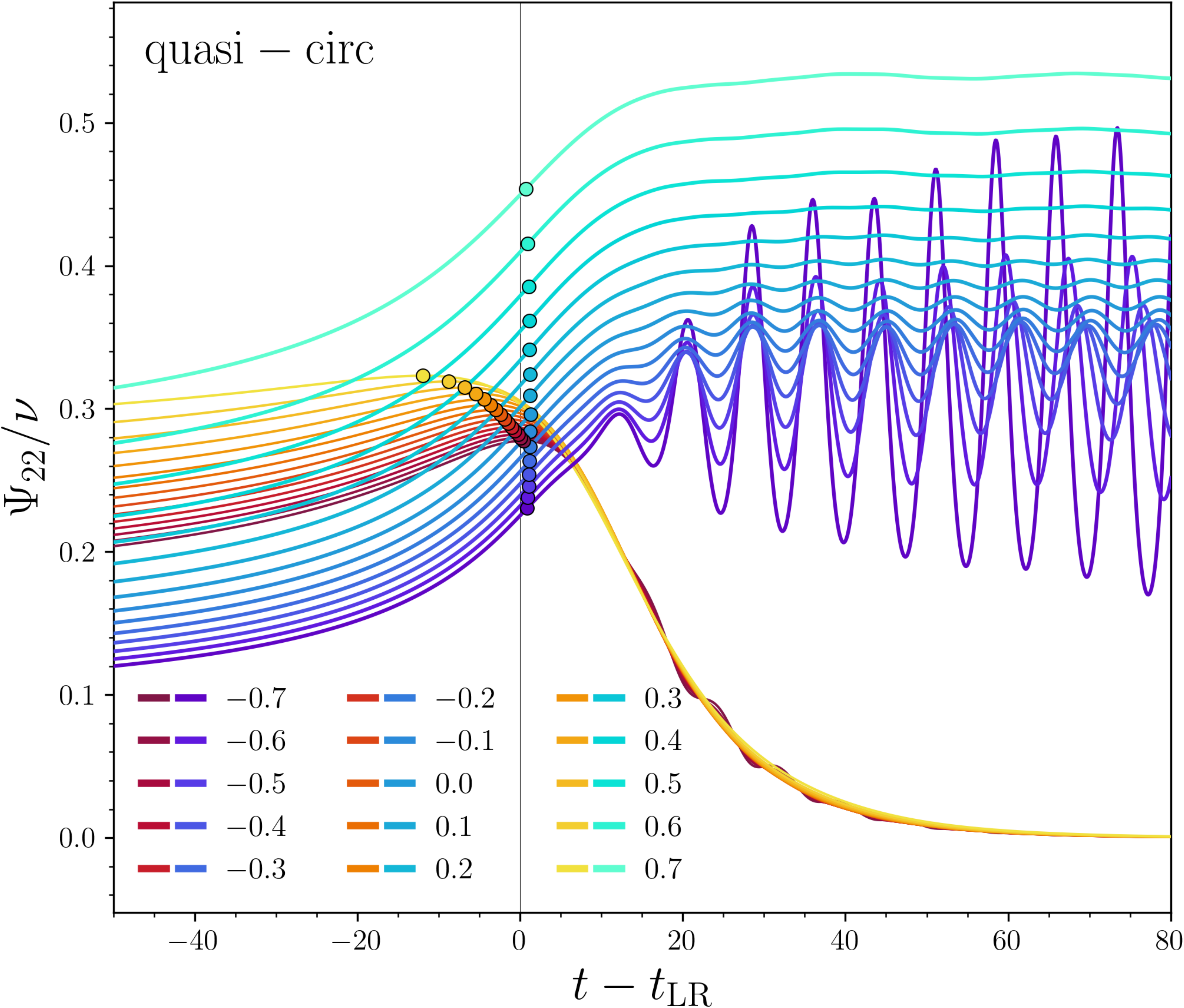}
	\includegraphics[width=0.325\textwidth]{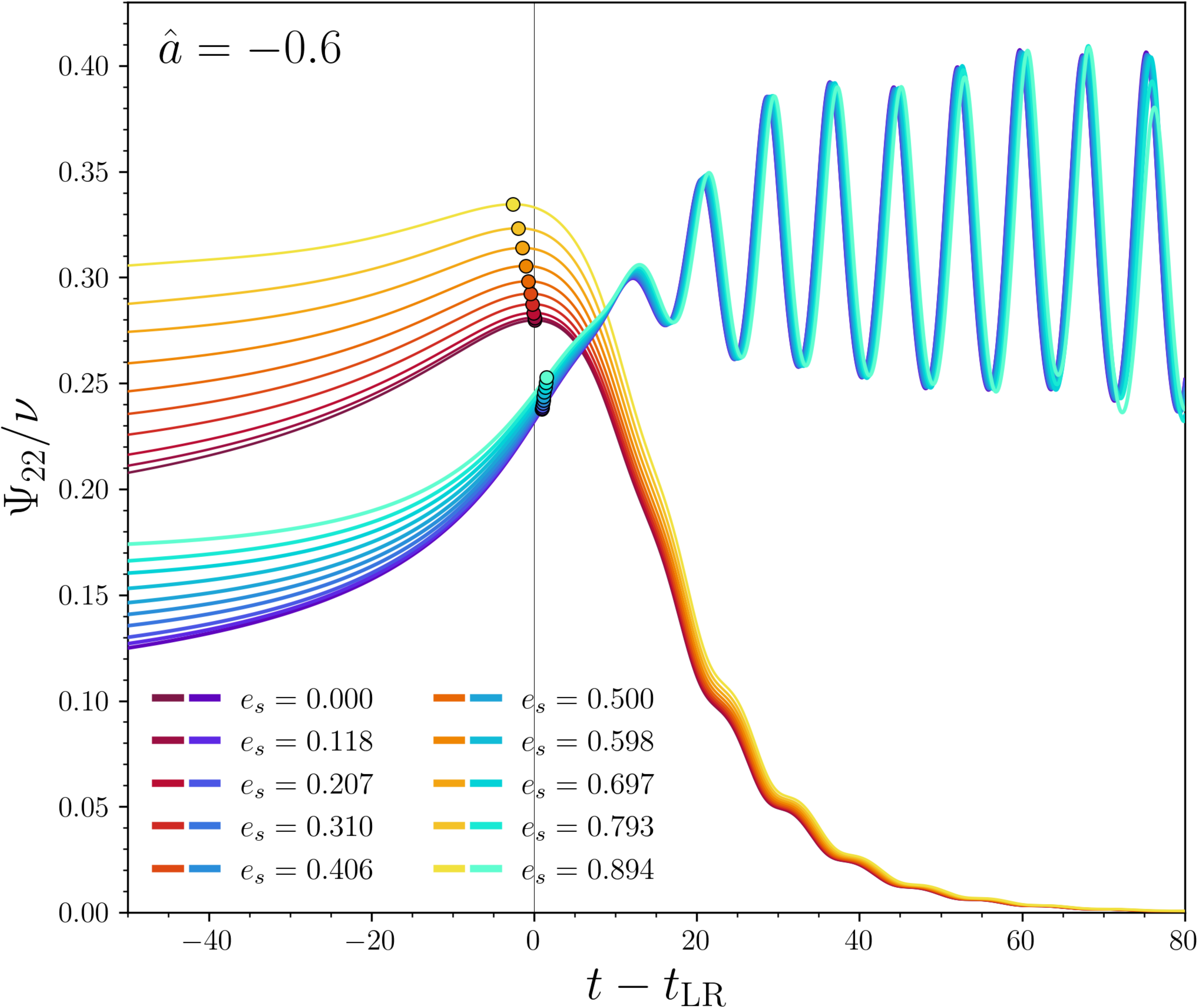}
	\includegraphics[width=0.322\textwidth]{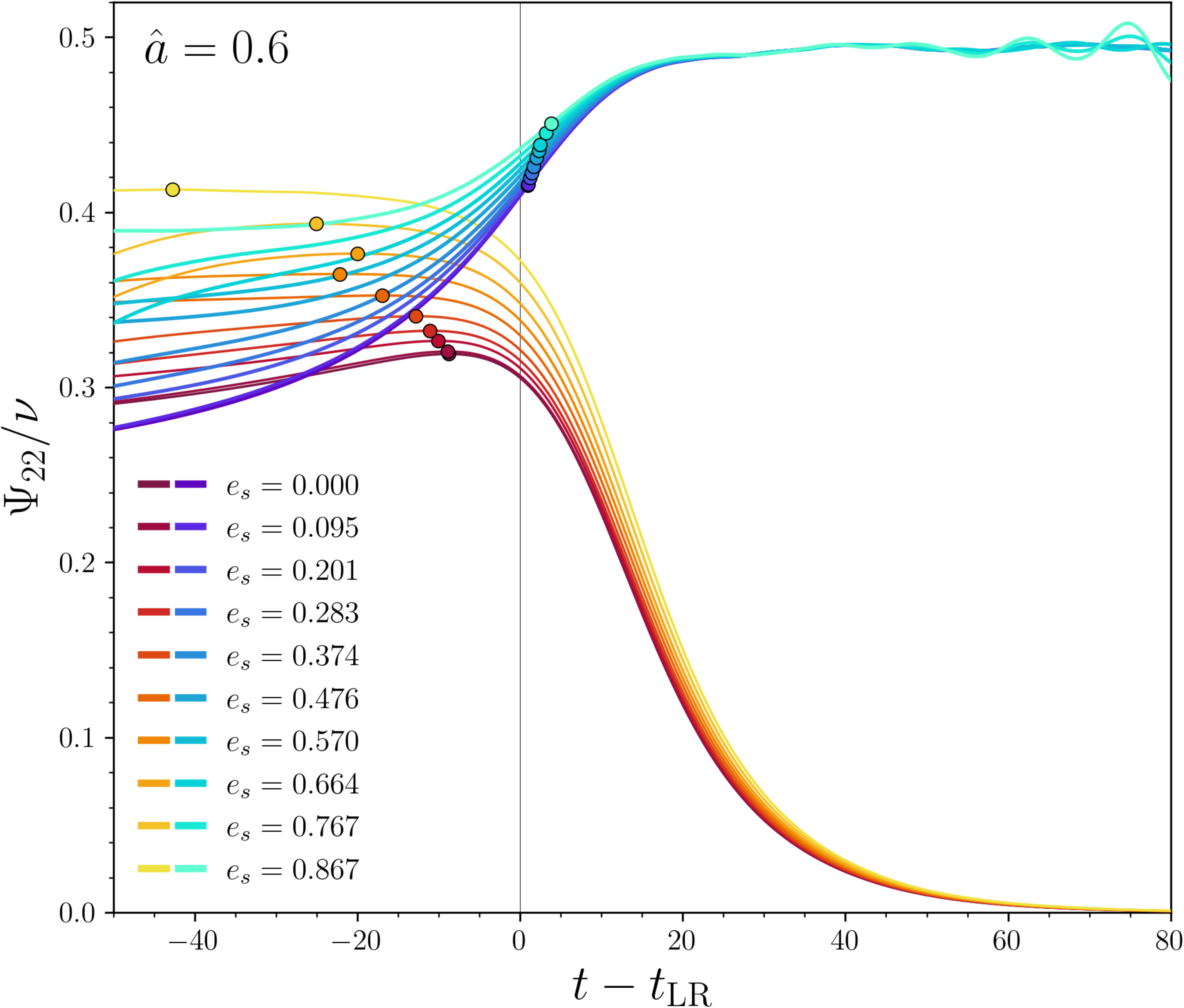}	
	\caption{\label{fig:num_waves}
	Left panel: amplitudes $A_{22}$ (warm colors) and frequencies $\omega_{22}$ (cold colors) for 
	RWZ-normalized waveforms $\Psi_{22}$ for quasi-circular systems
	with different Kerr spins, $\ha\in[-0.7,0.7]$ (reported in the legend).
	Dots on the amplitudes mark the locations of the maxima, $\tA22$,
	while dots on the frequencies mark the inflection point, $\td2omg0$.
	Waveforms aligned with respect to the \ac{lr} crossing $\tLR$.
	Middle panel: same quantities, but for $\ha=-0.6$ and eccentricities $\esep\in[0,0.9]$.
	Right panel: as the middle panel, but for $\ha=0.6$.}
	\end{center} 
\end{figure*}
We decompose the waveform strain in multipoles using the spin-weighted spherical harmonics
with weight $s=-2$,
\begin{equation}
\label{eq:strain}
h_+-{\rm i} h_\times = D_L^{-1} \sum_{\l=2}^{\infty} \sum_{m=-\l}^{\l} h_\lm \,\Ylm(\Theta, \Phi), 
\end{equation}
where $D_L$ is the luminosity distance, and $(\Theta,\Phi)$ are the angular variables of
a distant observer. We only consider multipoles up to $\l=8$ for the inspiral-plunge waveform 
and in the radiation reaction, and we explicitly provide a merger-ringdown model for modes 
up to $\l=5$, as detailed in Sec.~\ref{subsec:global}. We denote the amplitude and the waveform phase
as $h_\lm = A_\lm e^{- i \phi_\lm}$, and the waveform frequency as $\omega_\lm = \dot{\phi}_\lm$.

For any considered orbit, the corresponding waveform 
at linear order in the mass ratio can be obtained by solving numerically the \ac{rwz}
equations~\cite{Regge:1957td,Zerilli:1970se,Moncrief:1974am} in the non-spinning case, 
and more generally solving the Teukolsky equation~\cite{Teukolsky:1973ha}. In this work, we will focus 
on the spinning case, since the non-spinning scenario has been already
extensively discussed in Ref.~\cite{Albanesi:2023bgi}. 

The numerical setup that we use for obtaining Teukolsky waveform is the one discussed in Ref.~\cite{Albanesi:2021rby},
that we briefly recall here. We use the time-domain solver \Teuk{}~\cite{Harms:2014dqa}, which 
employs horizon-penetrating and hyperboloidal coordinates, allowing us to include the horizon and the future null infinity 
in the computational domain~\cite{Zenginoglu:2007jw,Zenginoglu:2009hd,Zenginoglu:2010cq}.
Exploiting the axial symmetry of the Kerr spacetime, the original 3+1 evolution equation is decomposed into 
a set of 2+1 equations, one for each azimuthal Fourier mode $m$. The solution provides the Weyl scalar $\psi_4$ at scri, 
defined as the contraction of the Weyl tensor with a suitable null tetrad (the Hawking-Hartle tetrad in our case),
that we decompose in $\l$-multipoles. 
The waveform multipoles are then reconstructed through a double time integration of the Weyl scalar multipoles, 
since we have, asymptotically, $\ddot{h}_\lm = \psi_4^\lm$.
If not specified otherwise, we compute the waveform using a resolution $N_r\times N_\theta = 3601\times 161$,
where $N_r$ and $N_\theta$ are the number of points in the radial and angular directions, respectively.
This resolutions has also been used in previous works~\cite{Albanesi:2021rby,Albanesi:2024fts}. We will however occasionally 
use an higher resolution, $N_r\times N_\theta = 5401\times 321$, to investigate small variations in the waveform induced by
slightly different dynamics. For a quantitative assessment of the numerical errors and comparisons with
other results in the literature, we point to Ref.~\cite{Albanesi:2021rby}.

We present representative examples for the dominant $(2,2)$ multipole in Fig.~\ref{fig:num_waves}.
For visualization purposes, we adopt the RWZ normalization
$\Psi_{22}=h_{22}/\sqrt{(\l+2)(\l+1)\l(\l-1)}$.
Starting from the left panel, we show amplitudes (warm colors) and frequencies (cold colors)
for quasi-circular systems with Kerr spins ranging from $\ha=-0.7$ to $\ha=0.7$.
All waveforms are aligned with respect to the \ac{lr} crossing.
Markers on the amplitudes indicate the location of the maxima, $\tA22$,
while dots on the frequencies denote their inflection points, $\td2omg0$.
As expected, increasing the spin leads to both larger amplitude peaks and higher final frequencies.
Notably, while the inflection points remain close to the \ac{lr} crossing for all spins,
the amplitude maxima are clearly anticipated as the spin increases.
This feature, already well known for quasi-circular test-mass systems~\cite{Taracchini:2014zpa,Harms:2014dqa},
will play a central role in our analysis, as discussed in the following sections.
Oscillations in both amplitude and frequency, generated by the beating between co-rotating and
counter-rotating \acp{qnm}, are particularly evident for retrograde orbits with large spin magnitudes,
as further discussed in Sec.~\ref{subsec:beating}.
In the middle panel of Fig.~\ref{fig:num_waves}, we show the same waveform quantities for systems
with fixed spin $\ha=-0.6$ and different eccentricity $\esep$, which we report in the legend.
Increasing the eccentricity leads to larger amplitudes at merger and to a slight anticipation
of $\tA22$ with respect to the \ac{lr}.
Frequencies prior to the \ac{lr} crossing are also higher for more eccentric configurations,
while the inflection points $\td2omg0$ are instead slightly delayed relative to the \ac{lr}.
No significant change is observed in the amplitude of the beating.
These effects are qualitatively similar to those observed in the Schwarzschild case
discussed in Ref.~\cite{Albanesi:2023bgi} (see, e.g., Fig.~4 therein),
and essentially related to the fact that the plunge
starts from lower radii for more eccentric systems.
Finally, in the right panel we consider eccentric configurations with $\ha=0.6$.
While the qualitative impact of eccentricity is similar, an important difference emerges:
the location of the amplitude peak $\tA22$ is more sensible to the eccentricity  and, crucially,
is no longer monotonic in the eccentricity.
In particular, $\tA22-\tLR$ is larger for $\esep=0.570$ than for $\esep=0.664$.
This behavior is related to the value of the relativistic anomaly at the separatrix crossing,
$\xisep$, as discussed further in Sec.~\ref{subsec:plunge}.
Although $\xisep$ strongly affects the timing of $\tA22$ for systems with positive spin,
we will argue that $\xisep$ can be neglected when constructing a post-merger waveform model,
provided a suitable anchoring point is chosen.
Specifically, we will show that an anchoring point closely tied to the \ac{lr} crossing
is preferable to the peak of the $(2,2)$ amplitude, see Sec.~\ref{subsec:anchoring}.
We finally note that, at late times, oscillations appear in the frequency for highly eccentric
systems.
These are caused by tail effects, which are enhanced by eccentricity and have been discussed
extensively in previous 
works~\cite{Albanesi:2023bgi,DeAmicis:2024not,Islam:2024vro,Becker:2025zzw,Islam:2025wci,DeAmicis:2024eoy,Ma:2024hzq}. 

%-------------------------------------------------------------------------------------------------------
\subsection{Transition from inspiral to plunge}
\label{subsec:plunge}
%-------------------------------------------------------------------------------------------------------
%
\begin{figure*}[t]
	\begin{center}
	\includegraphics[width=0.31\textwidth]{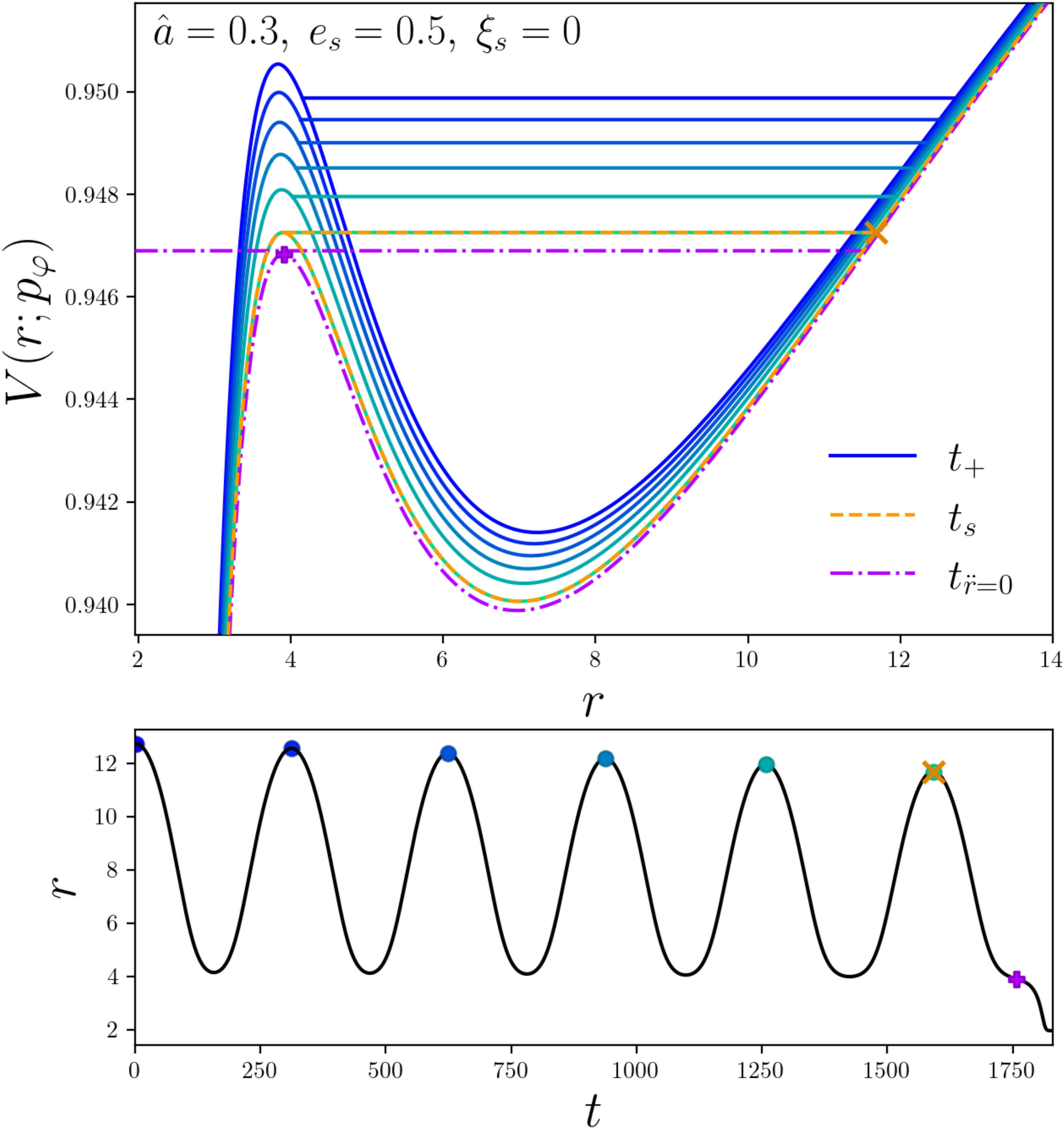}
	\hspace{1mm}
	\includegraphics[width=0.31\textwidth]{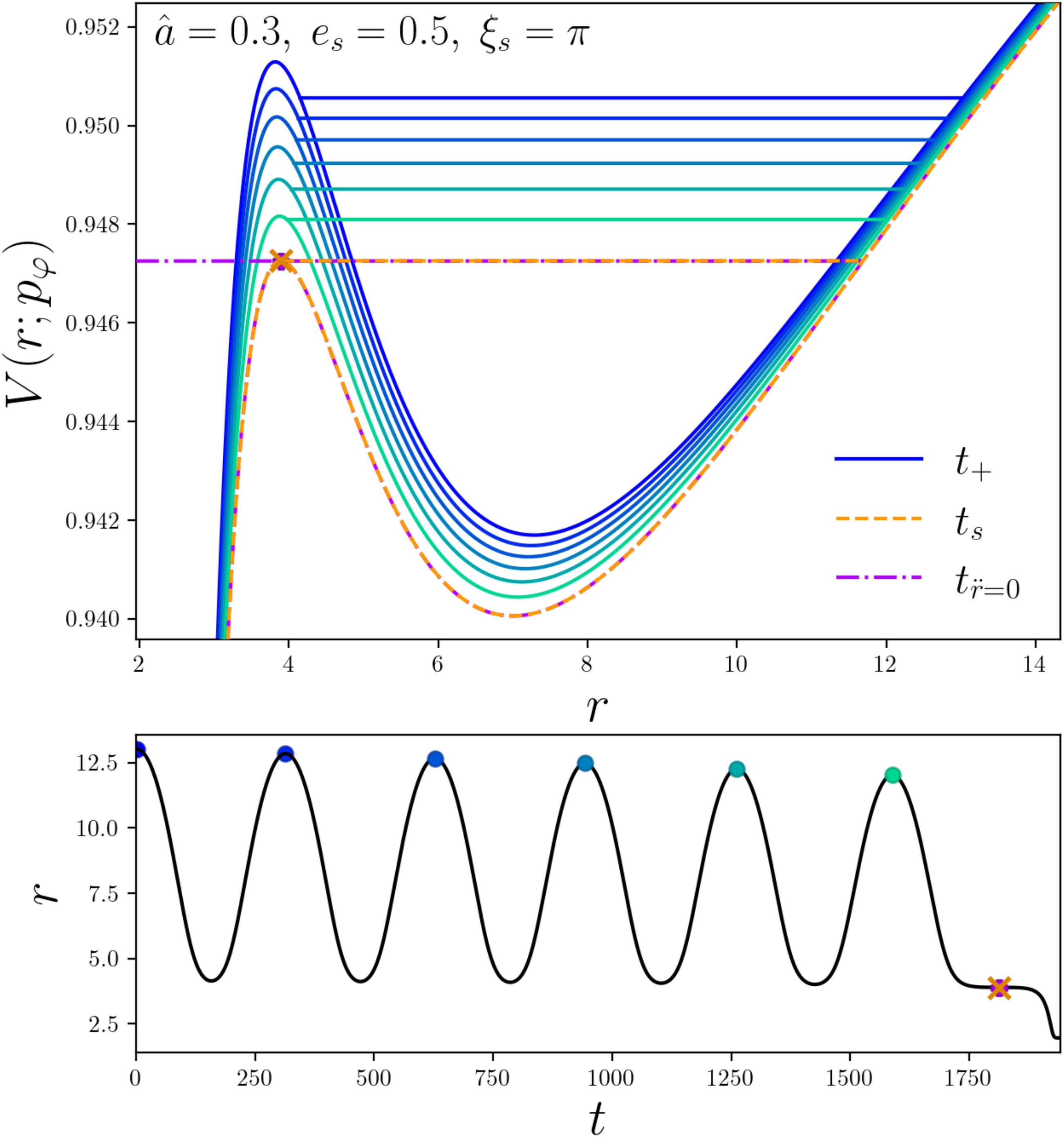}
	\hspace{1mm}
	\includegraphics[width=0.31\textwidth]{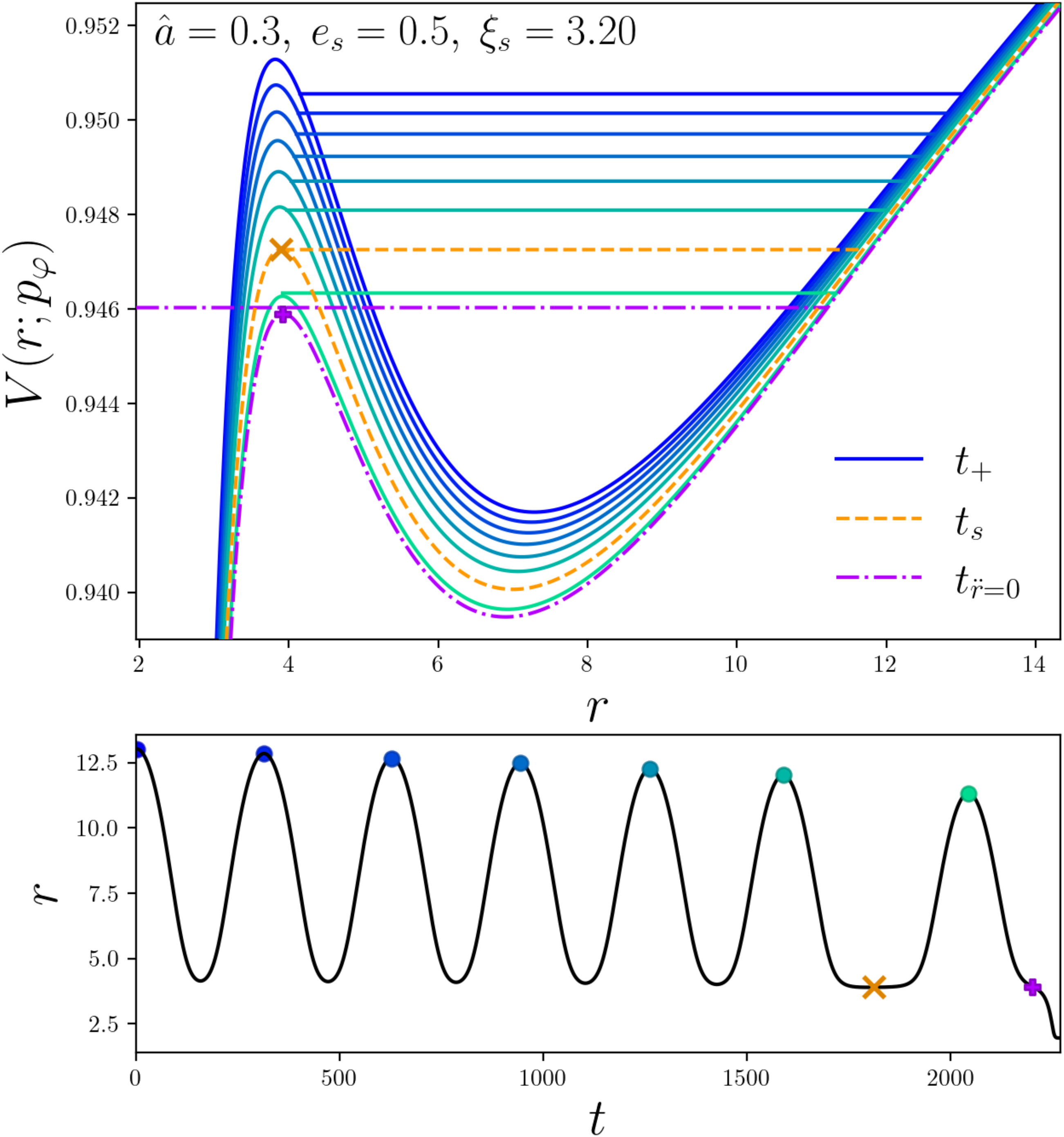}	
	\caption{\label{fig:potentials}
	 Potentials $V(r;p_\varphi)$ for three cases with
	 $\ha=0.3$, $\esep=0.5$ but different anomalies at the separatrix crossing: 
	 $\xisep = \left( 0, \pi, 3.20 \right)$. We show the potential at the 
	 apastron passages (blue/green solid lines), at $\tsep$ (dashed orange), and at
	 $\tplunge$ (dash-dotted purple). The orange cross marks the radial location
	 of the test-mass at $\tsep$, and the purple plus marks it at $\tplunge$. The same
	 markers are used on the radial evolution $r(t)$ reported in the bottom panels,
	 together with circles which mark the apastron (same color scheme of the potentials).
	 }
	\end{center} 
\end{figure*}
We now focus on the transition from the eccentric inspiral to the plunge. 
We start by discussing the dynamics, and we then move on the waveform properties. 
Note that a qualitatively similar discussion
for the Schwarzschild case has already been presented in Ref.~\cite{Albanesi:2023bgi}. 
Here, we briefly recall the main concepts for completeness, and make explicit use of the 
relativistic anomaly $\xi$ to better highlight some aspects.

%------------
% dynamics
%------------
\subsubsection{Dynamics and $\xisep$}
As previously outlined, the dynamics can be naturally understood in terms of 
a time-evolving effective potential $V(r;p_\varphi)$. The separatrix crossing occurs when the 
energy equals the maximum of this potential. However, such a crossing may take place at 
any point along the orbit, and therefore does not necessarily coincide with the beginning of the plunge, 
which instead occurs at a later time. To identify the latter, we search for an inflection point 
in the radial motion occurring after the last apastron passage, as proposed in Ref.~\cite{Albanesi:2023bgi}.

Some examples of effective potentials are shown in Fig.~\ref{fig:potentials}, where we consider 
three configurations with $\ha=0.3$ and $\esep=0.5$, but different separatrix-crossing times $\tsep$. 
To indicate the orbital phase at which the crossing occurs, we report the value of the relativistic 
anomaly at those times, $\xisep$. From a practical point of view, we choose as initial data 
$\left( e_0, p_0, \xi_0 \right)=\left( \esep, p_s, \xisep \right)$, 
perform a backward-in-time evolution for a prescribed number of orbits, and then evolve the system 
forward in time from the final point of the backward evolution, thereby also computing the portion of 
the dynamics after $\tsep$.
The bottom panels display the radial separation $r(t)$, with the separatrix-crossing time $\tsep$
marked by an orange cross and the onset of the plunge $\tplunge$ indicated by a purple marker. 
From these plots, it is clear that the start of the plunge can be heuristically interpreted as a ``missed'' 
periastron passage.

As shown in Fig.~\ref{fig:potentials}, the value of $\xisep$ can significantly affect the dynamics around $\tplunge$. 
In the leftmost case, the separatrix crossing occurs at the last apastron ($\xisep=0$); consequently, 
once the particle completes the orbit, it rapidly starts to plunge.
In contrast, when the separatrix crossing takes place at the periastron ($\xisep=\pi$), as shown in the 
middle panel, the particle undergoes a long-lived, quasi-circular whirl before plunging. This behavior 
arises because the orbital energy is close to the maximum of the effective potential, which corresponds to 
unstable circular orbits. We observe that, among the scenarios considered in this work, this is the closest 
to the critical plunge geodesics studied in Ref.~\cite{Faggioli:2025hff}, which focused on studying 
the transition from an eccentric inspiral to plunge in Kerr spacetime, but in the fully-conservative 
(geodesic) case. However, we remark that here the set up is different, since we are including a 
dissipative effects in the dynamics by means of a radiation reaction force $\F$.
Finally, in the rightmost case we consider $\xisep=3.20$, i.e. a value slightly larger than $\pi$, such 
that the separatrix crossing occurs when the particle is already moving away from the Kerr black hole. 
In this scenario, the particle must first complete an additional orbit, and only then plunge. 
Note that in this case the plunge is even more radial than for 
$\xisep=0$, since the difference $\hat{E}-V_{\rm max}$ at $\tplunge$ is larger, 
reflecting the fact that the system has had more time to radiate away \acp{gw}.
We conclude by noting that for $\xisep$ slightly larger than $\pi$, a long-lived, quasi-circular whirl can occur around $\tsep$, 
for the same reason that a similar whirl is observed at $\tplunge$ in the case $\xisep=\pi$, as discussed above.
We remark that the different phenomenologies that we observe at the beginning of the plunge, or 
at $\tsep$ in some cases, are a direct consequence of the time-evolution of the potentials or,
in other words, to the presence of a dissipative contribution in the dynamics. 

%------------
% wave
%------------
\subsubsection{Effects on the (2,2) numerical merger-ringdown waveforms}
\begin{figure}[t]
	\begin{center}
	\includegraphics[width=0.50\textwidth]{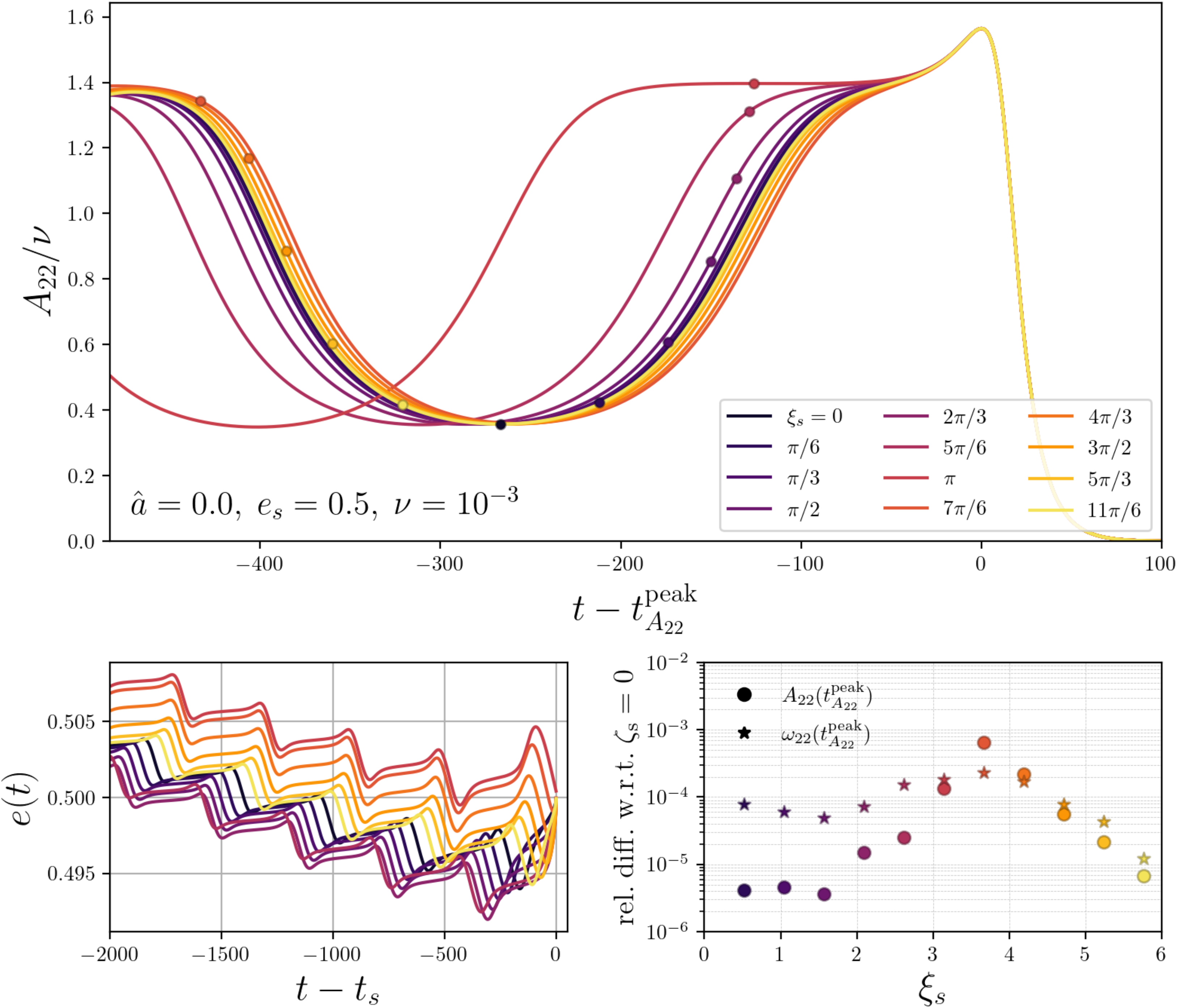}
	 \caption{\label{fig:anomaly_schw} Upper panel: \Teuk{} amplitudes of the (2,2) mode for configurations with 
	 $\ha=0$, $\esep=0.5$, $\nu=10^{-3}$, but different $\xisep$ (reported in the legend), obtained with
	 $N_r\times N_\theta=5401\times 321$ resolution. The dots mark $\tsep$.
	 Bottom left: corresponding eccentricities as a function of time. Bottom right: relative difference of 
	 $A_{22}$ and $\omega_{22}$ between the configurations with $\xisep>0$ and the one with $\xisep=0$.  
	 These differences are also shown in purple in Fig.~\ref{fig:anomaly_summary}. }
	\end{center} 
\end{figure}

We now turn our attention to how different separatrix-crossing times can affect the corresponding 
waveforms, and in particular the merger-ringdown portion, that we want to study in detail. 
We thus consider different configurations, all with same eccentricity at
the separatrix-crossing, but with different anomalies. We start by considering the non-spinning case with 
$\esep=0.5$, and the usual particle's mass $\mu=10^{-3}M$. We consider twelve different
values for $\xisep$, ranging from $0$ to $11\pi/6$. For this set, we consider
resolution $N_r \times N_\theta = 5401\times 321$. 
The resulting amplitudes for the (2,2) mode 
are reported in the upper panel of Fig.~\ref{fig:anomaly_schw}, aligned using the time 
of their peak $\tA22$. We remark that this time is often referred to as merger-time in
the comparable mass case, but in the test-mass limit this terminology can be confusing,
since $\tA22$ typically occurs before the \ac{lr} crossing for $\ha\gtrsim -0.5$.
(see e.g. Fig.~2 of Ref.~\cite{Albanesi:2023bgi} or Table~A3 of Ref.~\cite{Harms:2014dqa}).
Since the merger-ringdown waveforms obtained are visually indistinguishable, we better 
quantify the differences by computing the relative differences 
of the amplitude $A_{22}$ and the waveform frequency $\omega_{22}$ computed at $\tA22$
between the $\xisep>0$ configurations and the $\xisep=0$ case. These differences
are shown in the bottom right panel of Fig.~\ref{fig:anomaly_schw} and, as can be seen, 
the maximum discrepancies for amplitude and frequency are obtained for $\xisep=7\pi/6$,
and remain significantly below the $0.1\%$ threshold 
($|\delta A|/A \sim 0.07 \%$, $|\delta \omega|/\omega \sim 0.03 \%$). 
The differences for the other anomalies are even smaller, some of them also being
affected by the resolution considered. 
We can thus conclude that for $\ha=0$, $\esep=0.5$ and $\mu=10^{-3}M$ 
the merger-ringdown waveform does not depend significantly on $\xisep$. This is a remarkable result, because 
it allows us to fully characterize the merger-ringdown waveform using only one
parameter, $\esep$, as opposed to two parameters as expected, a priori, for generic planar orbits. 
The reason why this happens is that the the radiation reaction has a small effect on the evolution of 
potential, while it can have a significant effect on the early plunge (as the long-lived whirl previously mentioned), 
and this change in the potential is not enough large to considerably modify the latest part, and thus the merger-ringdown waveform. 
\begin{figure}[t]
	\begin{center}
	\includegraphics[width=0.50\textwidth]{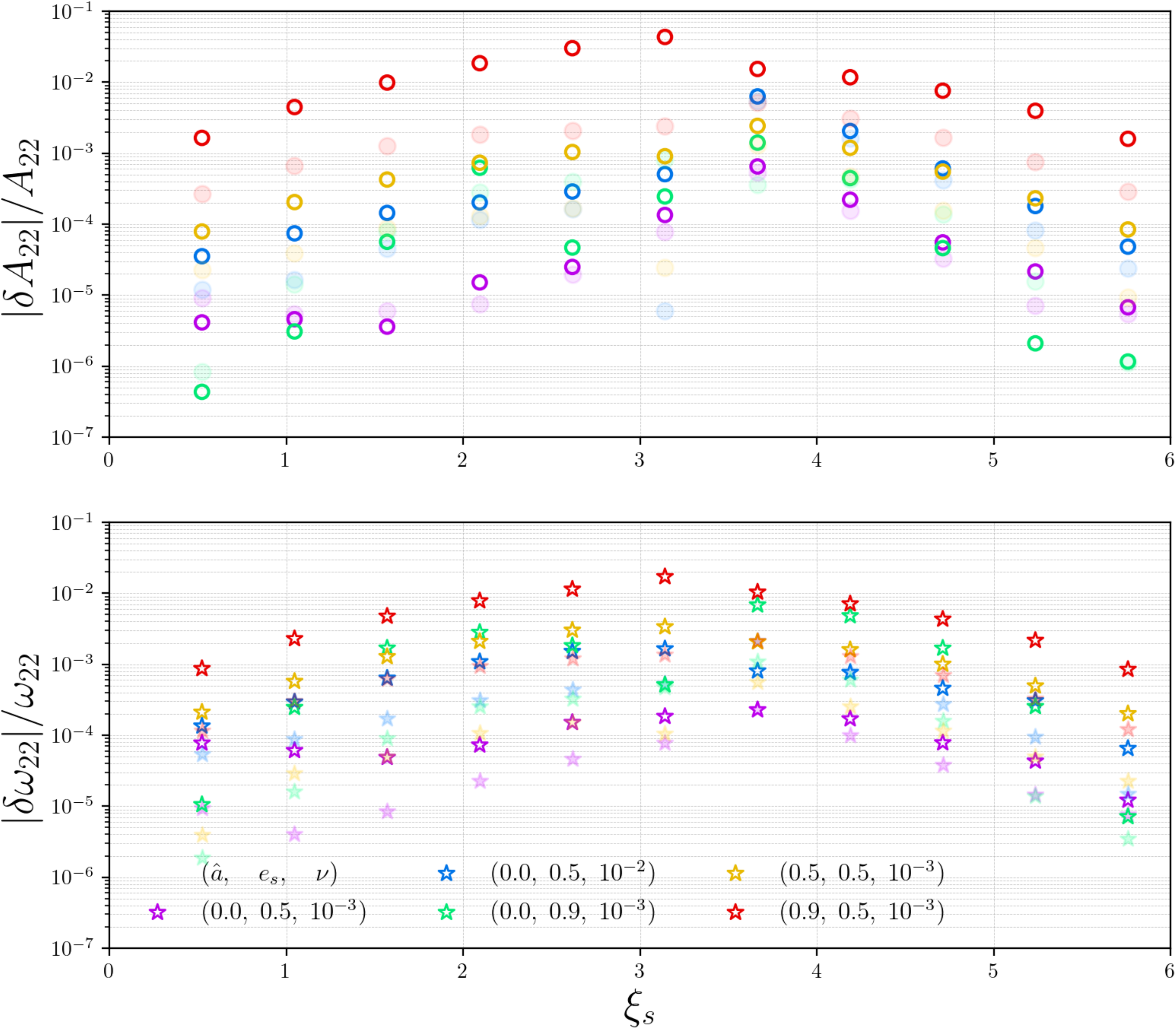}
	 \caption{\label{fig:anomaly_summary} Relative differences for the amplitude 
	 and frequency of the (2,2) mode between
	 configurations with $\xisep>0$ and the one with $\xisep=0$ computed at $\tA22$ (hollow tick markers) 
	 and at $\tLR$ (faint filled markers). Different colors refer to different
	 $(\ha,\esep,\nu)$-configurations, see legend. 
	 }
	\end{center} 
\end{figure}

Given the discussion above, it is natural to ask whether $\xisep$ becomes more relevant in regions 
of the parameter space where radiation reaction effects are stronger, for example for higher mass ratios,
higher eccentricities or higher (positive) spins.
As a first step in this direction, we recall that the radiation-reaction 
force $\F$ scales with the symmetric mass ratio $\nu$, and thus consider a larger
value, namely $\nu=10^{-2}$. While the resulting post-merger waveforms remain visually similar, 
we observe more pronounced differences in both amplitude and frequency at $\tA22$. These values are shown with blue 
hollow markers in Fig.~\ref{fig:anomaly_summary}. In this case, the relative difference in amplitude is $\sim 0.6\%$, 
approximately an order of magnitude larger than in the $\nu=10^{-3}$ case. Further increasing 
the mass ratio would lead to even larger differences; however, the test-mass approximation adopted in this 
work would no longer be valid. We note, nonetheless, that recent studies have reported a dependence of the 
amplitude on the mean anomaly in \ac{nr} simulations of comparable-mass binaries with fixed eccentricity~\cite{Nee:2025zdy}. 
The investigation of such effects within an EOB framework for comparable-mass systems is beyond the scope 
of this work and is therefore left for future studies.

Alternatively, instead of increasing the mass ratio, we can consider larger eccentricities, 
specifically $\esep=0.9$. In this case we still find that the maximum differences in amplitude and frequency 
remain small (hollow green markers in Fig.~\ref{fig:anomaly_summary}), 
at the level of $0.1\%$ and $0.7\%$, respectively.
Therefore, for $\ha=0$, the post-merger waveform can be safely assumed to be fully characterized 
by a single parameter, even at high eccentricity. This result is not unexpected, since a 
faithful EOB description of eccentric inspiral-merger-ringdown test-mass waveforms in 
Schwarzschild spacetime has been already presented in Ref.~\cite{Albanesi:2023bgi}, 
where eccentricities up to $e \simeq 0.95$ were considered, and the merger-ringdown waveform was 
modeled using a single independent variable. In particular, Ref.~\cite{Albanesi:2023bgi} employed 
an ``impact parameter" $\hat{b}$ rather than an eccentricity; we will return to this point 
in Sec.~\ref{subsec:global}.

Finally, we investigate how the spin affects this scenario. Since we are interested in
configurations where the radiation reaction acts more strongly, we focus on positive spins.
We start by considering the usual eccentricity $\esep=0.5$ and mass $\mu=10^{-3}M$, but 
Kerr spin $\ha=0.5$. For this case the amplitude and frequency differences with respect to 
the $\xisep=0$ case are reported with yellow markers in Fig.~\ref{fig:anomaly_summary}. While greater 
than in the Schwarzschild case for the same eccentricity and mass ratio, they are still quite small: 
$0.3\%$ for both amplitude and frequency at $\tA22$. The (2,2) amplitudes are indeed almost indistinguishable 
when aligned with respect to their peak, as show in 
the left panel of Fig.~\ref{fig:anomaly_kerr} of Appendix~\ref{app:anomaly}.
However, if we further increase the spin up to $\ha=0.9$, we immediately notice large discrepancies, 
especially in the amplitude.
The differences at $\tA22$, shown with red markers in Fig.~\ref{fig:anomaly_summary}, 
already highlight this effect: the amplitude ones are consistently above $0.1\%$, reaching up to 
$4.3\%$ for $\xisep=\pi$, while frequency differences can be as large as $1.7\%$. 
This effect becomes even more evident when inspecting the full merger-ringdown amplitudes aligned 
with respect to $\tA22$, as can be seen in the right panel of Fig.~\ref{fig:anomaly_kerr} 
of Appendix~\ref{app:anomaly}. 

%-------------------------------------------------------------------------------------------------------
\subsubsection{Amplitude peak and light-ring}
\label{subsubsec:zsep_anchorLR}
%-------------------------------------------------------------------------------------------------------
%
\begin{figure}[t]
	\begin{center}
	\includegraphics[width=0.49\textwidth]{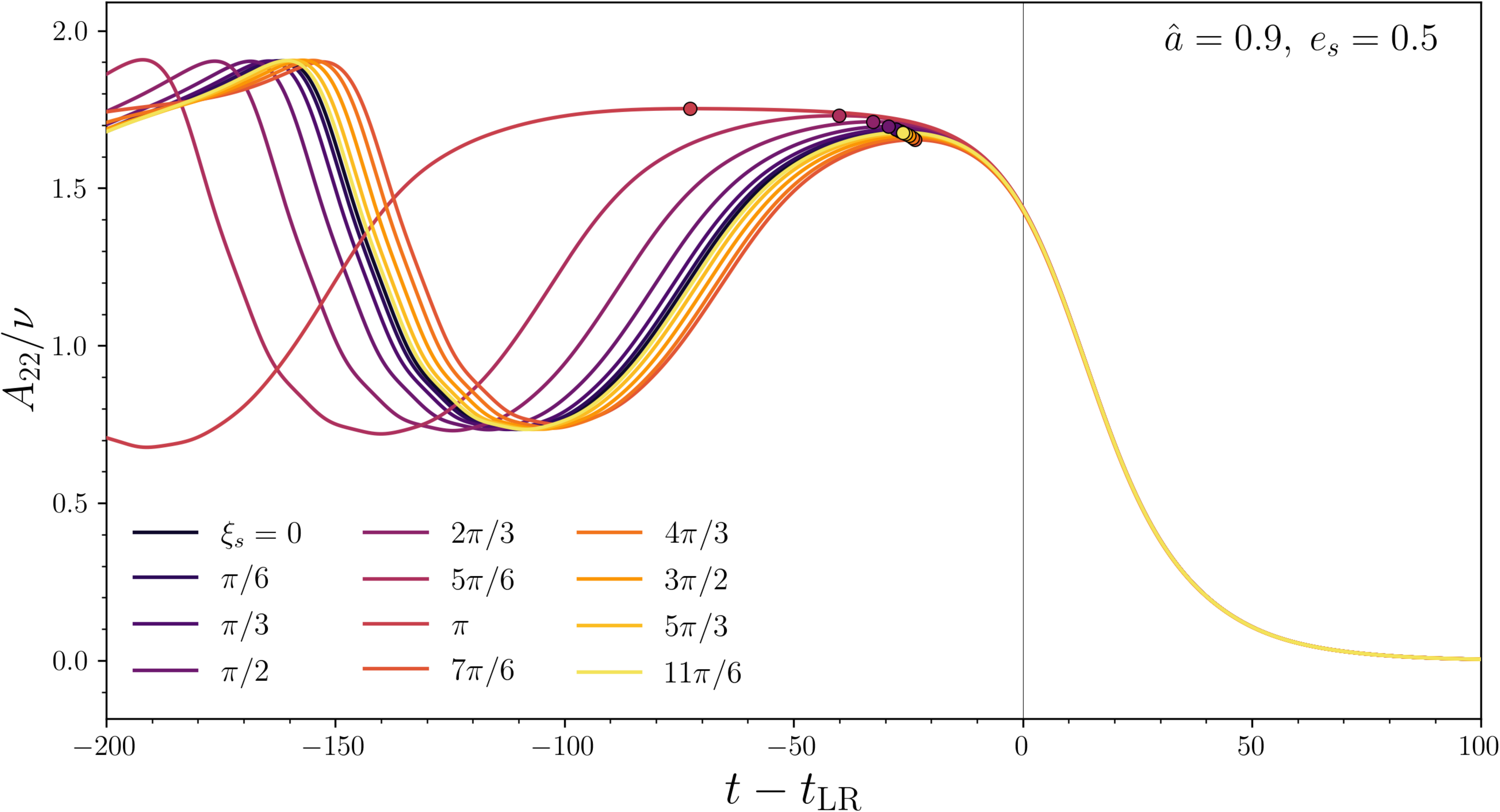}
	\includegraphics[width=0.49\textwidth]{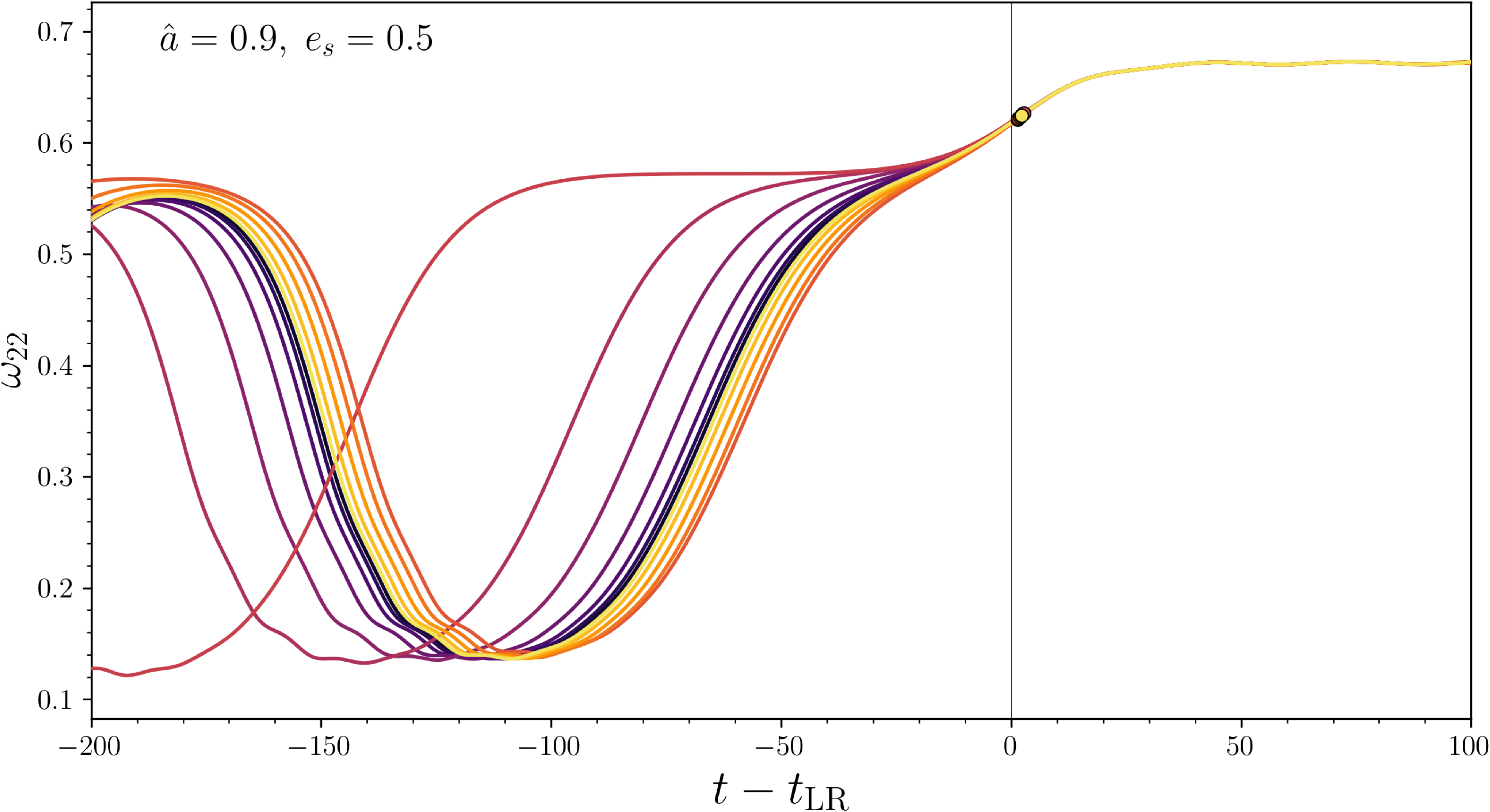}
	 \caption{\label{fig:a09_e05_tLR}
	 Amplitude (upper panel) and frequency (lower panel) of the (2,2) mode for 
	 $\ha=0.9$, $\esep=0.5$ and different $\xisep$, shifted with respect
	 to the \ac{lr} crossing (vertical line). The dots mark the amplitude maxima
	 and the inflection points of the frequency. 
	 The amplitudes aligned with respect to $\tA22$ 
	 rather than $\tLR$ are shown in Fig.~\ref{fig:anomaly_kerr}. 
	 }
	\end{center} 
\end{figure}
Given the previous discussion,
it is immediately clear that, if the ringdown waveform is defined as the portion of the 
signal with $t \geq \tA22$, then $\xisep$ can have a significant and non-negligible impact on it.
Consequently, attempting to model the waveform using a single parameter to 
quantify non-circularity (for instance $\esep$) would be inherently limited,
in particular in the high-spin regime where the influence of $\xisep$ is not small.
However, if one instead considers 
the waveform starting from the \ac{lr} crossing time $\tLR$, all signals with the 
same $(\ha,\esep)$ become very similar, independently of $\xisep$. Indeed, the differences 
in both amplitude and frequency decrease by approximately one order of magnitude when 
evaluated at $\tLR$ rather than at $\tA22$, as illustrated by the faint filled 
markers in Fig.~\ref{fig:anomaly_summary}. More quantitatively, 
the maximum differences are $0.6\%$ in amplitude and $0.1\%$ in frequency, with typical 
values well below these bounds.
To better highlight this property, we report in Fig.~\ref{fig:a09_e05_tLR} the amplitude
and frequencies of the configurations with $(\ha,\esep)=(0.9,0.5)$ but different 
$\xisep$. Note that these configurations correspond to the red markers in Fig.~\ref{fig:anomaly_summary},
and are also shown in the right panel of Fig.~\ref{fig:anomaly_kerr} in Appendix~\ref{app:anomaly}. 
As anticipated, the key point is that, when these waveforms are aligned with respect to the
\ac{lr} crossing, they all appear very similar. In the upper panel, we also mark the amplitude 
maxima with colored dots to indicate the times of peak amplitude, $\tA22$. These peaks occur 
well before $\tLR$ and exhibit a strong dependence on $\xisep$: the closer the separatrix 
crossing occurs to the maximum of the effective potential, the longer the particle whirls 
on an unstable circular orbit. This prolonged whirling anticipates the occurrence 
of $\tA22$ relative to the \ac{lr} crossing.
This behavior demonstrates that the amplitude peak is closely tied to the particle’s dynamics
and may occur well before the source term in the Teukolsky equation becomes negligible. 
In other words, the maximum is reached while the waveform is still significantly 
influenced by the source and therefore should not be described by a homogeneous solution,
such as a pure QNM-superposition. Note that the problem of {\it when} the description 
of the waveform as a superposition of QNMs starts to be valid is a crucial aspect of black 
hole spectroscopy~\cite{Berti:2025hly}. 

Given these considerations, it is natural to abandon the notion that the {\it post-merger} 
waveform should start at $\tA22$, especially for high-spin configurations. 
We remark that for lower spins the anticipation of $\tA22$ with respect to the \ac{lr} crossing 
is much less pronounced; for instance, in the Schwarzschild quasi-circular case one finds 
$\Delta t_{A_{22}}^{\rm LR} \equiv \tLR-\tA22 \simeq 2.4$. 
In such situations, defining the {\it post-merger} waveform as the portion of the signal following 
$\tA22$ is therefore reasonable. This picture, however, breaks down when highly eccentric 
and/or rapidly spinning configurations are considered. Indeed, a significant delay of the \ac{lr} 
crossing relative to $\tA22$ has already been observed in spinning quasi-circular test-mass systems~\cite{Taracchini:2013rva,Taracchini:2014zpa,Harms:2014dqa,Gralla:2016qfw}. In this section we have argued 
that non-circularity further amplifies this effect, demonstrating that the delay depends not only on 
the spin, but also on the eccentricity and, most crucially, on $\xisep$.

Consequently, when constructing a post-merger waveform model to complete an 
inspiral-plunge \ac{eob} waveform, careful consideration must be given to the choice of the anchoring time. 
For example, Ref.~\cite{Taracchini:2014zpa} completed test-mass quasi-circular \ac{eob} waveforms by 
attaching a (pseudo-)QNM description to the inspiral-plunge signal at a matching time delayed with 
respect to $\tA22$, but still preceding $\tLR$ (see also Ref.~\cite{Taracchini:2013rva} 
for a QNM-based approach employing an anchor point delayed with respect to the amplitude peak, 
but for generic mass ratios).
Modern \ac{eob} models, however, typically complete the waveform using the more general 
phenomenological framework introduced in Ref.~\cite{Damour:2014yha}, which we will describe in 
detail in Sec.~\ref{subsec:primary} (see also Ref.~\cite{Baker:2008mj} for a closely related approach). 
This prescription, which strongly improves a pure-QNM-based description of the post-peak waveform,
has been applied both to the dominant $(2,2)$ mode and to the most relevant subdominant modes, 
originally by attaching the model at $\tAlm$ for each multipole. 
Following the proposal of Ref.~\cite{Cotesta:2018fcv}, more recent implementations 
instead use $\tA22$ as the common attachment point also for some or all higher modes~\cite{Pompili:2023tna,Nagar:2023zxh}. 
In contrast, in this work we adopt a different strategy and choose the \ac{lr} crossing 
(or a point in its immediate vicinity) as the anchor for the post-merger description. 
The portion of the waveform between $\tA22$ and $\tLR$ is modeled using the standard inspiral-plunge 
\ac{eob} waveform, enhanced by \ac{nqc}. We will show that this choice is significantly 
more robust in the test-mass limit, particularly for high-spin configurations, while still yielding 
accurate waveforms. Moreover, this approach allows us to not consider $\xisep$ in the picture,
as clearly suggested by the examples reported in Fig.~\ref{fig:a09_e05_tLR}. 

We conclude this discussion by noting that while $\tA22$ can be unambiguously defined
both in the test-mass limit and in the comparable mass case, as it is a property
of the observable waveform, this is not the case for the \ac{lr} crossing, which is a property
of the dynamics. 
Indeed, when considering equal mass binaries in \ac{nr} simulations,
the \ac{lr} is not defined, at least not in a clear sense as in the test-mass limit. 
However, we observe that the inflection point of the (2,2) waveform frequency after the plunge, 
i.e. the first point at which $\ddot{\omega}_{22}=0$ after $\tplunge$, is really close to the \ac{lr} crossing
and does not strongly depends on $\xisep$. This is explicitly shown in the bottom
panel of Fig.~\ref{fig:a09_e05_tLR}, where this point is marked with circles which 
are almost indistinguishable, as opposed to the amplitude maxima
highlighted in the upper panel, which are instead clearly separated.
This suggests that for this inflection point of the frequency could be a suitable
anchor point for generic mass binaries. However, since in this work we deal 
only with the test-mass limit, we will consider a point strictly related to the \ac{lr}
as our default choice.
In Appendix~\ref{app:d2omg0}, we also briefly discuss the usage of $\td2omg0$ as anchoring point,
since it is closely related to the \ac{lr} crossing (see markers in Fig.~\ref{fig:num_waves}).

%=======================================================================================================
\section{Post-merger waveform modeling}
\label{sec:postmrg_waveform}
%=======================================================================================================
We now turn our attention on how to model the post-merger (or ringdown) waveform,
which is indeed the main topic of this article. We start by recalling
the QNM structure of the ringdown waveform in the linear stationary phase, 
and then discuss how we model the post-merger waveform by following the ideas
introduced in Ref.~\cite{Damour:2014yha}. 

%-------------------------------------------------------------------------------------------------------
\subsection{QNM structure and mode-mixing} 
\label{subsec:qnm_mixing}
%-------------------------------------------------------------------------------------------------------
As usually done in \ac{gw} physics, we decompose the waveform in multipoles over $\Ylm$,
the spin-weighted spherical harmonics base,
both with the numerical setup described in Sec.~\ref{subsec:num_waves} and in our EOB framework. 
However, this is not an optimal choice when studying Kerr perturbations. 
This is due to the fact that the Teukolsky equation 
is separable in the frequency-domain, thus splitting in radial and angular equations. The solutions of the latter, 
known as spheroidal harmonics with spin-weight $s$, are typically denoted as ${}_s S_{\lm}$ and provide a more natural
base for Kerr perturbations. 
Indeed, the ringing of a remnant generated by a black hole coalescence, is naturally described on this base
as a superposition of \ac{qnm} modes~\cite{Berti:2025hly},
\begin{align}
\label{eq:qnm_super}
D_L \left( h_+ - i \right. & \left. h_\times \right)  = 
 \sum_{\l,m,n,\pm}^{\infty} {}^S h_\lmn {}_{-2}S_{\lm n}(\Theta,\Phi,a\sigma) \\
= & \sum_{\l,m,n,\pm}^{\infty} C^{\pm}_{\lm n} e^{-\sigma^\pm_{\lm n}\tau} {}_{-2}S_{\lm n}(\Theta,\Phi,a\sigma) \nonumber
\end{align}
where ${}^S h_\lmn$ are the spheroidal modes, 
$\sigma_\lmn^\pm=\alpha^\pm_{\lm n}+i \omega^\pm_{\lm n}$ are the complex \ac{qnm} frequencies, which are fully determined
by the mass and spin of the remnant, and  $C^{\pm}_{\lm n}$ are source-dependent complex constants, not known a priori.
However, significant advancements in their analytical computation have been recently achieved for eccentric 
orbits in Schwarzschild~\cite{DeAmicis:2025xuh}, where the $C^{\pm}_{\lm n}$ are treated as dynamical functions
The time is shifted with respect to a certain reference time, $\tau = t-t_{\rm ref}$, from which this
QNM-description is valid. Indeed, since the $\sigma_\lmn^\pm$ are complex, the \ac{qnm} signal is exponentially damped, 
and therefore at later times the signal is dominated by tails, hereditary effects which behaves, 
asymptotically, as power laws~\cite{Price:1971fb,Leaver:1986gd,Andersson:1996cm}. Note that
for eccentric binaries the tail is enhanced and can have more complex behaviors, which are inherited from 
the inspiral~\cite{Albanesi:2023bgi,DeAmicis:2024not}. In Appendix~\ref{app:tails} we briefly discuss these effects,
but here we focus on the QNM-dominated part of the signal, which is the dominant one for the first few hundreds of $M$
after merger. In Eq.~\eqref{eq:qnm_super}, superscripts $\pm$ indicate rotating and counter-rotating \acp{qnm},
while the index $n=0$ denotes the fundamental mode and $n\geq 1$ its
overtones\footnote{Note that in the EOB literature, sometimes the indexing for $n$ starts with $n=1$
being the fundamental mode.}.
However, in the EOB framework we work with spherical multipoles, and we therefore
need the mapping between the two angular basis, which reads
\be
\label{eq:SYlm}
{}_{-2}S_{\l' m n} = \sum_{\l\geq m} \mu^*_{\l \l' m n}\; \Ylm,
\ee
where the complex quantities $\mu^*_{\l \l' m n}$ are the spin-dependent spherical-spheroidal mixing coefficients~\cite{Berti:2014fga,London:2018nxs}. 
In this work we will consider the fits provided by Ref.~\cite{Berti:2014fga}.
Combining Eqs.~\eqref{eq:strain}, \eqref{eq:qnm_super}, and \eqref{eq:SYlm}, 
the spherical multipoles can be expressed in terms of the spheroidal ones as~\cite{Pompili:2023tna}
\be
\label{eq:hS}
h_\lm = \sum_{\l'\geq|m|,n\geq 0} {}^S h_\lmn \mu^*_{\l \l' m n} \ .
\ee
This expression shows explicitly that a given spherical mode receives contributions 
from spheroidal modes with the same $m$ but different $\l'$. This phenomenon, 
known as {\it mode-mixing}, is particularly pronounced in certain higher-order modes, 
such as the $(3,2)$ mode, as first pointed out in Ref.~\cite{Buonanno:2006ui} and 
confirmed in numerous subsequent studies~\cite{Berti:2007fi,Kelly:2012nd,London:2014cma,Taracchini:2014zpa}.

Modes which are heavily effected by this mode-mixing are not monotonic 
in the amplitude and thus difficult to model with simple closed-form ans\"atze.
It is therefore more convenient to model the spheroidal modes ${}^S h_\lmn$, as proposed by Ref.~\cite{Pompili:2023tna},
which are instead monotonic (modulo beating between co-rotating and counter-rotating \acp{qnm}, 
as we will discuss later), and then reconstruct the spherical modes from Eq.~\eqref{eq:hS}.
However, written as it is, Eq.~\eqref{eq:hS} is clearly not invertible. We thus follow
the same approximations proposed by Ref.~\cite{Pompili:2023tna}:
i) we neglect the overtones $n\geq 1$, which decay faster than the fundamental $n=0$ mode,
ii) we neglect contributions from the $\l'>\l$ modes, since their contributions are 
small if compared to the $(\l,m,0)$ mode. Within this approximation, we can
then write spheroidal modes in terms of spherical ones by inverting Eq.~\eqref{eq:hS}.
Clearly, modes with $\l=2$ or $\l=m$ only differ by a complex constant in the two basis.
However, the other spheroidal modes have to be written as a sum of different spherical modes.
All other spheroidal modes considered in this work can be written as a sum of at most three spherical modes, 
namely\footnote{From this point onward we drop the $n=0$ index and write $\hS_\lm \equiv \hS_{\lm0}$.}
\be
\label{eq:sum3}
\hS_{\lm} =  c_\l h_\lm + c_{\l-1} h_{\l-1} + c_{\l-2, m} h_{\l-2,m},
\ee
where
\begin{align*}
c_\l       &= \frac{1}{\mu^*_{m\l\l0}}, \cr
c_{\l-1}   &= -\frac{\mu^*_{m\l,\l-1,0}}{\mu^*_{m\l\l0} \mu^*_{m,\l-1,\l-1,0} }, \cr
c_{\l-2}   &= \frac{\mu^*_{m,\l-1,\l-2,0}\mu^*_{m\l,\l-1,0}-\mu^*_{m,\l-1,\l-1,0} \mu^*_{m\l,\l-2,0}}{
\mu^*_{m,\l-2,\l-2} \mu^*_{m,\l-1,\l-1,0} \mu^*_{m\l\l0} }.
\end{align*}
For some modes, such as $(3,2)$, $(3,1)$, $(4,3)$, and $(5,4)$, the condition $\l\geq 2$ and $|m|\leq\l$ 
implies $c_{\l-2,m}=0$, so that only two spherical modes contribute (cf. Eqs.~(60a)-(60b) of Ref.~\cite{Pompili:2023tna}
for the $(3,2)$ and $(4,3)$ modes). In this work, however, we also consider the $(4,2)$, $(4,1)$, 
and $(5,3)$ modes, for which $c_{\l-2,m}\neq0$.

As a result, to accurately model the ringdown waveform, we directly use the spherical 
modes obtained from the numerical solution of the Teukolsky equation for the $\l=2$ and $\l=m$ modes. 
For all other multipoles, we first obtain the corresponding spheroidal modes from the numerical 
spherical modes using Eq.~\eqref{eq:sum3}, and then apply our modeling techniques.
The spherical modes are finally reconstructed from Eq.~\eqref{eq:hS}, and included in our EOB framework.

%-------------------------------------------------------------------------------------------------------
\subsection{Anchoring point}
\label{subsec:anchoring}
%-------------------------------------------------------------------------------------------------------
The description in Eq.~\eqref{eq:qnm_super} provides an accurate representation 
of the linear, stationary ringdown regime, but it is generally not applicable to 
the early ringdown, where source-driven effects remain relevant. 
Indeed, it is by now well established that, for comparable-mass binaries, a description of 
the signal solely in terms of quasi-normal modes becomes accurate only sufficiently far 
into the ringdown~\cite{Baibhav:2023clw}. Moreover, Ref.~\cite{Albanesi:2023bgi} argued that
this is the case also for a test-mass plunging in an Schwarzschild black hole. For these reasons, 
we model the post-merger waveform following the approach introduced in Ref.~\cite{Damour:2014yha}, 
also in the cases here considered. Before doing so, however, it is crucial to specify the time 
from which this modeling is applied. 

As previously argued, we want to model the waveform starting in the proximity 
of the \ac{lr} crossing, and not starting from the the peak of the (2,2) mode. 
However, rather than using $\tLR$ directly as the starting (or anchoring) 
point of the post-merger model, we instead choose for all modes
\be
\label{eq:anchor}
\to = \tLR - 2.4,
\ee 
where the constant offset corresponds to the time difference $\Delta t_{A_{22}}^{\rm LR} \equiv \tLR-\tA22$ 
in the Schwarzschild, quasi-circular case. With this choice, the anchoring time 
$\to$ coincides with $\tA22$ for the Schwarzschild quasi-circular dynamics, while this 
coincidence no longer holds for nonzero spin or eccentric systems.
This time is both the starting time for the ringdown model, and thus 
the matching time with the EOB inspiral-plunge waveform. 
At first sight, this offset may appear somewhat arbitrary. However, it is motivated by 
the procedure used to attach the inspiral-plunge \ac{eob} waveform to the post-merger 
signal described here. More specifically, it is closely related to the performance of 
the \ac{nqc} employed to ensure a smooth matching of the two solutions at $\to$. 
Indeed, choosing $\to=\tLR$ would still yield a reasonably accurate waveform, but the 
prescription in Eq.~\eqref{eq:anchor} results in a more accurate instantaneous 
frequency (see, e.g., Fig.~\ref{fig:different_anchor_qc}). 
This improvement stems from the enhanced effectiveness of the \ac{nqc} coefficients 
when evaluated at $\to$ rather than at $\tLR$. We will return to this point in 
Sec.~\ref{subsec:nqc} when discussing the construction of the complete EOB waveform.
We emphasize, however, that choosing $\to$ or $\tLR$ as the anchoring time does not 
significantly affect the accuracy of the post-merger modeling itself; the prescription 
of Eq.~\eqref{eq:anchor} becomes advantageous only when constructing the full \ac{imr} 
waveform. This would not be the case if $\tA22$ were instead adopted as the anchor point. In 
that scenario, the post-merger modeling itself would be substantially less accurate 
for highly spinning and eccentric configurations. 
These different choices for the anchoring point 
are tested for complete EOB waveforms in Sec.~\ref{subsubsec:anchoring_test_qc};
see also Fig.~\ref{fig:different_anchor_qc}.

%-------------------------------------------------------------------------------------------------------
\subsection{Phenomenological modeling}
\label{subsec:primary}
%-------------------------------------------------------------------------------------------------------
%
\begin{figure}[t]
	\begin{center}
	\includegraphics[width=0.50\textwidth]{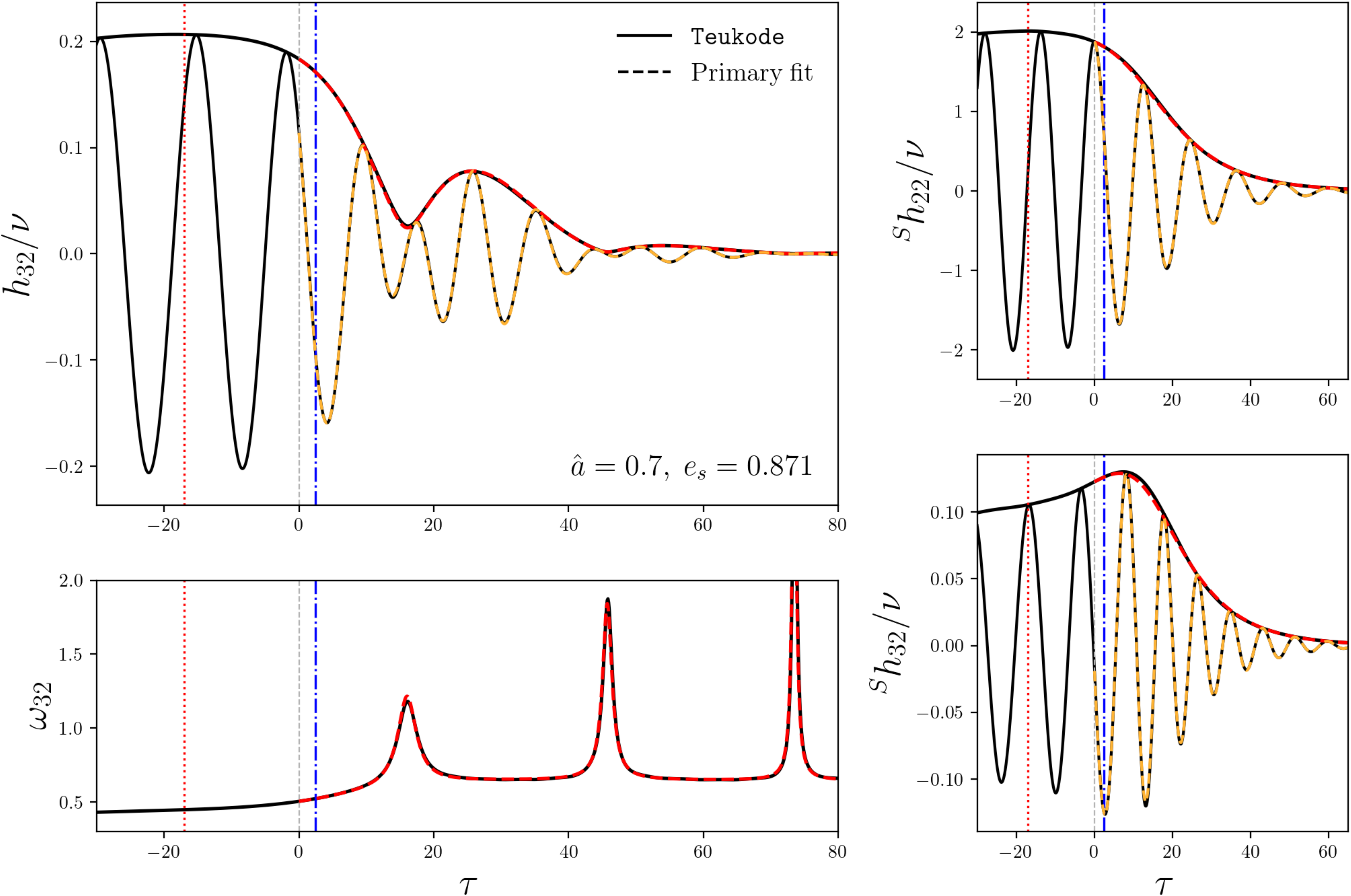}
	 \caption{\label{fig:pfit_a07_e09_l3m2} 
	 Primary fits for $\ha=0.7$ and $\esep=0.871$. Right panels: 
	 primary fits (dashed) performed on the numerical (2,2) and (3,2) spheroidal  modes 
	 obtained from Eq.~\eqref{eq:sum3} (solid black). Left:
	 amplitude and real part (top) and frequency (bottom) of the (3,2) spherical mode
	 obtained by summing the spheroidal fits with the appropriate
	 mode-mixing coefficients $\mu^*_{\l\l'm0}$ (dashed, red and orange), compared with the numerical result (solid black). 
	 The vertical lines mark $\tA22$ (dotted red), $\to=\tLR-2.4$ (dashed grey), and the \ac{lr} crossing (dash-dotted blue). 
	 }
	\end{center} 
\end{figure}
\begin{figure}[t]
	\begin{center}
	\includegraphics[width=0.50\textwidth]{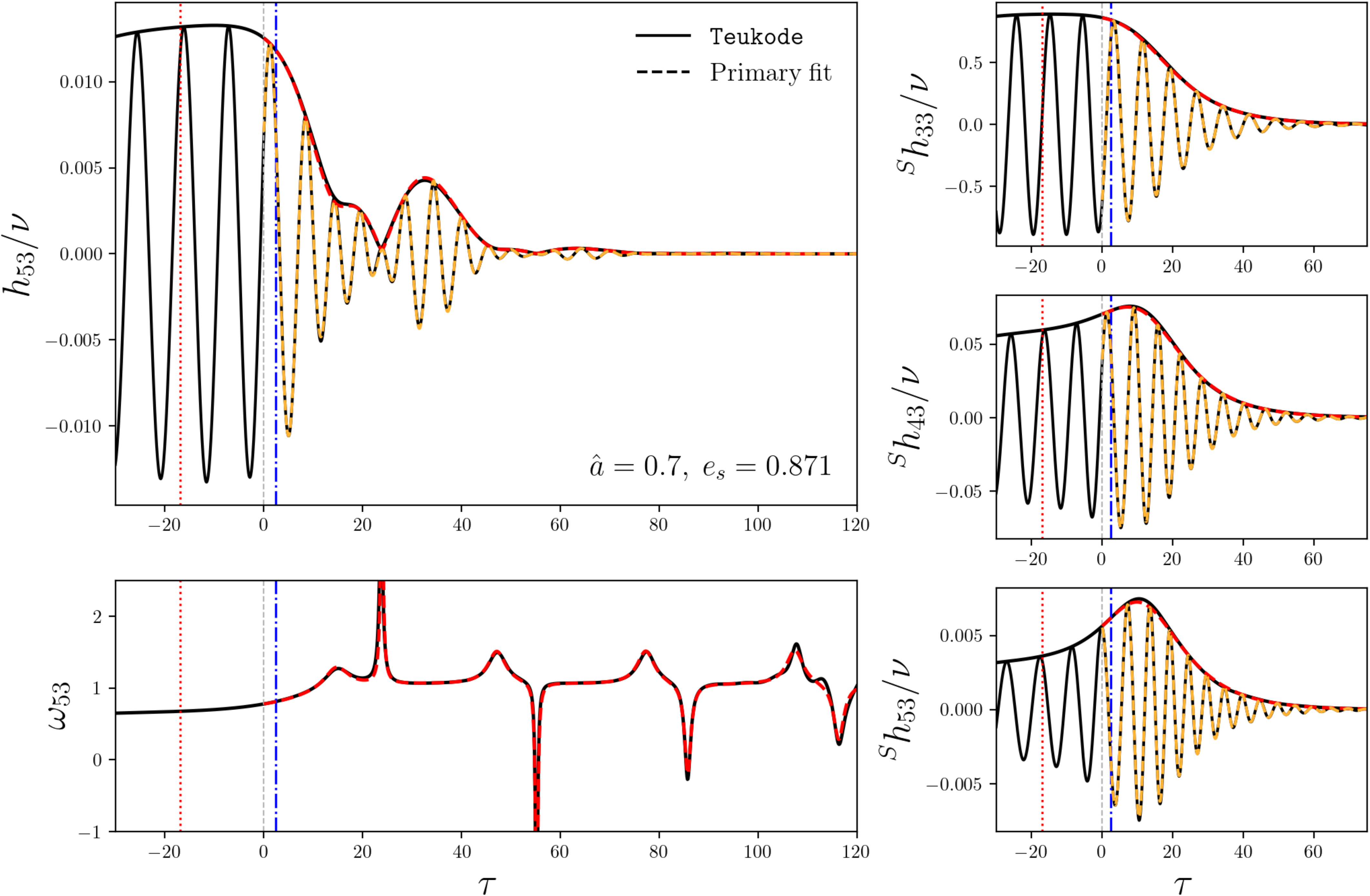}
	 \caption{\label{fig:pfit_a07_e09_l5m3} 
	 Analogous to Fig.~\ref{fig:pfit_a07_e09_l3m2}, but for the spherical (5,3) mode,
	 which is given by the mixing of the $m=3$, $\l\leq5$ spheroidal modes. 
	 }
	\end{center} 
\end{figure}
Having clarified our choice for the starting time of our description, we proceed to model
the post-merger waveform following an approach based on Ref.~\cite{Damour:2014yha}. 
The basic idea is to use phenomenological ans\"atze 
to describe the QNM-rescaled waveform $\bar{h}_\lm$, rather the post-merger waveform itself, defined as 
\be
\label{eq:hbar}
\barh_\lm(\tau) = e^{\sigma^+_{\lm 0} \tau + i \phi_\lm^{0}} h^{\rm rng}_\lm(\tau),
\ee
where $h^{\rm rng}_\lm$ is the numerical ringdown/post-merger waveform, $\sigma^+_{\lm0}$
the fundamental co-rotating \ac{qnm} frequency, $\tau$ is the time rescaled with respect to 
$\to$, $\tau\equiv t-\to$, and finally $\phi_\lm^{0}$ is the phase at $\tau=0$. In the following, 
we drop the $(\l,m)$-indices for simplicity, since the same procedures apply to all the multipoles
here considered. 
The QNM-rescaled waveform $\bar{h}_\lm(\tau)=A_{\barh}e^{i \phi_\barh}$ is then described using two
phenomenological ans\"atze for the amplitude and phase. With the rescaling of Eq.~\eqref{eq:hbar}, 
these two quantities look like activation functions, which saturate at certain constant values; 
see e.g. Fig.~5 of Ref.~\cite{Damour:2014yha} for a comparable
mass case, or Fig.~5 of Ref.~\cite{Albanesi:2023bgi} for some test-mass results. 
This occurs because the overtones decay faster than the fundamental \ac{qnm}, which is thus the only contribution 
left in in the late \ac{qnm} ringdown.

The templates that we use in this work are 
\begin{align}
A_\barh    & = c^A_1 \tanh{z} + c^A_4 + c^A_5 \sech{z},\quad z=c^A_2 \tau + c^A_3, \label{eq:Ah} \\
\phi_\barh & = -c_1^\phi \log{\left( \frac{ 1+c_3^\phi e^{-c_2^\phi \tau } }{1+c_3^\phi }\right)} \label{eq:phih}. 
\end{align}
The amplitude ansatz introduced here generalizes the one proposed in Ref.~\cite{Damour:2014yha}, 
to which it reduces in the limit $c_5^A=0$. Note that Ref.~\cite{Nishimura:2026nse} also recently introduced 
a different generalization of the template~\eqref{eq:Ah}.
The phase template is instead directly inspired by Ref.~\cite{Damour:2014yha}, with the 
additional simplification $c_4^\phi=0$
(cf. Eq.~(5) therein). We note that these two templates, corresponding to $c_5^A=0$ and, in general, 
$c_4^\phi\neq0$, are currently employed in both the \TEOB{}~\cite{Nagar:2024oyk,Albanesi:2025txj} 
and \SEOB{}~\cite{Pompili:2023tna,Gamboa:2024hli} families. 
 
As usual, we impose continuity conditions which allow us to fix part of the coefficients.
In this work, we aim to enforce continuity between the inspiral-plunge \ac{eob} waveform and the post-merger 
description by using the quantities $(A_0,\dot{A}_0,\ddot{A}_0)$ and 
$(\omega_0,\dot{\omega}_0,\ddot{\omega}_0)$ when determining the \ac{nqc} 
coefficients (see Sec.~\ref{subsec:nqc} for details).
As a consequence, these quantities can be used to constrain several of the post-merger 
parameters without introducing any additional numerical input.
For the amplitude template of Eq.~\eqref{eq:Ah}, three parameters can be fixed in this way, yielding
\begin{subequations}
\label{eq:Ah_constraints}
\begin{align}
c^A_5 =& -\frac{\cosh{c^A_3}}{{c^A_2}^2} \left[\ddot{A}_0 + \alpha_0^+\left(2 \dot{A}_0 + A_0 \alpha^+_0\right)\right. \cr
&\left. + 2 c^A_2 \left(\dot{A}_0 + A_0  \alpha^+_0 \right)\tanh{c^A_3}\right],\\
c^A_4 =& A_0 - \frac{\cosh{c^A_3}}{c^A_2}\left[c^A_2 c^A_5  \right. \cr
&\left. + \left(\dot{A}_0 + A_0 \alpha^+_0\right) \sinh{c^A_3}\right],\\
c^A_1 =& \left(A_0 - c^A_4 - c^A_5 \sech{c^A_3}\right)\coth{c^A_3}.
\end{align}
\end{subequations}
From a heuristic standpoint, one might expect $c^A_2 = \alpha_{10}/2$, where 
$\alpha_{10}\equiv \alpha^+_1-\alpha^+_0$ is the difference between the damping rates of the 
fundamental \ac{qnm} and the first overtone~\cite{Damour:2014yha}.
However, imposing this additional constraint, and thus leaving only $c^A_1$ as a free parameter, 
leads to a significant degradation of the fit quality for several configurations.
Introducing the $c^A_5$-term, and thus having two free parameters, 
allows us to straightforwardly impose $C^2$-continuity conditions on the amplitude 
while still achieving reliable fits.
The choice of the $\sech(z)$ functional form is particularly convenient in this
context, since it leads to algebraic equations for the $C^2$-continuity conditions that are
simple to solve.
We also note that the $c^A_5$ term vanishes for large $\tau$, so that the correct asymptotic
behavior is preserved: at late times the waveform naturally reduces to the sole
contribution of the fundamental \ac{qnm}.
We thus only treat ($c^A_2,c^A_3$) as free parameters, which are 
determined by fitting numerical waveforms. We note, however, that the fitted $c^A_2$
does not differ drastically from the heuristic expectation mentioned above.

We now turn our attention to the phase template of Eq.~\eqref{eq:phih}. In this case, we
impose only a single constraint using the instantaneous frequency $\omega_0$, namely
\be
c_1^\phi = \frac{1+c_3^\phi}{c_2^\phi c_3^\phi}\Delta\omega_0,
\ee
where $\Delta\omega_0 \equiv \omega^+_0-\omega_0$, and $\omega^+_0$ denotes, as usual, 
the oscillation frequency of the fundamental \ac{qnm}, while $\omega_0$ is the waveform
frequency at $\tau=0$.
Also in this case, we find that leaving two parameters free significantly improves the quality
of the waveform fits, and we therefore do not enforce further constraints.
For instance, adopting the heuristic prescription $c_2^\phi=\alpha_{10}$~\cite{Damour:2014yha}, or alternatively
fixing $c_2^\phi$ by imposing continuity with $\dot{\omega}_0$, leads to a visibly worse agreement
with the numerical phase evolution.
We also explored the use of the phase template introduced in Ref.~\cite{Damour:2014yha}, which
extends Eq.~\eqref{eq:phih} by including an additional parameter $c_4^\phi$.
However, even after imposing extra constraints to retain one or two free coefficients,
we find that the simpler ansatz adopted here provides the best compromise between flexibility,
robustness, and overall accuracy. The same template and constraint for the phase has been used in
Ref.~\cite{Nishimura:2026nse} for the quasi-circular test-mass case. 

Summarizing, we constrain $(c_1^A,c_4^A,c_5^A,c_1^\phi)$ using 
$(A_0,\dot{A}_0,\ddot{A}_0, \omega_0)$, and leave as free parameters $(c_2^A,c_3^A,c_2^\phi,c_3^\phi)$,
which are determined by fits of numerical waveforms, denoted as {\it primary} fits.
As extensively discussed in Sec.~\ref{subsec:qnm_mixing}, for the $\l\geq3$ with $m<\l$ it is convenient
to perform these fits on spheroidal modes, and then reconstruct the spherical modes through Eq.~\eqref{eq:hS}. 
For the $\l=2$ or $\l=m$ fitting spherical or spheroidal modes is equivalent, since within our approximation
they only differ by one complex constant, $\mu^*_{m\l\l0}$. 
Two examples for $\ha=0.7$ are reported 
in Fig.~\ref{fig:pfit_a07_e09_l3m2} and~\ref{fig:pfit_a07_e09_l5m3}. We consider a configuration 
with high positive spin, since the mode-mixing between modes with same $m$ but different $\l$ 
is particularly relevant for these cases. On the other hand, 
the beating between co-rotating and counter-rotating \acp{qnm}, that we still have to
discuss, is not relevant. We pick an eccentricity of $\esep\simeq0.9$, but this is not of primary
relevance for this discussion. 
In the right panels of Fig.~\ref{fig:pfit_a07_e09_l3m2} we show the primary fits
performed with Eqs.~\eqref{eq:Ah} and~\eqref{eq:phih} on $\hS_{22}$ and $\hS_{32}$.
We also highlight with vertical lines the \ac{lr} crossing (dash-dotted, blue), 
the starting point of our fits $\to$ (dashed, grey), and the peak of the (2,2) amplitude (dotted, red). 
The first thing that we notice is that $\tA22$ strongly anticipates $\tLR$,
as expected. Secondly, we notice that the spheroidal modes which we fit are indeed
monotonic in the amplitude and phase, and we are thus able to reproduce
them using our ans\"atze. The closed-form spherical mode reconstructed from the 
spheroidal fits matches well the numerical (3,2) spherical mode, as show in the left
panels, for amplitude, real part and frequency. The peaks observed in 
the latter correspond to the modulation observed the amplitude, and they grow overtime
(see also discussion in Ref.~\cite{Taracchini:2014zpa}).
However we recall that at later time, the signal is dominated by the tail and not by a 
\ac{qnm} signal. Unfortunately, for the (3,2) multipole we do not observe a clean tail with
our current numerical setup. The reason of this  can be understood by considering, for a moment, the 
dominant (2,2) mode. In this case, the transition from \ac{qnm} and tail
occurs between $\sim 100 - 150 M$ after the \ac{lr} crossing, and at the end of this transition 
the amplitude of the tail is approximately $A^{\rm tail}_{22}/\nu\simeq 8\cdot 10^{-5}$. At later times,
around $\tau\simeq 450$, the amplitude of the (2,2) tail reaches $10^{-5}$, and starts
to show numerical artefacts. For the (3,2) mode, instead, when the transition between \ac{qnm}
and tail occurs, the waveform amplitude is already approximately $A^{\rm tail}_{32}/\nu\simeq 10^{-6}$, and we are
not thus able to accurately resolve the tail signal. 

In Fig.~\ref{fig:pfit_a07_e09_l5m3} we consider the same physical configuration, but we
focus on the spherical (5,3) mode and the $m=3$ spheroidal modes. Also in this case,
the spheroidal primary fits (right panels) yield an accurate description of the spherical
mode (left). Here the amplitude and the frequency exhibit more complicated behaviors,
since the $h_{53}$ is given by the sum of three different spheroidal modes. 
This aspect is particularly evident in the instantaneous frequency $\omega_{53}$ reported
in the bottom left panel. The mixing between the three modes causes the frequency to have
small bumps, the first being at $\tau\simeq 15$ and the other following periodically after
$\Delta \tau \simeq  30$, each of them followed by larger spikes. The first spike reaches
a positive value, while the others reach negative values which decrease
(in absolute value) over time. Also in this case, the maxima/minima in the frequency can
be linked to the modulations observed in the amplitude.%, as expected from Eq.~\eqref{eq:freq}. 

%-------------------------------------------------------------------------------------------------------
\subsection{Beating between co-rotating and counter-rotating \acp{qnm}}
\label{subsec:beating}
%-------------------------------------------------------------------------------------------------------
%
\begin{figure*}[t]
	\begin{center}
	\includegraphics[width=0.49\textwidth]{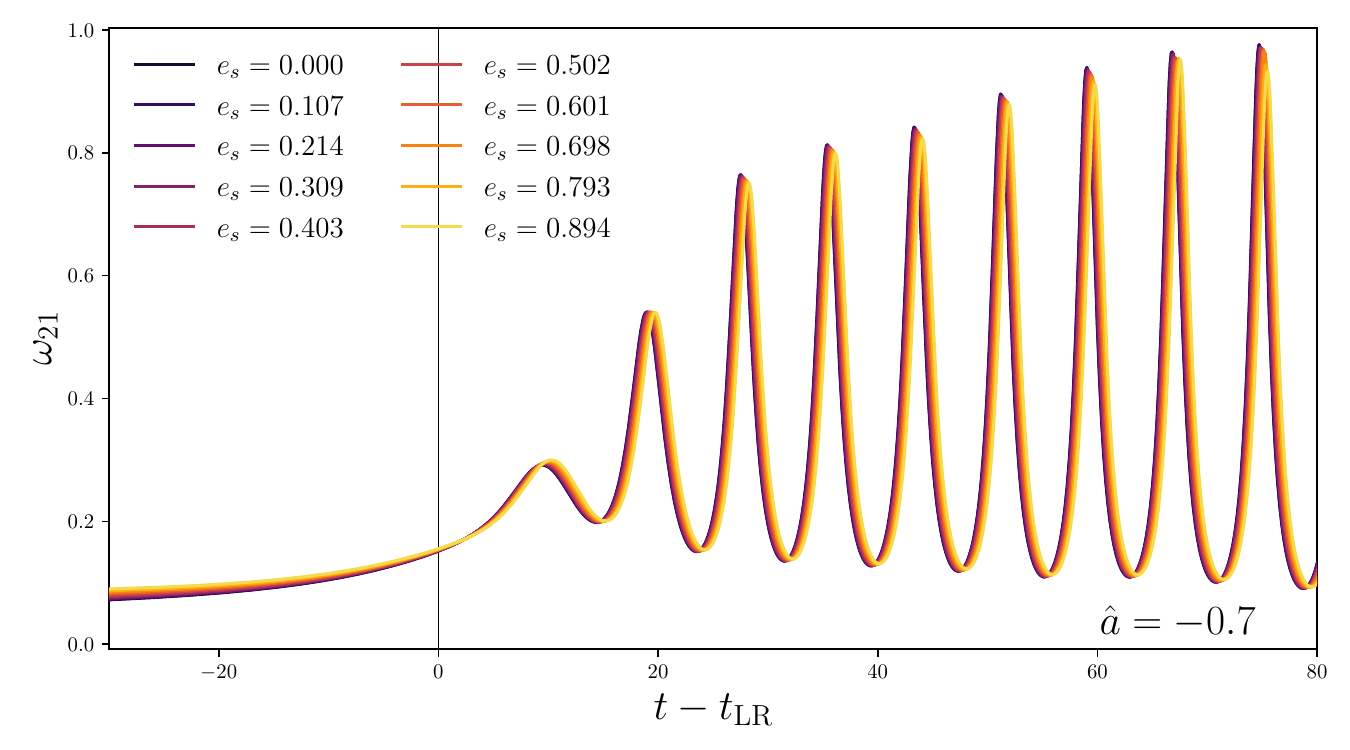}
	\includegraphics[width=0.49\textwidth]{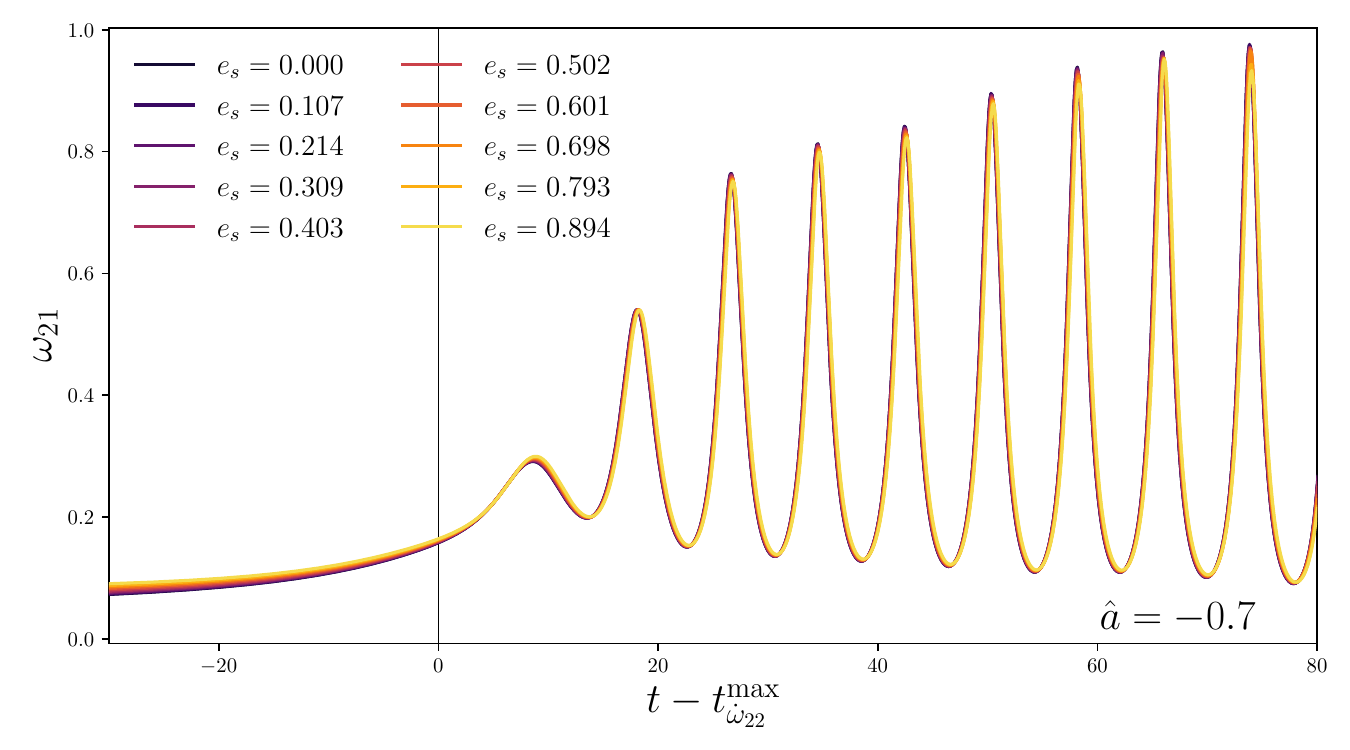}
	 \caption{\label{fig:beating_a-07} Instantaneous frequencies for the (2,1) mode, $\omega_{21}$, 
	 for $\ha=-0.7$ and different eccentricities, up to $e_s\simeq 0.9$. Left:
	 alignment with respect to the \ac{lr} crossing. Right: alignment with respect to the
	 inflection point of the (2,2) frequency during the plunge, $\td2omg0$.
	 }
	\end{center} 
\end{figure*}
As shown by Eq.~\eqref{eq:qnm_super}, both co-rotating and counter-rotating \acp{qnm} contribute
to the waveform signal. The beating between the two dominant corresponding modes gives raise 
to oscillations in the amplitude and frequency of the waveform, which are particularly evident 
for retrograde orbits, as shown in Fig.~\ref{fig:num_waves}. 
This is due to the fact that counter-rotating modes are more excited during more radial plunges,
while they are suppressed for more circular dynamics, as in the case of high and positive spins.
For example, this effect is not visible in Fig.~\ref{fig:pfit_a07_e09_l3m2} and~\ref{fig:pfit_a07_e09_l5m3},
where $\ha=0.7$. 

This effect is particularly visible in the frequency, which we recall being defined by 
$\omega_\lm = - \Im(\dot{h}_\lm/h_\lm)$. 
It is useful to focus, for a moment, on  the simpler Schwarzschild case, where the frequency obtained
by ignoring overtones is given by~\cite{Bernuzzi:2010ty}
\be
\omega^{\rm schw}_{\lm 0} = \frac{ (1 - \hat{A}_{\ell m 0}^2) \omega_{\ell 0} }{ 1 +   
\hat{A}_{\ell m 0}^2 + 2 \hat{A}_{\ell m 0}^2 \cos(2 \omega_{\ell 0} \t + \hat{\theta}_{\ell m 0}) },
\label{eq:beating_schw}
\ee
where $\omega_{\ell 0}$ are the fundamental \ac{qnm} oscillatory frequency, which is
the same for co-rotating and counter-rotating modes modulo a sign, since in Schwarzschild we have 
$\sigma_{\l 0}^- = (\sigma_{\l 0}^+)^*$, and thus $\sigma_{\l 0}^\pm = \alpha_{\l 0} \pm i \omega_{\l 0}$. 
Note that there are no $m$-indices in the \ac{qnm} frequencies, since we are in a spherically symmetric spacetime. 
The other quantities appearing are $\hat{A}_{\lm 0}$ and $\theta_{\lm 0}$, defined as
$\hat{a}_{\lm 0} \equiv C^-_{\lm 0}/C^+_{\lm 0}\equiv \hat{A}_{\lm 0} e^{i \theta_{\lm 0}}$.
The quantity $\hat{A}_{\lm 0}$ is a measure of the relevance of the counter-rotating 
modes with respect to the co-rotating ones. Indeed, if $\hat{A}_{\lm 0}=0$, then there is no beating
and the waveform frequency is simply given by $ \omega_{\ell 0}$. For non-zero values, there is instead
a modulation, whose amplitude is strictly related to $\hat{A}_{\lm 0}$. 

In the more general Kerr case, we have $\omega_{\lm 0}^+\neq -\omega^-_{\lm 0}$, since
the relations between co-rotating and counter-rotating \ac{qnm} frequencies involves the $m$-number, 
i.e. $\sigma^-_{\lm n} = (\sigma_{\l -m n}^+)^*$. In this case, the frequency given by fundamental \acp{qnm}
can be written as 
\be
\label{eq:qnm_freq_kerr}
\omega_{\lm} = \Im\left( \sigma_{\lm 0}^+  \frac{
1 + \hat{\sigma}_{\lm 0} \hat{a}_{\lm 0}  e^{\Delta \hat{\sigma}_{\lm 0} \tau} }{ 
1 +                      \hat{a}_{\lm 0}  e^{\Delta \hat{\sigma}_{\lm 0} \tau}}\right),
\ee 
where $\hat{a}_{\lm 0}= \hat{A}_{\lm 0} e^{i \theta_{\lm 0}}$ as above, 
$\hat{\sigma}_{\lm 0}\equiv \sigma^-_{\lm 0}/\sigma^+_{\lm 0}$, and
$\Delta\hat{\sigma}_{\lm 0}\equiv \sigma^+_{\lm 0} - \sigma^-_{\lm 0}$.
So even if the structure is more complicated than before, we still have a 
complex coefficients $\hat{a}_{\lm 0}$ which quantifies the relevance of the beating. 
Note that, in general, the amplitude  of these oscillations in the frequency is not constant in time
(see also Eq.~(6) of Ref.~\cite{Taracchini:2014zpa} and discussion therein). 
We model this effect with a straightforward generalization of the approach proposed by Ref.~\cite{Albanesi:2023bgi}
for Schwarzschild. We therefore introduce an additional factor in the ringdown waveform
\be
\label{eq:beating_factor}
h^{\rm rng}_{\lm} \rightarrow  h^{\rm rng}_{\lm} \left( 1+ \hat{A}_{\lm 0} e^{\Delta\hat{\sigma}_{\lm 0} \bar{\t}+ i \theta_{\lm0}} \mathscr{S}(\bar{\t}-\tau_0)   \right),
\ee 
where $\bar{\t} = t-\td2omg0$, and $\mathscr{S}(x)=1/\left(1+e^{-x}\right)$ 
is a sigmoid which switches-on the beating correction. 
Note that we do not know a priori $\hat{a}_{\lm 0}=\hat{A}_{\lm 0} e^{i \theta_{\lm 0}}$, and therefore we extract
these two quantities by fitting the waveform frequency with Eq.~\eqref{eq:qnm_freq_kerr} in the late ringdown
when overtones have already decayed. Note that recently Ref.~\cite{Nishimura:2026nse} considered a similar modeling 
for the beating, introduced a more refined sigmoid and extracting the beating coefficients with 
\texttt{qnmfinder}~\cite{Mitman:2025hgy}.
For the Schwarzschild case, Ref.~\cite{Albanesi:2023bgi} noted that the coefficient 
$\hat{A}_{\lm 0}$ does not strongly depends on the eccentricity (see Fig.~4 therein).
However, when aligning the waveform frequencies with respect to the \ac{lr} crossing for different eccentricities,
a small but systematic dephasing of the beating oscillations was observed.
In this work we find that
(i) the amplitude coefficient $\hat{A}_{\lm 0}$ is independent of the eccentricity also in the Kerr case, and
(ii) the phase $\theta_{\lm 0}$ is likewise only weakly dependent on the eccentricity, provided it is extracted
using an appropriate reference time. In particular, we find that measuring $\theta_{\lm 0}$ at the inflection
point of the frequency, rather than at the \ac{lr} crossing, removes this residual dependence.
This effect is illustrated in Fig.~\ref{fig:beating_a-07}, where we show the frequencies of the $(2,1)$ mode
for several configurations with $\ha=-0.7$ and different eccentricities.
As shown in the left panel, aligning the waveforms using $\tLR$ leads to a small but visible misalignment
in the beating pattern across different eccentricities.
By contrast, when the frequencies are aligned using the time corresponding to the inflection point of the
$(2,2)$ frequency during the plunge, denoted by $\td2omg0$, the beatings overlap remarkably well,
as shown in the right panel.
We have verified that this behavior persists for other values of the spin and for the other multipoles.
Therefore, when $\theta_{\lm 0}$ is defined with respect to $\td2omg0$, it becomes effectively independent
of the eccentricity, in close analogy with the behavior of $\hat{A}_{\lm 0}$.
In other words, the beating between co-rotating and counter-rotating modes is controlled by the Kerr spin
alone and does not retain memory of the orbital eccentricity.
As a consequence, when constructing global fits over the parameter space, we can safely adopt
quasi-circular fits for both $\hat{A}_{\lm 0}$ and $\theta_{\lm 0}$ without any appreciable loss of accuracy.
Finally, for modes affected by mode mixing, all such fits are performed using the spheroidal-mode
frequencies ${}^S \omega_{\lm}$.

%-------------------------------------------------------------------------------------------------------
\subsection{Global fits}
\label{subsec:global}
%-------------------------------------------------------------------------------------------------------
In the previous sections we have described how we model the post-merger waveform, and in particular the 
ans\"atze that we employ. We thus proceed to apply this description to several waveforms 
on the parameter space, and then provide {\it global}  fits of these quantities in terms of $(\ha,e_s)$. Once that
we have the global fits of these quantities, we can then provide a closed-form description
of for the post-merger waveform across the parameter space.
To perform these fits, we consider 137 configurations with $\ha\in[-0.8, 0.9]$ and
$\esep \leq 0.9$. Note that in the global fits we do not consider $\ha<-0.8$ due to the
heavy mode-mixing observed that contaminates the primary fits, but the post-merger waveform 
obtained by extrapolating our model to lower spins up to $\ha=-0.9$ is still reliable
(see, e.g., the left panel of Fig.~\ref{fig:a09_e09} for an analytical/numerical comparison with this spin).
The models however loses reliability for spins $|\ha|>0.9$; for retrograde orbits, 
in particular, due to the heavy mixing between co-rotating and counter-rotating modes. 
A quasi-circular case with $\ha=0.95$ is however discussed in Appendix~\ref{app:nqc_derivs}. 

Before proceeding, it is also useful to motivate the choice of our independent variables.
While the Kerr spin $\ha$ is an obvious choice, $\esep$ is way less straightforward. 
Ref.~\cite{Albanesi:2023bgi} indeed used an impact parameter for Schwarzschild, 
defined as $b\equiv p_\varphi/E$, which was evaluated at the peak of the orbital 
frequency $\Omega_{\rm pk}$ (which corresponds to the \ac{lr}) and shifted with the corresponding quasi-circular value, 
$\hat{b}^{\rm schw}_{\rm LR} = b^{\rm schw}_{\rm LR} - b_{\rm LR}^{\rm schw, QC}$.
A similar but more general approach has also been used in Ref.~\cite{Carullo:2023kvj} for
non-circularized comparable mass binaries, thus showing its reliability even beyond
the test-mass limit.
The definition can be easily generalized to Kerr as 
$\hat{b}_{\rm LR} = b_{\rm LR}  - b_{\rm LR} ^{\rm QC, \ha}$,
where now the value that we subtract is the quasi-circular one
for the corresponding Kerr spin $\ha$. Note that one could directly use $b_{\rm LR}$, but it is convenient
to have an independent variable which is zero in the quasi-circular limit, so that we can easily build
hierarchical fit. 
\begin{figure}[t]
	\begin{center}
	\includegraphics[width=0.49\textwidth]{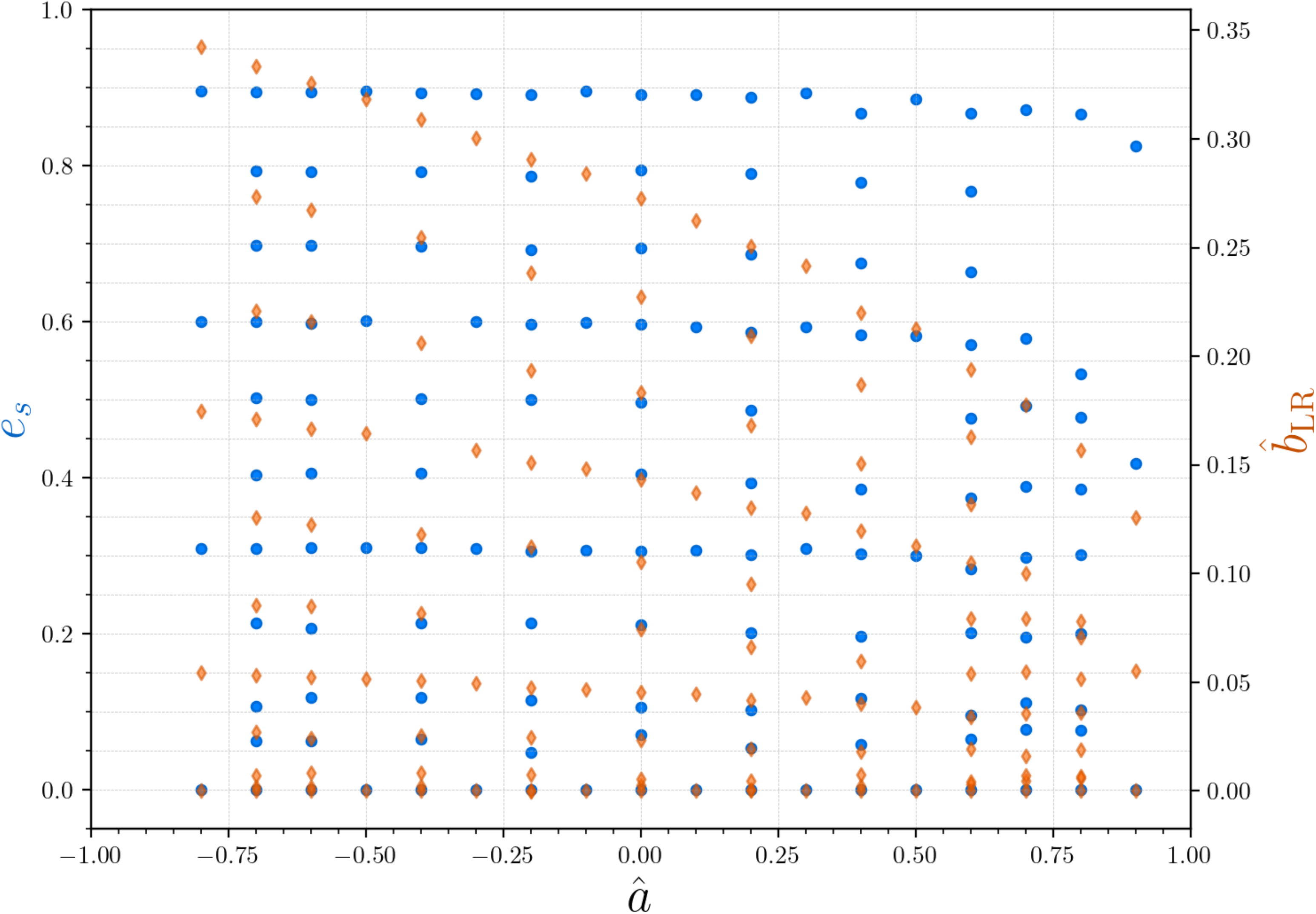}
	 \caption{\label{fig:parspace_e_vs_b} Configurations considered in the global fits:
	 blue dots for $(\ha,\esep)$, orange diamonds for $(\ha,\hat{b}_{\rm LR})$.
	 }
	\end{center} 
\end{figure}
However, despite this possible choice, we still decide to use $\esep$ for our global fits.
The reason is twofolds. First, the maximum value allowed for $\hat{b}_{\rm LR}$ is spin dependent,
while $e_s^{\rm max}=1$ for all the spins. This is shown in Fig.~\ref{fig:parspace_e_vs_b}, where
we report the configurations considered in the global fits: blue dots for $(\ha,\esep)$,
orange diamonds for $(\ha,\hat{b}_{\rm LR})$.
Therefore, not only the fits are easier to obtain, but it is
also easier to visually inspect them over the whole parameter space. Secondly, even if $\esep$ is obtained
way before the \ac{lr} crossing for some configurations, we are still able to smoothly fit all the needed quantities in term of
this variable. Nonetheless, it should be noted that this choice is possible in the test-mass limit, but
when dealing with generic mass ratio, $\hat{b}$ (evaluated at a suitable time) could be more reliable choice, 
as already argued by Ref.~\cite{Carullo:2023kvj}. We leave this investigation to future works. 
We finally remark that we span the parameter space by selecting $\ha$ and $e_0$, together with an
initial semilatus rectum which yields at least 3-4 orbits before the plunge. Therefore, $\esep$ is typically 
close to $e_0$ and, most importantly, we do not select a priori any $\xisep$. Indeed, 
we expect our approach to be reliable independently of $\xisep$, as extensively discussed
in Sec.~\ref{subsec:plunge}. Note that this is not be the case when considering $\to=\tA22$ as anchoring point.
Indeed, in that case we observe scattered points for $A_0$ and the other quantities
in the high-spin and high-eccentricity corner of the parameter space, due to dependance on $\xisep$. 
Instead, we do not have such scattered quantities with $\to=\tLR$ or $\to=\tLR-2.4$.

Having clarified our choice of variables, we proceed to discuss the templates that we use for the global
fits. Since we use a hierarchical approach, we start by performing one-dimensional fits 
along $x_1$ and $x_2$. We use rational templates of the kind
\begin{equation}
\label{eq:template_1d}
R_d^n(x) = \frac{a_0 + b_1 x + \dots + b_n x^n}{1+ c_1 x + \dots + c_d x^d},
\end{equation}
where $a_0$ is set to be equal to the value which corresponds
to $(x_1,x_2)=(0,0)$, while the $b_i$ and $c_i$ coefficients are instead obtained with 
1-dimensional fits. The 2-dimensional template is then obtained as 
\begin{equation}
\label{eq:template_2d}
F_2(x_1,x_2) = a_0 R_{12} R_{21}
\end{equation}
where $R_{12}$ is a rational function in $(x_1,x_2)$ which reduces
exactly to the $x_1$-fit for $x_2=0$, and $R_{21}$ similarly.
The notation can be made compact using the permutations $p=\lbrace 12,21\rbrace$. We denote the $x_{p_2}$-deformation of a parameter $b_i^{p_1}$ found with a $x_{p_1}$-fit as
\be
\label{eq:template_calB}
{\cal B}_p\left(x_{p_2}; b_i^{p_1} \right) = \frac{b_i^{p_1} + \sum\limits_{j=1}^{n_{p,p_2}} \alpha^{b_i^{p_1}}_j x_{p_2}^j}{1 + \sum\limits_{j=1}^{d_{p,p_2}} \beta^{b_i^{p_1}}_j x_{p_2}^j}
\ee
where $p_i$ marks the $i$th-variable of the permutation, e.g. if $p=21$, then $p_1=2$ and $p_2=1$. 
We can thus write $R_{12}$ and $R_{21}$ explicitly as
\be
R_{p} = \frac{1 + \sum\limits_{i=1}^{n_{p,{p_1}}} a_0^{-1} {\cal B}_{p}\left(x_{p_2}; b_i^{p_1} \right) x^i_{p_1}}{
 1 + \sum\limits_{i=1}^{d_{p,{p_1}}} {\cal B}_{p}\left(x_{p_2}; c_i^{p_1} \right) x^i_{p_1} }.
\ee
The functional form of $R_p$ depends on 4 parameters: $(n_{p,p_1}, d_{p,p_1})$ are the two parameters 
which give the orders of the numerator and denominator of the rational template~\eqref{eq:template_1d} used for the 
1-dimensional fit over $x_{p_1}$, while $(n_{p,p_2}, d_{p,p_2})$ are the orders of numerators
and denominators for the $x_{p_2}$-deformations of the parameters obtained with Eq.~\eqref{eq:template_calB}. 
Therefore, the functional form of Eq.~\eqref{eq:template_2d} depends on 8 parameters.
We call this approach hierarchical since the  coefficient $a_0$ is fixed with the $(x_1,x_2)=(0,0)$ value, 
the $\lbrace b_i,c_i\rbrace$ coefficients in the two $R_p$ terms are found with 1-dimensional fits over $x_{1,2}$, 
and finally $\lbrace\alpha_j,\beta_j\rbrace$ are found with a 2-dimensional fit over $(x_1,x_2)$.
In our specific case, we consider $(x_1, x_2)=(\ha,\esep)$, and therefore $a_0$ is fixed to the Schwarzschild quasi-circular value.
In the cases in which $a_0=0$ (or even $a_0\simeq 0$), the template is clearly ill-posed, but we can just rescale the data 
points with a constant $\bar{c}$ (typically $\bar{c}=1$), perform the fits and obtain the physical 
quantity by subtracting $\bar{c}$ from the obtained fit.  

Therefore, for each multipole, we use the template of Eq.~\eqref{eq:template_2d} to fit the numerical quantities 
computed at the anchoring point $\to$, $(A_0,\dot{A}_0,\ddot{A}_0)$ and $(\omega_0, \dot{\omega}_0,  \ddot{\omega}_0)$,
and the coefficients obtained with primary fits, $(c_2^A,c_3^A,c_2^\phi,c_3^\phi)$.
We provide these fits for all the modes with $m\geq 1$ and $\l\leq4$, plus the 
(5,5), (5,4) and (5,3) modes. Note that the quantities at $\to$ are always obtained from spherical modes,
while for mode-mixed multipoles the coefficients $c_{2,3}^{A,\phi}$ are obtained on spheroidal modes.
Moreover, we model the (2,0) mode with the approach 
proposed in Ref.~\cite{Albanesi:2024fts}, as detailed in Sec.~\ref{subsec:l2m0}.

For the beating instead, we use only quasi-circular fits as argued in Sec.~\ref{subsec:beating}, and we thus use the template
of Eq.~\eqref{eq:template_1d} for the phase $\theta_{\lm0}$. For $\hat{A}_{\lm0}$ instead,
we employ an exponentiation of Eq.~\eqref{eq:template_1d} without fixing $a_0$, 
with the only exception of the (4,1) spheroidal mode in which we avoid the exponentiation.
Moreover, since Eq.~\eqref{eq:beating_factor} requires the inflection point of the $(2,2)$ frequency,
$\td2omg0$, we provide a fit for the difference $\td2omg0-\tLR$ using Eq.~\eqref{eq:template_2d}.
Note that, in principle, $\td2omg0$ could be directly extracted from the full EOB waveform
before applying Eq.~\eqref{eq:beating_factor}.
Although we tested this possibility, we found that fitting $\td2omg0-\tLR$ yields a more robust
and stable procedure. We also stress that this is a single additional fit, which is then
used consistently for all multipoles.
By contrast, in our previous study of eccentric dynamics in Schwarzschild spacetime
(Ref.~\cite{Albanesi:2023bgi}), we did not provide a global fit for $\td2omg0-\tLR$.
Instead, we fitted each phase $\theta_{\lm}$ directly as a function of an impact parameter $\hat{b}$.
However, as illustrated by the example in Fig.~\ref{fig:beating_a-07}, this procedure becomes
unnecessary once a suitable reference time is chosen for the correction in
Eq.~\eqref{eq:beating_factor}, namely $\td2omg0$.
Moreover, in Ref.~\cite{Albanesi:2023bgi} the anchoring points were taken to be the amplitude
peaks $\tAlm$ of each multipole. As a consequence, additional fits were required: i) $\tA22$
with respect to the peak of the orbital frequency, and ii) $\tAlm$ 
with respect to $\tA22$.
None of these quantities are needed in the present approach, since we adopt as anchoring point
the dynamical time $\to=\tLR-2.4$, as discussed in Sec.~\ref{subsec:anchoring}.
In addition, Ref.~\cite{Albanesi:2023bgi} employed an extra free parameter in the phase template,
i.e., three phase parameters instead of the two used here.

We therefore conclude that the model presented in this work requires significantly fewer fits
of physical quantities, despite being applicable to the more general Kerr case and providing
a more accurate description of higher multipoles even in the Schwarzschild limit, as we will
show below. However, almost all the physical quantities (with the exception of the beating) are fitted 
over the two-dimensional parameter space $(\ha,\esep)$, rather than in terms of a single parameter 
encoding the non-circularity of the dynamics. We conclude by noting that some quantities
are weakly dependent on the eccentricity, such as $c_2^A$ and $c_2^\phi$, which are indeed heuristically 
expected to be $\propto\alpha_{1}^+-\alpha_{0}^+$, or $\ddot{A}_0$ and $\omega_0$ (but notably, not $\dot{\omega}_0$).
However, when we observe a clear trend in the datapoints, we still provide a $\esep$-fit, even
if subdominant with respect to the $\ha$-fit and thus negligible for practical purposes.
In other cases, the data points show no systematic dependence on the eccentricity and instead appear scattered 
around an approximately constant value, with the variations being consistent with numerical noise. 
In these situations, we therefore restrict to quasi-circular fits, as done, for instance, 
for the parameter $c_2^\phi$ of the $(5,4)$ mode. The global fits for our default model ($\to=\tLR-2.4$) 
are reported in an ancillary file, together with an implementation of the templates~\eqref{eq:template_2d}.

%=======================================================================================================
\section{Effective-one-body waveform}
%=======================================================================================================
\label{sec:eob}
Having obtained a model for the post-merger waveform $h_\lm^{\rm rng}$, we can match it to the EOB solution
for the inspiral-plunge as 
\be
\label{eq:eob_imr}
h_\lm = \theta(\to-t) h_\lm^{\rm inspl} \hNQC + \theta(t-\to) h_\lm^{\rm rng},
\ee
where $\Theta$ is the Heaviside step-function, $h_\lm^{\rm inspl}$ is the EOB analytical 
solution discussed in Sec.~\ref{subsec:inspiralwave}, and $\hNQC$ are
the \ac{nqc} presented in Sec.~\ref{subsec:nqc}. The latter ensure a smooth
connection between the inspiral-plunge EOB waveform and the ringdown model.
The matching is performed at $\to$, which is also the starting time of our ringdown
modeling (see Sec.~\ref{subsec:anchoring}).

%------------------------------------------------------------------------
\subsection{Inspiral EOB waveform}
%------------------------------------------------------------------------
\label{subsec:inspiralwave}
The standard factorization and resummation of the \ac{pn} expanded
multipoles for quasi-circular systems is~\cite{Damour:2008gu},
\be
\label{eq:DIN}
h_{\lm}^{\rm inspl}= h_{\lm}^{(N, \epsilon)_{\rm c}} \hat{h}^{(\epsilon)_{\rm c}}_{\lm} = h_{\lm}^{(N, \epsilon)_{\rm c}} \hat{S}^{(\epsilon)} \hat{h}_\lm^{\rm tail} (\rho_{\lm})^{\l} , 
\ee
where $\epsilon$ denotes the parity of the multipole, $h_{\lm}^{(N, \epsilon)_{\rm c}}$ is the Newtonian circular 
contribution and $\hat{h}^{(\epsilon)_{\rm c}}_{\lm}$ is the circular PN correction.
In the latter, $\hat{S}^{(\epsilon)}$ 
is the effective-source term, i.e.~the energy if $\epsilon=0$ or the Newtonian-normalized
angular momentum if $\epsilon=1$, $\hat{h}_\lm^{\rm tail}=T_{\lm} e^{i \delta_{\lm}}$ is 
the tail factor and $\rho_{\lm}$  is the residual amplitude corrections. Note that while
$\hat{S}^{(\epsilon)}$ and $T_\lm$ are written in closed form, the residual 
phase and amplitude corrections $\delta_\lm$ and $\rho_\lm$ are \ac{pn}-expanded series. 
Therefore, additional resummations, especially for the $\rho_\lm$, are often employed to make these results more 
reliable in the strong-field regime~\cite{Nagar:2016ayt,Messina:2018ghh,Nagar:2024oyk}.
Recently, more sophisticated factorization and resummations have been proposed, 
either based on the mapping of the Teukolsky equation into a confluent Heun equation~\cite{Aminov:2020yma,Fucito:2023afe,Cipriani:2026xmx} or
on effective field theory techniques~\cite{Ivanov:2025ozg}, thus proposing improvements to 
the paradigm of Eq.~\eqref{eq:DIN}. 
However, in this paper our main focus is the post-merger waveform, and we thus leave to future work a more 
in depth study of the inspiral waveform and on how the results of Ref.~\cite{Cipriani:2026xmx} can be extended
to spinning and non-circular dynamics.

The waveform of Eq.~\eqref{eq:DIN} can be generalized to non-circular dynamics including corrections 
that are known up to 3PN~\cite{Chiaramello:2020ehz,Khalil:2021txt,Placidi:2021rkh,Albanesi:2022ywx,
Albanesi:2022xge,Placidi:2023ofj,Grilli:2024lfh,Faggioli:2024ugn,Gamboa:2024imd}.
In particular, Ref.~\cite{Chiaramello:2020ehz} proposed to generalize the waveform of Eq.~\eqref{eq:DIN} to
generic orbits by including the non-circular corrections obtained from the Newtonian source multipoles,
\be
\label{eq:genericNP}
h_{\lm}^{(N, \epsilon)_{\rm c}} \rightarrow h_\lm^{(N,\epsilon)_{\rm c}} \hat{h}_\lm^{(N,\epsilon)_{\rm nc}}.
\ee
For example, for the (2,2) mode this correction reads
\be
\label{eq:NP22}
\hat{h}_{22}^{(N,0)_{\rm nc}} = 1 - \frac{\ddot{r}}{2 r \Omega^2} - \frac{\dot{r}^2}{2 r^2 \Omega^2} 
+ \frac{2 \mathrm{i} \dot{r}}{r \Omega} + \frac{\mathrm{i} \dot{\Omega}}{2 \Omega^2}.
\ee 
We compute $\dot{r}$, $\ddot{r}$, and $\dot{\Omega}$ from Hamilton’s equations and their 
time derivatives, and proceed analogously for $\dddot{r}$ and $\ddot{\Omega}$.
In evaluating the latter, we neglect terms proportional to $\dot{\F}$, which are found to be numerically small.
Note that these derivatives are also the ones entering the radiation-reaction force considered
here~\cite{Chiaramello:2020ehz,Albanesi:2021rby}; in that context, they are computed 
following the procedure detailed in Appendix~A of Ref.~\cite{Nagar:2024oyk}.
Non-circular corrections for higher-order waveform multipoles (which are not included in the 
radiation reaction) likewise require higher-order derivatives (see, e.g., Appendix~B of 
Ref.~\cite{Albanesi:2021rby}), which we compute using a fourth-order centered finite-difference stencil
scheme.

While the correction in Eq.~\eqref{eq:genericNP} is clearly essential for eccentric inspirals, we 
argued in Ref.~\cite{Albanesi:2023bgi} that it also improves the analytical-numerical agreement 
during the plunge even in the quasi-circular case.
For this reason, we include it throughout the entire evolution of the binary.
By contrast, the 2PN non-circular corrections considered in previous works~\cite{Placidi:2021rkh,Albanesi:2022xge} 
were shown to improve the waveform phase during the inspiral but were then switched-off during the plunge.
Since the present work focuses on the late stages of the dynamics, we do not include these 2PN corrections here.

%------------------------------------------------------------------------
\subsection{Next-to-Quasi-Circular corrections}
%------------------------------------------------------------------------
\label{subsec:nqc}
The \ac{nqc} needed to improve the waveform during the plunge phase can be written as 
\begin{equation}
\label{eq:hnqc}
\hat{h}^{\rm NQC}_\lm = \left( 1 + \sum_{i=1}^{3}  a_i^\lm n_i \right) \exp{ \left( i  \sum_{j=1}^{3} b_j^\lm n_{j+3} \right) },
\end{equation}
where the $n_i$ functions are combinations of quantities that, for circularized systems,
are only relevant during the plunge and negligible during the quasi-circular inspiral,
hence the name. The base that we adopt in this work for all the modes is 
\begin{equation}
\label{eq:nqc_base}
\begin{alignedat}{2}
n_1 & = \frac{p_{r_*}^2}{r^2\Omega^2}, \qquad &
n_4 & = \frac{p_{r_*}}{r \Omega}, \\
n_2 & = \frac{\ddot{r}}{r \Omega^2}, \qquad &
n_5 & = n_4 \, p_{r_*}^2, \\
n_3 & = n_1 \, p_{r_*}^2, \qquad &
n_6 & = n_5 \, p_{r_*}^2.
\end{alignedat}
\end{equation}
Note that no special choice is done for the (2,2), as opposed to previous
works~\cite{Albanesi:2023bgi}, since we are now considering the same anchoring point 
and prescriptions for all the modes.
The coefficients $a_i$ and $b_i$ are determined at a specific time, generally $\tNQC$,
using the quantities $(A,\dot{A},\ddot{A})$ and $(\omega,\dot{\omega},\ddot{\omega})$.
In this work we consider $\tNQC=\to$, so that we can re-use some of these quantities in 
our ans\"atze for the amplitude and phase of Eqs.~\eqref{eq:Ah} and~\eqref{eq:phih}.
To obtain the $\lbrace a_i,b_j \rbrace$-coefficients for each multipole, we thus solve 
the two linear systems 
\begin{equation}
\label{eq:nqc_system}
\begin{alignedat}{2}
            A_{\rm EOB}(\to) & =            A_{0}, \qquad &
       \omega_{\rm EOB}(\to) & =        \omega_{0}, \\
      \dot{A}_{\rm EOB}(\to) & =       \dot{A}_{0}, \qquad &
 \dot{\omega}_{\rm EOB}(\to) & =  \dot{\omega}_{0}, \\
     \ddot{A}_{\rm EOB}(\to) & =      \ddot{A}_{0}, \qquad &
\ddot{\omega}_{\rm EOB}(\to) & = \ddot{\omega}_{0},
\end{alignedat}
\end{equation}
where on the left-hand sides the amplitude, frequency and corresponding time-derivatives 
are computed from $h_{\rm EOB} = h_\lm^{\rm inspl} \hat{h}_\lm^{\rm NQC}$.
In \TEOB{} only the first derivatives of $A$ and $\omega$ are considered,
so that $n_3$ and $n_6$ are not actually used. The impact 
of using the second time-derivatives $\ddot{A}$ and $\ddot{\omega}$ in Eqs.~\eqref{eq:nqc_system}
is discussed in Appendix~\ref{app:nqc_derivs}.

By construction, the \ac{nqc} are negligible during quasi-circular inspirals, 
but they are not negligible in eccentric inspirals, since $p_{r_*}$ is not small, nor is $\ddot{r}$.
This however unwanted, since these corrections only need to act on the plunge. We thus
switch-off the NQC corrections during the eccentric inspiral using a sigmoid, 
\be
\hNQC \rightarrow \hNQC \mathscr{S}(t-\tplunge)
\ee
where $\mathscr{S}(x)=1/\left(1+e^{-\alpha^s x}\right)$.
The natural time shift for this sigmoid is $\tplunge$, since we argued in Sec.~\ref{subsec:plunge}
that at this time the particle is close to an (unstable) circular orbit, and thus the base~\eqref{eq:nqc_base}
vanishes.
With this choice, the precise value of $\alpha^s$ is not critical; throughout this work we adopt
$\alpha^s = 0.2$, as already done in Ref.~\cite{Albanesi:2023bgi}.
Note that for quasi-circular systems or configurations with very low eccentricity,
the radial separation $r(t)$ may remain concave ($\ddot{r}<0$) throughout the late inspiral and plunge,
so that an inflection point of $r(t)$ is not defined.
In such cases, we instead use the shifted sigmoid $\mathscr{S}(t-\tLSO)$,
where $\tLSO$ denotes the time of the \ac{lso} crossing~\cite{Ori:2000zn}
(i.e., the separatrix crossing in the quasi-circular limit).
This choice has no significant impact on the results, since in these configurations
the \ac{nqc} are not relevant during the inspiral, thereby obviating the
need for a sigmoid. 

%=======================================================================================================
\section{Analytical/numerical comparisons}
\label{sec:comparisons}
%=======================================================================================================
%
\begin{figure*}[t]
	\begin{center}
	\includegraphics[width=0.325\textwidth]{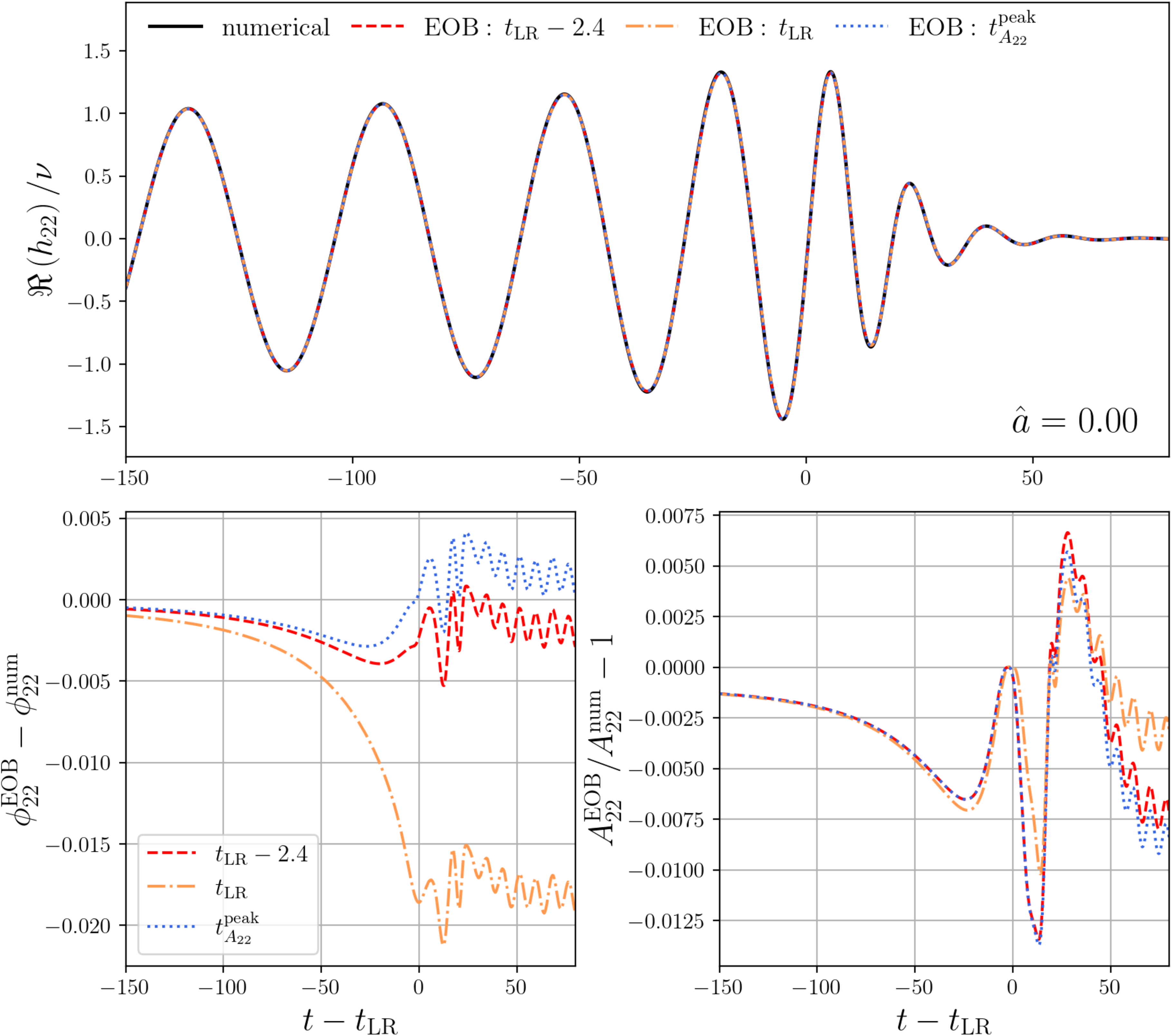}
	\includegraphics[width=0.325\textwidth]{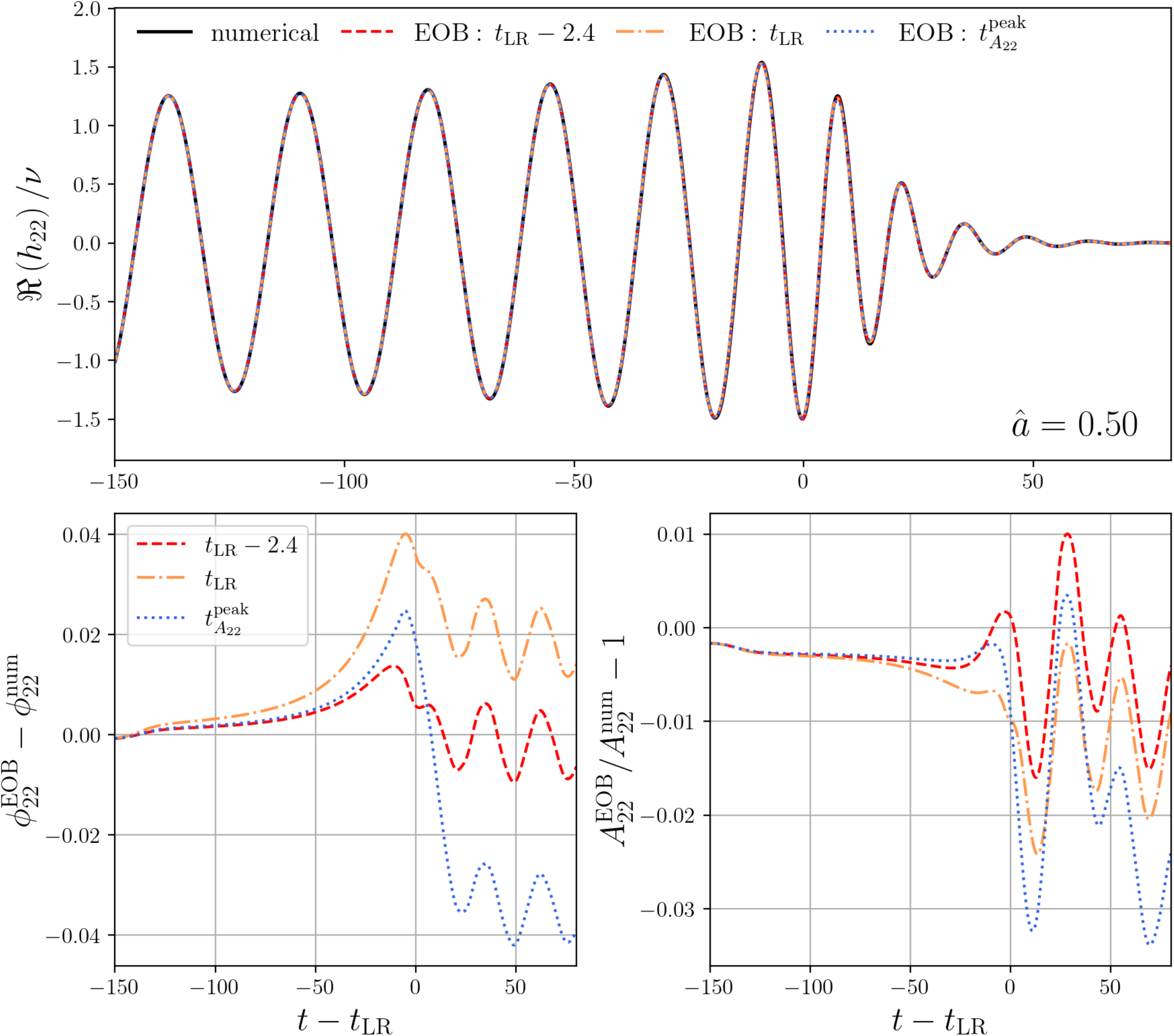}
	\includegraphics[width=0.325\textwidth]{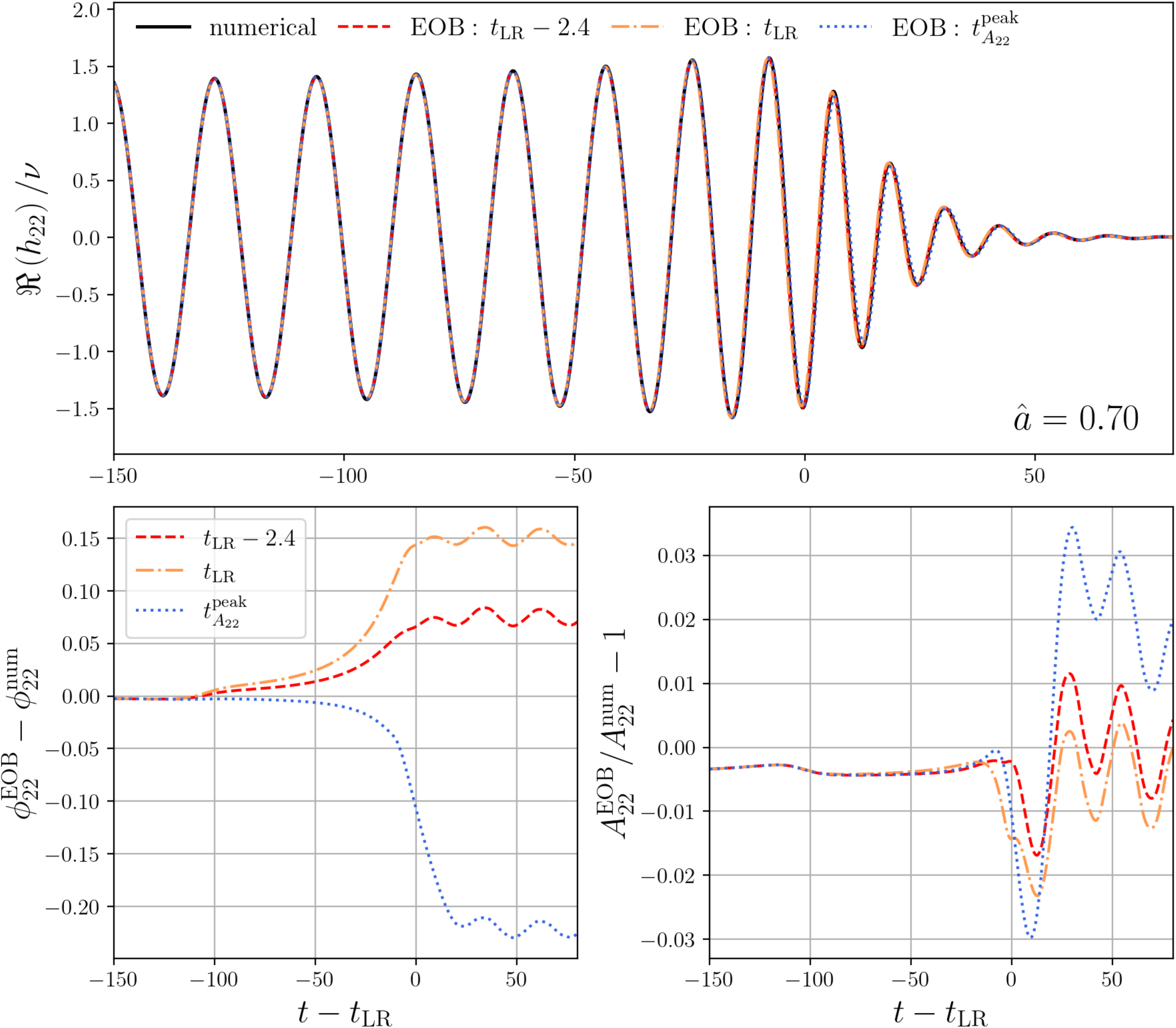}
	 \caption{\label{fig:different_anchor_qc} Analytical/numerical comparisons 
	 for quasi-circular configurations with spins $\ha=\lbrace0,0.5,0.7\rbrace$ 
	 (from left to right). For each case,
	 we show the real parts in the upper panel, and the phase and relative amplitude
	 difference in the two smaller bottom panels. The default model
	 considered in this work ($\to=\tLR-2.4$) is shown with dashed red lines.
	 Two additional models with $\to=\tLR$ (dash-dotted orange) and $\to=\tA22$
	 (dotted blue) are also reported. See text for more details.}
	\end{center} 
\end{figure*}
In this section we test the accuracy of the EOB waveform, focusing on
the plunge and post-merger signal, which is the main topic of this work.
Improvements for the late-inspiral analytical waveforms, which are likely needed
for high eccentricity and high spin, are postponed to future work.  

%------------------------------------------------------------------------
\subsection{Quasi-circular systems}
%------------------------------------------------------------------------
\label{subsec:comparisons_qc}
Since the novelty here introduced yields improvements
also in the quasi-circular case, we start by focusing on this scenario and
postpone the discussion of eccentric dynamics to Sec.~\ref{subsec:comparisons_ecc}.

%........................................................
\subsubsection{Testing different anchoring points}
%........................................................
\label{subsubsec:anchoring_test_qc}
We begin by demonstrating that the choice of anchoring point discussed in
Sec.~\ref{subsec:anchoring} is the optimal one among those explored in this work,
even for the quasi-circular case.
Specifically, we consider the dominant (2,2) multipole and 
we compare our default choice against two alternatives: using the \ac{lr}
crossing time $\tLR$ directly as anchoring point, and using the peak of the dominant
mode amplitude, $\tA22$.
For the latter case, we restrict the global fits to $\ha \leq 0.8$, since for higher spins
even the primary fits cease to be reliable.
Moreover, when adopting $\to=\tA22$, we additionally employ a fit for the time difference
between the peaks of the amplitude and of the orbital frequency, $t_{\Omega_{\rm orb}}^{\rm peak}$, following
Ref.~\cite{Albanesi:2023bgi}.
The results for three representative configurations with
$\ha=\lbrace0,0.5,0.7\rbrace$ are shown in Fig.~\ref{fig:different_anchor_qc}.
In the nonspinning case (left panel), all three models perform well.
Nevertheless, the model anchored at $\to=\tLR$ (dash-dotted orange curve) exhibits a
larger late-time dephasing, reaching $\Delta\phi_{22}\simeq 0.02$ rad.
The other two prescriptions, $\to=\tLR-2.4$ (dashed red) and $\to=\tA22$ (dotted blue),
perform comparably better, with $|\Delta\phi_{22}|\lesssim 0.005$ rad.
No significant differences are observed in the amplitude, and all three models remain highly
accurate in this regime.
More pronounced discrepancies arise for $\ha=0.5$ (middle panel), particularly in the phase.
In this case, the $\to=\tLR-2.4$ prescription clearly outperforms the others, yielding
$|\Delta\phi_{22}|\lesssim0.01$ rad.
Even larger differences are observed for $\ha=0.7$ (right panel), where
the model with anchoring point at the peak yields $\Delta\phi_{22}\simeq-0.22$ rad, 
while our default choice gives $\Delta\phi_{22}\simeq 0.07$ rad.
Increasing the spin further decrease the accuracy of the $\tA22$ model.
Indeed, while for $\ha\leq 0.8$ one finds
$-3 \lesssim \tA22 - t_{\Omega_{\rm orb}^{\rm peak}} \lesssim 6$,
this quantity grows dramatically at higher spins, reaching  $\sim 50$
for $\ha=0.9$ and even a few hundreds for higher spins (see Table~A3 of Ref.~\cite{Harms:2014dqa}).
For this reason, we restrict the global fit of
$\tA22 - t_{\Omega_{\rm orb}^{\rm peak}}$ to $\ha \leq 0.8$.
Moreover, as discussed earlier, for very high spins even the primary fits anchored at $\tA22$
break down due to the presence of long-lived amplitude plateaus.
By contrast, both approaches that anchor the model near the \ac{lr} crossing yield physically
consistent waveforms\footnote{Recently, Ref.~\cite{Nishimura:2026nse}
introduced a model for quasi-circular test-mass configurations using the amplitude peak
as matching time and still yielding accurate results for spins as high as $\ha=0.9$.}.
The analytical/numerical comparison of our model with $\to=\tLR-2.4$ for $\ha=0.9$
is instead reported in the middle panel of Fig.~\ref{fig:nqc2} in Appendix~\ref{app:nqc_derivs}.
We observe a dephasing that never
exceeds $0.03$ rad and a relative amplitude difference remaining below $\sim 1\%$.
The model anchored directly at $\tLR$ (not shown in Fig.~\ref{fig:nqc2}) performs worse, reaching
$\Delta\phi_{22}\simeq 0.2$ rad and a relative amplitude difference of $\sim 2.5\%$.
Finally, we note that for retrograde orbits all the three anchoring points explored in this section
yield consistent waveforms, but our default choice of Sec.~\ref{subsec:anchoring} typically performs slightly better. 

%........................................................
\subsubsection{Multipoles}
%........................................................
\label{subsubsec:comparisons_qc}
\begin{figure*}[t]
	\begin{center}
	\includegraphics[width=0.300\textwidth]{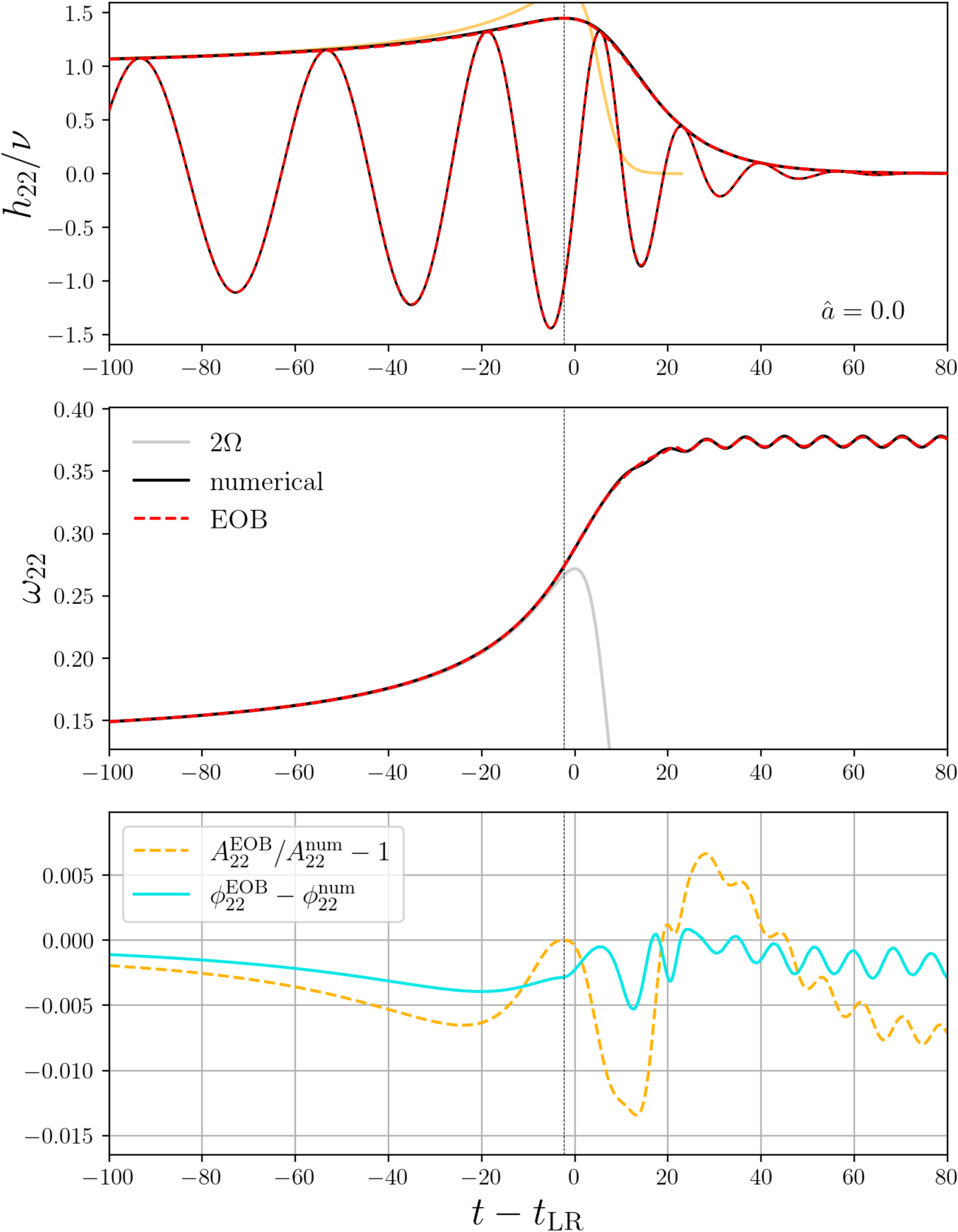}
	\includegraphics[width=0.301\textwidth]{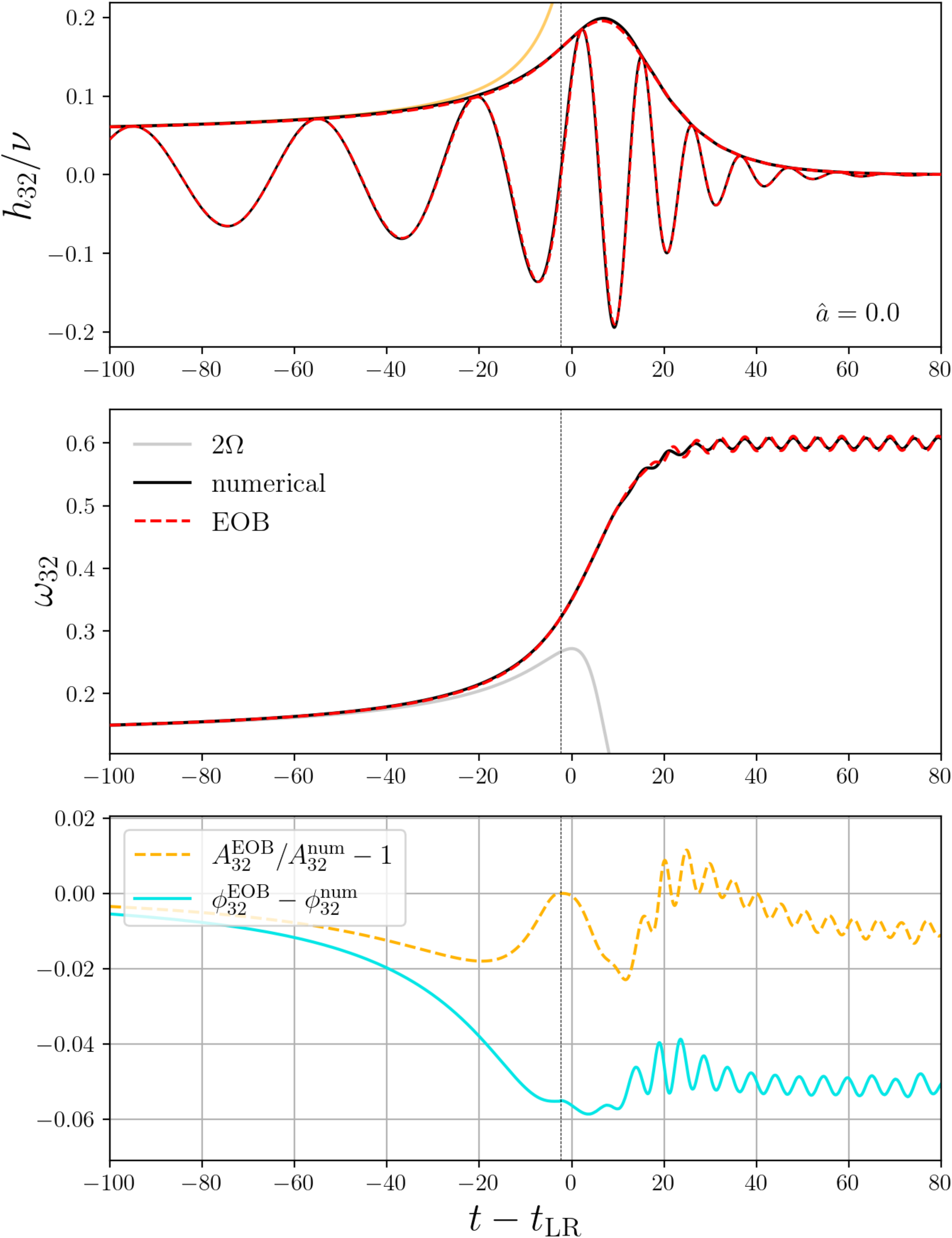}
	\includegraphics[width=0.303\textwidth]{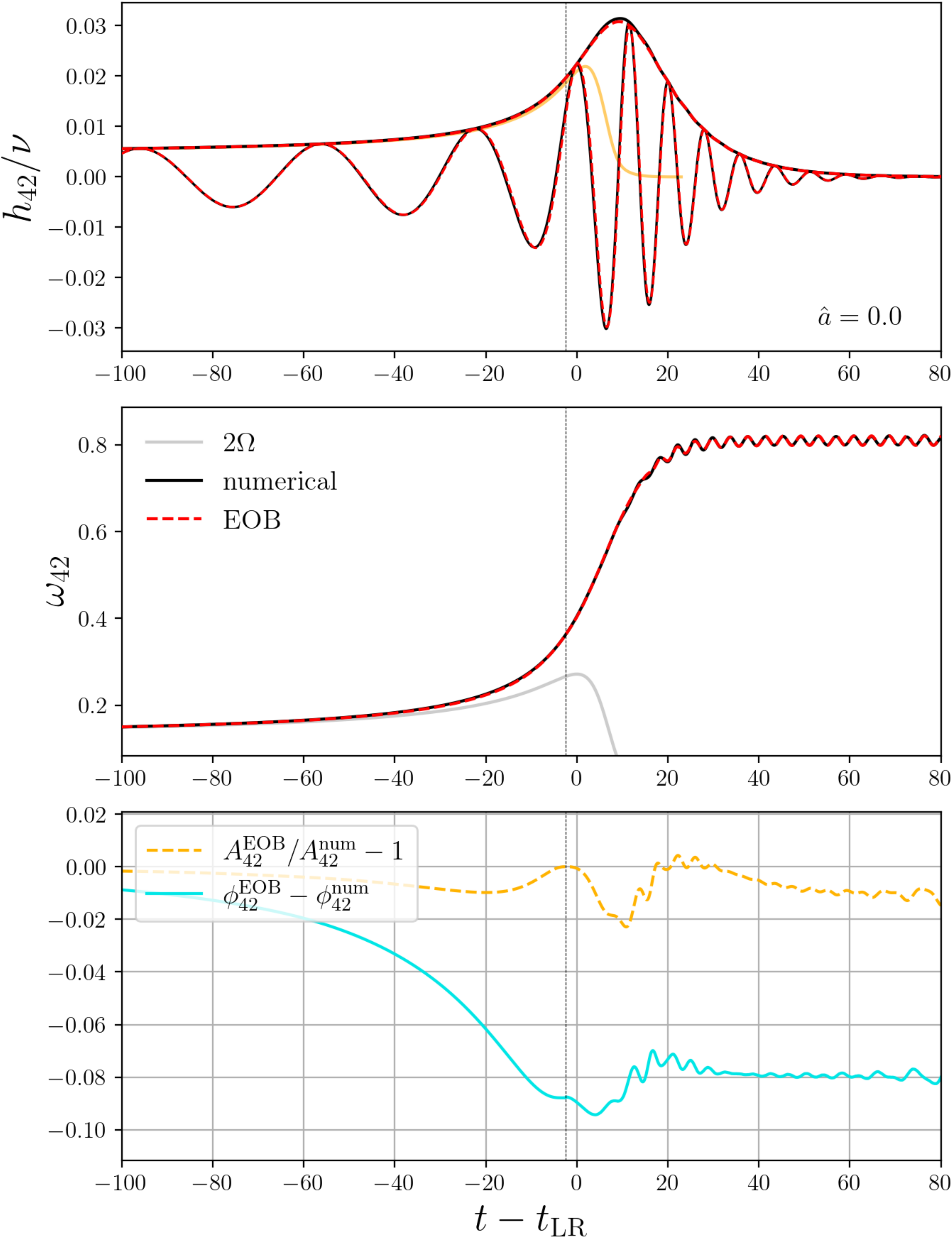}\\
	\hline
	\vspace{0.2cm}
	\includegraphics[width=0.301\textwidth]{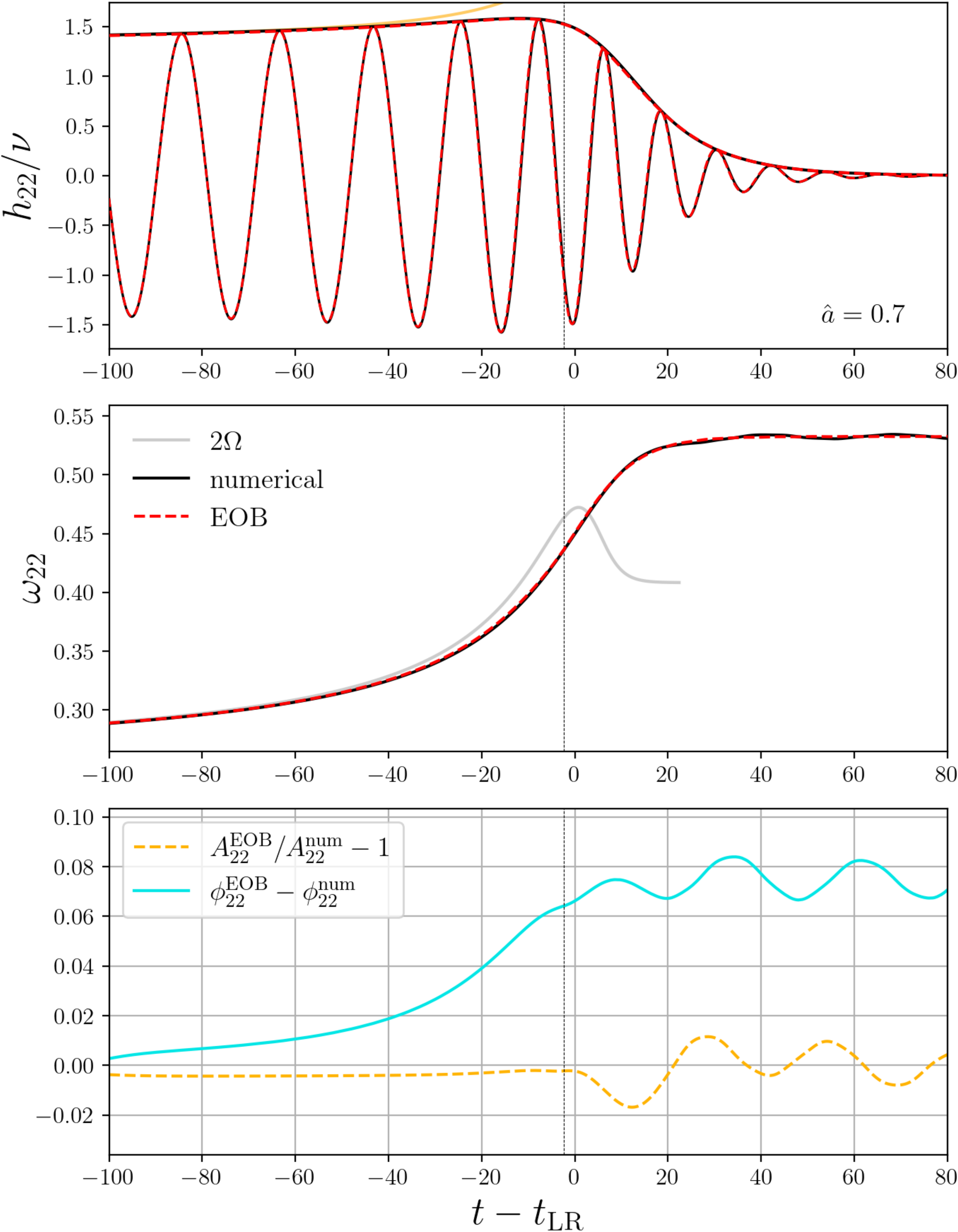}
	\includegraphics[width=0.305\textwidth]{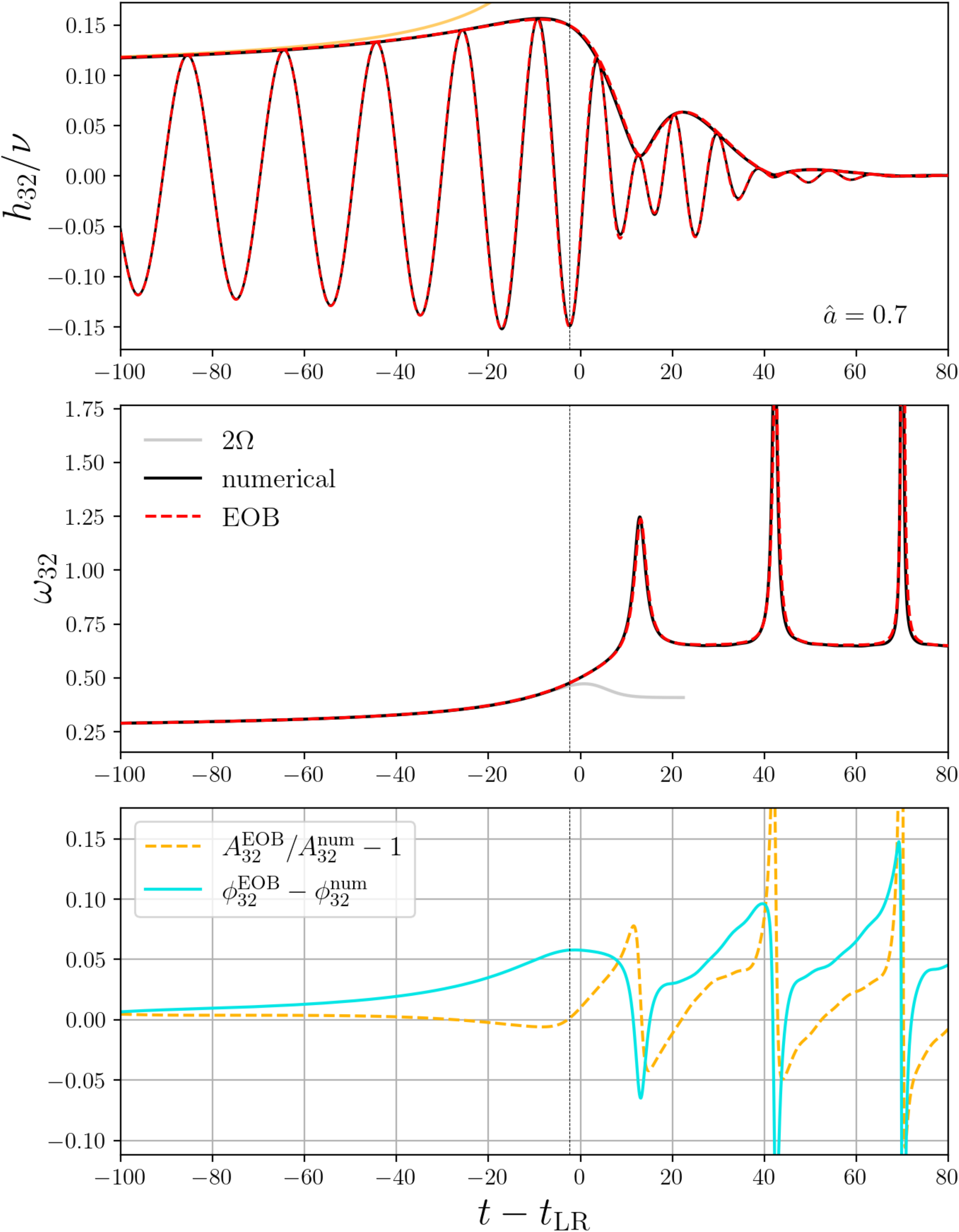}
	\includegraphics[width=0.303\textwidth]{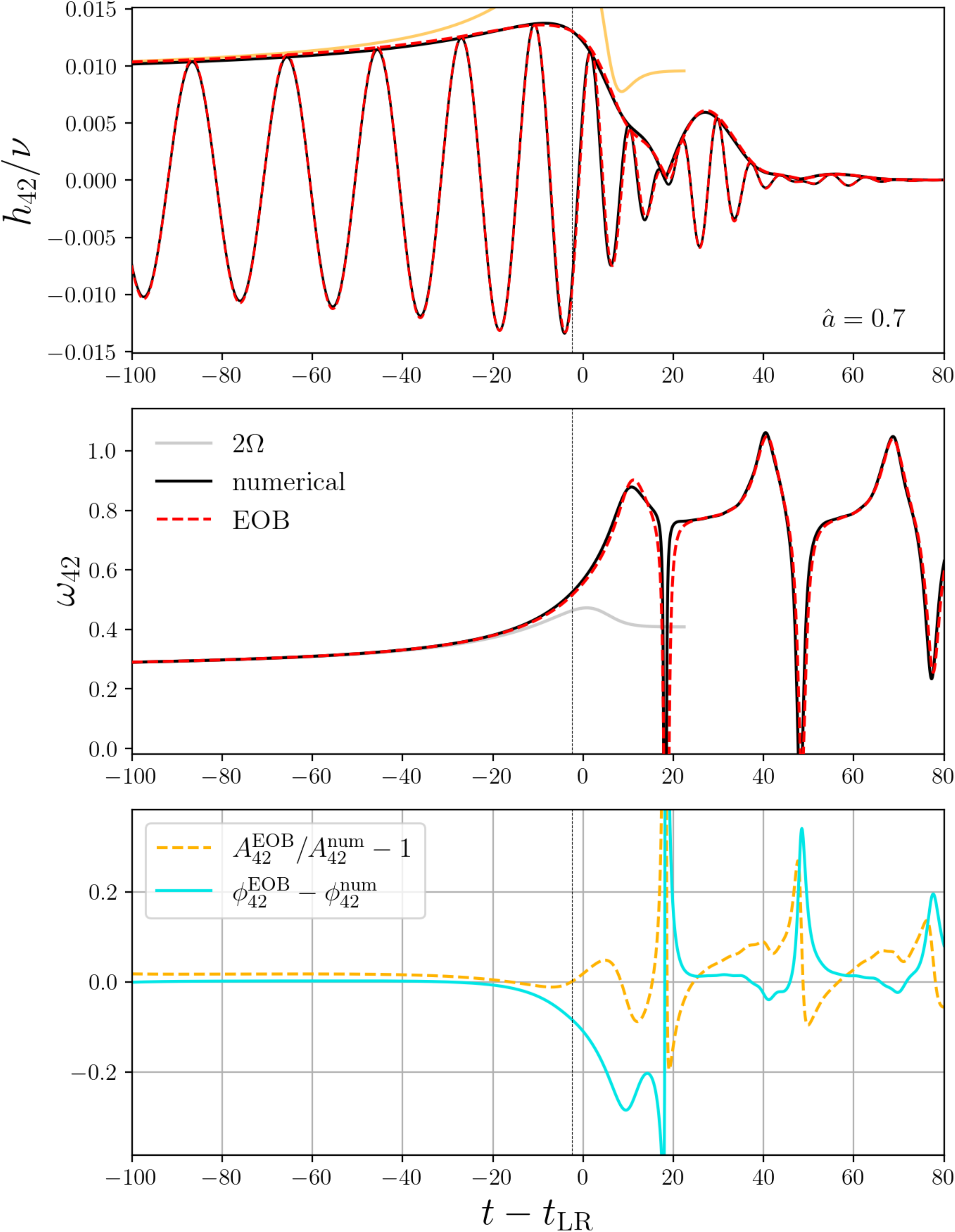}\\
	\hline
	\vspace{0.2cm}
	\includegraphics[width=0.301\textwidth]{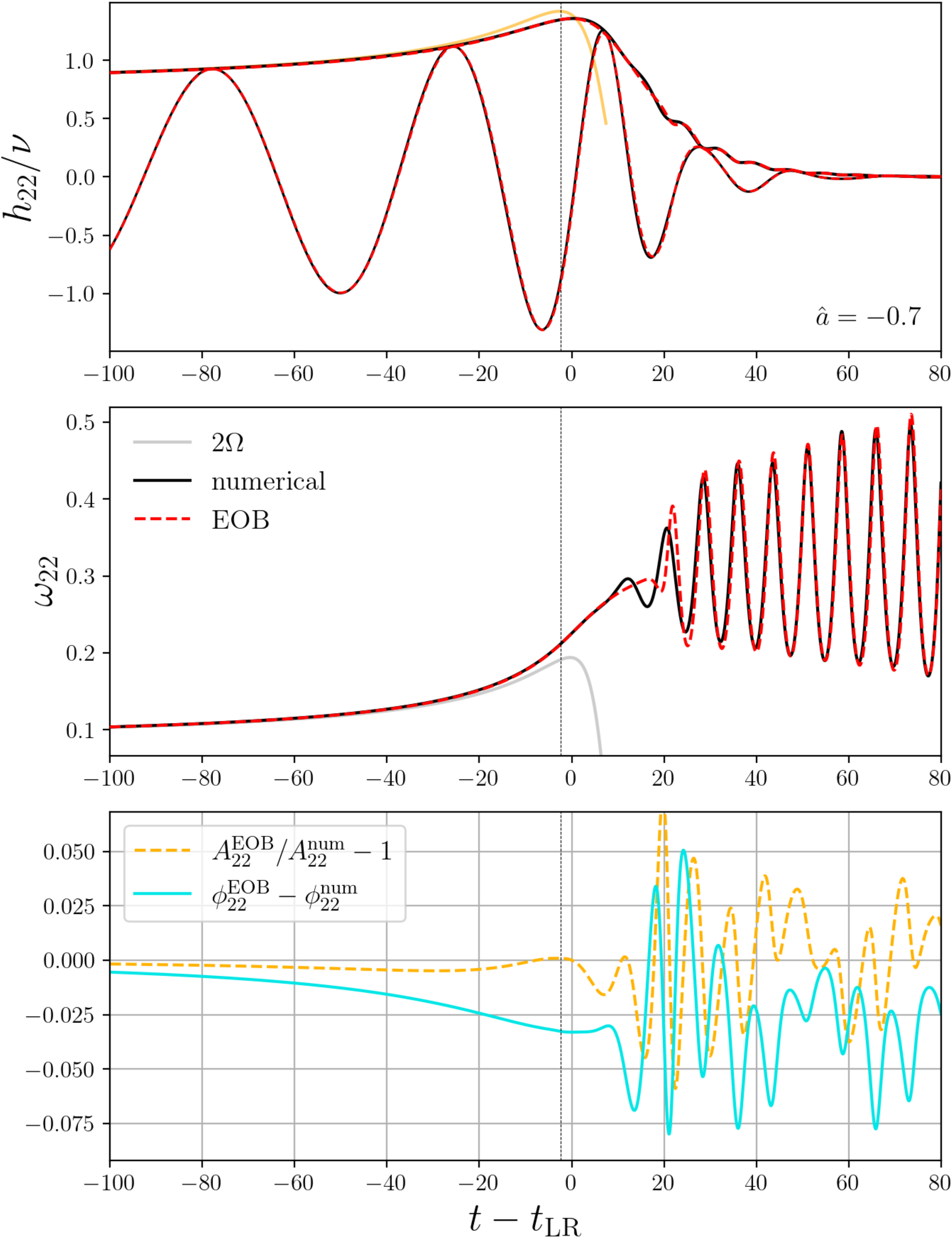}
	\includegraphics[width=0.303\textwidth]{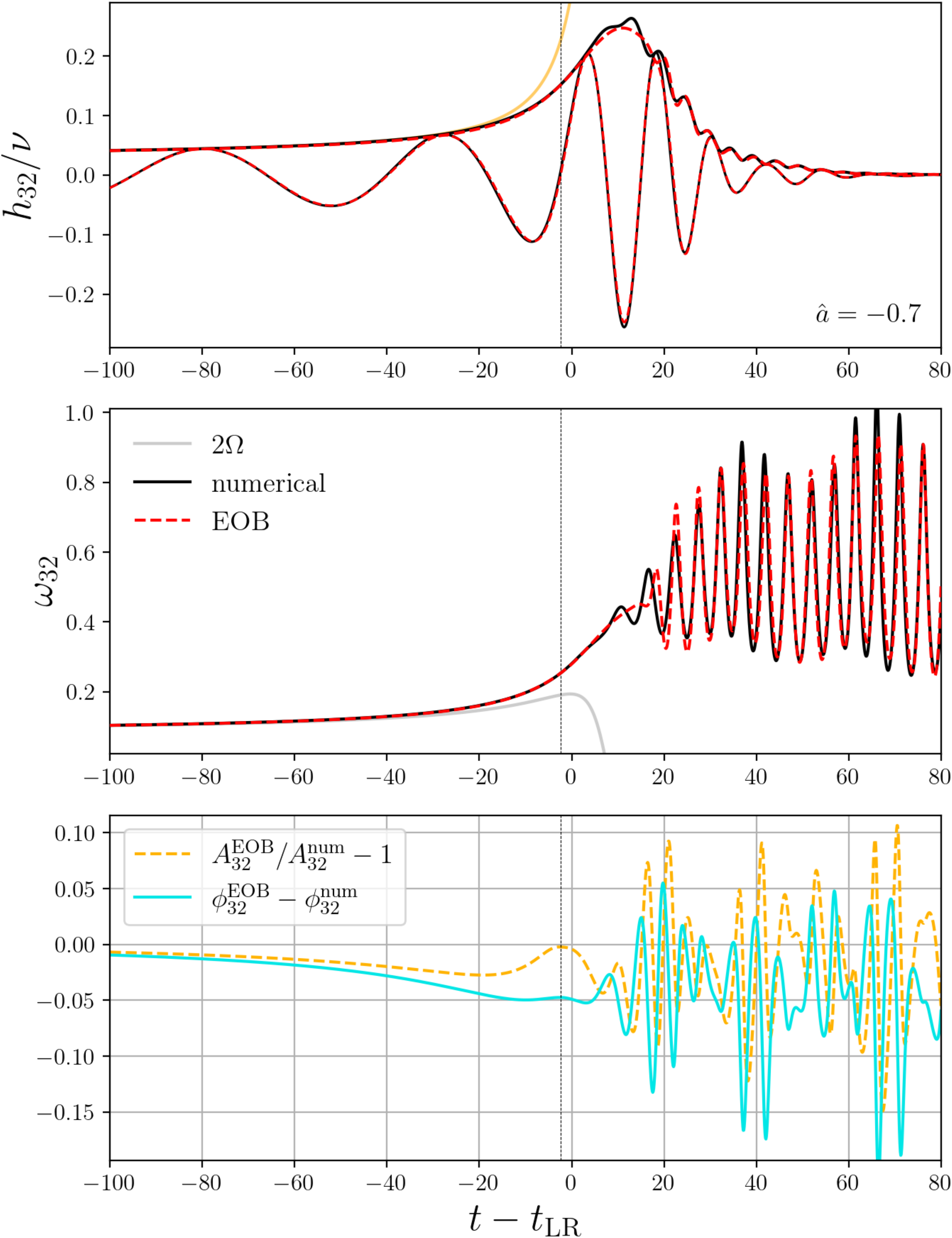}
	\includegraphics[width=0.300\textwidth]{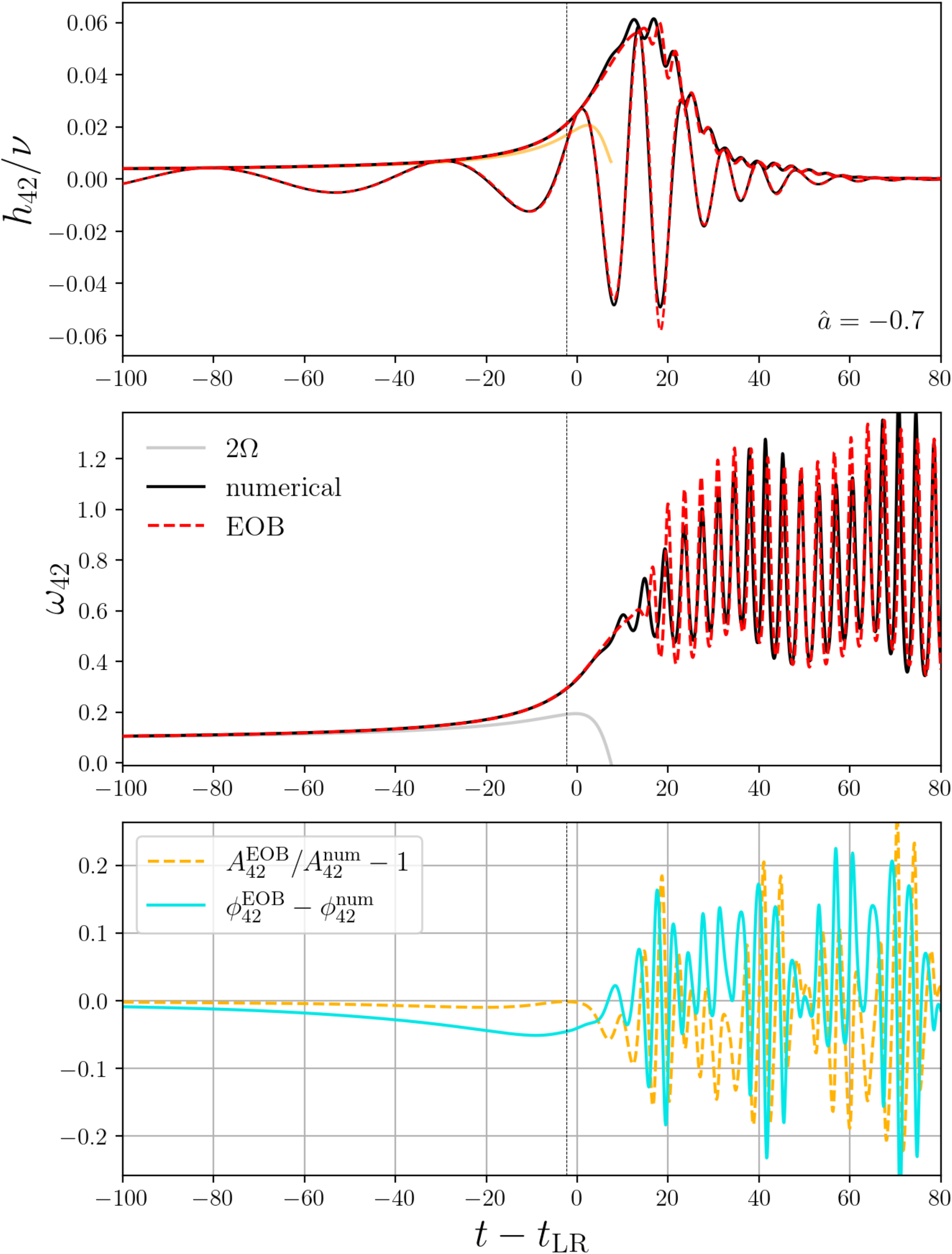}
	 \caption{\label{fig:comparisons_m2_qc} Analytical/numerical comparisons for quasi-circular systems with 
	 $\ha=\lbrace 0,\pm 0.7 \rbrace$ for the $m=2$ modes considered in this work. Complete EOB waveforms are highlighted 
	 in dashed red, while numerical results in black. We also report the inspiral-plunge
	 amplitude without NQC corrections in the upper panels (orange) and $2 \Omega$ in the middle
	 panels. For each case, we show phase (azul) and relative amplitude differences (dashed orange).
	 The dashed vertical line marks the anchoring time $\to=\tLR-2.4$.}
	\end{center} 
\end{figure*}
We now discuss how the accuracy of our default model varies for different spins. 
We report in Fig.~\ref{fig:comparisons_m2_qc} three quasi-circular configurations with
different spins, $\ha=\lbrace 0,\pm0.7 \rbrace$. We focus on the modes 
with $m=2$ considered in this work: $(2,2)$, $(3,2)$, and $(4,2)$. We
will discuss the other higher modes directly when addressing eccentric configurations. 
In the first row of Fig.~\ref{fig:comparisons_m2_qc}, 
we report the multipoles for the non-spinning case. In this case 
the post-peak amplitudes are monotonic since there is no spherical-spheroidal mode-mixing and 
the beating between co-rotating and counter-rotating \acp{qnm} is not so relevant, even if some
oscillations can be clearly seen during the stationary phase in the frequency,
which we report in the middle panels. These oscillations are however well-modeled by the approach	
discussed in Sec.~\ref{subsec:beating}.
While the dominant (2,2) mode is the one which is best-described by our model, with an
instantaneous analytical/numerical phase difference of $|\Delta \phi_{22}|\lesssim 0.005$ rad,
the model remains accurate also for the higher modes, with $\Delta \phi_{32}\simeq -0.06$ rad and
$\Delta \phi_{42}\simeq -0.010$ rad. The amplitude relative differences are instead around 
the $\sim 1-2\%$ for boh higher modes. Notably, the accuracy of the phase for higher modes 
has been significantly increased with respect to Ref.~\cite{Albanesi:2023bgi}, where 
we were using $\to=\tAlm$, rather than $\to=\tLR-2.4$. Indeed,
in that case we had $\Delta \phi_{32}\simeq -0.11$ and $\Delta \phi_{42}\simeq 0.7$ rad (cf. Fig.~10 therein).
It should be noted, however, that even using $\tA22$ as anchoring point instead of $\tAlm$ 
for the higher modes would strongly improve the phase for the non-spinning case,
since in that case the large phase differences were linked to the \ac{nqc} rather than to the post-peak
waveform modeling itself.
In Fig.~\ref{fig:comparisons_m2_qc} we also report the amplitude of $h_\lm^{\rm inspl}$ (solid orange, upper panels).
As can be seen, the amplitude \ac{nqc} start to be relevant from $\sim 40\,M$ before $\tLR$. We also report 
$2\Omega$ (grey, middle panels), which is the leading quasi-circular Newtonian contribution to the 
waveform frequency $\omega_\lm$. For the (2,2) mode in particular, we have $\omega_{22}\simeq 2 \Omega$
until the \ac{lr} crossing, thus implying that the \ac{nqc} only introduce a small correction
to the EOB waveform frequency and thus yield a very accurate phase. For the higher modes,
the differences between orbital and waveform frequencies are instead higher.

We now move our attention to the quasi-circular case with $\ha=0.7$, 
reported in the middle row of Fig.~\ref{fig:comparisons_m2_qc}. 
In this case, a visible discrepancy between $2 \Omega$ and $\omega_{22}$ can be observed starting
from $\tLR-40$. While this is partially accounted by the \ac{pn} circular
corrections included in $h_{22}^{\rm inspl}$, this still results in a larger phase difference, 
which however reaches at most $\sim 0.08$ rad.  
The amplitude is instead accurate at the $\sim 2\%$ level. When moving to the higher modes,
we clearly see the effect of the spherical-spheroidal mode-mixing, both in the amplitude and in the
frequency. We have already discussed this effect in Sec.~\ref{subsec:primary} when discussing the primary fits on 
spheroidal modes. Here we just highlight that the attachment of these post-merger to the inspiral-plunge
waveform yield remarkably accurate complete waveforms. For the (3,2) and (4,2) modes, we  
have $\Delta \phi_{32}\simeq 0.06$ and $\Delta \phi_{42}\simeq 0.1$ rad at the \ac{lr}. Amplitudes are 
also accurate at $\tLR$, having relative differences around $\sim 1\%$ and $\sim 2\%$ for the two
higher modes. However, at later times we observe larger discrepancies, which are linked
to the modulation in amplitude and phase. However, on average, amplitude and phase remain reliables, 
as can be better seen by the direct comparison between amplitudes and real parts reported in the upper panels.
It should be noted that, for higher spins, larger discrepancies are observed in the higher modes,
while the (2,2) mode remains more accurate: $\Delta \phi_{22}\lesssim 0.14$ and $\Delta \phi_{22}\lesssim 0.03$ 
rad for $\ha=0.8$ and $0.9$, respectively (for $\ha=0.9$, see also middle panels of Fig.~\ref{fig:nqc2}).
We conclude the discussion of this prograde orbit by noting that for the (4,2) mode, we have
a $1.7\%$ analytical/numerical relative difference in the amplitude before the \ac{nqc} kick-in. 
This highlight that the resummation of the $\rho_\lm$ here adopted, which strictly follows
Ref.~\cite{Messina:2018ghh} for this higher mode
starts to be inaccurate for high spins. For higher spins, this difference grows to $\sim 9\%$ 
and $\sim 14\%$ for $\ha=0.90$ and $0.95$, respectively. Similar effects 
are observed for the other multipoles, especially with high $\l$. However, this has been already mentioned
in Sec.~\ref{subsec:inspiralwave}, and the improvement of the analytical prescription for the inspiral
waveform (and thus the radiation reaction) with more up-to-date methodologies~\cite{Cipriani:2026xmx,Ivanov:2025ozg}
is postponed to future works.

We finally discuss the retrograde orbit with $\ha=-0.7$, which is shown 
in the last row of Fig.~\ref{fig:comparisons_m2_qc}. The main differences with the
previous case is that the contribution from the beating is now quite large. We recall
that the beating modeling adopted in this paper, discussed in Sec.~\ref{subsec:beating}, 
includes a factor which only takes into account the beating between co-rotating and counter-rotating
for the fundamental mode, and it is therefore accurate only in the stationary \ac{qnm} regime. 
Indeed, for times prior to $t\simeq \tLR + 20$, there is also a beating contribution from the overtones 
which we are not modeling. Therefore, the prescription for the beating becomes 
accurate only after those times. This discrepancy is particularly evident when looking at
the analytical/numerical comparison for the frequencies. 
However, at later times, the beating is reasonably well described.
For the (2,2) mode (leftmost panel), there is no spherical-spheroidal mode-mixing, and 
therefore the beating is the only kind of mixing. For the two higher modes (middle and rightmost panels),
the spherical-spheroidal mixing is also relevant, and therefore we observe two
modulations in $\omega_\lm$: the one with higher frequency is given by the 
beating, while the one with lower frequency given by the spheroidal-spherical mixing. 
The accuracy of beating description increases for milder spins, and gets worse
for faster rotating Kerr black holes, as a priori expected. 
However, the matching between the inspiral-plunge waveform and the post-merger model
yield reliable results. As can be seen from the $\ha=-0.7$ example reported in
Fig.~\ref{fig:comparisons_m2_qc}, the analytical/numerical phase differences $\Delta \phi_\lm$
at the \ac{lr} are approximately 0.035, 0.05, and 0.04 rad for the (2,2), (3,2), and
(4,2) modes, respectively.  

We conclude this section by mentioning that the model produces physical waveforms up to 
$\ha=0.95$ for the (2,2) mode, despite the global fits being performed up to 
$\ha=0.9$. We briefly discuss this high-spin case in Appendix~\ref{app:nqc_derivs},
when discussing the relevance of second-time derivatives in the \ac{nqc}.
Indeed, it should be noted that all the phase difference here discussed are dependent
on the prescriptions adopted for the \ac{nqc}, and they are not only related to the 
accuracy of the post-merger modeling itself. On the other hand, the relative differences in amplitude
are less heavily dependent on the \ac{nqc} choice, since they are not cumulative, as opposed to the phase ones. 
We also mention that for spin $\ha=0.99$ the model stops producing physically
robust waveforms, since artefacts related to the \ac{nqc} appear.

%------------------------------------------------------------------------
\subsection{Eccentric systems}
%------------------------------------------------------------------------
\label{subsec:comparisons_ecc}
\begin{figure*}[t]
	\begin{center}
	\includegraphics[width=0.31\textwidth]{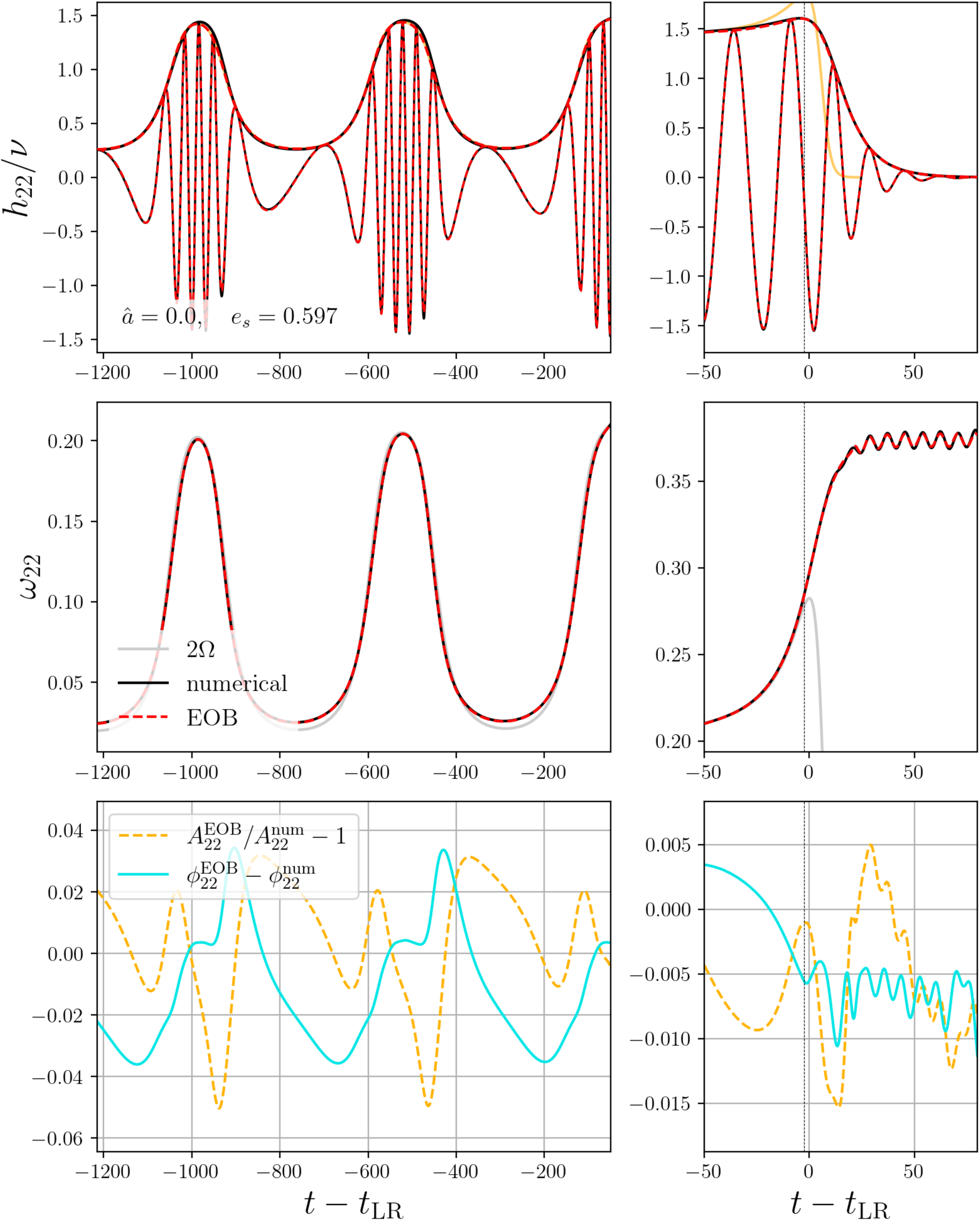}
	\includegraphics[width=0.31\textwidth]{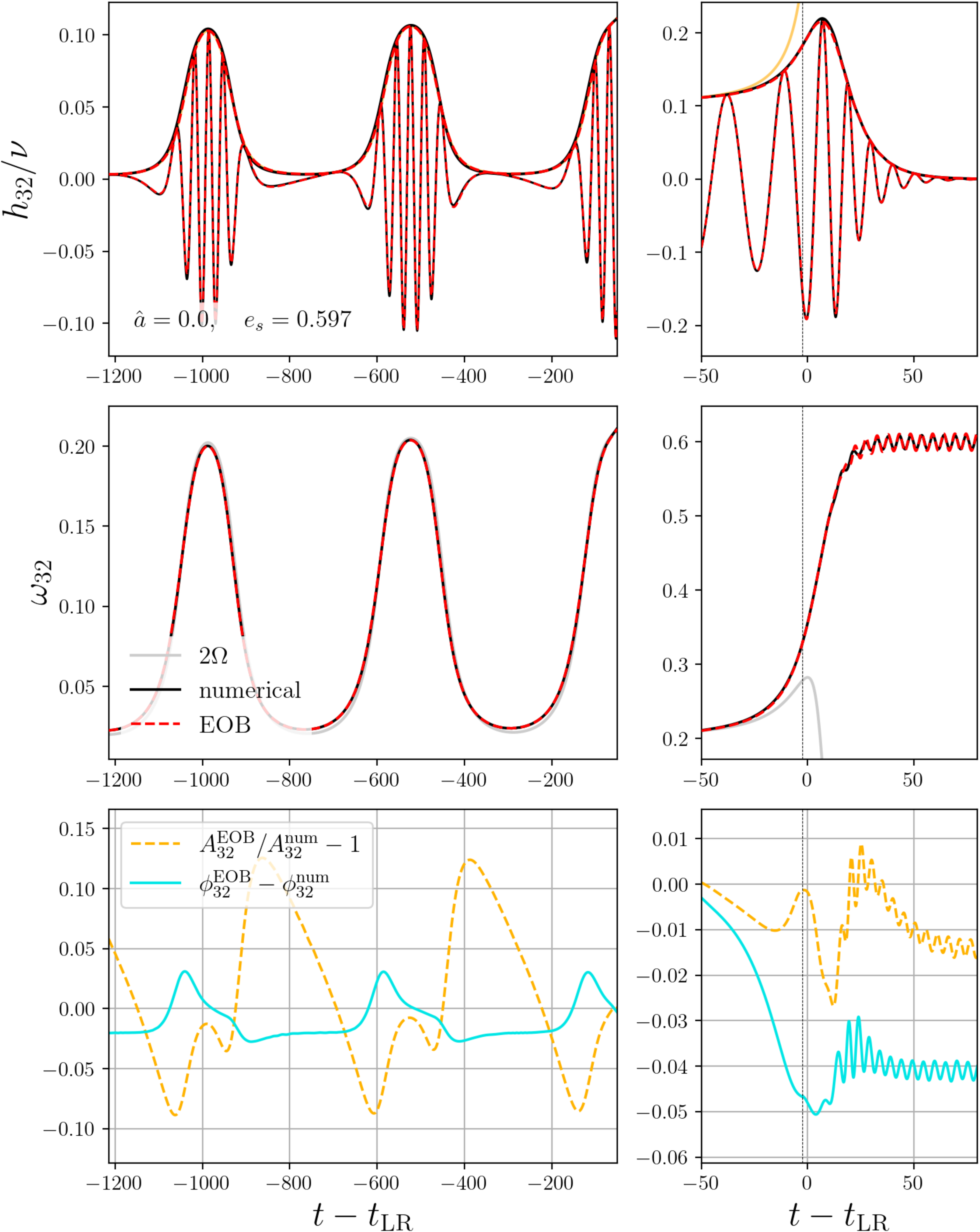}
	\includegraphics[width=0.31\textwidth]{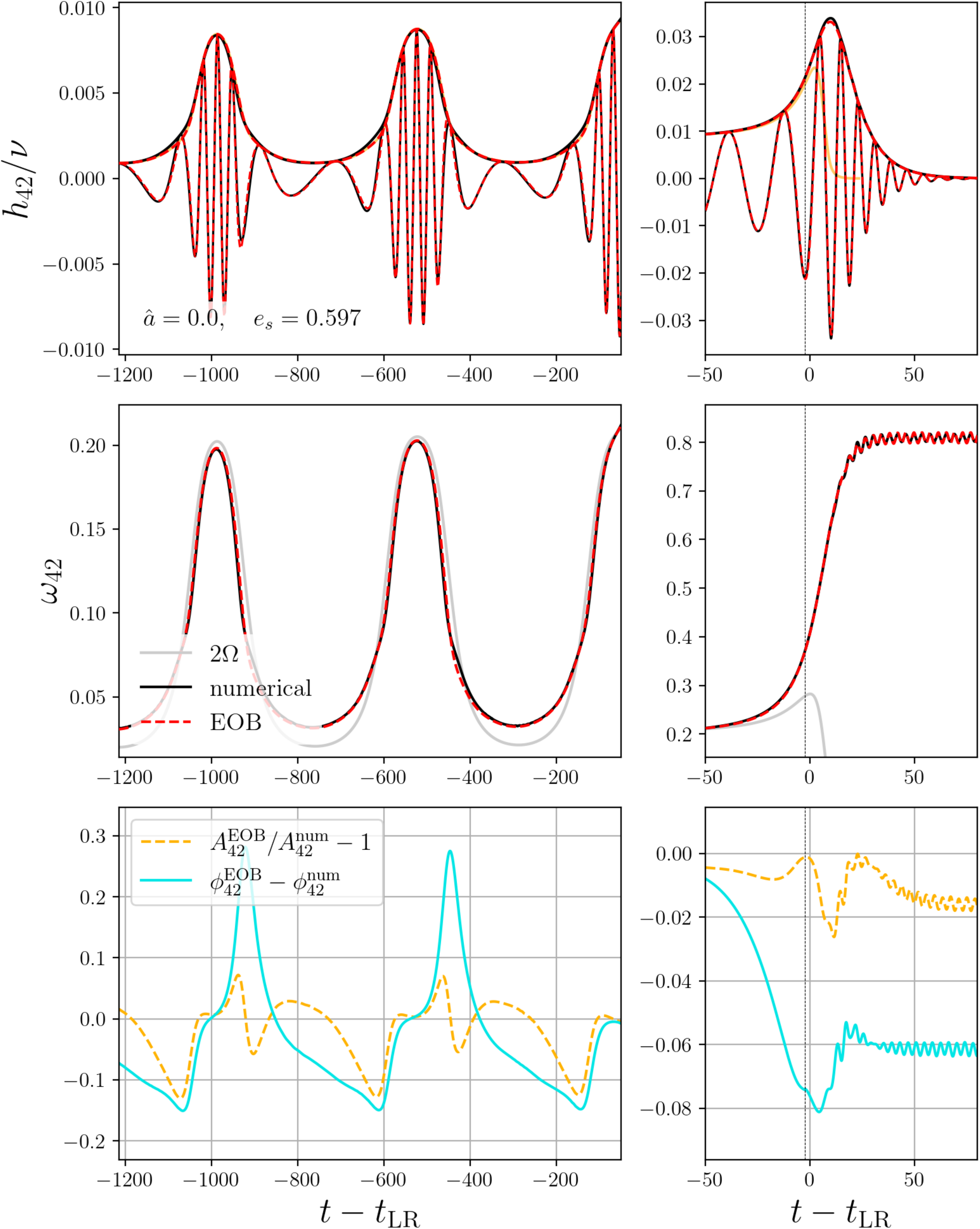}\\
	\hline
	\vspace{0.2cm}
	\includegraphics[width=0.31\textwidth]{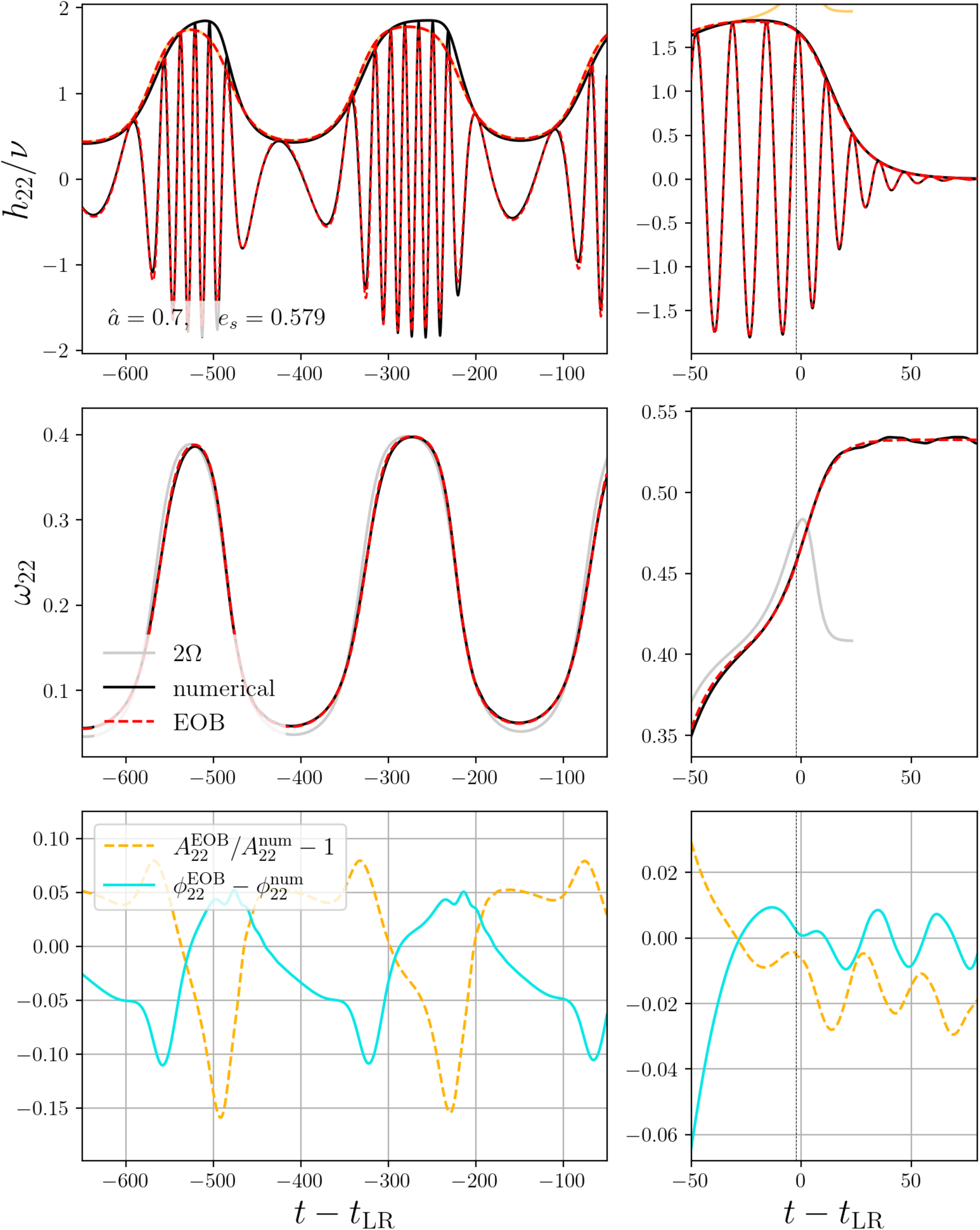}
	\includegraphics[width=0.31\textwidth]{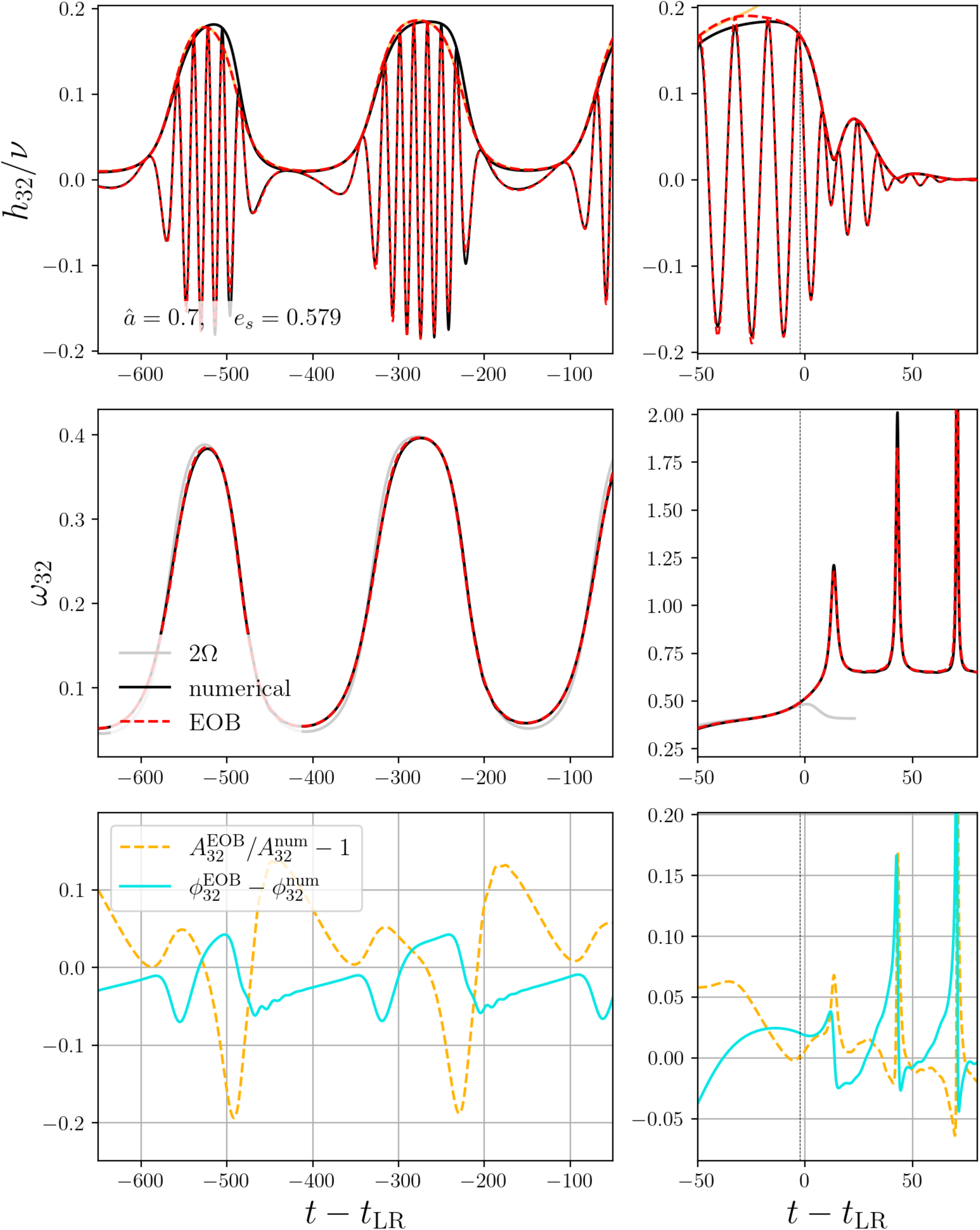}
	\includegraphics[width=0.31\textwidth]{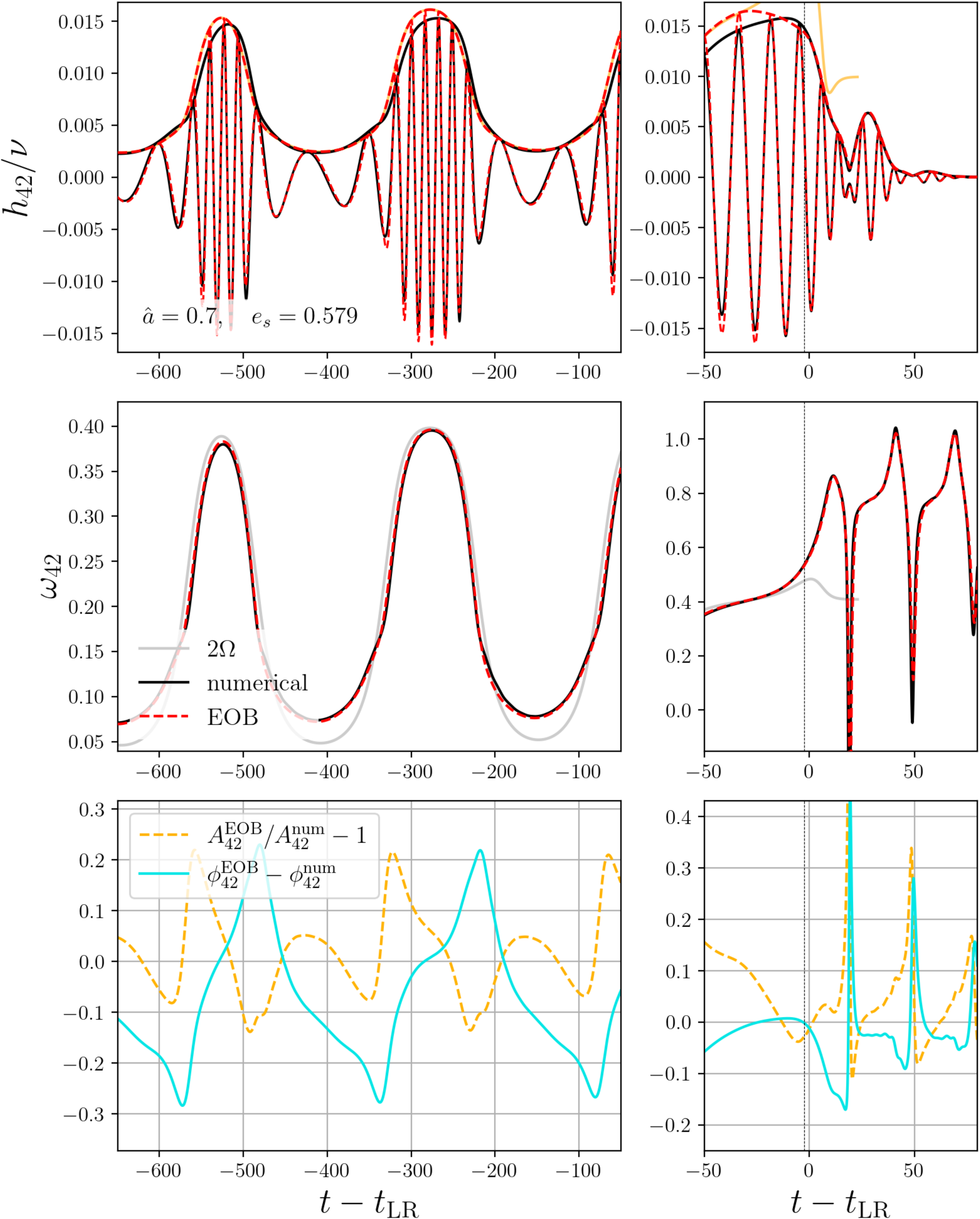}\\
	\hline
	\vspace{0.2cm}
	\includegraphics[width=0.31\textwidth]{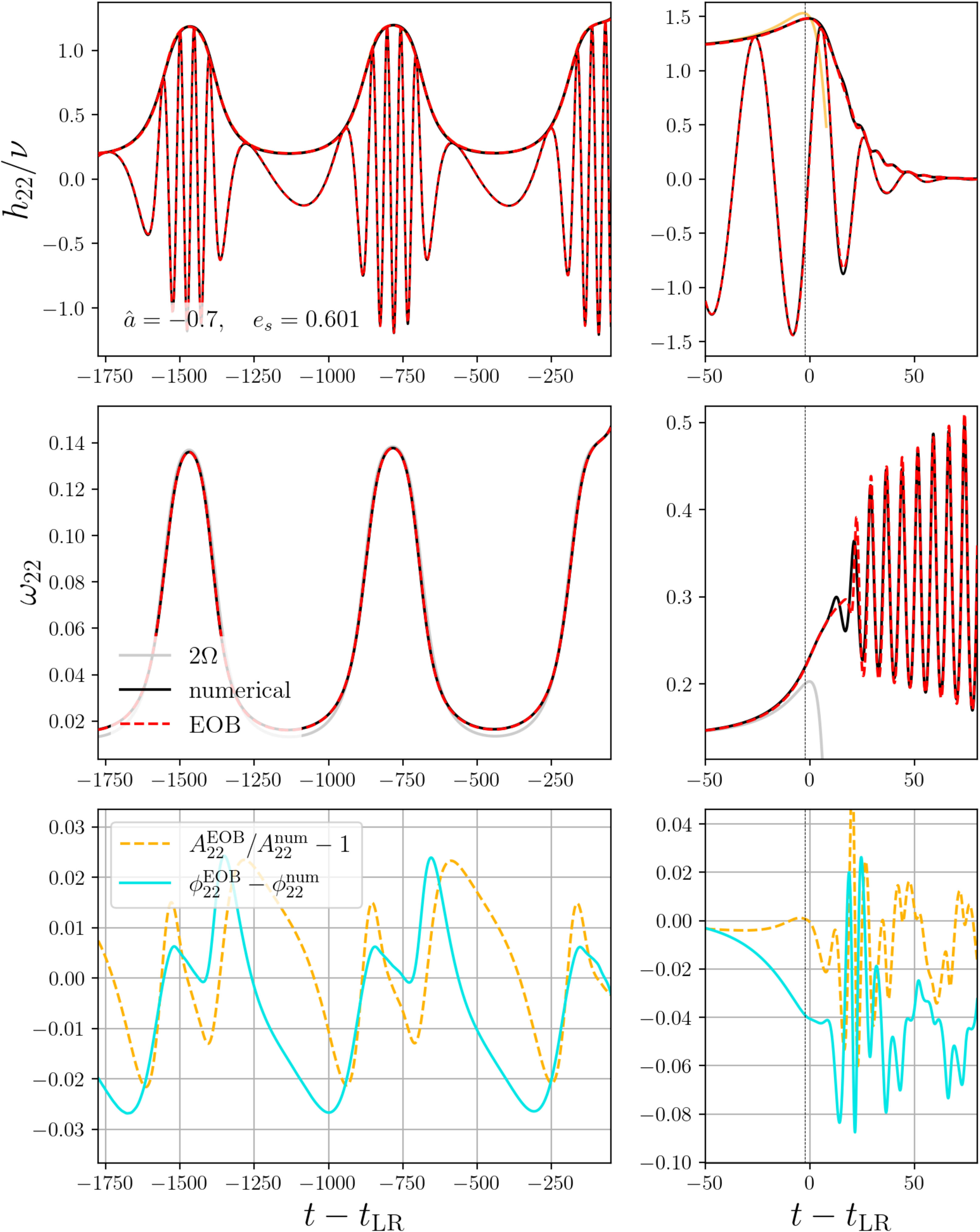}
	\includegraphics[width=0.31\textwidth]{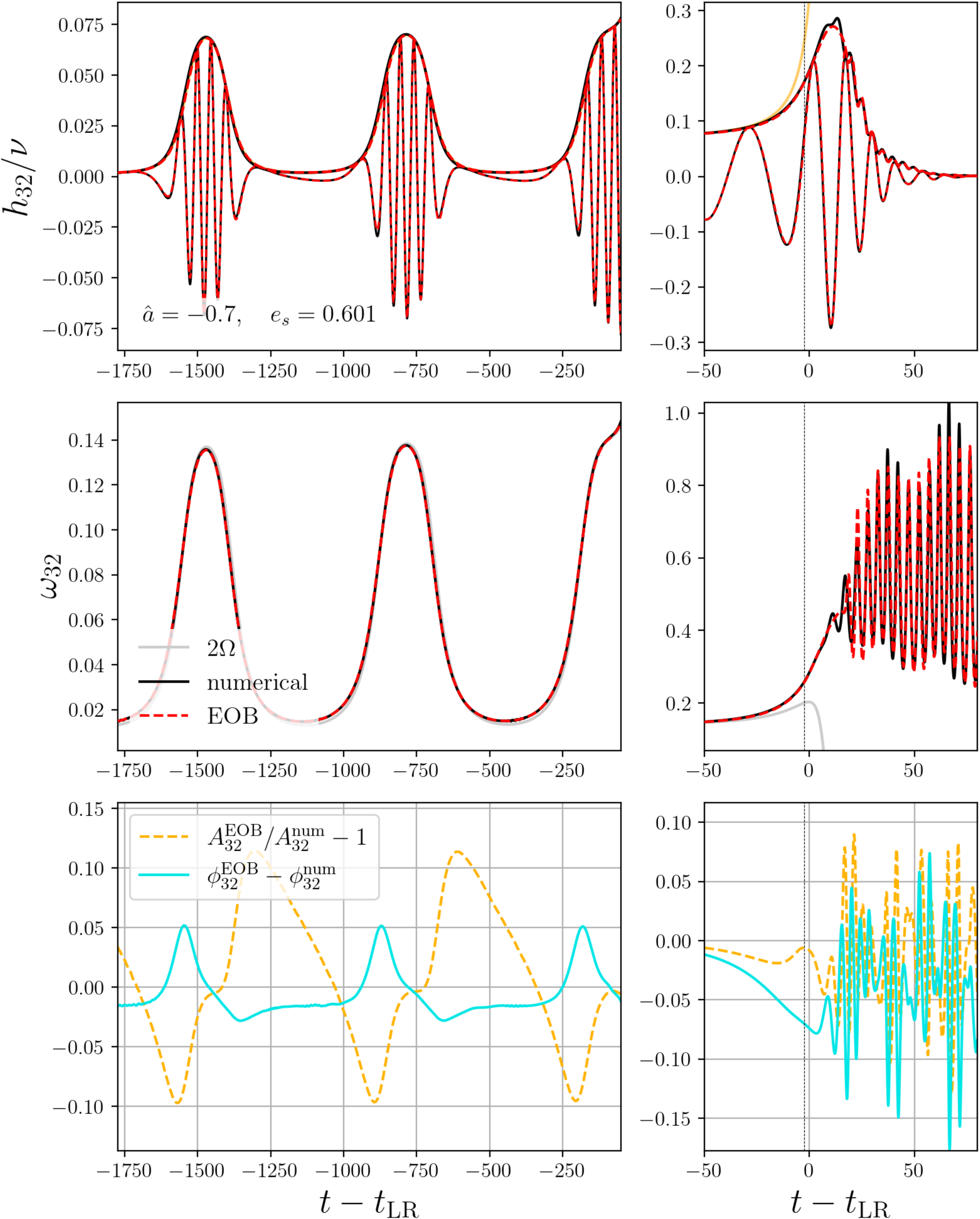}
	\includegraphics[width=0.31\textwidth]{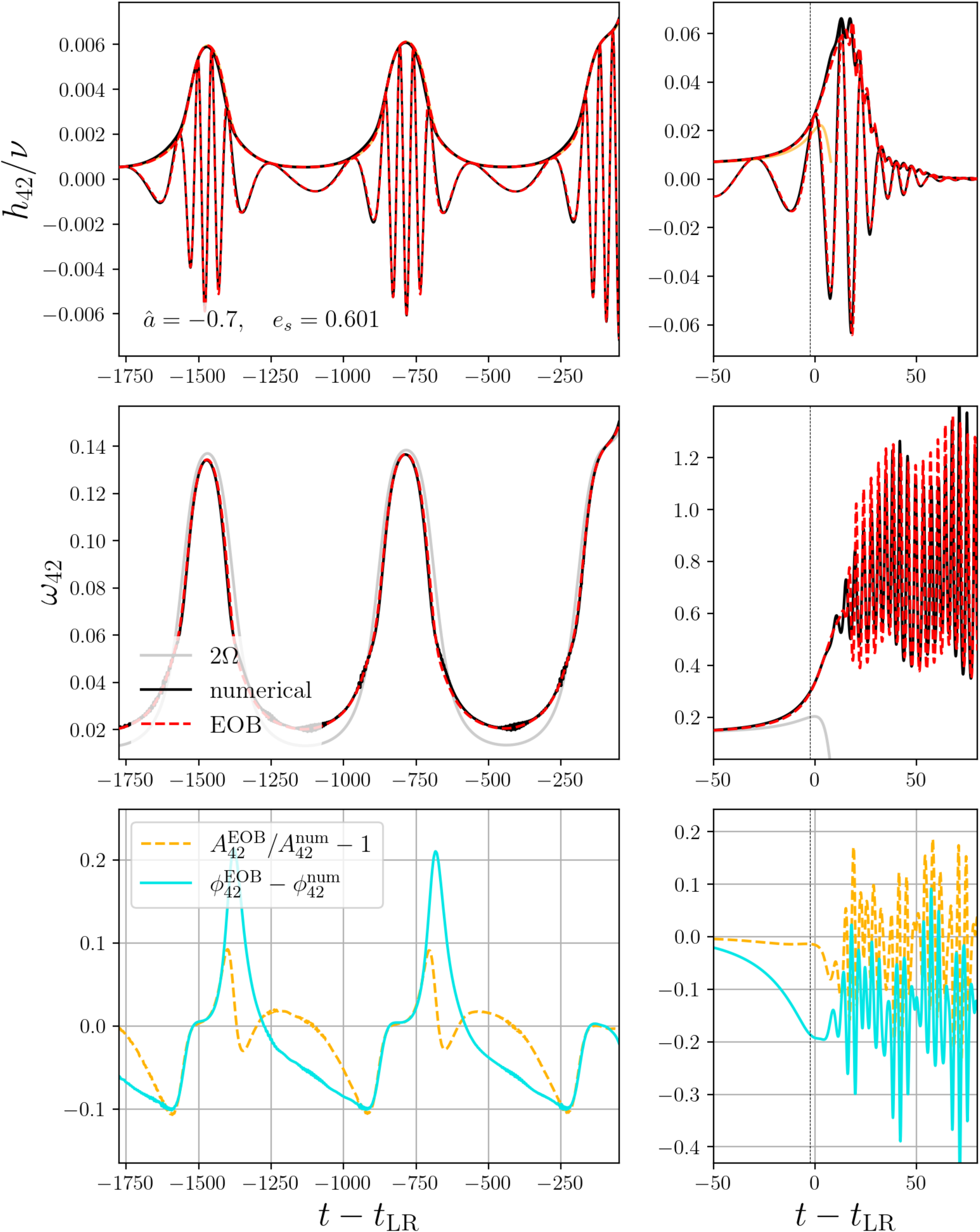}
	 \caption{\label{fig:comparisons_m2_e06} Analytical/numerical comparisons for quasi-circular systems with 
	 $\ha=\lbrace 0,\pm 0.7\rbrace$ and $\esep\simeq 0.6$ for the $m=2$ modes considered in this work. 
	 The EOB complete waveforms are reported 
	 in dashed red, while numerical results in black. We also show the inspiral-plunge
	 amplitude without NQC corrections in the upper panels (orange) and $2 \Omega$ in the middle
	 panels. For each case, we report phase (azul) and relative amplitude differences (dashed orange).
	 The dashed vertical line marks the anchoring time $\to=\tLR-2.4$.}
	\end{center}
\end{figure*}
Having discussed the quasi-circular scenario, we now move to the eccentric case.
Since we are in the test-mass limit, the remnant is not affected by the dynamics, 
and therefore the \ac{qnm} content is left unchanged by eccentricity. However,
as  already highlighted in Sec.~\ref{subsec:num_waves}, eccentricity causes an increase
in the amplitude peak and the waveform frequency before merger, thus influencing 
the matching procedure between the inspiral-plunge and post-merger waveforms. 
Moreover, eccentric dynamics can significantly move the peak of the amplitude, depending 
on the relativistic anomaly at the separatrix crossing, $\xisep$, as extensively discussed 
Sec.~\ref{subsec:plunge}. However, at $\tA22$ the waveform is still strongly source-driven
and can therefore be modeled by the inspiral-plunge EOB waveform, which is completed with
the post-merger waveform in the proximity of the \ac{lr} crossing. Note that this last fact
was already relevant for quasi-circular cases with high spin.
Moreover, in the eccentric inspiral we can be observed \ac{qnm} excitations known 
as wiggles~\cite{Kojima:10.1143,Rifat:2019fkt,Thornburg:2019ukt,Albanesi:2021rby},
generated during close passages. This effect can be seen, for example, 
in the amplitude and frequency oscillations observable before the apastron passages
in Fig.~\ref{fig:a09_e05_tLR}. We do not provide a model for wiggles,
but their modelization and impact on the inspiral radiation reaction should be studied 
in future works. 

%------------------------------------------------------------------------
\subsubsection{$m>0$ modes}
%------------------------------------------------------------------------
\label{subsec:ecc_mpositive}
We start by considering the same Kerr spins and multipoles discussed in Sec.~\ref{subsec:comparisons_qc}
for quasi-circular systems, but we now consider $e_0=0.6$. Since the simulations are only
a few thousands $M$ long, the eccentricity at the separatrix crossing is still $\esep\simeq 0.6$. We report the 
(2,2), (3,2), and (4,2) multipoles for spins $\ha=\lbrace 0,\pm 0.7\rbrace$ and $\esep\simeq 0.6$ in 
Fig.~\ref{fig:comparisons_m2_e06}. The modes with $m=3$ and $m=1$ are shown in Figs.~\ref{fig:comparisons_m3_e06} 
and~\ref{fig:comparisons_m1_e06}, which we report at the end of the paper to not overload the main text. 
For each spin-configuration, we show two full orbits before the plunge-merger, in order to also showcase the performance
of $h^{\rm inspl}_\lm$ during the eccentric inspiral. We remind however, that a more careful discussion 
on the accuracy of the inspiral waveform (and radiation reaction) that we use here has been carried out in Ref.~\cite{Albanesi:2021rby},
and 2PN corrections have been also considered in subsequent 
works~\cite{Placidi:2021rkh,Albanesi:2022ywx,Albanesi:2022xge,Placidi:2023ofj}. 
Since here we focus on the plunge and post-merger waveform, we switch off these 2PN corrections, 
which are only relevant for the inspiral phase~\cite{Placidi:2021rkh}.

Regarding the post-merger waveform, we find that the accuracy is comparable to the
quasi-circular case and, in some configurations, even better.
In particular, in the Schwarzschild case the final phase differences of the $m=2$ modes
are slightly smaller than in the quasi-circular limit.
This can be understood by noting that, for eccentric configurations, the plunge
frequencies are higher (see also the discussion in Sec.~\ref{subsec:plunge}) but must
ultimately converge to the same final \ac{qnm} frequency.
As a consequence, the \ac{nqc} corrections required to enforce this convergence are smaller.
For the $\ha=0.7$ case, the phase accuracy of the $(2,2)$ mode is substantially improved,
with $|\Delta\phi_{22}|<0.01$ rad throughout the post-merger phase, whereas in the
quasi-circular case the phase difference oscillates around $0.07$ rad.
A mild improvement in phase accuracy is also observed for the $(3,2)$ and $(4,2)$ modes.
On the other hand, the post-merger amplitudes are slightly less accurate in the eccentric case.
The largest amplitude discrepancies, however, are visible during the inspiral
and plunge phases.
In particular, for the $(3,2)$ and $(4,2)$ modes, a noticeable bump appears in the analytical
waveform at the beginning of the final burst of radiation.
This feature is instead related to the factorization and resummation procedures employed
in the inspiral EOB waveform (see Sec.~\ref{subsec:inspiralwave}) and is therefore beyond
the scope of the present work.
For retrograde orbits with $\ha=-0.7$, the inspiral waveform is significantly more accurate
than the $\ha=0.7$ case. This is expected since the dynamics probes a less strong-field regime 
farther from the central black hole.
In this case, we observe a slight degradation of the phase accuracy around the \ac{lr}
crossing: for instance, in the quasi-circular configuration we find
$\Delta\phi_{22}\simeq -0.03$ rad, while for $\esep\simeq 0.6$ this increases modestly to
approximately $-0.04$ rad. The amplitude accuracy remains comparable, as does the modeling 
of the beating between co-rotating and counter-rotating \acp{qnm}.
This further confirms that tying the beating phase to $\td2omg0$ and performing fits
depending only on the spin is a robust and reliable modeling choice, as discussed
in Sec.~\ref{subsec:beating}.

\begin{figure*}[t]
	\begin{center}
	\includegraphics[width=0.32\textwidth]{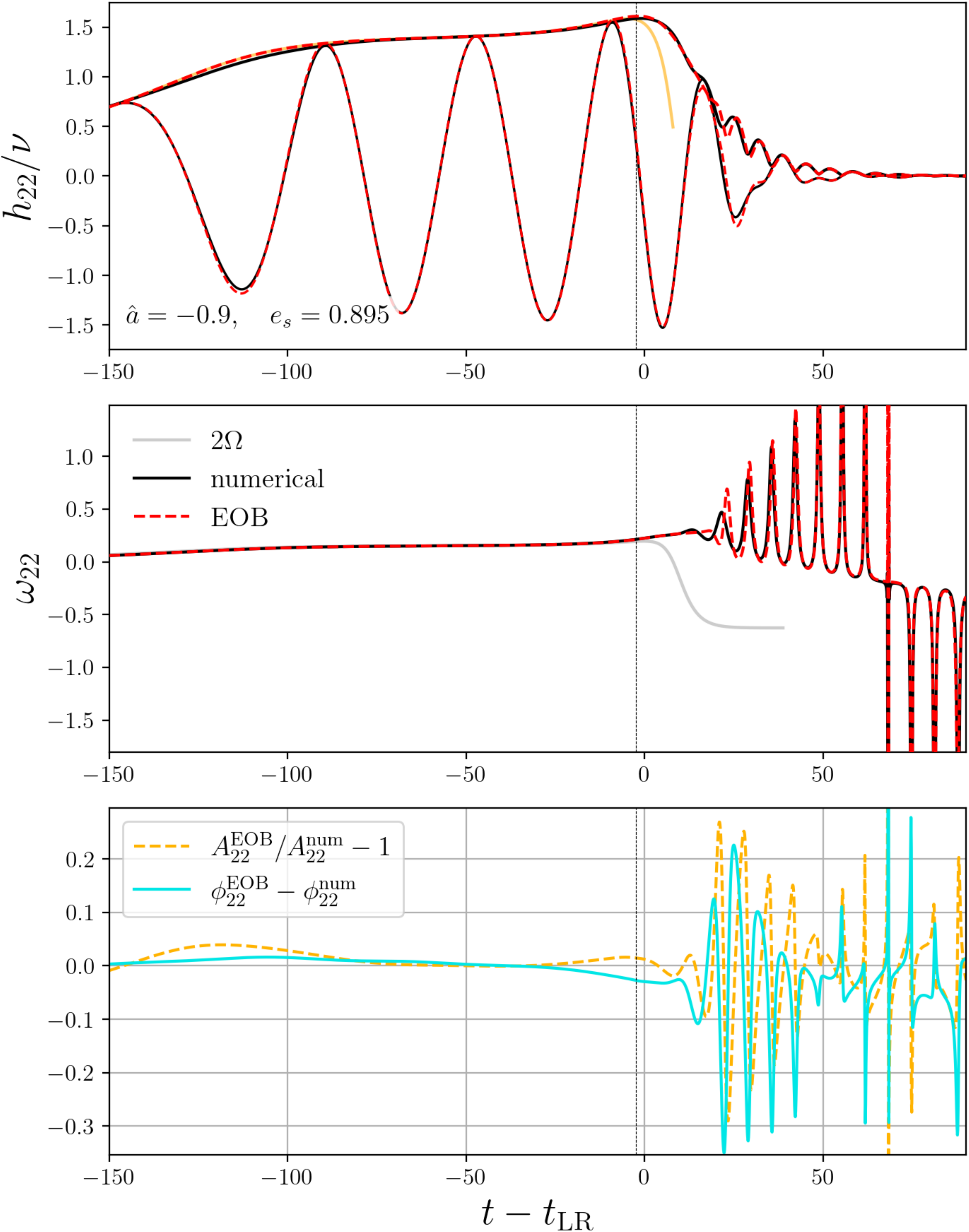}
	\includegraphics[width=0.32\textwidth]{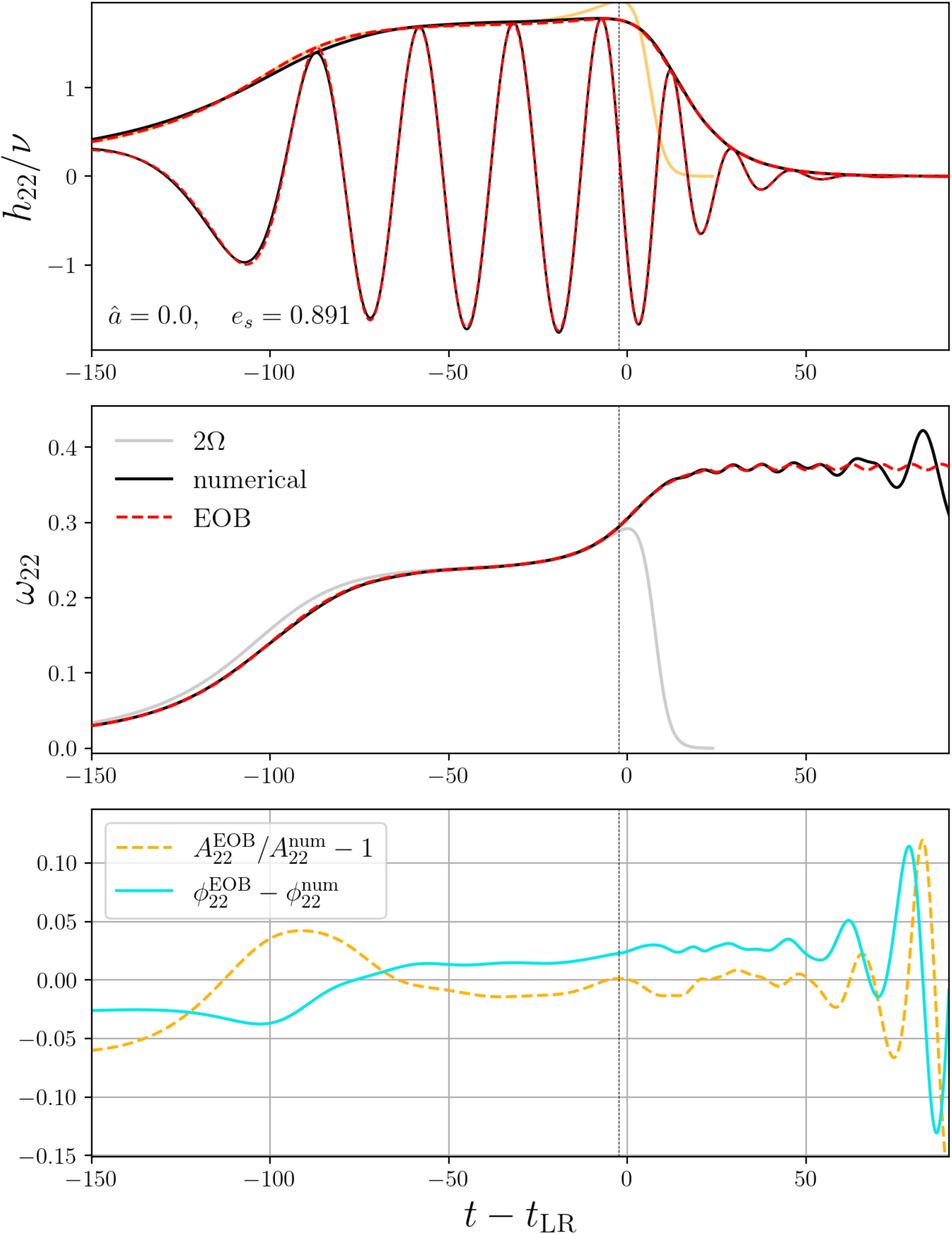}
	\includegraphics[width=0.32\textwidth]{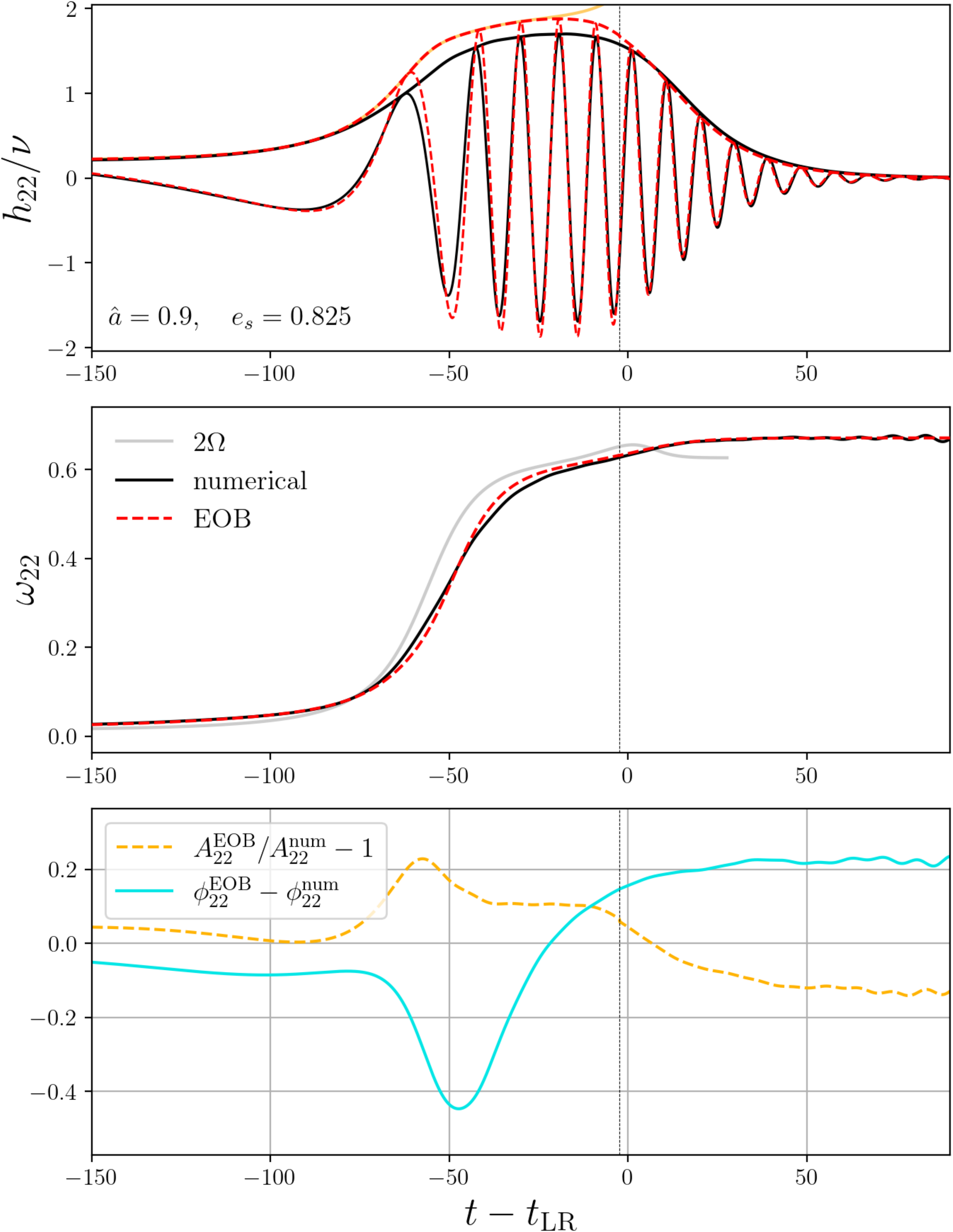}
	 \caption{\label{fig:a09_e09} 
	 Analytical/numerical comparisons of the (2,2) mode for the most eccentric elliptic-like configurations 
	 with $\ha=\lbrace0,\pm0.9\rbrace$ considered in this work.
	 The EOB complete waveforms are reported 
	 in dashed red, while numerical results in black. We also show the inspiral-plunge
	 amplitude without NQC corrections in the upper panels (orange) and $2 \Omega$ in the middle
	 panels. For each case, we report phase (azul) and relative amplitude differences (dashed orange).
	 The dashed vertical line marks the anchoring time $\to=\tLR-2.4$.}
	\end{center}
\end{figure*}

For these eccentric cases, we also show analytical/numerical comparisons for 
the $m=1$ and $m=3$ modes in Figs.~\ref{fig:comparisons_m1_e06} and~\ref{fig:comparisons_m3_e06}, respectively. 
The (4,1) mode is the less accurate multipole modeled in this work, but it is also
the less relevant in the \ac{gw} strain. Even in the numerical cases,
the $(4,1)$ mode is affected by numerical noise, especially for waveforms which correspond
to dynamics at large radii. However, the post-merger waveform is always well resolved,
and therefore we do not run these numerical simulations at higher resolutions.
Both the (3,1) and (4,1) modes have frequencies which strongly deviate from the
Newtonian quasi-circular contribution $\Omega$. These deviations, which are a 
consequence of the non-circularity of the motion, are qualitatively
well-reproduced by the prescription introduced in Ref.~\cite{Chiaramello:2020ehz} 
and discussed in Sec.~\ref{subsec:inspiralwave}, but there are still relevant quantitative differences.
The other modes, especially the ones with $\l=m$, have instead frequencies closer to 
$m \Omega$, as expected.
Overall, the post-merger waveform itself retains its accuracy for the most 
relevant modes for $\ha\leq 0.8$, with the exception of the (4,1) mode. 
It should be noted, however, that higher modes affected by spheroidal-spherical 
mode-mixing become less accurate at $\ha=0.9$. Moreover, if high spins are coupled with 
high eccentricity, the inspiral shows evident asymmetries in the waveform generated
at the periastron passages, which might be linked to both \ac{qnm} emission, as previously
mentioned, and to non-circular hereditary effects which are here not modeled here.
These asymmetries are already visible for the $\ha=0.7$, $\esep\simeq 0.6$ case 
considered in this discussion. 

In Fig.~\ref{fig:a09_e09} we finally report the analytical/numerical comparisons for the (2,2) mode in the cases with 
$\ha=\lbrace 0, \pm 0.9\rbrace$ and highest eccentricities considered in this work (except
for the dynamical captures, which we will discuss in Sec.~\ref{subsec:comparisons_hyp}). 
The discussion of these results is qualitatively similar to the previously discussed ones.
In the retrograde case with $\ha=-0.9$ and $\esep\simeq 0.895$ (left panel), the largest inaccuracy are obtained
around $10-20\,M$ after $\tLR$, where our beating modeling is less accurate since it only takes into account
the fundamental frequencies. The beating description becomes however more reliable at later times.
For the prograde orbit with $\ha=0.9$ and $\esep\simeq0.825$ (right panel), a consistent
discrepancy is observed in the pre-merger waveform, due to the inaccuracy of the analytical inspiral/plunge
waveforms, but the post-merger waveform agrees remarkably well with the numerical data, especially considering that
this is the most ``extreme" configuration for the elliptic-like cases considered in this work.
The Schwarzschild case with $\esep\simeq 0.891$ (middle panel of Fig.~\ref{fig:a09_e09}) 
is the one with the best overall agreement, as expected. In the numerical 
waveform frequency of this Schwarzschild case it is also evident that the tail begins to be  
relevant already at $\sim 50\,M$ after the \ac{lr} crossing.

%------------------------------------------------------------------------
\subsubsection{(2,0) modes}
%------------------------------------------------------------------------
\label{subsec:l2m0}
\begin{figure}[t]
	\begin{center}
	\includegraphics[width=0.48\textwidth]{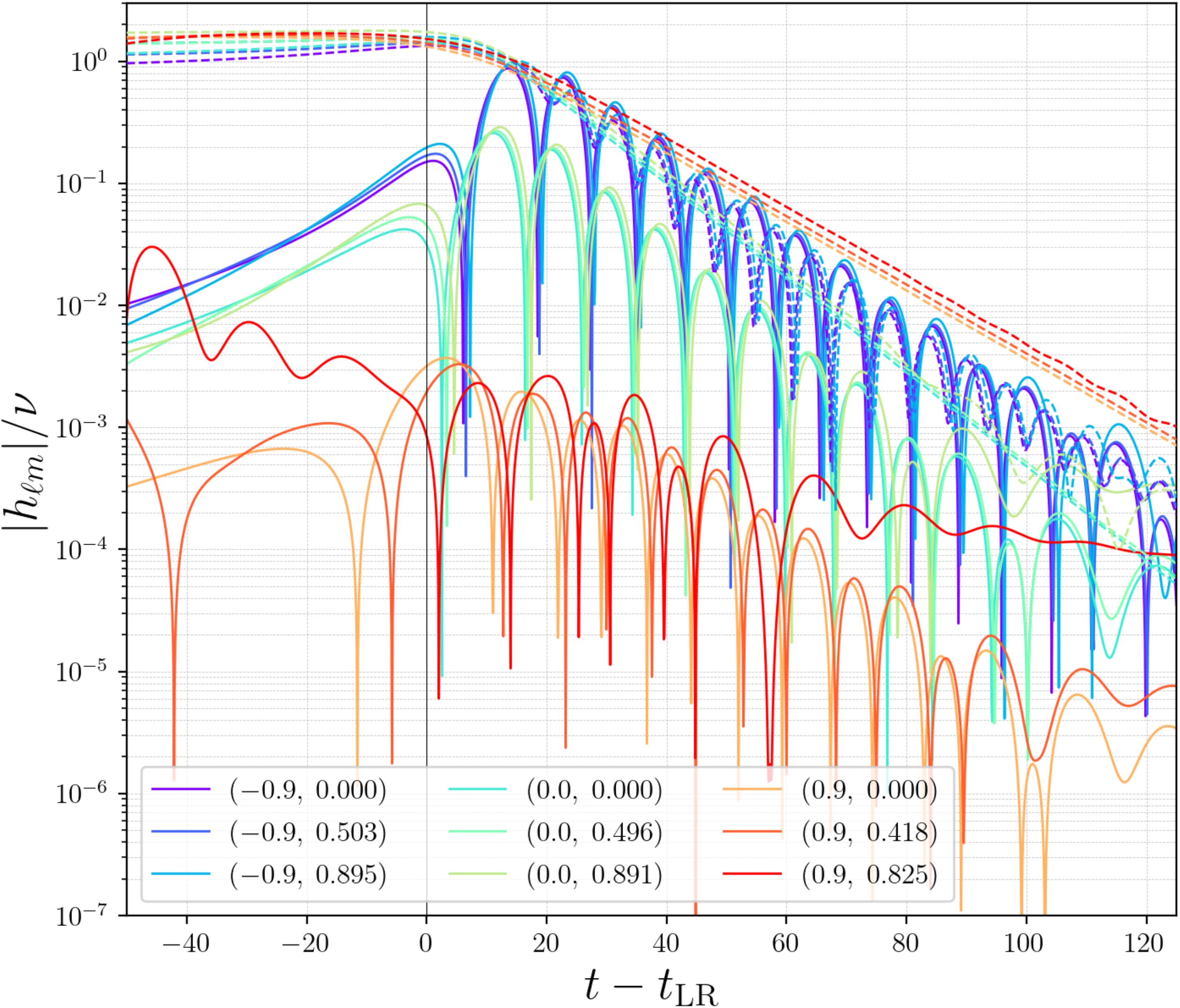}
	 \caption{\label{fig:teuk_l2m0} Numerical multipoles for different $(\ha,\esep)$ configurations,
	 as reported in the legend. We show the (2,2) and (2,0) modes with dashed and solid lines.}
	\end{center}
\end{figure}
\begin{figure*}[t]
	\begin{center}
	\includegraphics[width=0.32\textwidth]{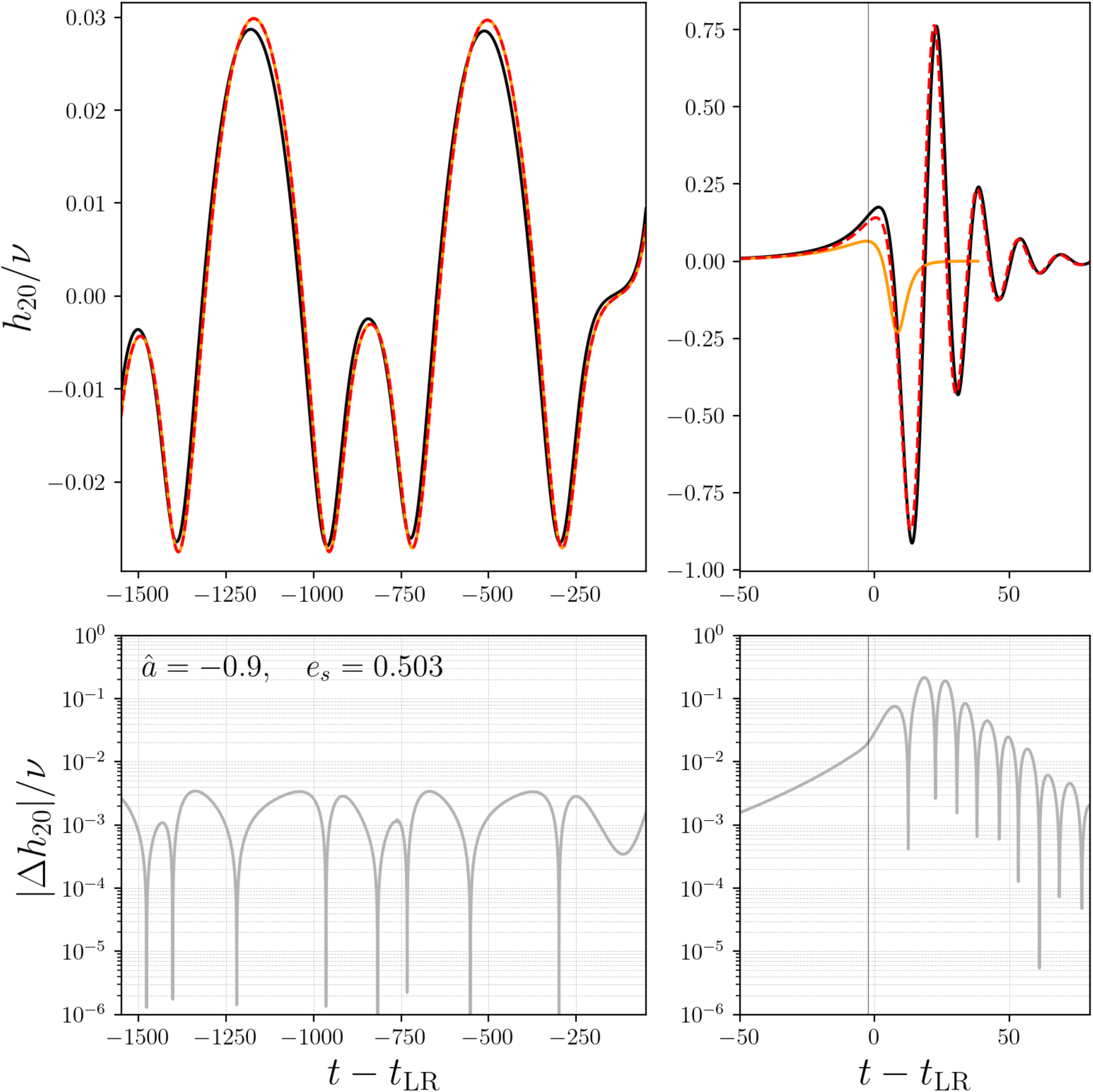}
	\includegraphics[width=0.32\textwidth]{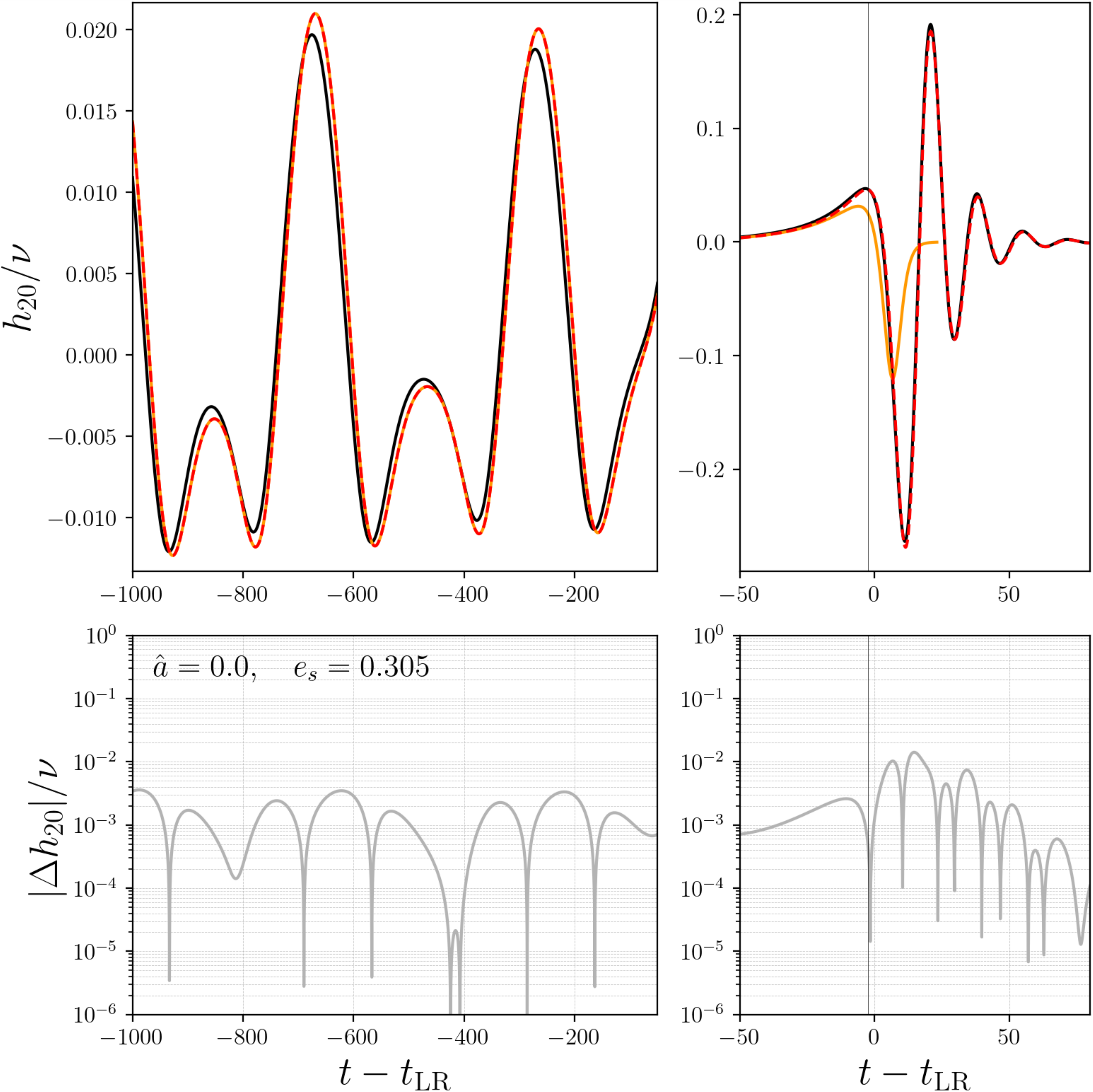}
	\includegraphics[width=0.32\textwidth]{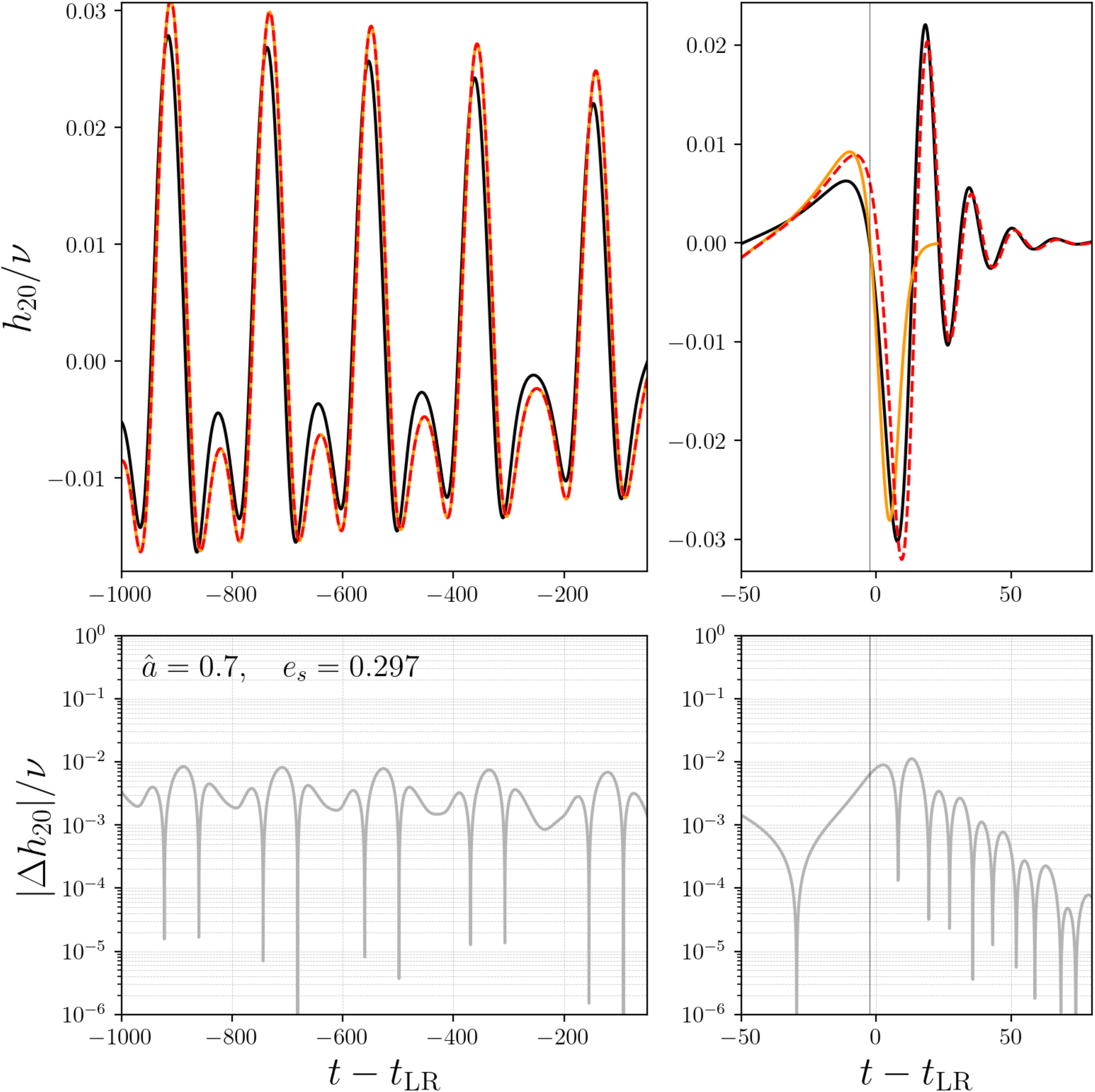}
	 \caption{\label{fig:comparisons_l2m0} Analytical/numerical comparisons for the (2,0) mode
	 for three different $(\ha,\esep)$ configurations. In the upper panel we report the numerical 
	 multipoles (black), the inspiral waveforms given by Eq.~\eqref{eq:h20_inspl} (solid orange), 
	 and the complete EOB waveforms (dashed red). In the bottom panels we report the 
	 residuals $|\Delta h_{20}|\equiv |h_{20}^{\rm Teuk} - h_{20}^{\rm EOB}|$ 
	 in absolute value. Logarithmic vertical scale.
	 The dashed vertical line marks the anchoring time $\to=\tLR-2.4$.}
	\end{center}
\end{figure*}
We now discuss the $(2,0)$ mode, which is treated differently from the $m>1$ 
multipoles due to its intrinsically real (rather than complex) nature. However, it is convenient 
to work with complex quantities, so that amplitude and phase are well defined and remain 
non-oscillatory during the plunge-merger-ringdown evolution.
We therefore follow the procedure introduced in Ref.~\cite{Albanesi:2024fts}, 
which consists of complexifying the numerical waveform with a Hilbert transform before performing the 
primary fits. The coefficients obtained in this way are 
then interpolated across the parameter space, as explained in Sec.~\ref{subsec:global}, 
and the complex post-merger model is attached to a Hilbert-complexified version of the 
real inspiral-plunge waveform. For the latter, we consider the Newtonian expression~\cite{Chiaramello:2020ehz}
\be
\label{eq:h20_inspl}
h_{20}^{\rm inspl} = 4 \sqrt{\frac{2 \pi}{15}} \nu \left(r \ddot{r} + \dot{r}^2\right).
\ee
The physical $(2,0)$ mode is then obtained by taking the real part of the complexified \ac{imr} signal.

Since the $m=0$ modes are purely radial, they are negligible during quasi-circular inspirals 
but become relevant during the plunge, merger, and ringdown, especially for retrograde orbits.
This occurs because the radius of the plunge starts at larger radii for retrograde orbits than for prograde 
ones. 
This spin-hierarchy is explicitly shown in Fig.~\ref{fig:teuk_l2m0}, where we report
configurations with spins $\ha=\lbrace0,\pm0.9\rbrace$, each of them with 
eccentricities $\esep\simeq\lbrace0,0.5,0.9\rbrace$.
For large negative spins, the contribution of the $(2,0)$ mode becomes comparable to that 
of the $(2,2)$ mode. For instance, in the quasi-circular case with $\ha=-0.9$, 
we find $A_{22}^{\rm peak}/\nu\simeq1.34$, while the peak of the $(2,0)$ mode reaches 
$h_{20}^{\rm peak}/\nu\simeq -0.87$. The opposite occurs for large positive spins and 
prograde orbits: for $\ha=0.9$, we obtain $A_{22}^{\rm peak}/\nu\simeq1.34$ but 
$h_{20}^{\rm peak}/\nu\simeq -0.004$.

However, when eccentric configurations are considered, radial variations become important already during the inspiral, and the $(2,0)$ 
mode provides a non-negligible contribution also during the early dynamics. For prograde orbits with 
high spins and high eccentricities, the inspiral can be more relevant than the merger-ringdown
portion. On the other hand, the eccentricity has a comparable small effect on the amplitude 
of the merger-ringdown signal, as shown in Fig.~\ref{fig:teuk_l2m0}.

The quasi-circular case has already been studied in Ref.~\cite{Albanesi:2024fts}, and more 
recently Ref.~\cite{Nishimura:2026nse} incorporated this modeling within their framework, 
also focusing on quasi-circular test-mass binaries. Here we thus focus on the eccentric case.
We begin by noting that Eq.~\eqref{eq:h20_inspl} represents only the Newtonian contribution, 
and therefore becomes progressively less accurate 
for high spins and high eccentricities. Indeed, even for moderate eccentricities combined 
with mild-to-high spins, the analytical/numerical agreement during the inspiral deteriorates 
significantly, as illustrated in the right panel of Fig.~\ref{fig:comparisons_l2m0}, where we 
consider $\ha=0.7$ and $\esep\simeq0.3$. A significant worse agreement is obtained for larger spins 
and/or larger eccentricities. In particular, while the same eccentricity in Schwarzschild
yields an acceptable agreement (middle panel of Fig.~\ref{fig:comparisons_l2m0}), eccentricities 
$\gtrsim0.6$ lead to significant phase and amplitude discrepancies that render the model 
unreliable even in the non-spinning case for such high eccentricities. 
However, this limitation is linked to the low accuracy of the simple inspiral-plunge 
waveform employed here, rather than to the post-merger modeling itself. Indeed, we find that the 
primary-fit procedure used for the post-merger signal remains applicable for all the configurations 
considered in this work.
Finally, in the left panel of Fig.~\ref{fig:comparisons_l2m0} we show a configuration with $\esep=0.5$ 
and $\ha=-0.9$. A slight dephasing is visible in the merger-ringdown waveform, although it becomes 
significantly smaller for configurations with $\ha=-0.8$ and similar or lower eccentricities. For 
instance, while in the $\ha=-0.9$ case of Fig.~\ref{fig:comparisons_l2m0} we find a maximum residual 
$|\Delta h_{20}|/\nu\sim0.21$, for the configuration with $\ha=-0.8$ and $\esep\simeq0.6$ the 
residual decreases to $|\Delta h_{20}|/\nu\sim0.05$.

We conclude by remarking that, although the post-merger approach of Ref.~\cite{Albanesi:2024fts} 
extends naturally to eccentric configurations within the framework presented here, the construction 
of a fully reliable \ac{imr} $(2,0)$ mode ultimately requires a more robust analytical prescription 
for the inspiral-plunge waveform. This is particularly important for high positive spins and high 
eccentricities. As mentioned throughout the paper, improving the accuracy of the multipolar inspiral waveform 
(and correspondingly of the fluxes) will be the focus of future work.

%------------------------------------------------------------------------
\subsection{Dynamical captures}
%------------------------------------------------------------------------
\label{subsec:comparisons_hyp}
\begin{figure*}[t]
	\begin{center}
	\hspace{0.6cm}
	\includegraphics[width=0.2\textwidth]{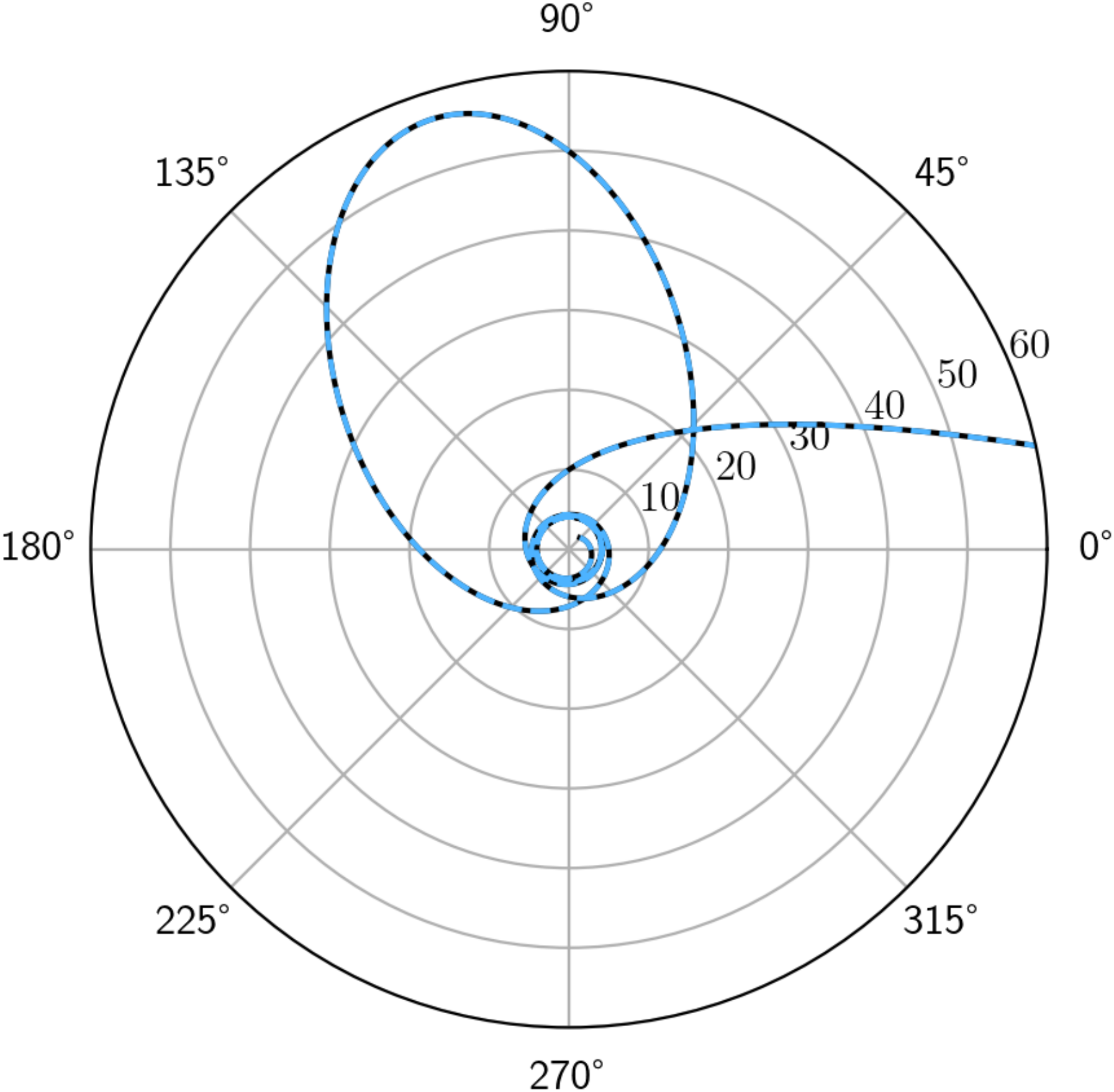}
	\hspace{2.0cm}
	\includegraphics[width=0.2\textwidth]{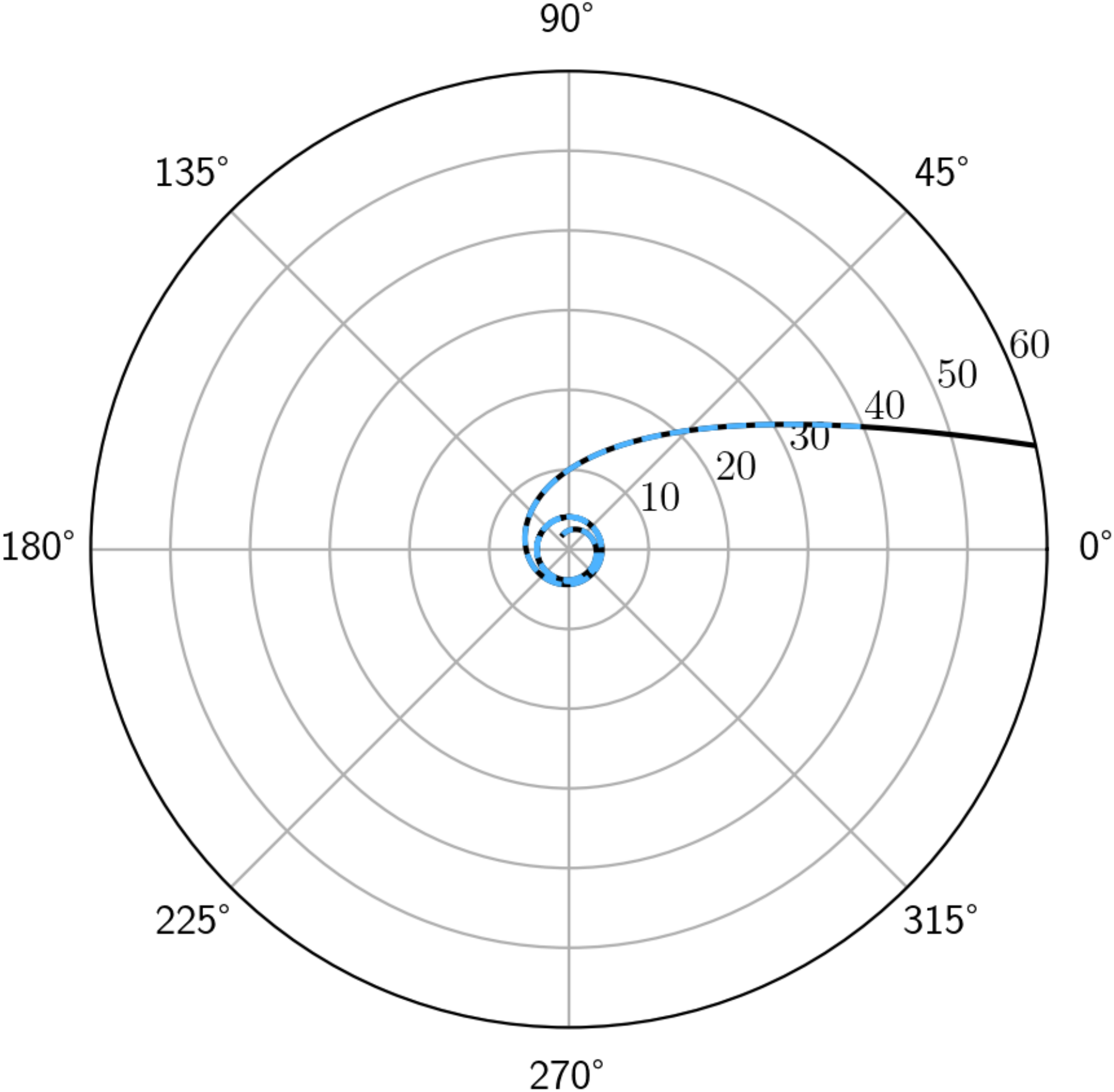}
	\hspace{2.0cm}
	\includegraphics[width=0.2\textwidth]{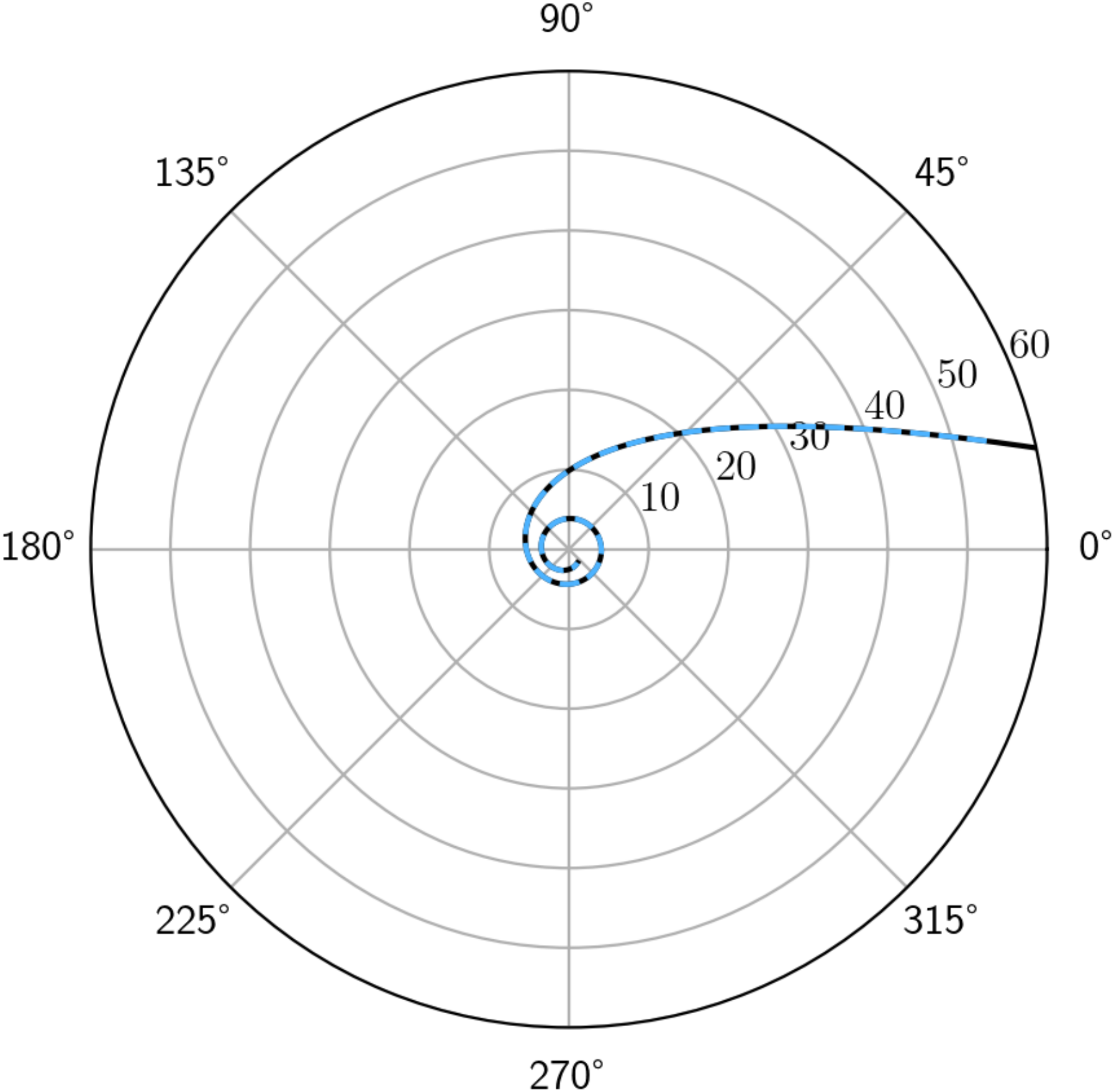}
	\hspace{0.6cm}\\
	\includegraphics[width=0.320\textwidth]{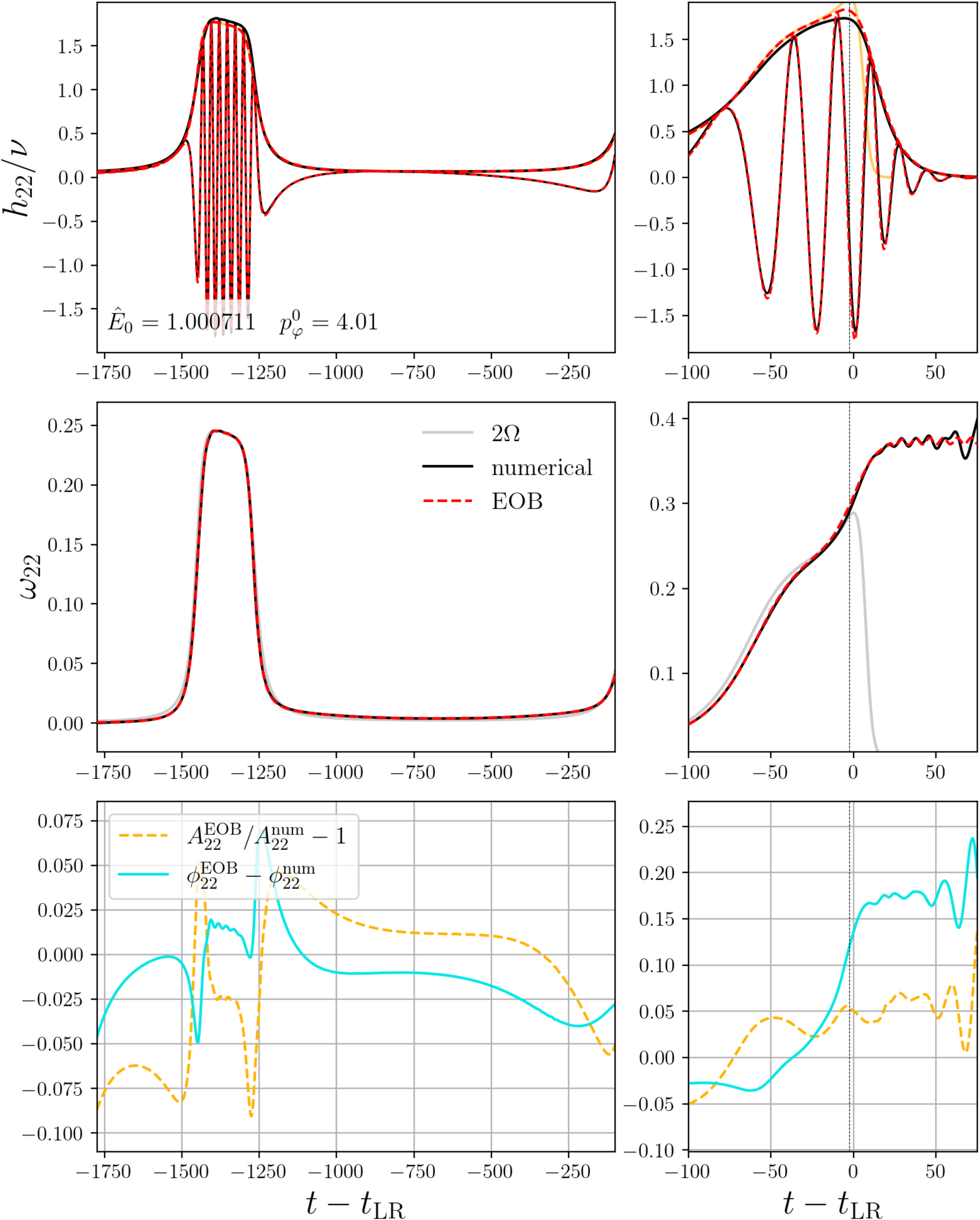}
	\includegraphics[width=0.309\textwidth]{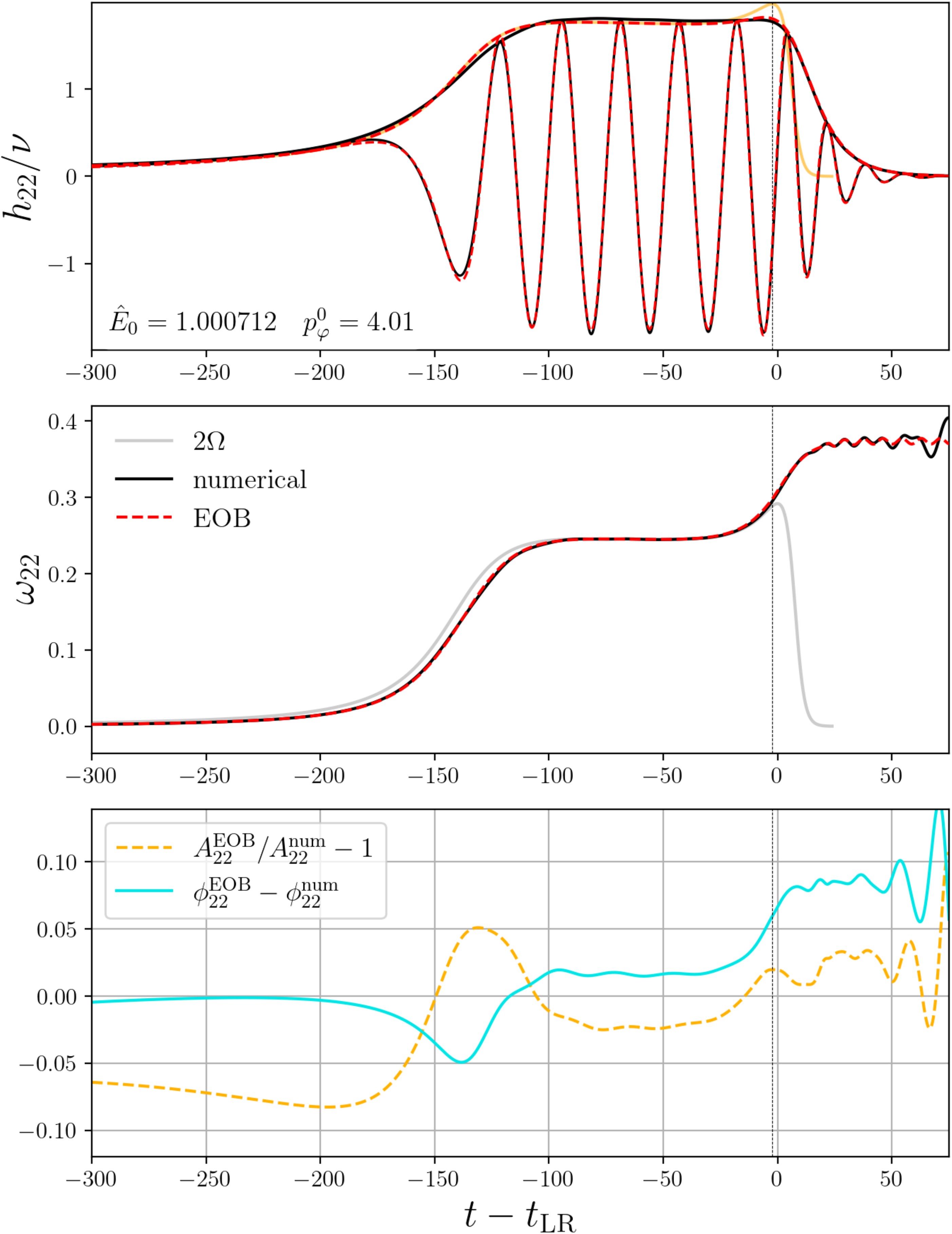}
	\includegraphics[width=0.310\textwidth]{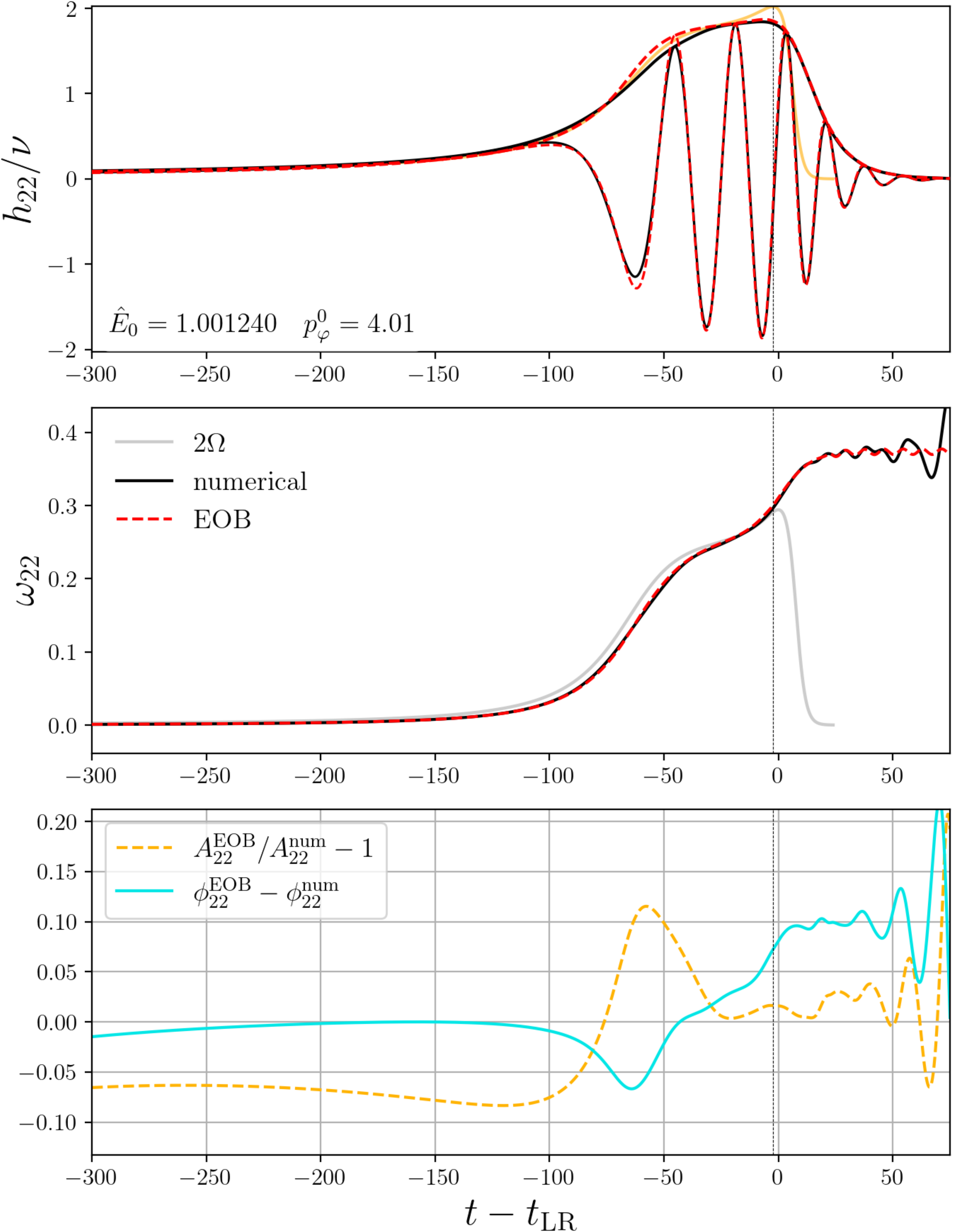}
	 \caption{\label{fig:comparisons_hyp_schw} Dynamical captures in Schwarzschild considered in Ref.~\cite{Albanesi:2021rby}.
	 Initial energies and angular momenta are reported in the panels. All systems have $\nu=10^{-2}$.
	 We also report the trajectory in black, highlighting with blue dashed lines the portions
	 which correspond to the waveforms of the bottom panels.
	 The numerical (2,2) modes are shown in black, while the EOB waveform and its frequency in dashed red. 
	 We also report the amplitude of the inspiral-plunge waveform without \ac{nqc} with solid orange lines in the upper panels.
	 In the bottom panels we show phase differences (azul) and relative amplitude differences (dashed orange). 
	 The dashed vertical line marks the anchoring time $\to=\tLR-2.4$.}
	\end{center} 
\end{figure*}
We now discuss the accuracy of the model in dynamical capture scenario. Indeed,
due to the presence of radiative effects, systems which are initially unbound 
($\hat{E}\geq 1$) can become bound ($\hat{E}<1$) due to the \ac{gw} emission. 
Such systems are more astrophysically relevant for comparable mass binaries,
where the impact of the radiation reaction is stronger and could thus lead
to captures in dense astrophysical environments~\cite{Samsing:2013kua,Rodriguez:2016kxx,Belczynski:2016obo,Samsing:2017xmd,
Rodriguez:2018pss,Mukherjee:2020hnm,DallAmico:2021umv,DallAmico:2023neb}. For example, this scenario 
might have been at the origin of GW190521~\cite{LIGOScientific:2020iuh,Gamba:2021gap}, and for this reason
previous works were devoted to the study of these configurations~\cite{Nagar:2020xsk,Andrade:2023trh,Albanesi:2024xus}.

In Ref.~\cite{Albanesi:2021rby} we also studied three captures in the nonspinning test-mass limit
with symmetric mass ratio $\nu=10^{-2}$.
In the same work, we also presented a rudimental ringdown waveform model with anchoring point $\tAlm$,
and tried to apply it to these dynamical captures for the dominant (2,2) mode. 
The model failed to reproduce the numerical data
(grey lines in Fig.~14 therein), but could be cured by inserting in the ringdown model  
values extracted directly from the numerical data: numerical values of amplitude and frequency at $\tA22$,
the value of $\tA22$ itself, and the primary coefficients. 
These ad-hoc prescriptions were shown with red lines in Fig.~14 of Ref.~\cite{Albanesi:2021rby}.
However, a robust model should be able to describe these scenarios without any additional ad-hoc
information. We now argue that the model introduced in this paper is able to do so, without including 
any additional information or any additional modification. 
Indeed, even if we consider orbits which are initially unbounded, if there is a merger 
there must be a time in which both the apastron and periastron are defined, 
and we therefore can use our definitions~\eqref{eq:ep_definition} to compute $\esep$ 
and thus evaluate our ringdown model. 

\begin{figure*}[t]
	\begin{center}
	\hspace{0.6cm}
	\includegraphics[width=0.2\textwidth]{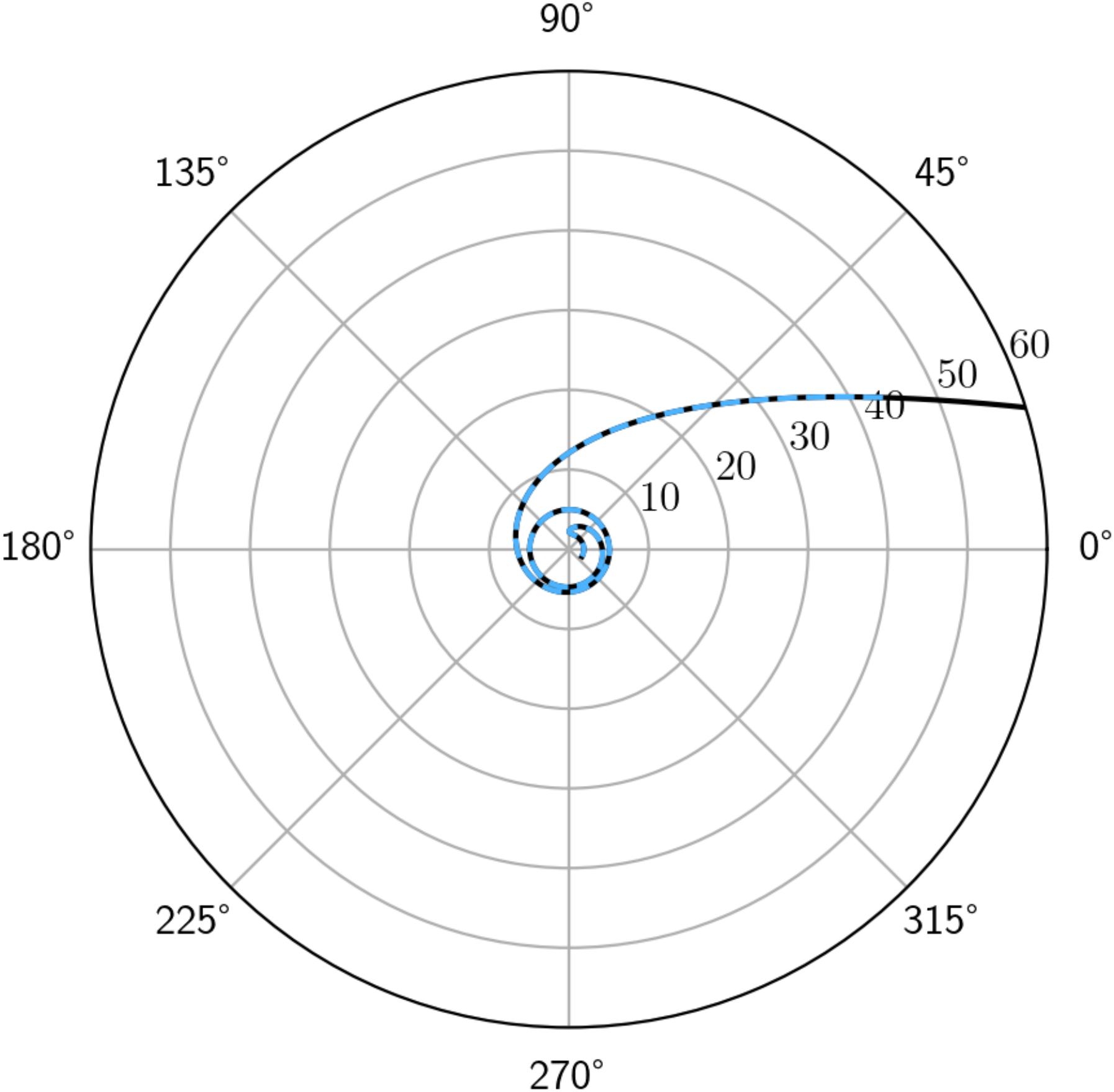}
	\hspace{2.0cm}
	\includegraphics[width=0.2\textwidth]{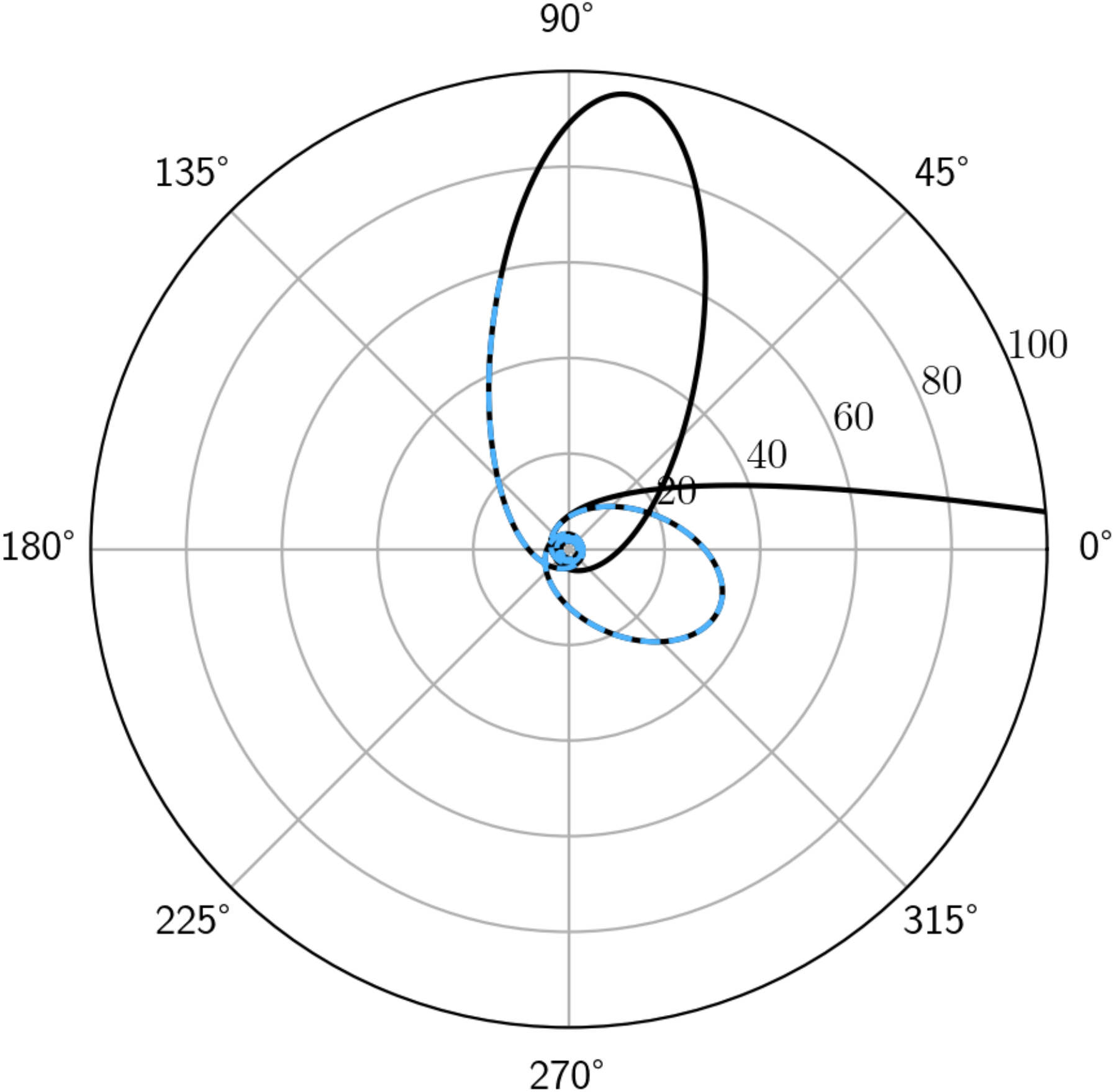}
	\hspace{2.0cm}
	\includegraphics[width=0.2\textwidth]{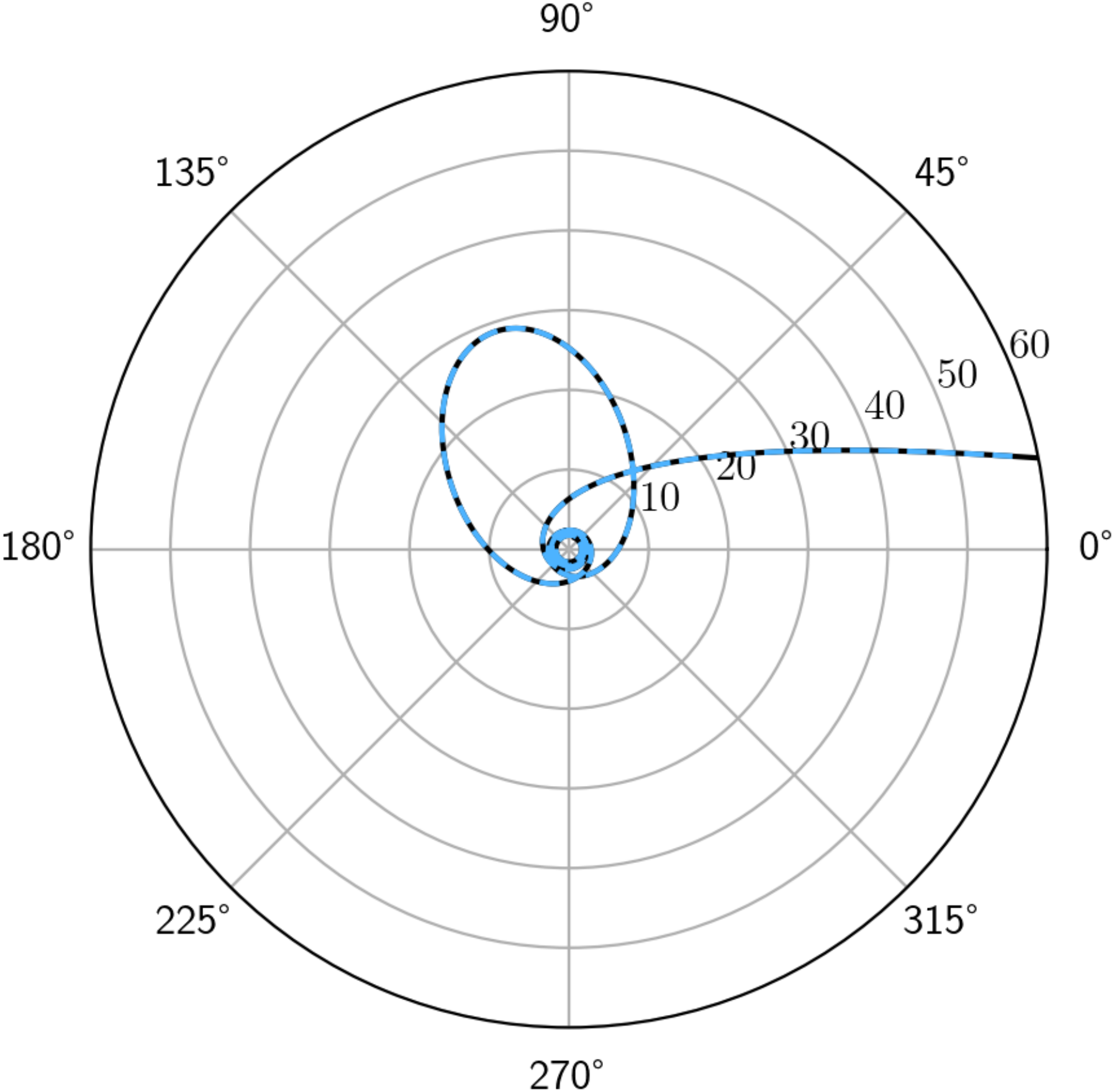}
	\hspace{0.6cm}\\
	\includegraphics[width=0.310\textwidth]{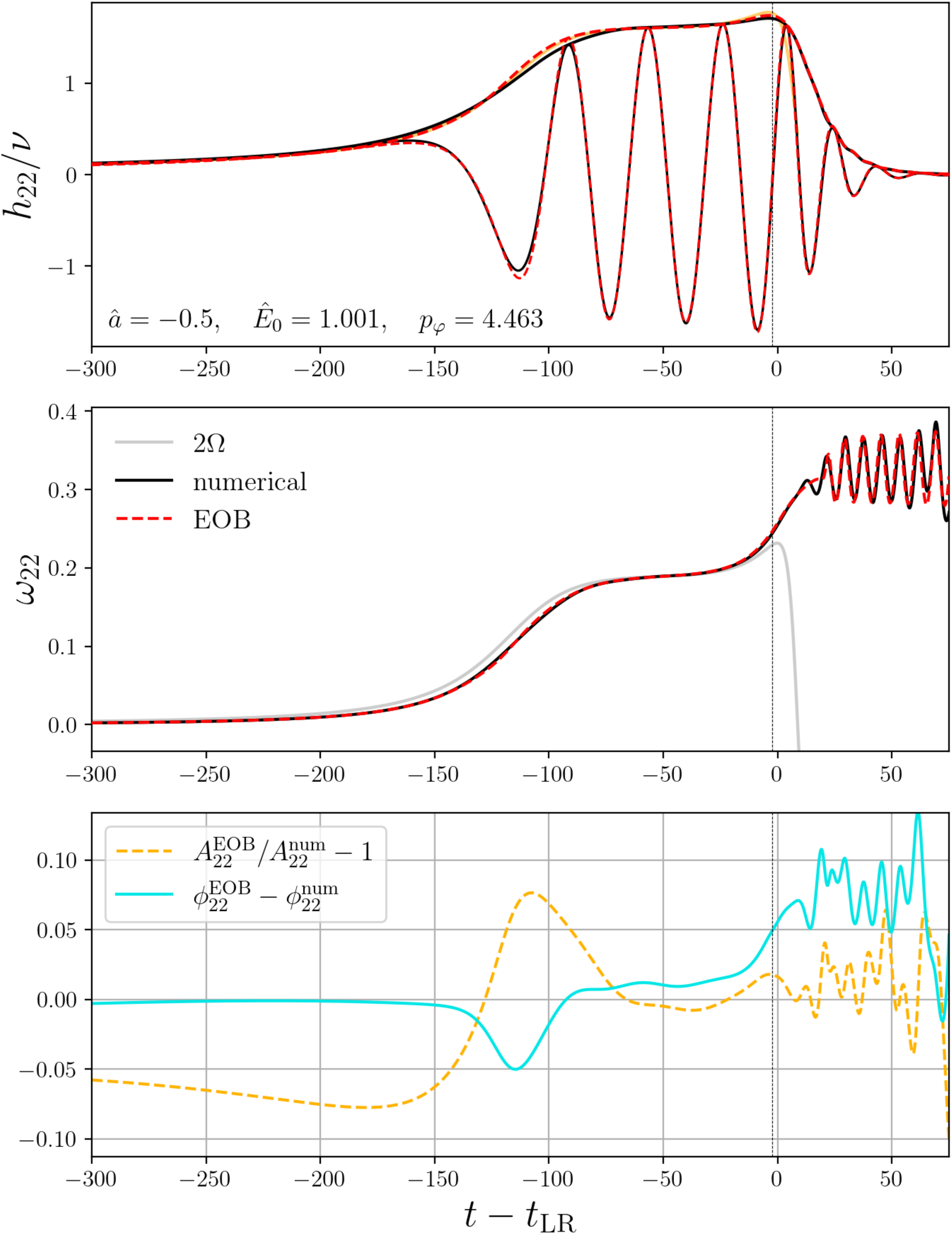}
	\includegraphics[width=0.320\textwidth]{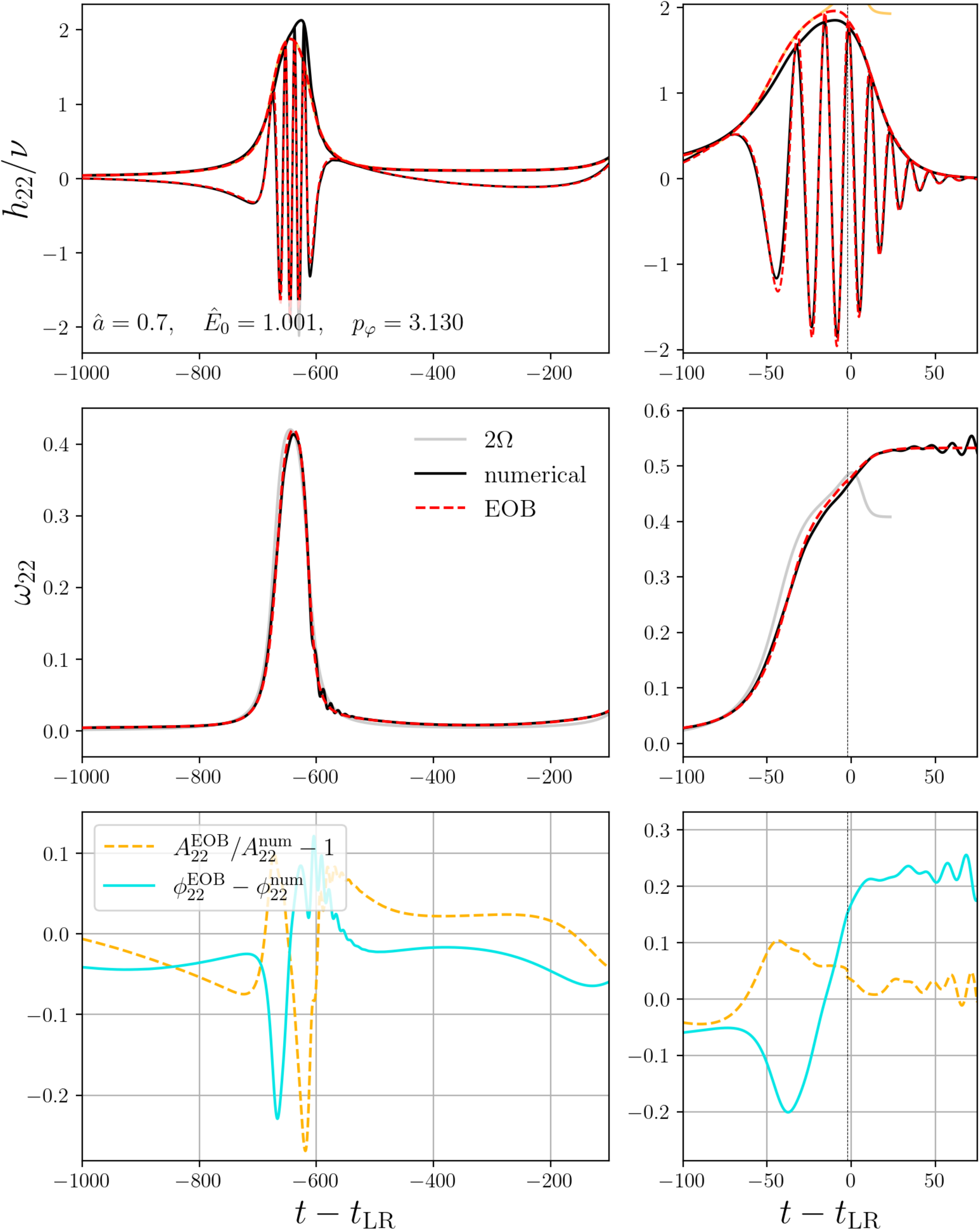}
	\includegraphics[width=0.320\textwidth]{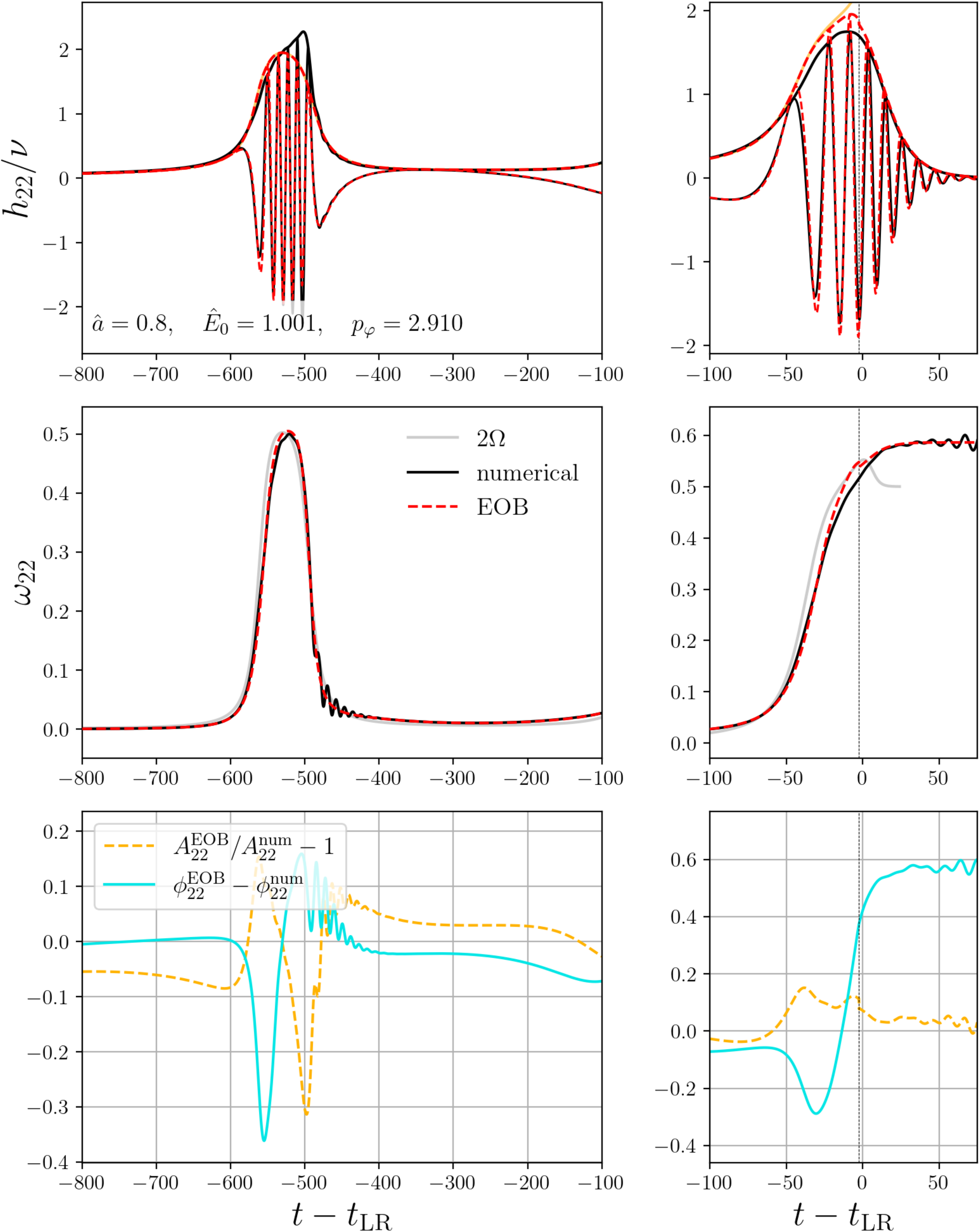}
	 \caption{\label{fig:comparisons_hyp_kerr} 
	 Dynamical captures for three configurations in Kerr spacetime.
	 Initial energies and angular momenta are reported in the panels. All systems have $\nu=10^{-2}$.
	 We also report the trajectory in black, highlighting with blue dashed lines the portions
	 which correspond to the waveforms of the bottom panels.
	 The numerical (2,2) modes are shown in black, while the EOB waveform and its frequency in dashed red. 
	 We also report the amplitude of the inspiral-plunge waveform without \ac{nqc} with solid orange lines in the upper panels.
	 In the bottom panels we show phase differences (azul) and relative amplitude differences (dashed orange).
	 The dashed vertical line marks the anchoring time $\to=\tLR-2.4$.
	 }
	\end{center} 
\end{figure*}
The results for the same physical configurations considered in Ref.~\cite{Albanesi:2021rby} 
are shown in Fig.~\ref{fig:comparisons_hyp_schw}. We report amplitudes, real parts and frequencies
for the (2,2) multipole as usual, but we also report the trajectories in the upper panels, where
the dashed blue lines highlight the portions that correspond to the waveforms shown below.
For these systems, we fix the initial
angular momentum to $p_\varphi^0=4.01$ and we considered different values of energy 
$\hat{E}_0 = \lbrace 1.000711, 1.000712, 1.001240\rbrace$. For the angular momentum considered, the peak
of the radial effective potential reaches $V_{\rm max}^0=1.001250$. Therefore, if no radiation
reaction is considered, all these configurations are scatterings. However,
once dissipative effects are included in the dynamics, they result in captures, and we
thus have merging systems. 
In the leftmost panel we consider the case with $\hat{E}_0=1.000711$, which corresponds to a double encounter, i.e. 
the particle gets captured at the first close encounter, completes an elliptic-like orbit afterwards,
and then finally plunges. In this case, the phase difference for the (2,2) mode reaches
$0.18$ rad, while the post-merger amplitude stays around $5-6\%$. After $50 M$ from the \ac{lr} crossing,
evident oscillations in the phase/amplitude difference appear. 
These oscillations are due to the tail starting to dominate the numerical signal.
Tail effects are also evident in the (2,2) frequency, where the oscillations related to the beating
between co-rotating and counter-rotating \acp{qnm} leave place to growing oscillations
at $t\sim \tLR + 50$, which are instead linked to the \ac{qnm}-tail transition regime. 
A clean tail regime without evident \ac{qnm} contamination starts at approximately $170 M$ after
$\tLR$, whit an initial amplitude of $A_{22}/\nu\simeq 1.7\cdot 10^{-4}$. For
the other two configurations the tail is even more enhanced, as expected. For comparison, 
at $170 M$ we have $A_{22}/\nu\simeq 2.3\cdot 10^{-4}$ for $\hat{E}_0 = 1.000711$
and $A_{22}/\nu\simeq 3.6\cdot 10^{-4}$ for $\hat{E}_0 = 1.001240$.
Regarding the phase differences $\Delta \phi_{22}$ in the stationary \ac{qnm} regime after merger, 
for the other two configurations of Fig.~\ref{fig:comparisons_hyp_schw}, which are direct captures, we get 
$0.08$ and $0.10$ rad. Note that the best phase agreement is obtained for the configuration in the middle
panel, which corresponds to a case with a circular whirl before the plunge. 
For these two direct captures, the analytical/numerical relative amplitude differences 
in the post-merger reach the few percents, but it is a bit higher ($\sim 6-7\%)$ for the double 
encounter in the leftmost panel. However, the amplitude before the plunge is less accurate for the most
direct capture (rightmost panel), where it reaches $11\%$, while for the other two cases
it is around $5\%$. 

We now briefly discuss the Kerr case. We do not perform a systematic exploration of the parameter space, 
but we rather select a few configurations and we discuss the capability and limitations of
our model. In Fig.~\ref{fig:comparisons_hyp_kerr} we report three configurations with $\nu=10^{-2}$ and 
same initial energy $\hat{E}_0=1.0010$, but
different spins, $\ha=\lbrace-0.5,0.7,0.8\rbrace$, and initial angular momenta (reported in the panels). 
For the retrograde orbits reported in the leftmost panel, we consider $p_\varphi^0=4.463$, which yields
$V_{\rm max}^0=1.0013$. Therefore, also this configuration would be a scattering in the fully conservative case. 
The model retains its accuracy for this case, having a final phase difference of approximately
$0.1$ rad, and post-merger amplitude accurate to the $\sim 5\%$ level. However, also in this case
we observe a larger amplitude discrepancy before the burst of radiation at $t\simeq \tLR-120$, which
yields $\delta A/A \simeq 7.5 \%$. In this case the tail signal is even more enhanced, overcoming the
\ac{qnm} regime at $t\sim \tLR+150$ with $A_{22}/\nu \simeq 3.0\cdot 10^{-4}$. The beating between
co-rotating and counter-rotating \acp{qnm} before
the kick-in of the tail is well reproduced by our model, as can be seen from the plot of the frequency. 

In the middle panel we consider a configuration with $\ha=0.7$ and $p_\varphi^0=3.13$. In this case we obtain 
a triple encounter, which has two elliptic-like orbits before the plunge. In Fig.~\ref{fig:comparisons_hyp_kerr} 
we show only the two bursts of radiation linked to the second encounter and to the merger. As already observed 
for elliptic orbits in Sec.~\ref{subsec:comparisons_ecc}, we have bursts of \ac{qnm} radiation after
the periastron passages. This yields a strong asymmetry in the \ac{gw} burst, and to visible oscillations
in the frequency. For this configuration the phase agreement is lower, having $\Delta \phi_{22}\simeq0.2$ rad 
during the post-merger (and also before the \ac{lr} crossing). The behaviour of the amplitude is
similar to what has been observed before: a $\sim 10\%$ inaccuracy in the first part of the burst, but 
the accuracy improves to a few percents during the post-merger phase. 
We note that the phase agreement decreases for more direct captures with the same 
spin, as previously discussed for the Schwarzschild case. 
For example, if we consider $p_\varphi^0=3.10$ and $3.11$, we get a double encounter
with $\Delta \phi_{22}\simeq 0.33$ rad and a direct capture with $\Delta \phi_{22}\simeq 0.25$ rad. 
If we further decrease the initial angular momentum to $p_\varphi^0=3.0$, we get a very large 
dephasing, which reaches $1.6$ rad. 

Finally, when considering the last configuration  
of Fig.~\ref{fig:comparisons_hyp_kerr} with $\ha=0.8$ and $p_\varphi^0=2.91$ (right panels), the situation remains
qualitatively similar but with a further decrease in the phase accuracy, which now reach $0.6$ rad. The 
\ac{qnm} bursts during the periastron passages are also now even more relevant, as expected.
For $p_\varphi^0=\lbrace 2.93,2.95,3.00\rbrace$ we get configurations
with multiple close encounters before merger, with phase  differences in the range $0.30-0.45$ rad.
If we instead decrease the angular momentum to $p_\varphi^0=2.90$, we still get a 
final phase difference of approximately $0.45$ rad, but we also observe unphysical ``bumps" 
in the analytical amplitude and frequency  $50 M$ before the \ac{lr} crossing.
We also explored some configurations with $\ha=0.9$, finding physical waveforms for 
captures with multiple encounters, but unphysical ones for some direct captures. 

We thus conclude that the waveform model presented in this work 
applies in the dynamical capture scenario for
spins $\ha\lesssim 0.7$ and captures which have multiple encounters or that which are not ``too direct".
To better quantify this last statement, we require $\hat{E}_0 < V^0_{\rm max}$. 
Note that the condition $1\leq\hat{E}_0\leq V^0_{\rm max}$
corresponds to unbound geodesics, and therefore in our case they are captures driven by the radiation
reaction. On the other hand, $\hat{E}_0>V^0_{\rm max}>1$ corresponds to configurations which would plunge  
even in the conservative case. 
We also mention that using an anchoring point based on the inflection point of the frequency 
as discussed in Appendix~\ref{app:d2omg0} results in lower phase differences,
as shown in Fig.~\ref{fig:different_anchor_d2omg}. However, this prescription yields a worse
amplitude agreement in the cases with high spins. 
Finally, we remark that we are working with a setup in which the dynamics is used both
for the numerical and analytical waveforms, and therefore inaccuracies linked to the radiation reaction are
not taken into account. We are thus only testing the prescription for the waveform, especially for the plunge-merger-ringdown 
phase. This is particularly relevant for captures, since in general small inaccuracies in the radiation reaction can
strongly influence the dynamics and thus the number of encounters.  

%=======================================================================================================
\section{Conclusions}
\label{sec:conclusions}
%=======================================================================================================
In this paper we studied the plunge, merger, and ringdown of a non-spinning particle falling into a Kerr black 
hole after an eccentric inspiral, thereby generalizing and improving the Schwarzschild analysis presented in 
Ref.~\cite{Albanesi:2023bgi}. The dynamics were obtained by solving Hamilton’s equations including an \ac{eob} 
radiation reaction at leading order in the symmetric mass ratio $\nu$.
The corresponding waveforms were computed by numerically solving the Teukolsky equation~\cite{Teukolsky:1973ha} 
using the 2+1 time-domain code \Teuk{}~\cite{Harms:2014dqa}.

We first investigated several phenomenological aspects of the plunge and merger.
In particular, we showed explicitly that the relativistic anomaly at the separatrix crossing $\xisep$, 
i.e. the transition from stable to unstable motion, does not have a significant impact on the peak of 
the $(2,2)$ amplitude in the eccentric Schwarzschild case.
However, the influence of the anomaly becomes increasingly relevant in configurations where radiation reaction plays a stronger role.
For instance, for $\ha=0.9$ and $\esep=0.5$, we found that the dependence 
on the relativistic anomaly $\xisep$ of the $(2,2)$ amplitude peak location $\tA22$
measured relative to the \ac{lr} crossing is remarkably large.
This would in principle require $\xisep$ to be taken into account in any post-peak waveform characterization.
Nevertheless, as shown in Fig.~\ref{fig:a09_e05_tLR}, the waveforms at fixed $\ha$ and $\esep$ become practically 
indistinguishable shortly after the \ac{lr} crossing. This demonstrates that the post-\ac{lr} portion of the signal 
can be fully characterized in terms of $(\ha,\esep)$ alone, thus neglecting $\xisep$.
It also implies that the amplitude peak can occur while the waveform is still strongly source-driven, 
making a \ac{qnm}-based description starting from $\tA22$ inadequate.
Notably, an anchoring point tied to the \ac{lr} is advantageous also in the quasi-circular limit at high spins, 
where the amplitude peak can precede the \ac{lr} crossing by tens to hundreds of $M$, rendering the post-peak 
modeling less practical.

Motivated by these considerations, we modeled the ringdown waveform starting from a point closely related to the \ac{lr},
$\to=\tLR-2.4$, where the numerical small offset is chosen such that $\to\equiv\tA22$ in the quasi-circular, non-spinning case.
This choice enables a closed-form description of the ringdown waveform based on the phenomenological framework introduced in 
Ref.~\cite{Damour:2014yha}, while employing a different template for the amplitude and, crucially, avoiding the use of amplitude peaks as reference points.
The same procedure is applied to all multipoles with $m\geq1$ and $\l\leq4$, as well as to the $(5,5)$, $(5,4)$, and $(5,3)$ modes.
The $(2,0)$ multipole is modeled analogously by extending the approach introduced in Ref.~\cite{Albanesi:2024fts}.
We incorporate spheroidal-spherical mode mixing for modes with $\l\geq3$ and $m<\l$ by extending the method 
introduced in Ref.~\cite{Pompili:2023tna}. This approach consists of applying the phenomenological description to approximate 
spheroidal modes and subsequently converting back to the spherical basis using spheroidal-spherical mixing coefficients~\cite{Berti:2014fga}.
In addition, we model the beating between fundamental co-rotating and counter-rotating \acp{qnm}, extending and improving the 
treatment presented in Ref.~\cite{Albanesi:2023bgi}.

The resulting ringdown model is then used to complete an \ac{eob} inspiral-plunge waveform for non-circular motion, as discussed in Sec.~\ref{sec:eob}.
A smooth transition between the inspiral-plunge and ringdown regimes is ensured through the inclusion of \ac{nqc} corrections.
The full \ac{eob} waveform accurately reproduces the numerical \ac{imr} signal in a wide range of configurations up to $\ha=0.9$, as shown 
in Sec.~\ref{sec:comparisons}. Despite being more general than the Schwarzschild model of Ref.~\cite{Albanesi:2023bgi} and relying 
on fewer fitted physical quantities, the present model provides a more accurate description of higher-order modes even in the Schwarzschild limit.
For highly eccentric and fast spinning configurations, the accuracy decreases. We find that the degradation in phase 
accuracy at merger for $\ha>0.9$ 
is primarily associated with the \ac{nqc} corrections rather than with the ringdown modeling itself 
(see Appendix~\ref{app:nqc_derivs}). Finally, the ringdown model naturally extends to dynamical capture scenarios 
without any further modification, as discussed in more detail in Sec.~\ref{subsec:comparisons_hyp}.
A more accurate EOB modeling for the full evolution of EMRIs with high eccentricity and high primary spins requires 
further improvements in the inspiral phase, related in particular to the accuracy of the radiation reaction,
which are however postponed to future works. 

While the modeling introduced here makes direct use of the \ac{lr}, a quantity that is not defined in the comparable-mass case, 
the results of this work nevertheless point toward promising directions for extending the approach beyond the test-mass limit.
Indeed, the \ac{lr} is closely related to the inflection point of the $(2,2)$ frequency during the plunge, $\td2omg0$
(see, e.g., the markers on the frequency curves in Fig.~\ref{fig:num_waves}).
This dynamical feature can therefore be used as an alternative anchoring point for a ringdown model.
We explore this possibility in Appendix~\ref{app:d2omg0}, where we show that anchoring the model at $\to=\td2omg0-3.6$ yields 
results that are broadly comparable to those obtained using $\to=\tLR-2.4$. Also in this case, the numerical offset is chosen such
that $\to\equiv\tA22$ in the quasi-circular, non-spinning case.
In particular, this prescription leads to an improved phase agreement for highly eccentric and rapidly 
spinning configurations, albeit at the cost of a slightly less accurate amplitude description.
Moreover, since an eccentricity directly tied to the dynamics cannot be defined in a
gauge-invariant way, for the comparable mass case one should be more careful in which variables 
to use in the global fits to parametrize the non-circularity of the systems. While different
works proposed definitions of eccentricity based on the 
observable waveform~\cite{Ramos-Buades:2019uvh,Ramos-Buades:2022lgf,Shaikh:2023ypz,Shaikh:2025tae}, an alternative approach would
be to use an impact parameter written in terms of energy and angular momentum, as mentioned in 
Sec.~\ref{subsec:global}.
This has already been done in previous works, 
both for the test-mass limit~\cite{Albanesi:2023bgi} and the comparable mass case~\cite{Carullo:2023kvj}. 
A systematic investigation of the applicability of this strategy to the comparable-mass regime 
is also left to future work. 

{\it Addendum -} During the final stages of this work, we became aware of a study of eccentric coalescences 
in the test-mass limit being carried out by Faggioli et al.~\cite{Faggioli:ecckerr}, which presents a complementary 
analysis for the systems studied in this work.

%=======================================================================================================
\section{Acknowledgment}
%=======================================================================================================
  The authors are grateful to R. Gamba for useful comments and suggestions on the draft.
  S.A. acknowledges support from the Deutsche Forschungsgemeinschaft (DFG) project ``GROOVHY'' 
  (BE 6301/5-1 Projektnummer: 523180871). 
  S.A. also expresses gratitude for the hospitality of IHES, 
  where part of this work was conducted. This visit was supported by the
  {\it ``2021 Balzan Prize for Gravitation: Physical and Astrophysical Aspects”}, 
  awarded to T. Damour.
  S.A. and A.N. thank G. Faggioli for useful discussions during the workshop 
  {\it ``EOB@Work25: 10 Years of Gravitational Wave Detections''}, 
  hosted by the INFN Section of Torino.
  S.A. is also grateful to E. D'Erme for support during the latest stages of this work.
  S.B. knowledges support by the EU Horizon under ERC Consolidator Grant,
  no. InspiReM-101043372.  
  Numerical simulations were performed 
  the Virgo ``Tullio'' server in Torino, supported by INFN,
  and on the ARA cluster
  at Friedrich Schiller University Jena. 
  The ARA cluster is funded in part by DFG grants INST
  275/334-1 FUGG and INST 275/363-1 FUGG, and ERC Starting Grant, grant
  agreement no. BinGraSp-714626.

\appendix

\begin{figure*}[t]
	\begin{center}
	\includegraphics[width=0.49\textwidth]{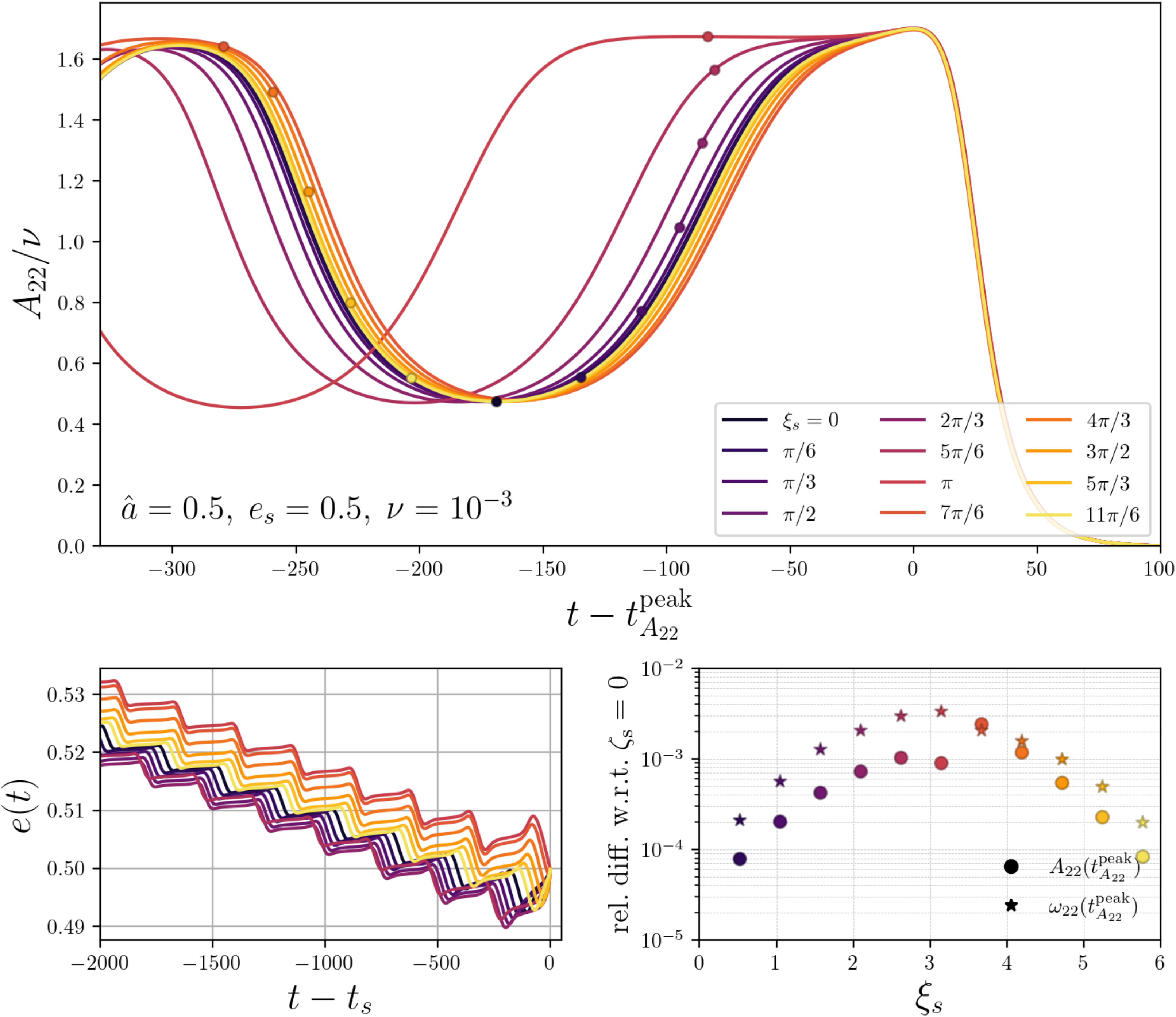}
	\includegraphics[width=0.49\textwidth]{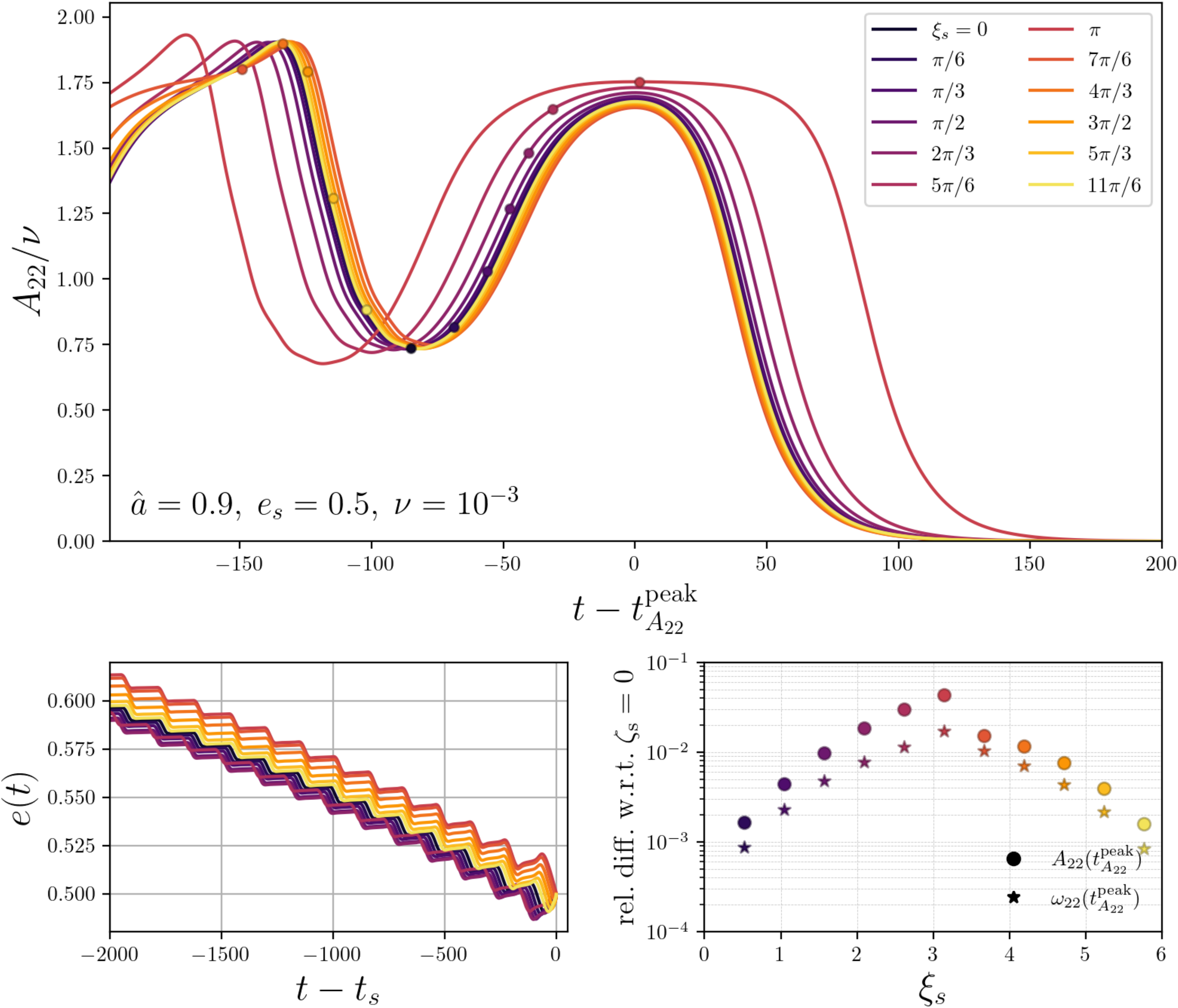}
	 \caption{\label{fig:anomaly_kerr} 
	 Upper panels: \Teuk{} amplitudes of the (2,2) mode for configurations with 
	 fixed $\ha$, $\esep=0.5$, $\nu=10^{-3}$, but different $\xisep$ (reported in the legend). 
	 The dots mark the separatrix-crossing time.
	 In the two small bottom panels we show the corresponding eccentricities as a function of time (left)
	 and the relative difference of $A_{22}$ and $\omega_{22}$ between the configurations with $\xisep>0$ and the one with $\xisep=0$
	 (right). 
	 In the leftmost group of panels we consider $\ha=0.5$, while on the right $\ha=0.9$. 
	 Figure analogous to Fig.~\ref{fig:anomaly_schw}, but for spinning cases. 
	 The amplitudes and frequencies of the $\ha=0.9$ case are also shown in Fig.~\ref{fig:a09_e05_tLR}, 
	 where they are aligned with respect to $\tLR$ rather than $\tA22$.
	 }
	\end{center} 
\end{figure*}

%=======================================================================================================
\section{Impact of $\xisep$ in the presence of spin}
%=======================================================================================================
\label{app:anomaly}
As extensively discussed in Sec.~\ref{subsec:plunge}, the relativistic anomaly at the separatrix crossing
$\xisep$ can have a strong impact on the location of the (2,2) amplitude peak with respect to the \ac{lr}
crossing. This effect is enhanced in configurations where the radiation reaction has a stronger impact
on the dynamics. In this Appendix we briefly discuss the series of configurations with 
$\esep=0.5$ and spins $\ha=\lbrace 0.5,0.9\rbrace$. The relative differences in amplitude and frequency
among configurations with $\xisep>0$ and the $\xisep=0$ case are shown with yellow and red markers
in Fig.~\ref{fig:anomaly_summary}. The corresponding amplitudes and evolutions of the eccentricities
are reported in Fig.~\ref{fig:anomaly_kerr}. When aligning the amplitudes with respect to the time 
of the (2,2) amplitude peak, $\tA22$, for the $\ha=0.5$ case, it is already possible to visually distinguish 
some configurations, even if the amplitudes overlap reasonably well starting from a few $M$ before the peak. 
When moving to the $\ha=0.9$ case, the post-peak amplitudes aligned using $\tA22$ start to differ drastically, 
thus making impossible a modelization based on solely $(\ha,\esep)$. However, as highlighted in the main text
and in particular in Fig.~\ref{fig:a09_e05_tLR}, the waveforms after the \ac{lr} crossing 
become indistinguishable once that they are aligned using $\tLR$, thus showing that an anchoring 
point related to $\tLR$ rather than $\tA22$ allows us to model the last portion of the waveform
using only $(\ha,\esep)$, and thus not taking into account $\xisep$. 

%=======================================================================================================
\section{Relativistic anomaly and post-merger tails}
\begin{figure}[t]
	\begin{center}
	\includegraphics[width=0.50\textwidth]{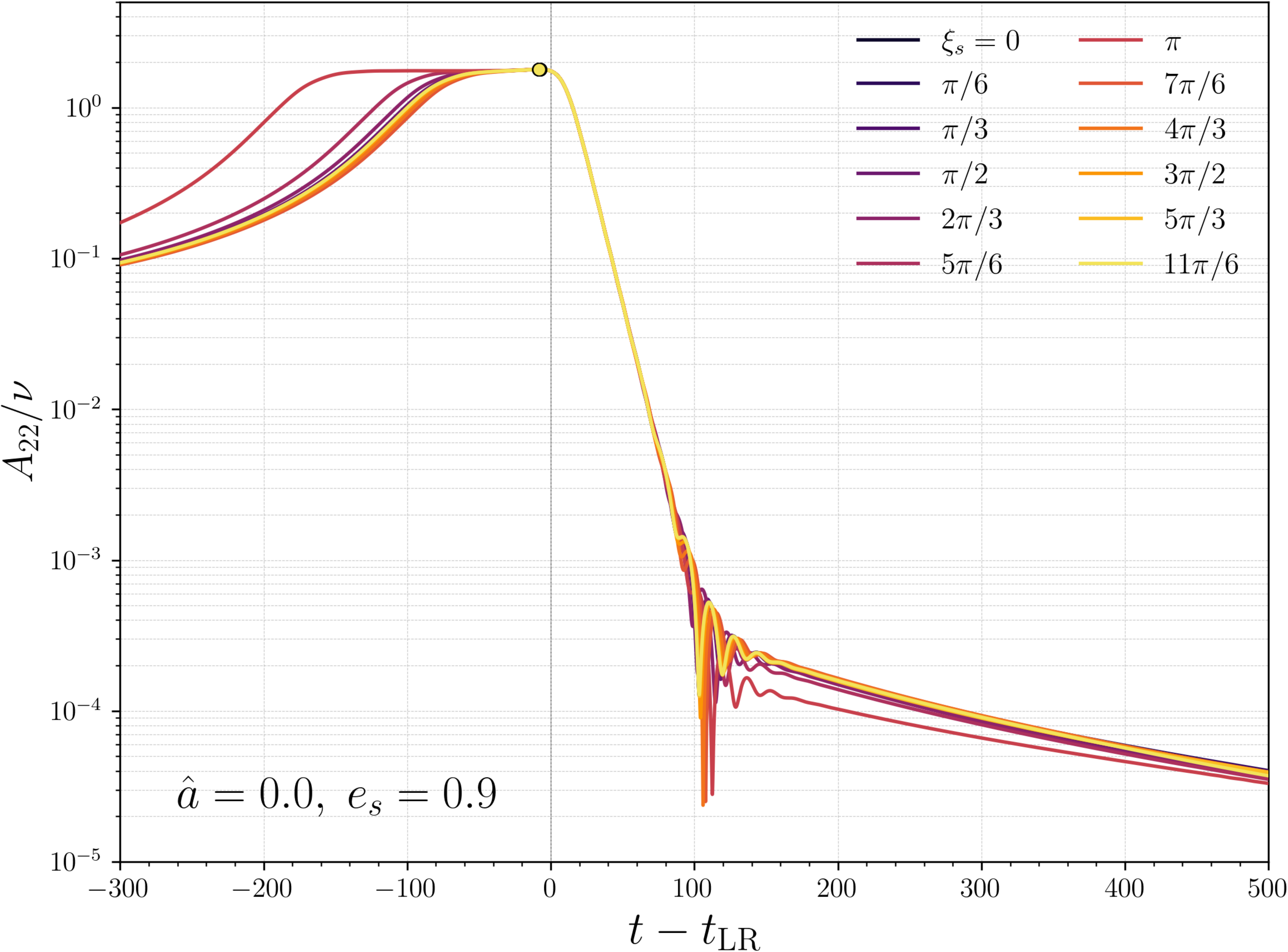}
	 \caption{\label{fig:tails_e09_a00_teuk} Amplitude of the (2,2) mode for 
	 $\ha=0$, $\esep=0.9$ and different $\xisep$, shifted with respect
	 to the \ac{lr} crossing (vertical line). We use a vertical log-scale
	 to highlight the post-merger tails. The (indistinguishable) dots mark $\tA22$ for each
	 configuration.}
	\end{center} 
\end{figure}

%=======================================================================================================
\label{app:tails}
Recent works have studied the impact of orbital eccentricity on the post-merger tails
at intermediate times generated by the infall of particles in a Schwarzschild black 
hole~\cite{Albanesi:2023bgi,DeAmicis:2024not}, showing that the tail phenomenology 
is indeed inherited from the inspiral and can be more articulate than a simple 
power law (see e.g. Fig.~3 of Ref.~\cite{DeAmicis:2024not}). 
These phenomena have been studied also in Kerr~\cite{Islam:2024vro,Becker:2025zzw,Islam:2025wci} 
and observed in full \ac{nr} simulations for comparable mass systems~\cite{DeAmicis:2024eoy,Ma:2024hzq}.
Since we are able to compute tails also within our numerical setup with \Teuk{}, 
we briefly discuss how $\xisep$ influences this hereditary contribution. We start by noting
that we obtain reliable numerical results only for high eccentricity, approximately for
$e\gtrsim 0.6$, where this value also depends on the spin (the amplitudes of the tails are indeed larger for 
retrograde orbits). We thus focus on an highly 
eccentric case ($\esep=0.9$) in Fig.~\ref{fig:tails_e09_a00_teuk}. As can be seen, the tail is slightly
suppressed for $\xisep=\pi$, that corresponds to the case in which the particle has a long-lived whirl before plunging
because the separatrix crossing occurs at periastron.
This is indeed expected, since the tail is generated by the long-range potential, and thus the peak-tail 
emission occurs at the apastron passages. If the last apastron occurs way before the \ac{lr} crossing,
then the corresponding tail is slightly suppressed with respect to the case with a more 
direct plunge. 

We also mention that we observe more relevant tails for dynamical captures than for elliptic-like orbits,
as expected and already discussed in Sec.~\ref{subsec:comparisons_hyp}. 
However, a more careful study of the Kerr tails using \Teuk{} is however postponed to future works.
%

%=======================================================================================================
\section{Second time-derivatives in the \ac{nqc}}
%=======================================================================================================
\label{app:nqc_derivs}
\begin{figure*}[t]
	\begin{center}
	\includegraphics[width=0.325\textwidth]{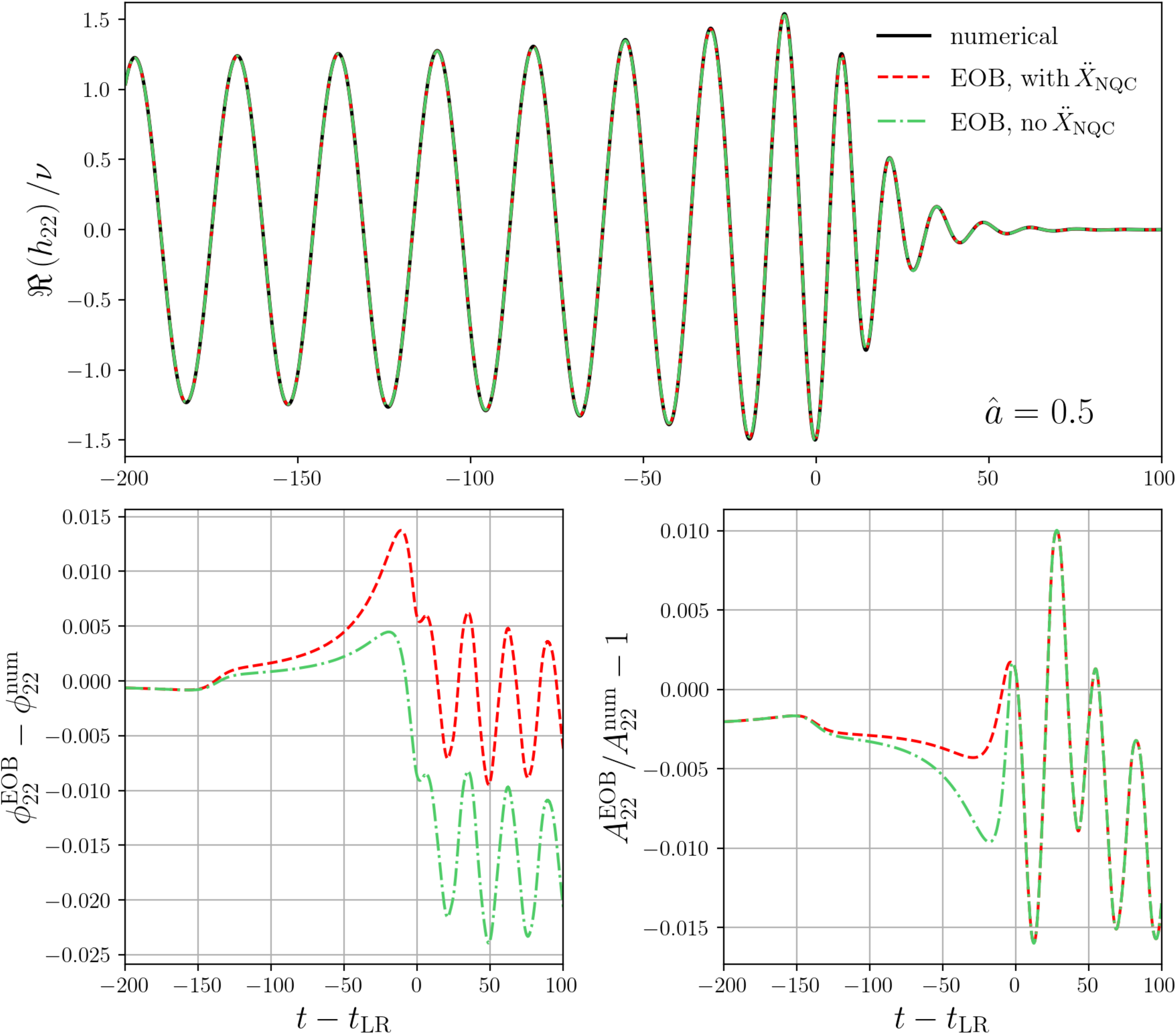}
	\includegraphics[width=0.325\textwidth]{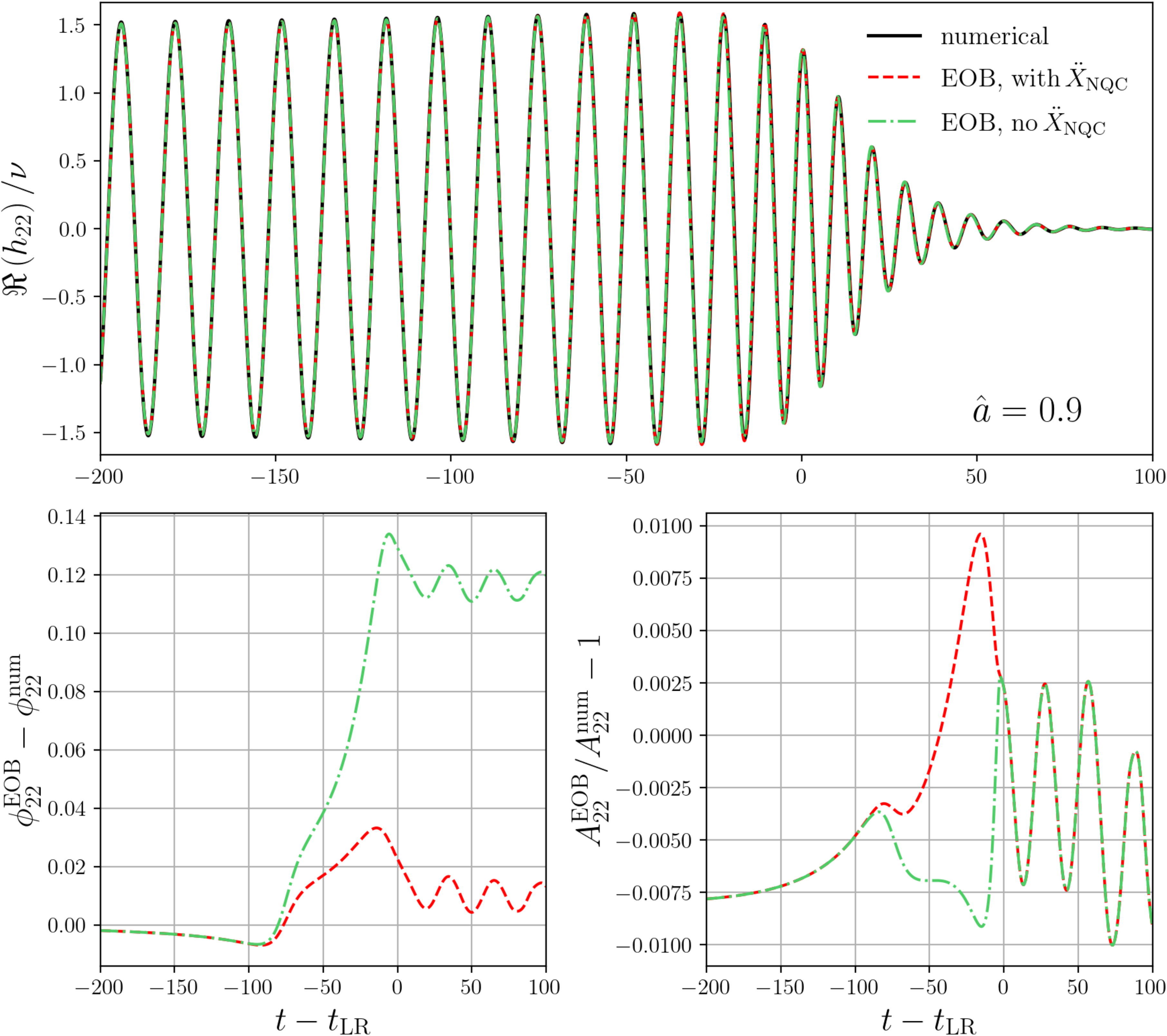}
	\includegraphics[width=0.325\textwidth]{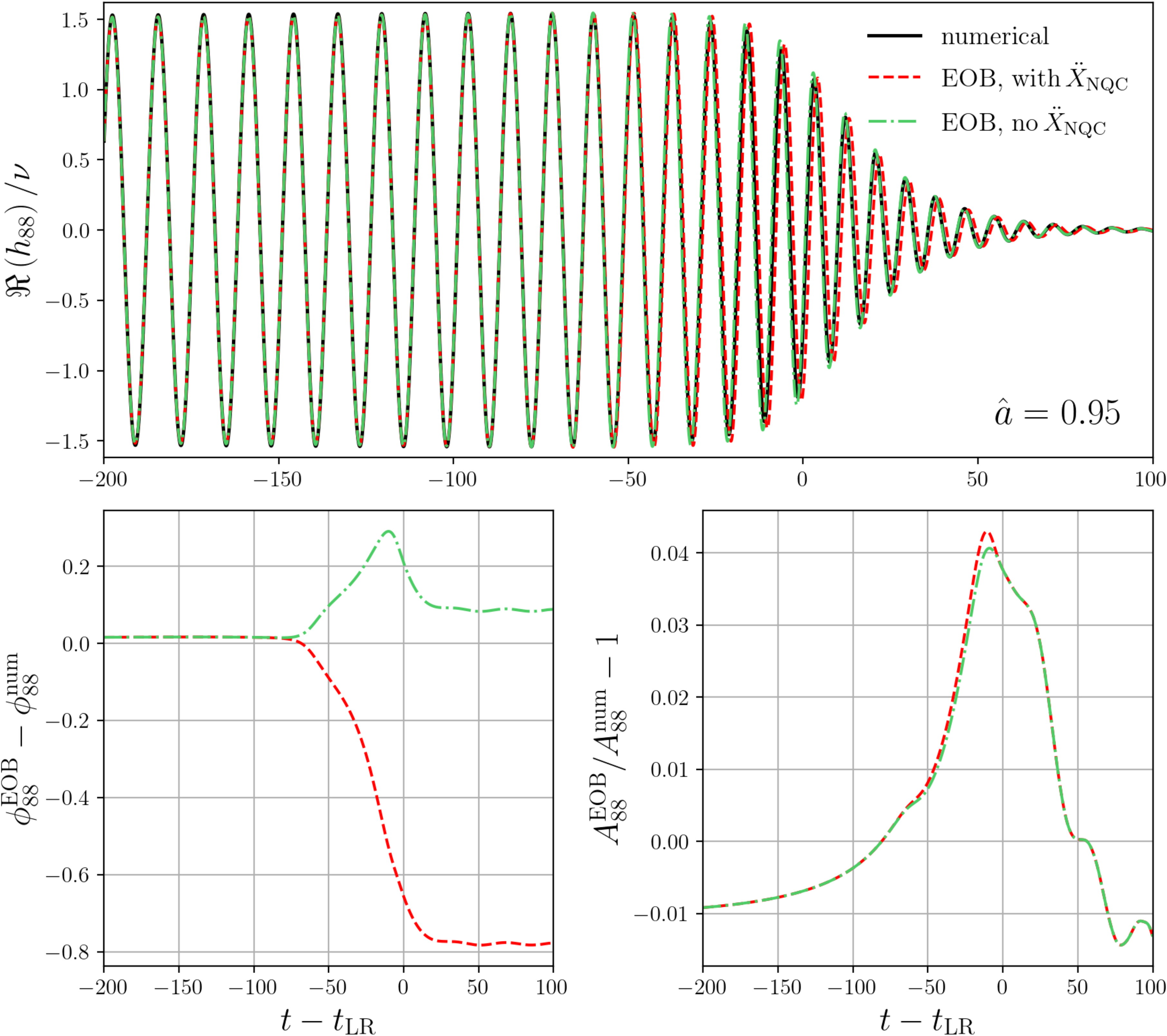}
	 \caption{\label{fig:nqc2} Analytical/numerical comparisons for the complete
	 EOB waveforms in three quasi-circular cases with $\ha=\lbrace 0.5,0.9,0.95 \rbrace$. 
	 We consider in dashed red the standard EOB waveform considered in this work, together
	 with the waveform obtained without considering the second-time derivatives in
	 the \ac{nqc} (dash dotted green). Phase and relative amplitude differences are reported
	 in the bottom panels, with same color scheme.}
	\end{center} 
\end{figure*}
In this Appendix, we briefly discuss the impact of including second-time derivatives
$\ddot{X}_{\rm NQC}$ in the \ac{nqc} of Eqs.~\eqref{eq:nqc_system} for the dominant $(2,2)$ mode.
Relevant quasi-circular examples are shown in Fig.~\ref{fig:nqc2}, where we compare the numerical waveform (solid black)
with the standard EOB prescription adopted in this work (dashed red), and with an EOB waveform constructed
in the same way but excluding second-time derivatives in the \ac{nqc},
i.e., by setting $a_3^{22}=b_3^{22}=0$ in Eq.~\eqref{eq:hnqc} (dash-dotted green).
In the leftmost panel, we show a quasi-circular configuration with $\ha=0.5$.
In this case, our default prescription yields both a better overall phase agreement and an improved
amplitude prior to the matching time $\to$.
Since the post-merger waveform is identical in the two models, the amplitudes coincide exactly after $\to$.
The improvement obtained by including $\ddot{X}_{\rm NQC}$ in the \ac{nqc} is a general feature
for all quasi-circular configurations with $\ha \leq 0.5$ considered in this work, which motivates our
default modeling choice.
For larger spins, however, the situation changes. For spins $\ha=\lbrace0.6, 0.7\rbrace$,
the prescription without $\ddot{X}_{\rm NQC}$ performs a slightly better,
while for $\ha=0.8$ the two prescriptions are very similar.
For $\ha=0.9$ however, the prescription with $\ddot{X}_{\rm NQC}$, perform 
significantly better, as shown in the middle panel of Fig.~\ref{fig:nqc2}.
For even higher spins, such as $\ha=0.95$ (right panel of Fig.~\ref{fig:nqc2}), including
$\ddot{X}_{\rm NQC}$ in the \ac{nqc} causes instead a significant phase disagreement,
with $\Delta \phi_{22}\simeq -0.8$ rad, compared to a phase difference of at most 
$|\Delta \phi_{22}|\simeq 0.25$ rad when second-time derivatives are excluded.
As mentioned above, this degradation in accuracy occurs only for very large spins.
Therefore, we retain the $\ddot{X}_{\rm NQC}$ terms in our model, since they yield better overall
performance across the parameter space.
We also note that for even higher spins, such as $\ha=0.99$, the model fails to produce physically
robust waveforms regardless of whether second-time derivatives are included.

Finally, we remark that the improved accuracy of the $\ddot{X}_{\rm NQC}$ prescription also holds
for low to mild eccentricities, while for large eccentricities the two prescriptions yield
essentially equivalent results.

%=======================================================================================================
\section{Anchoring at $\td2omg0$}
%=======================================================================================================
\label{app:d2omg0}
\begin{figure*}[t]
	\begin{center}
	\includegraphics[width=0.325\textwidth]{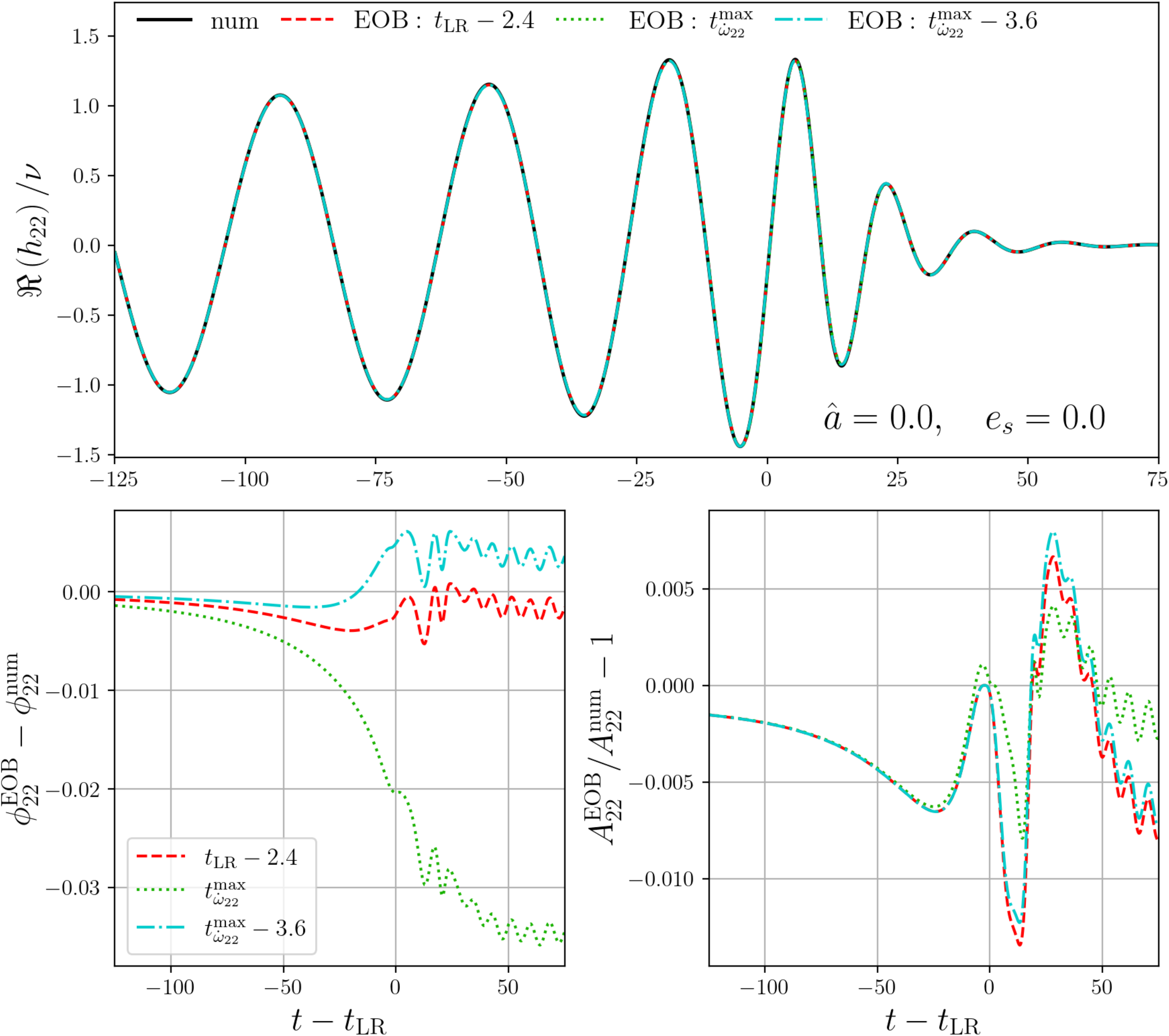}
	\includegraphics[width=0.325\textwidth]{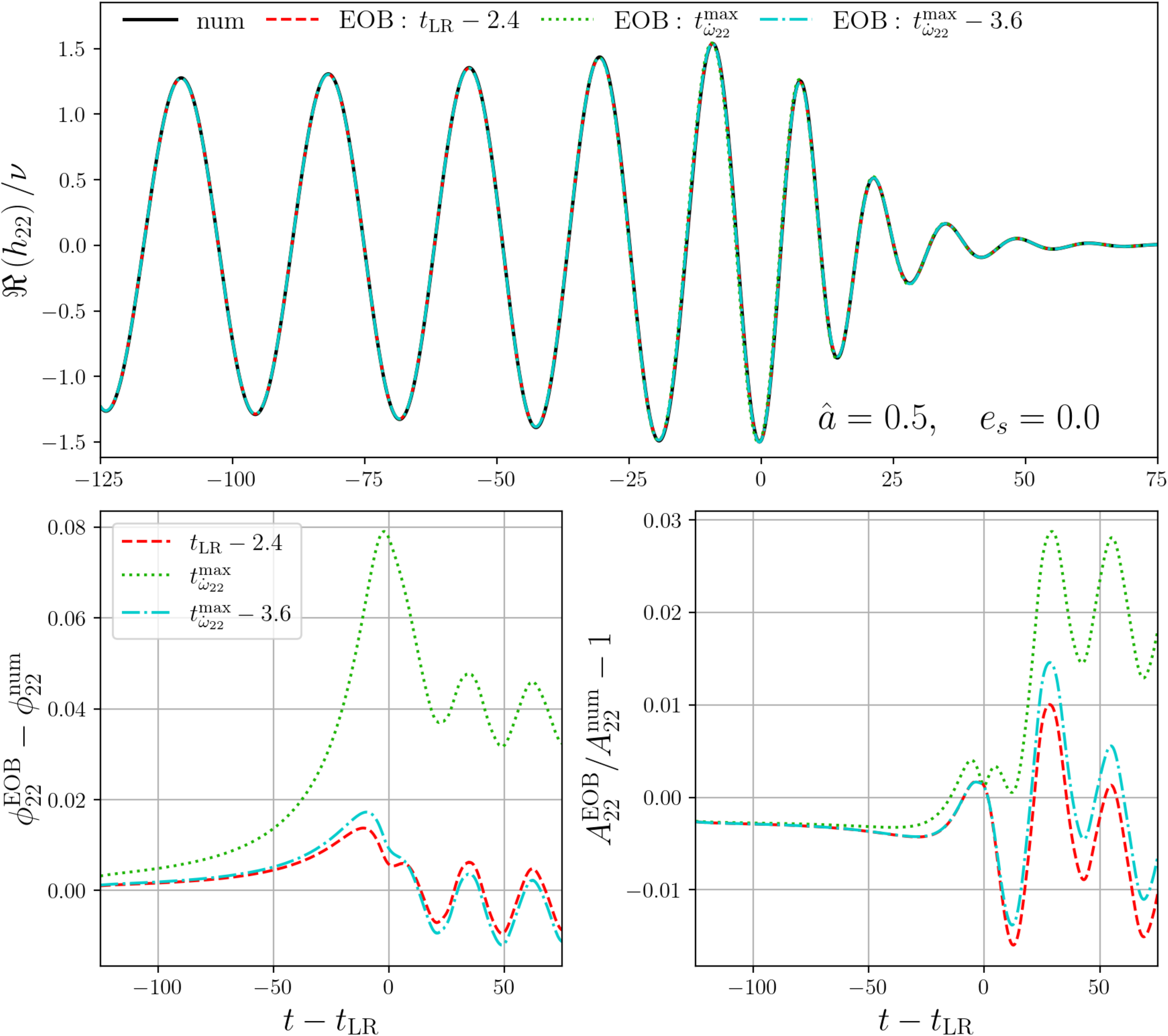}
	\includegraphics[width=0.325\textwidth]{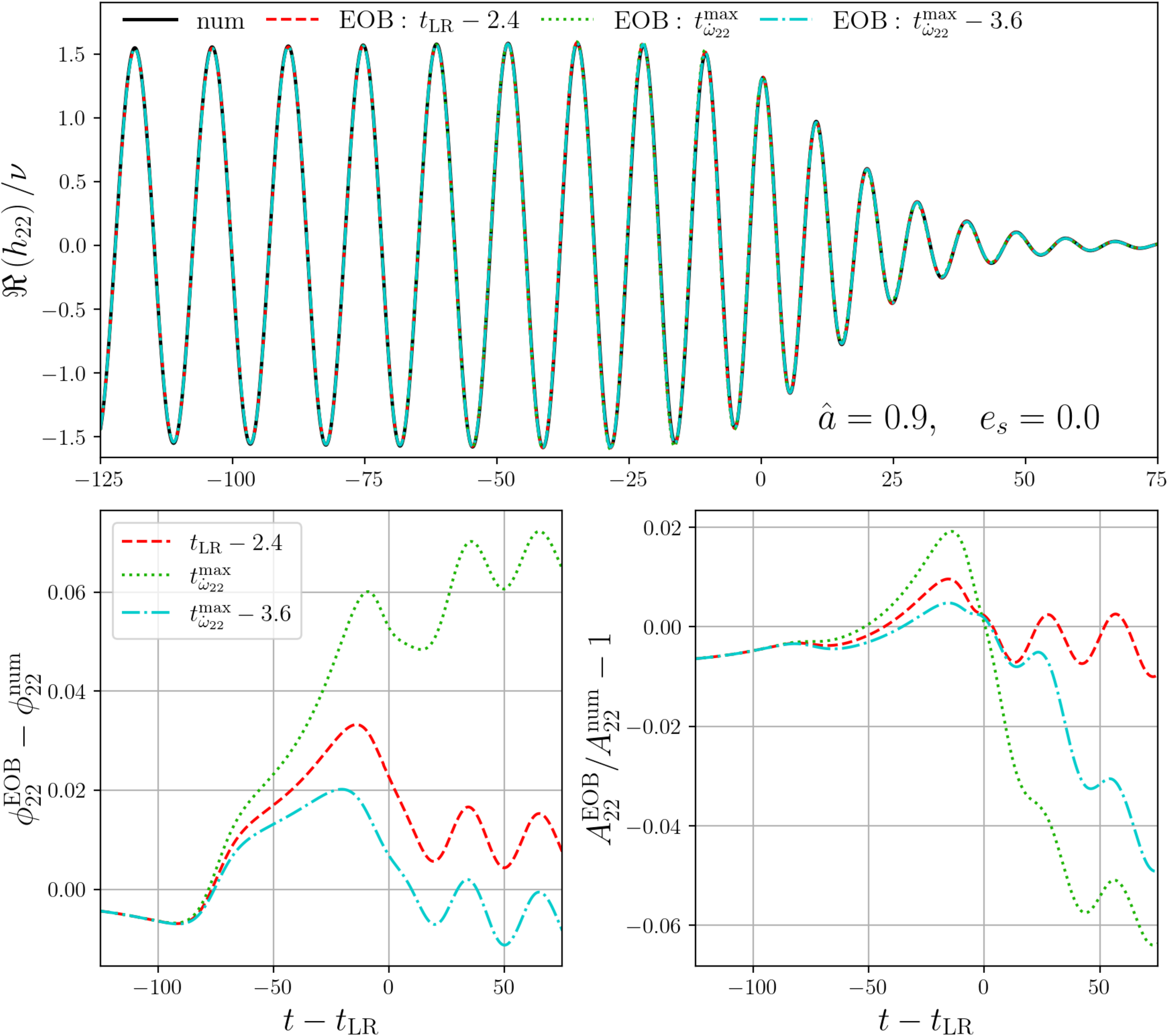}\\
	\vspace{-0.10cm}
	\hline
	\vspace{+0.25cm}
	\includegraphics[width=0.325\textwidth]{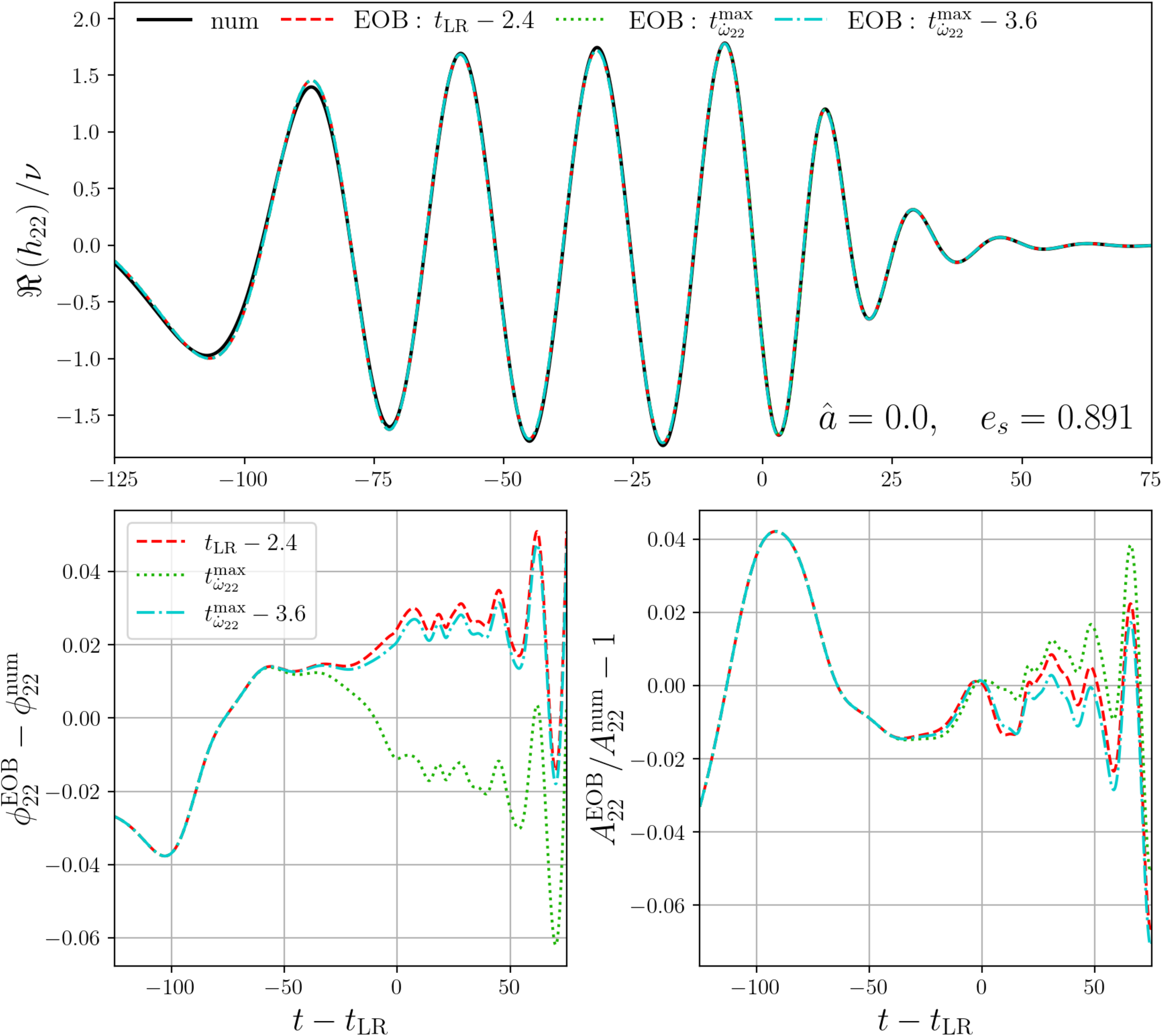}
	\includegraphics[width=0.325\textwidth]{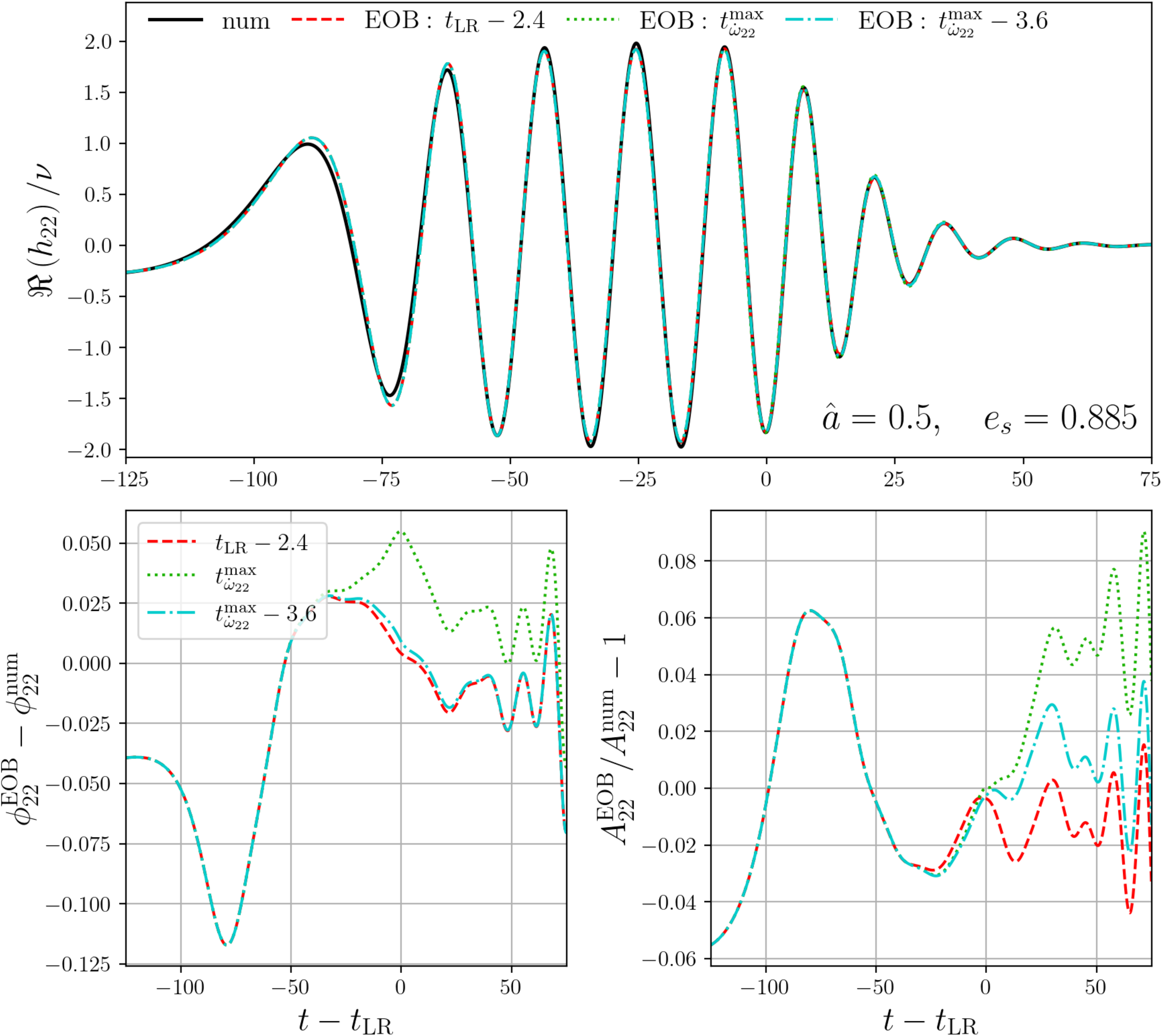}
	\includegraphics[width=0.325\textwidth]{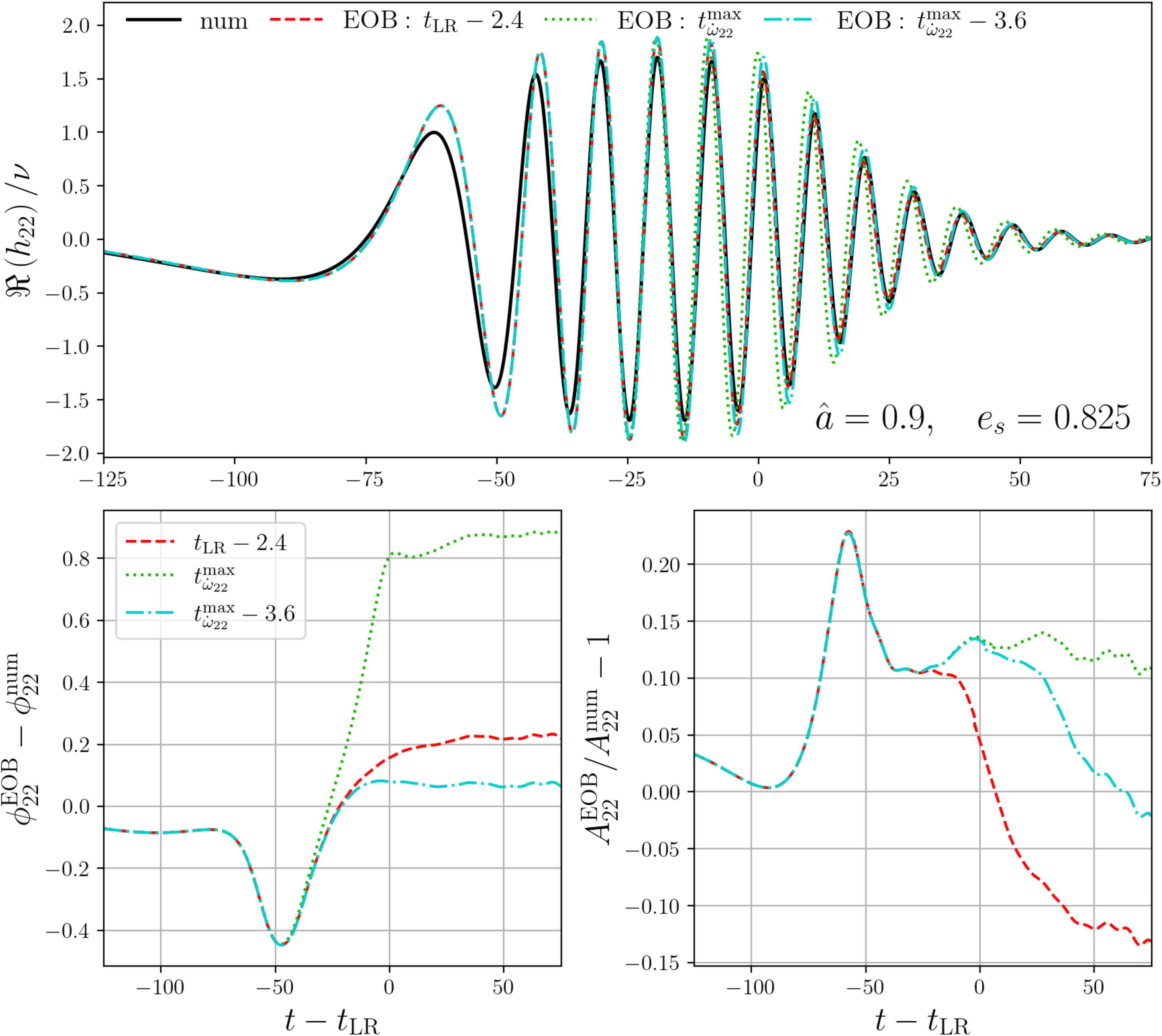}\\
	\vspace{-0.10cm}
	\hline
	\vspace{+0.25cm}
	\includegraphics[width=0.325\textwidth]{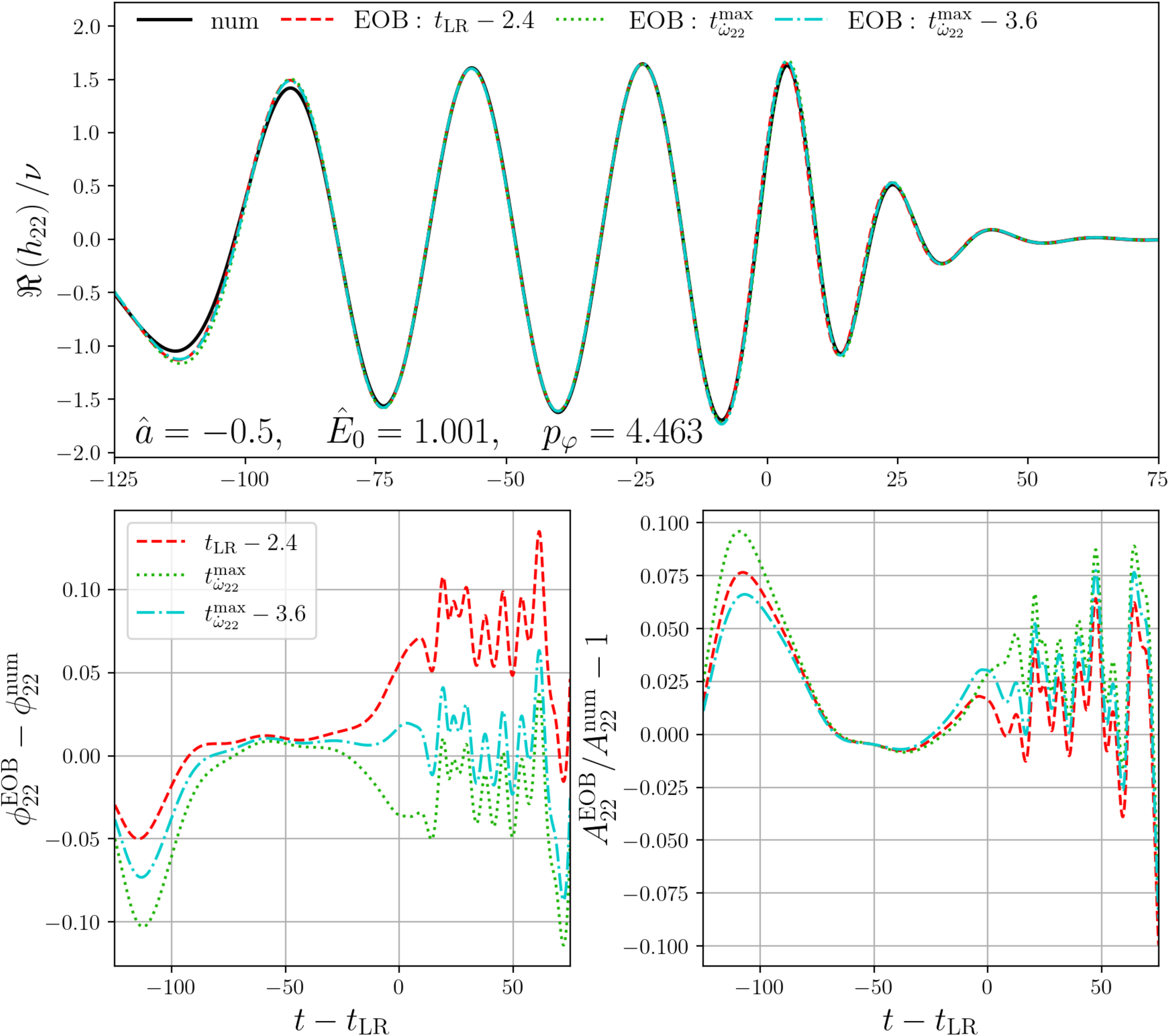}
	\includegraphics[width=0.325\textwidth]{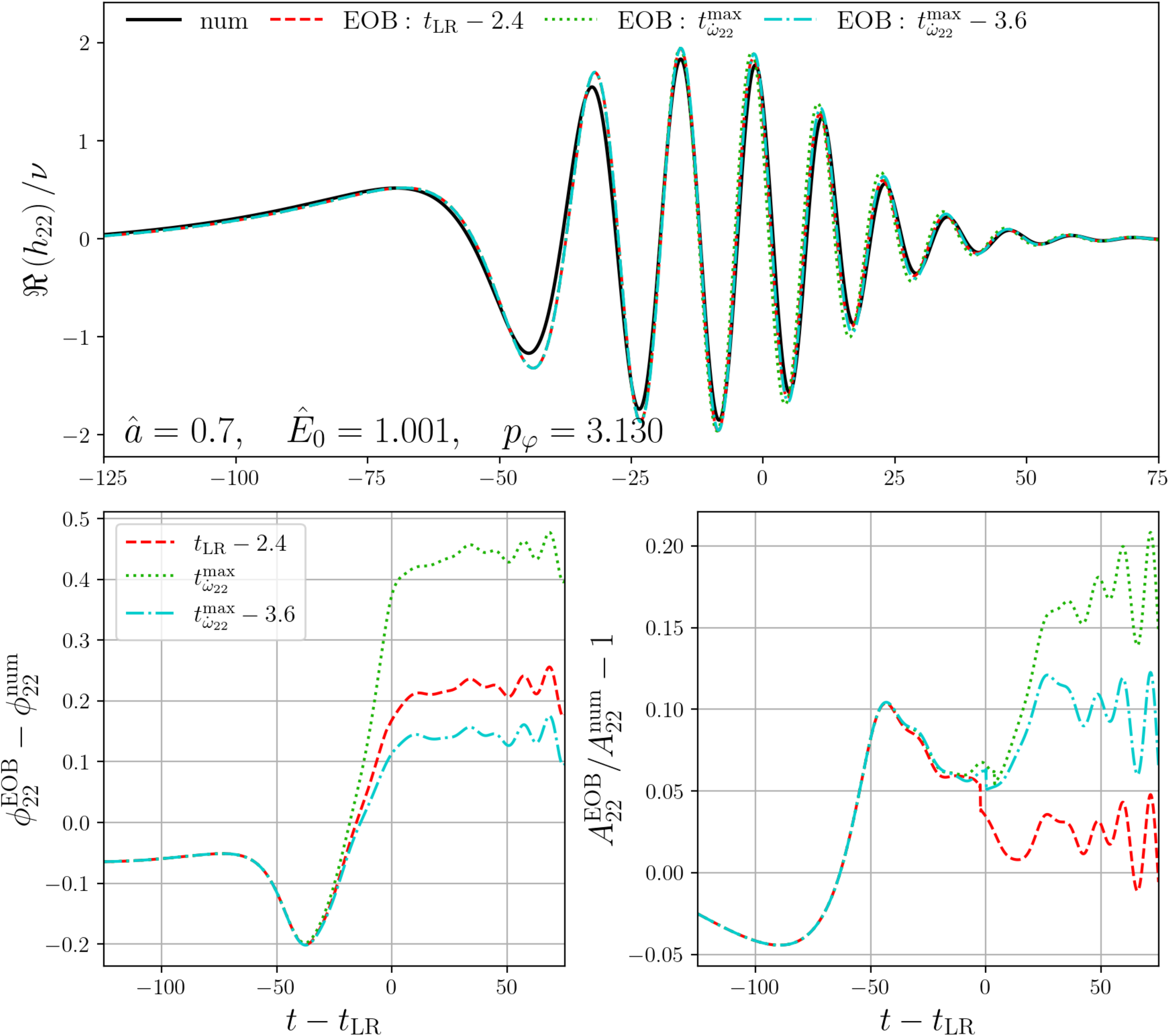}
	\includegraphics[width=0.325\textwidth]{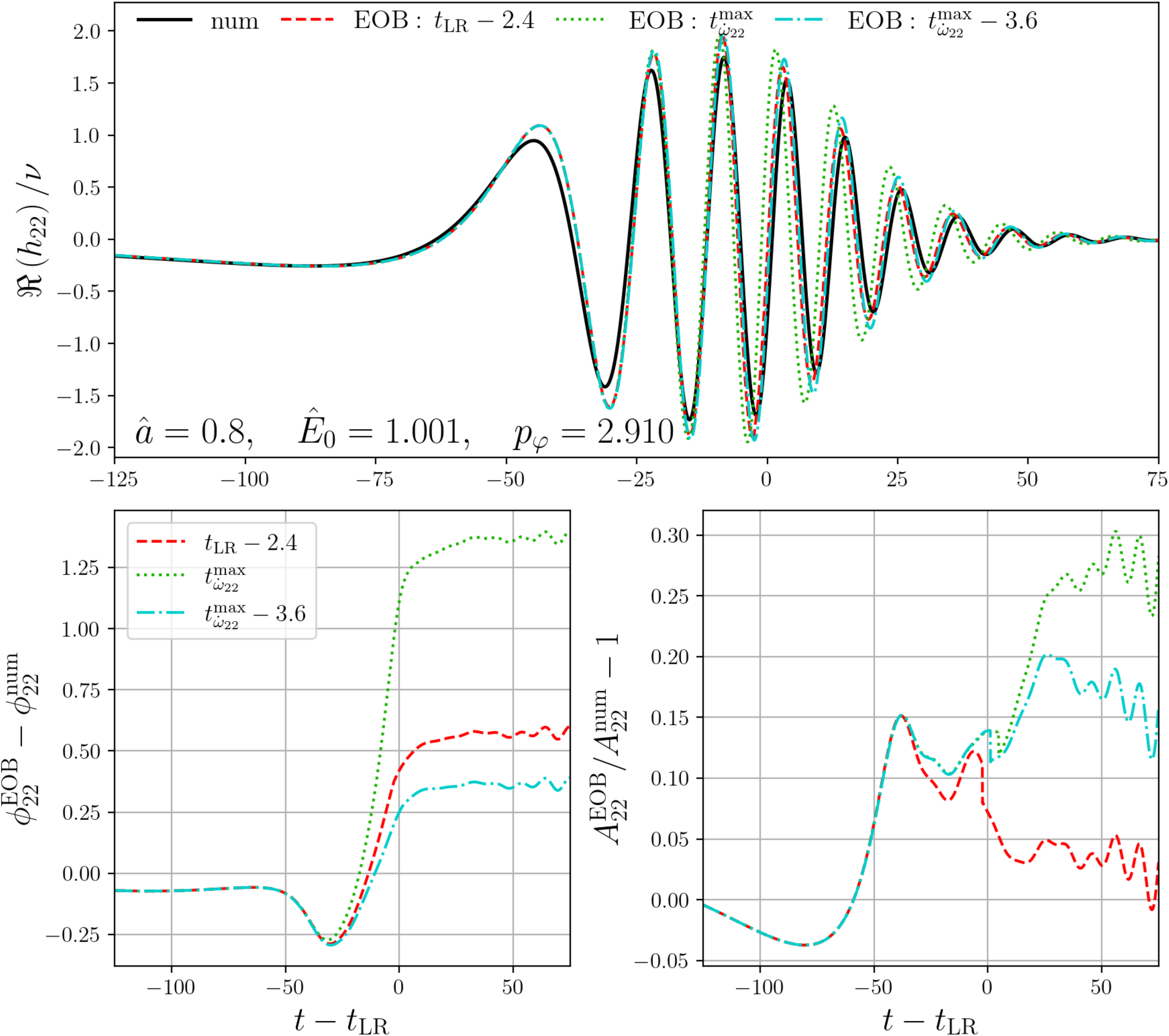}\\
	 \caption{\label{fig:different_anchor_d2omg} Top row: analytical/numerical comparisons 
	 for quasi-circular configurations with spins $\ha=\lbrace0,0.5,0.9\rbrace$ 
	 (from left to right). For each case,
	 we show the real parts in the upper panel, and the phase and relative amplitude
	 difference in the two smaller bottom panels. The default model 
	 considered in this work ($\to=\tLR-2.4$) is shown with dashed red lines.
	 Two additional models with $\to=\td2omg0$ (dotted green) and $\to=\td2omg0-3.6$ (dash-dotted azul)
	 are also reported. See Appendix~\ref{app:d2omg0} for more details.
	 Middle row: ibidem, but for $e_s\simeq 0.8-0.9$. Bottom row: ibidem, but for the
	 dynamical captures of Fig.~\ref{fig:comparisons_hyp_kerr}.}
	\end{center} 
\end{figure*}
In this Appendix, we discuss the modeling of the ringdown waveform using the inflection point of the $(2,2)$ 
frequency as anchoring time. More specifically, we explore two different scenarios:
i) using exactly $\to=\td2omg0$, and ii) using $\to=\td2omg0-3.6$.
In the latter case, the numerical value is chosen such that $\to$ coincides with $\tA22$ in the quasi-circular, 
non-spinning case. This second choice is inspired by the fact that, as shown in Sec.~\ref{subsubsec:anchoring_test_qc}, 
the time $\to=\tLR-2.4$ was more accurate than simply using $\tLR$ (see in particular Fig.~\ref{fig:different_anchor_qc}). 
Aside from the different anchoring points, these two models based on $\td2omg0$ use the same prescription discussed in 
the main text. We implement them only for the $(2,2)$ mode.

Some analytical/numerical comparisons for these two models against the standard model proposed in the main text are 
reported in Fig.~\ref{fig:different_anchor_d2omg}. In the left panel of the upper row, we report the Schwarzschild 
quasi-circular case. As can be seen, the largest analytical/numerical phase difference $\Delta \phi_{22}$ is obtained 
with the model using $\to=\td2omg0$ (dotted green). This is expected, because we already argued in the main text that 
using $\tLR$ is too late to allow effective \ac{nqc} corrections. Since $\td2omg0>\tLR$, we expect a similar effect for 
this anchoring time. However, if we consider $\to=\td2omg0-3.6$ (dash-dotted light blue), the phase difference becomes 
much smaller, reaching at most $|\Delta \phi_{22}|\simeq 0.006$ rad, close to the default model of this work (dashed red), 
which nevertheless still performs better. On the other hand, no significant differences are observed in the amplitudes.
If we increase the spin to $\ha=0.5$ (first row, middle panel), the situation remains qualitatively unchanged: the model 
with $\to=\td2omg0$ performs worse than the other two, which instead behave quite similarly. Increasing the spin further to 
$\ha=0.9$ (first row, right panel), we observe that $\to=\td2omg0-3.6$ yields the most accurate phase. An improvement is also 
obtained for $\ha=0.95$ (not shown in Fig.~\ref{fig:different_anchor_d2omg}), where $\to=\td2omg0-3.6$ yields $\Delta \phi_{22}\simeq -0.5$ rad, 
compared to $-0.8$ rad obtained with $\to=\tLR-2.4$. The model anchored exactly at the $(2,2)$ frequency inflection instead exceeds $-1.2$ rad. 
However, it is clear that for $\ha>0.9$ no model yields accurate phases unless the second-time derivatives in the \ac{nqc} corrections are 
switched off, as shown in Fig.~\ref{fig:nqc2}.
Moving now to the eccentric cases reported in the middle row, 
we see that $\to=\td2omg0-3.6$ yields an improved phase when high eccentricity and spin are combined. 
For the $\esep\simeq0.9$ non-spinning case (second row, left panel), this model performs similarly to the default one. However, when the 
spin is increased, this alternative choice leads to smaller phase differences, as shown for $\ha=0.5$ and $\ha=0.9$ in the middle and 
right panels of the second row. The improvement in phase accuracy is retained also in the dynamical capture scenario, as shown in the 
last row of Fig.~\ref{fig:different_anchor_d2omg}, where we report the same physical configurations considered in 
Fig.~\ref{fig:comparisons_hyp_kerr}. This improvement is particularly evident in the phases for $\ha=0.7$ and $\ha=0.8$. However, in both 
cases we also observe a degradation in the amplitude accuracy. This may be related to the global fits adopted for the $\to=\td2omg0-3.6$ 
model rather than to the \ac{nqc} corrections themselves. Indeed, switching on or off the \ac{nqc} terms related to $\ddot{A}$ does not 
significantly change the amplitude, whereas the second-time derivative of the frequency in the \ac{nqc} strongly affects the phase. For 
dynamical captures in Schwarzschild, instead, the model with $\to=\td2omg0-3.6$ performs similarly to the default one, as already noted 
for the highly eccentric elliptic-like configurations with moderate spins.

We have thus shown that choosing an anchoring point strictly related to the inflection point of the $(2,2)$ frequency, $\to=\td2omg0-3.6$, yields 
results comparable to the default choice adopted in this work, $\to=\tLR-2.4$. This is encouraging for the modeling of comparable-mass binaries, 
where the \ac{lr} cannot be used since it is not defined. However, a detailed study of the comparable-mass regime is left for future work.

\begin{figure*}[t]
	\begin{center}
	\includegraphics[width=0.32\textwidth]{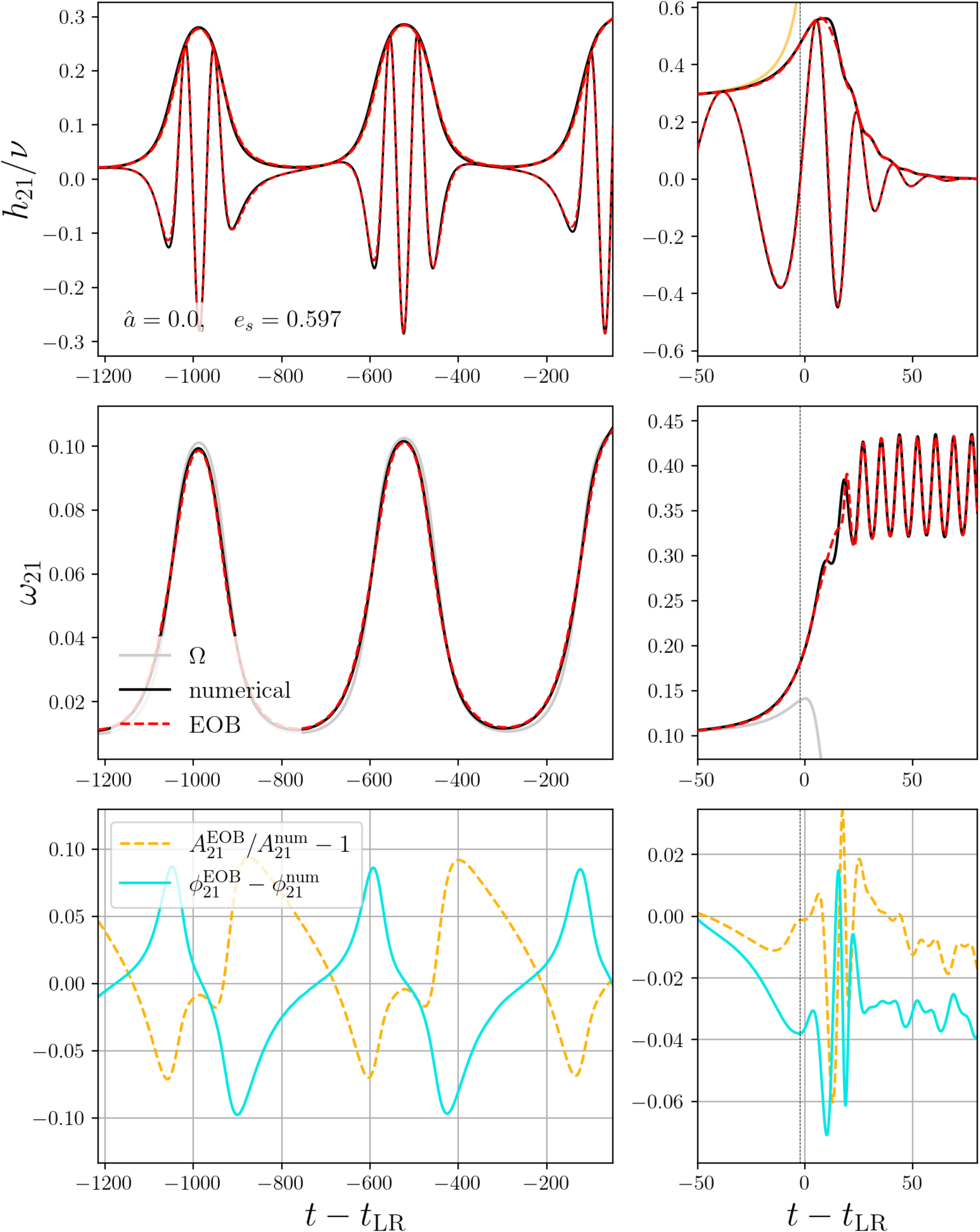}
	\includegraphics[width=0.32\textwidth]{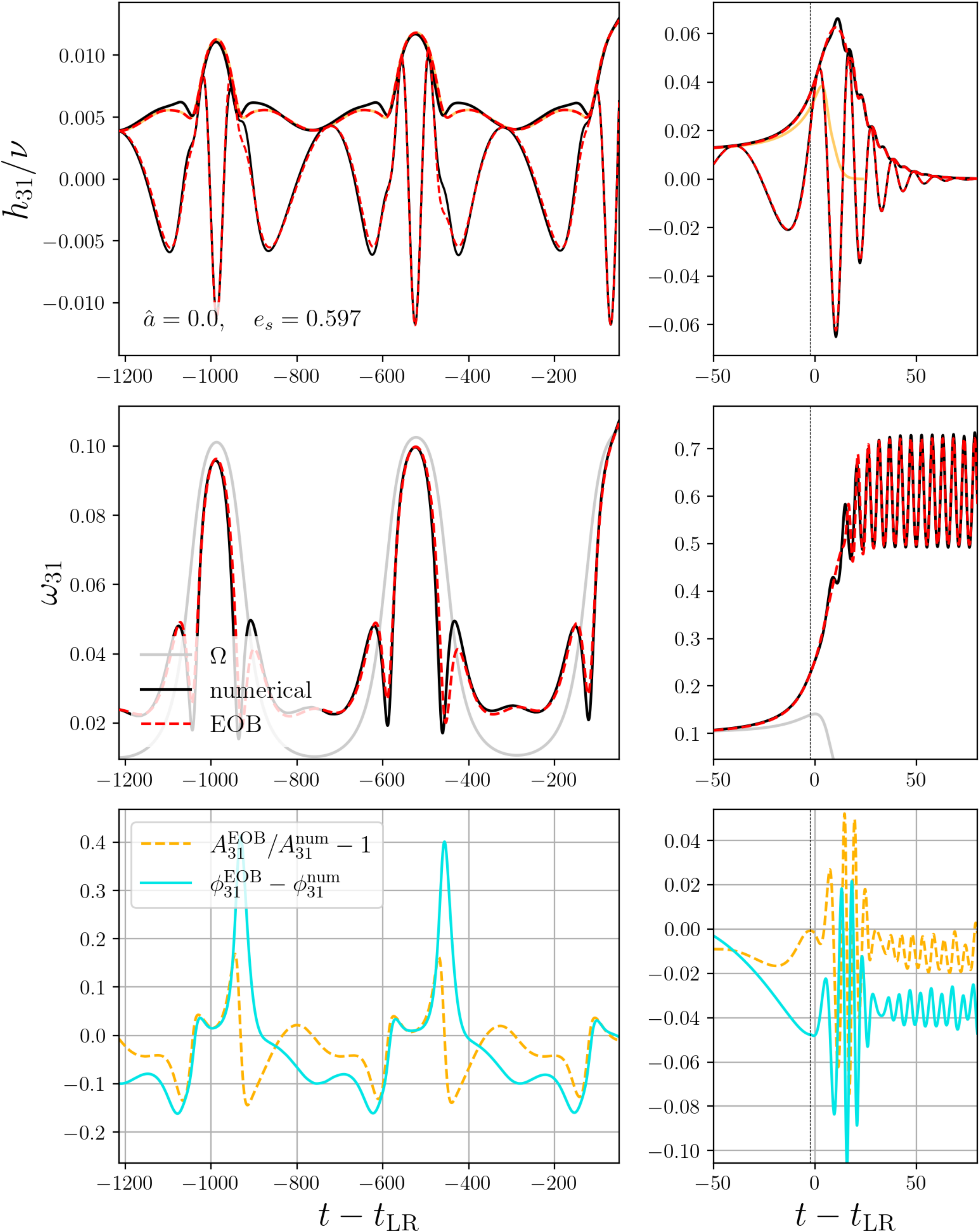}
	\includegraphics[width=0.32\textwidth]{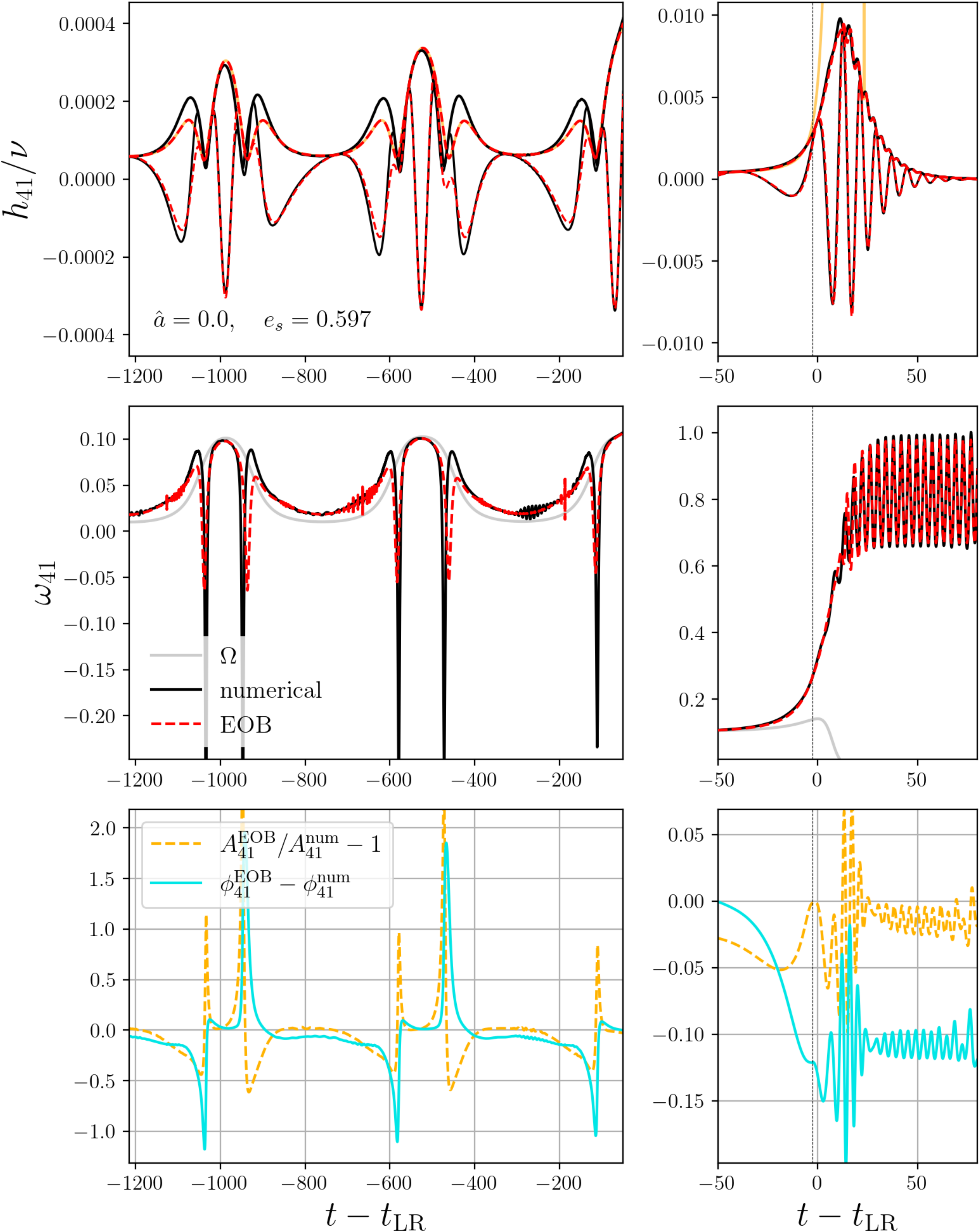}\\
	\hline
	\vspace{0.3cm}
	\includegraphics[width=0.32\textwidth]{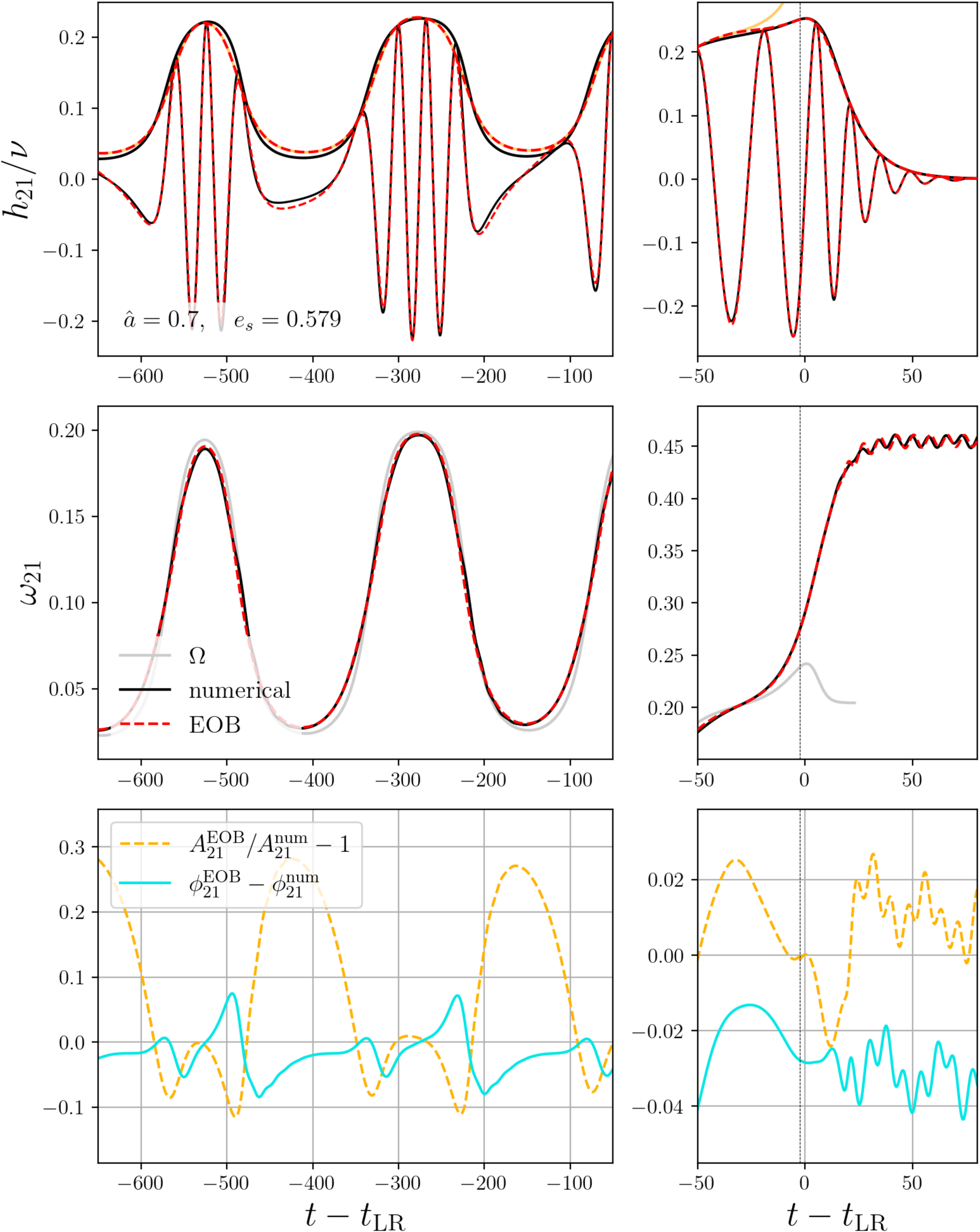}
	\includegraphics[width=0.32\textwidth]{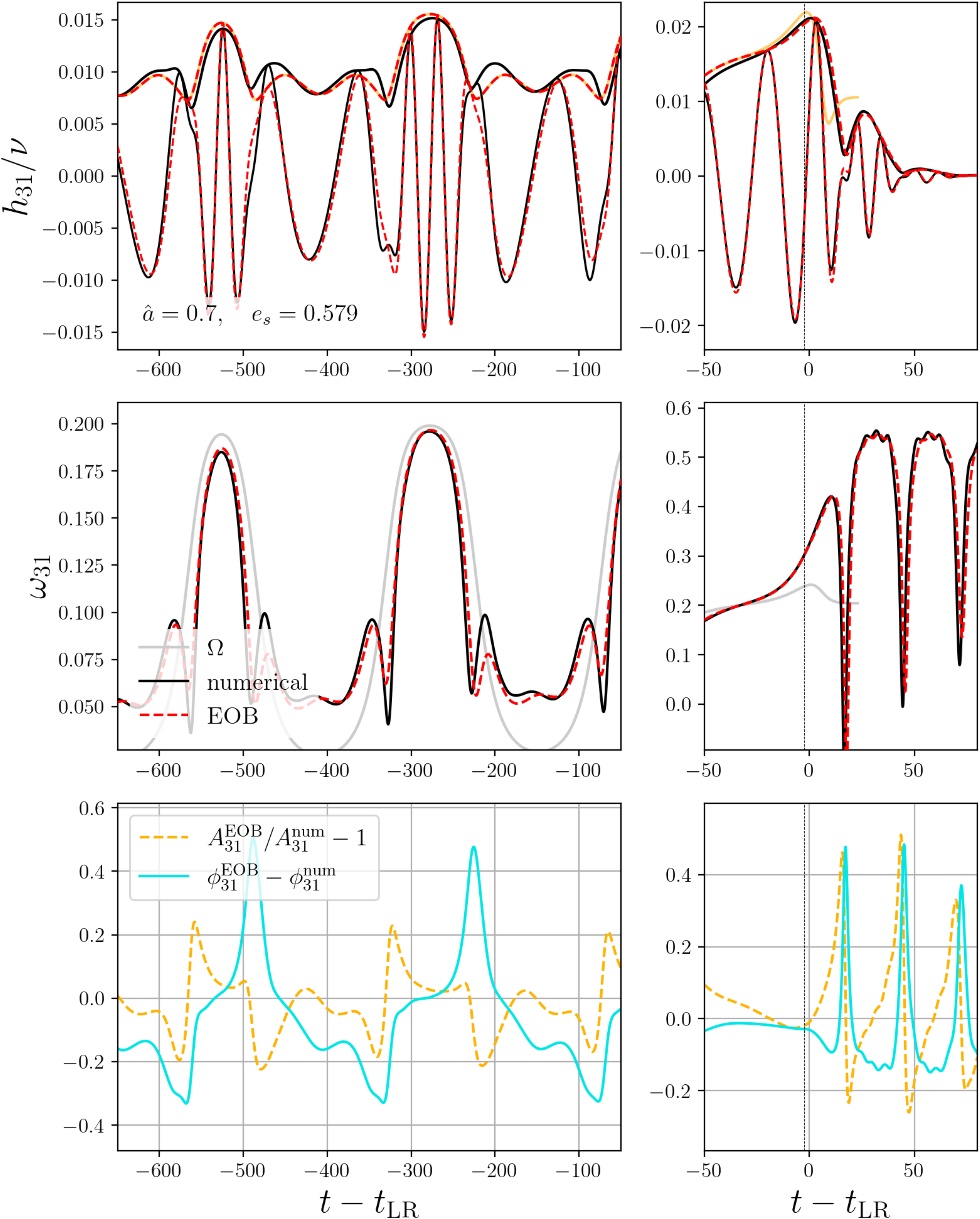}
	\includegraphics[width=0.32\textwidth]{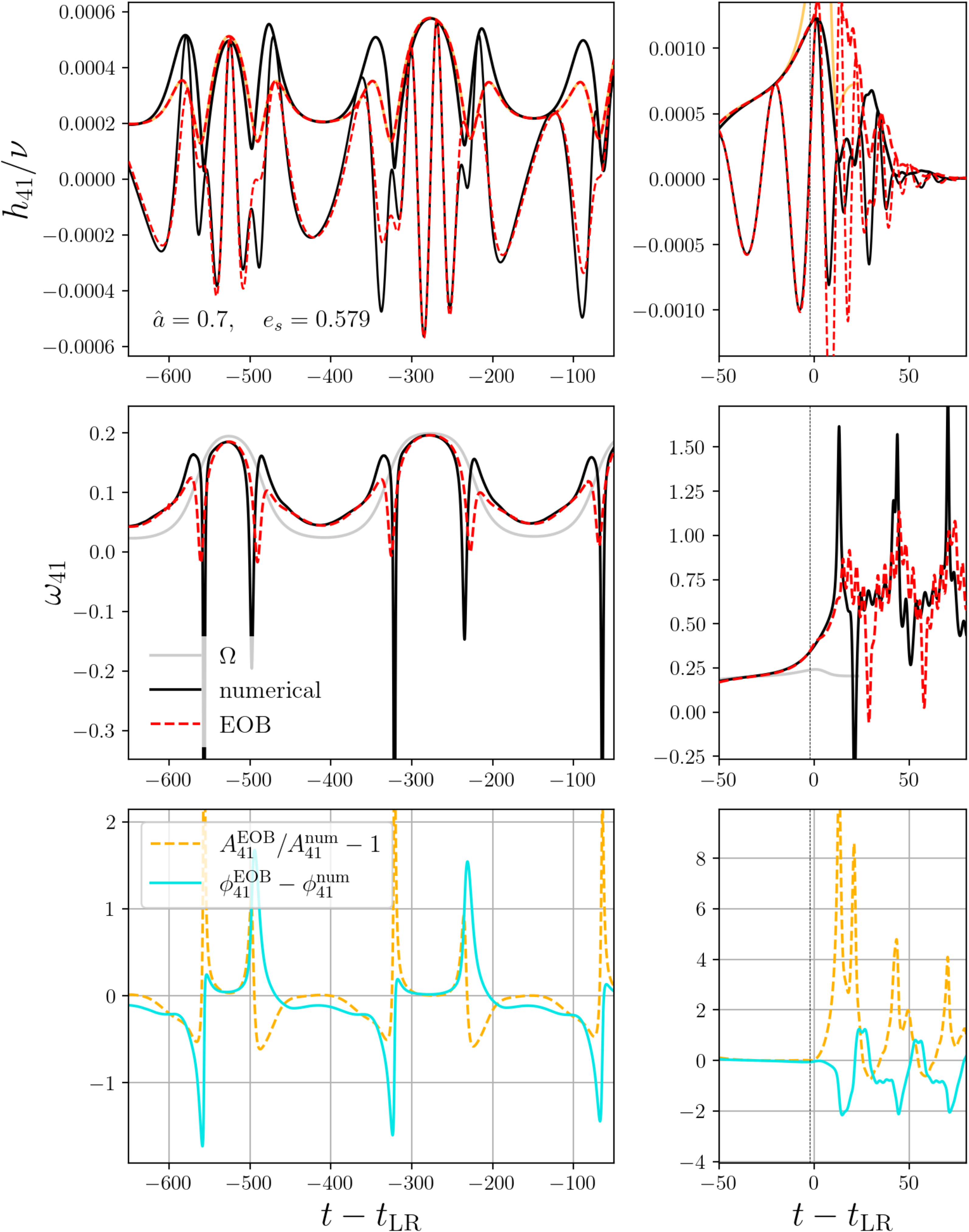}\\
	\hline
	\vspace{0.3cm}
	\includegraphics[width=0.32\textwidth]{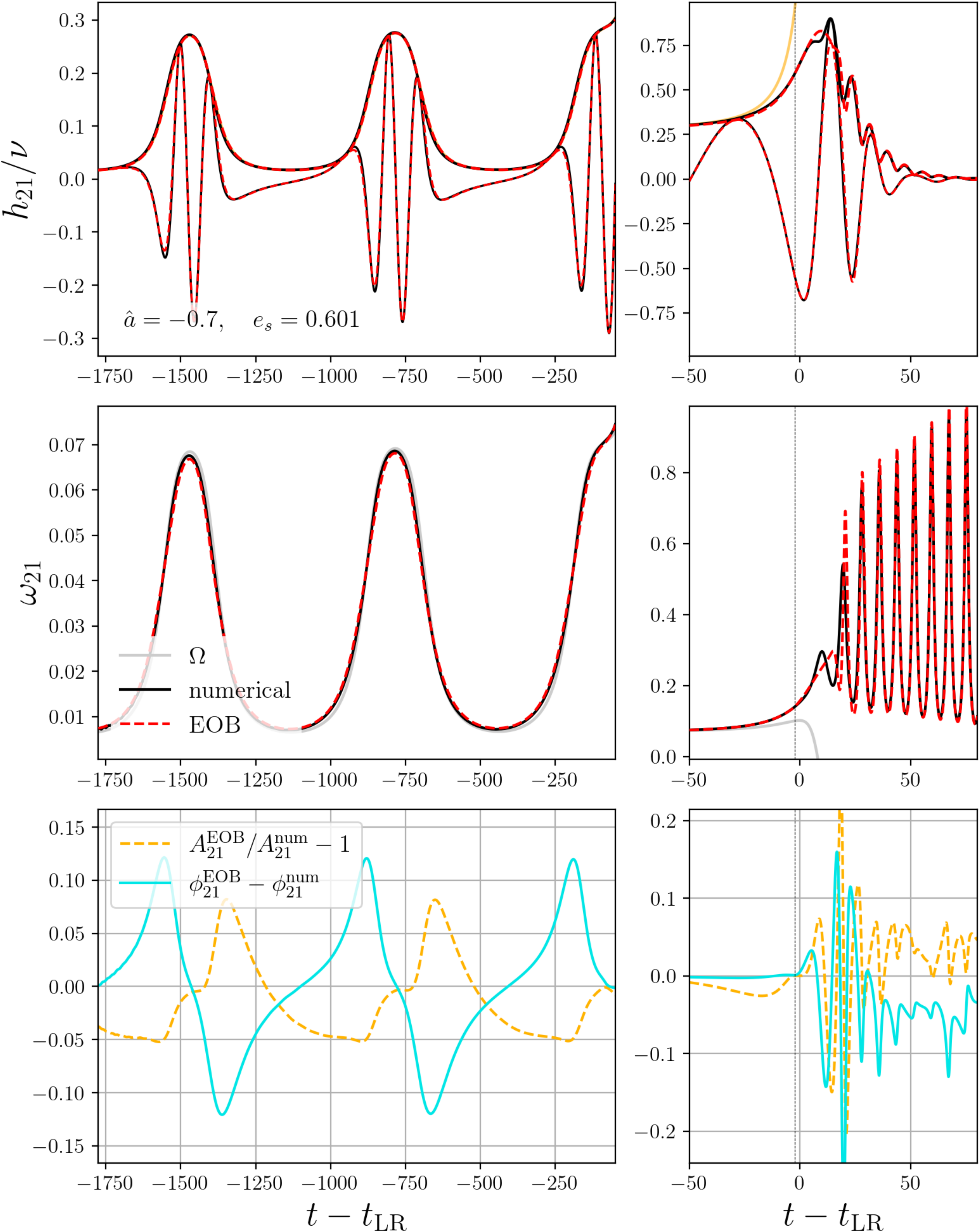}
	\includegraphics[width=0.32\textwidth]{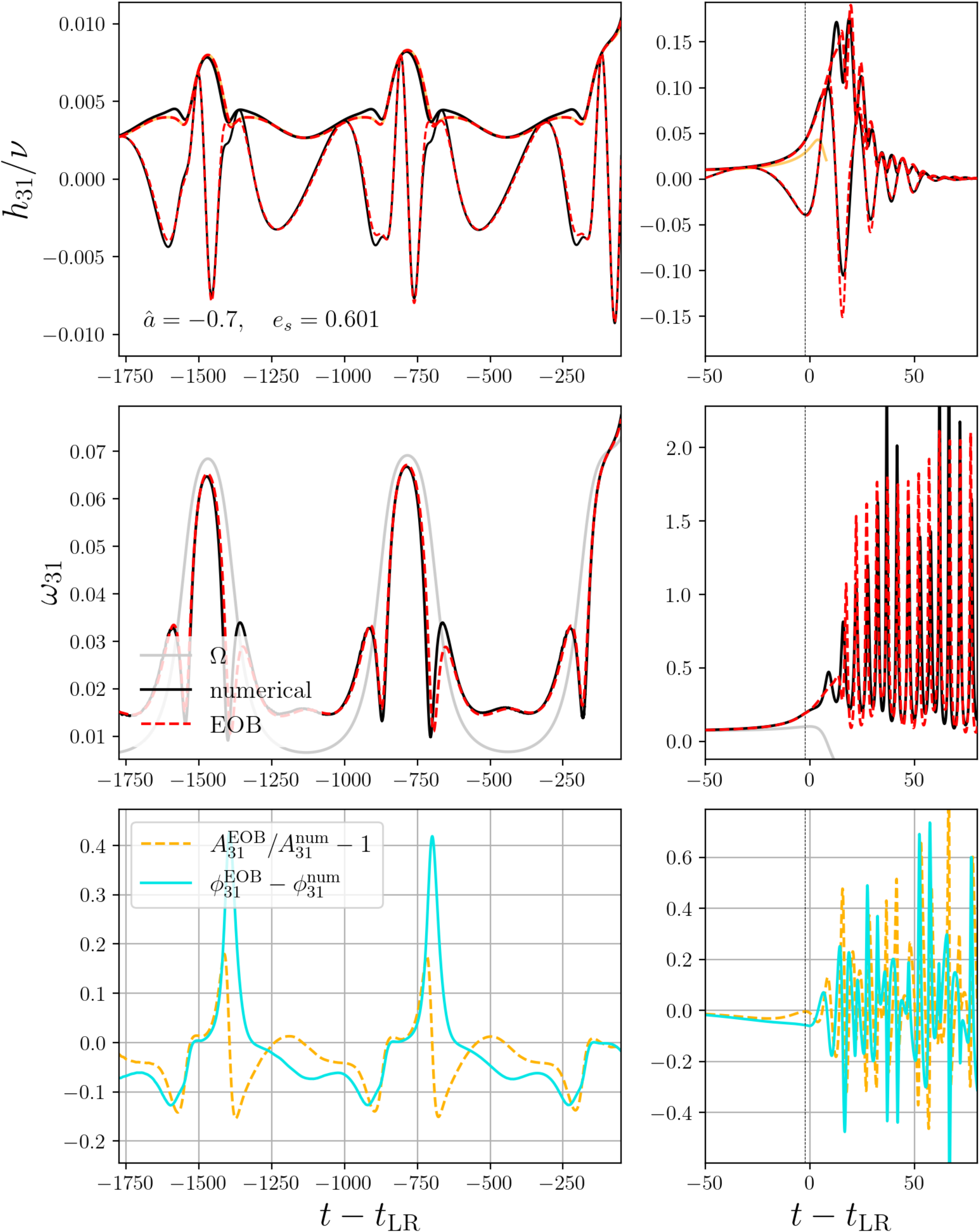}
	\includegraphics[width=0.32\textwidth]{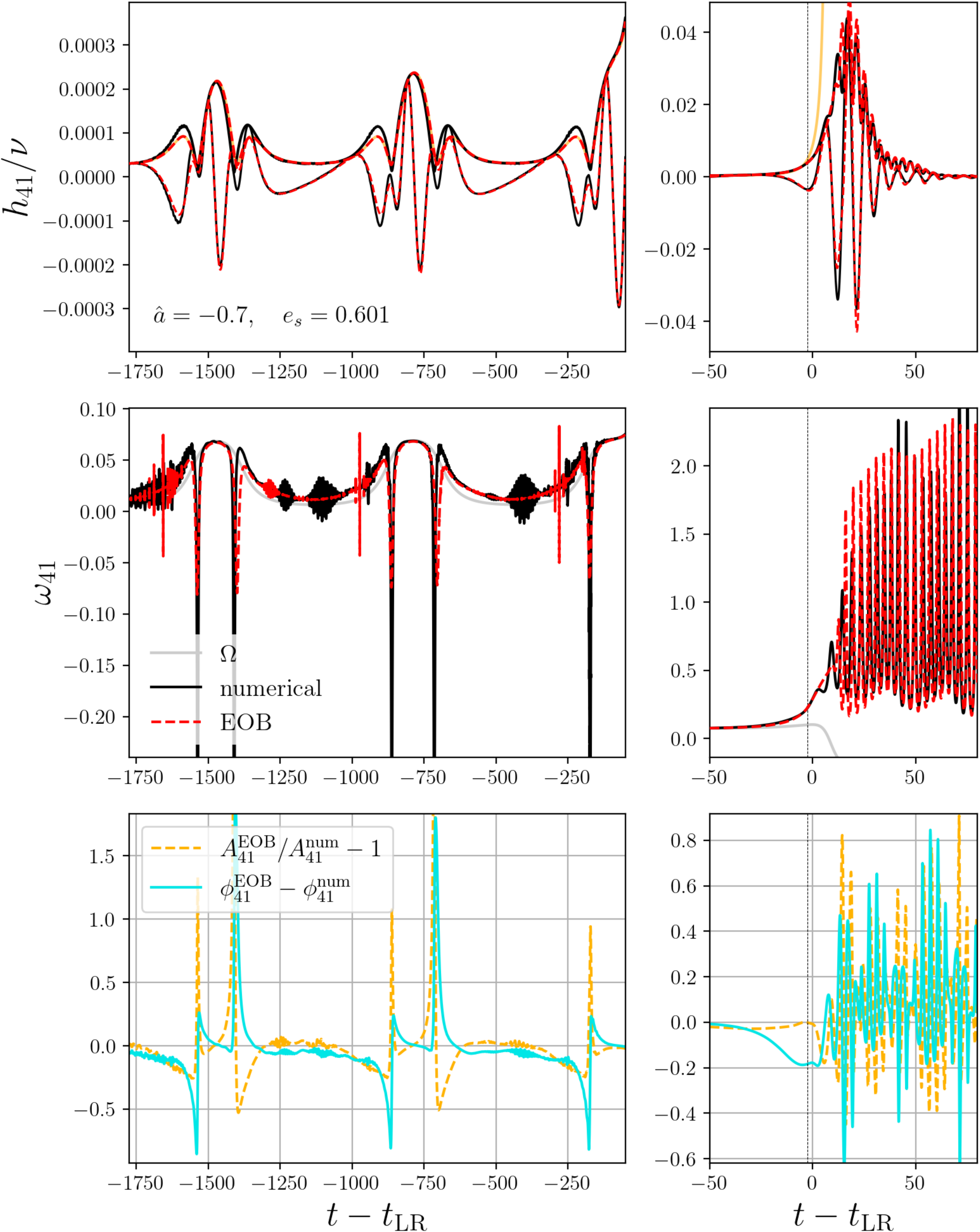}
	 \caption{\label{fig:comparisons_m1_e06} Analytical/numerical comparisons for quasi-circular systems with 
	 $\ha=0,\pm 0.7$ and $\esep\simeq 0.6$ for the $m=1$ modes considered in this work. 
	 The EOB complete waveforms are highlighted 
	 in dashed red, while numerical results in black. We also report the inspiral-plunge
	 amplitude without NQC corrections in the upper panels (orange) and $1 \Omega$ in the middle
	 panels. For each case, we show phase (azul) and relative amplitude differences (dashed orange).}
	\end{center} 
\end{figure*}
\begin{figure*}[t]
	\begin{center}
	\includegraphics[width=0.32\textwidth]{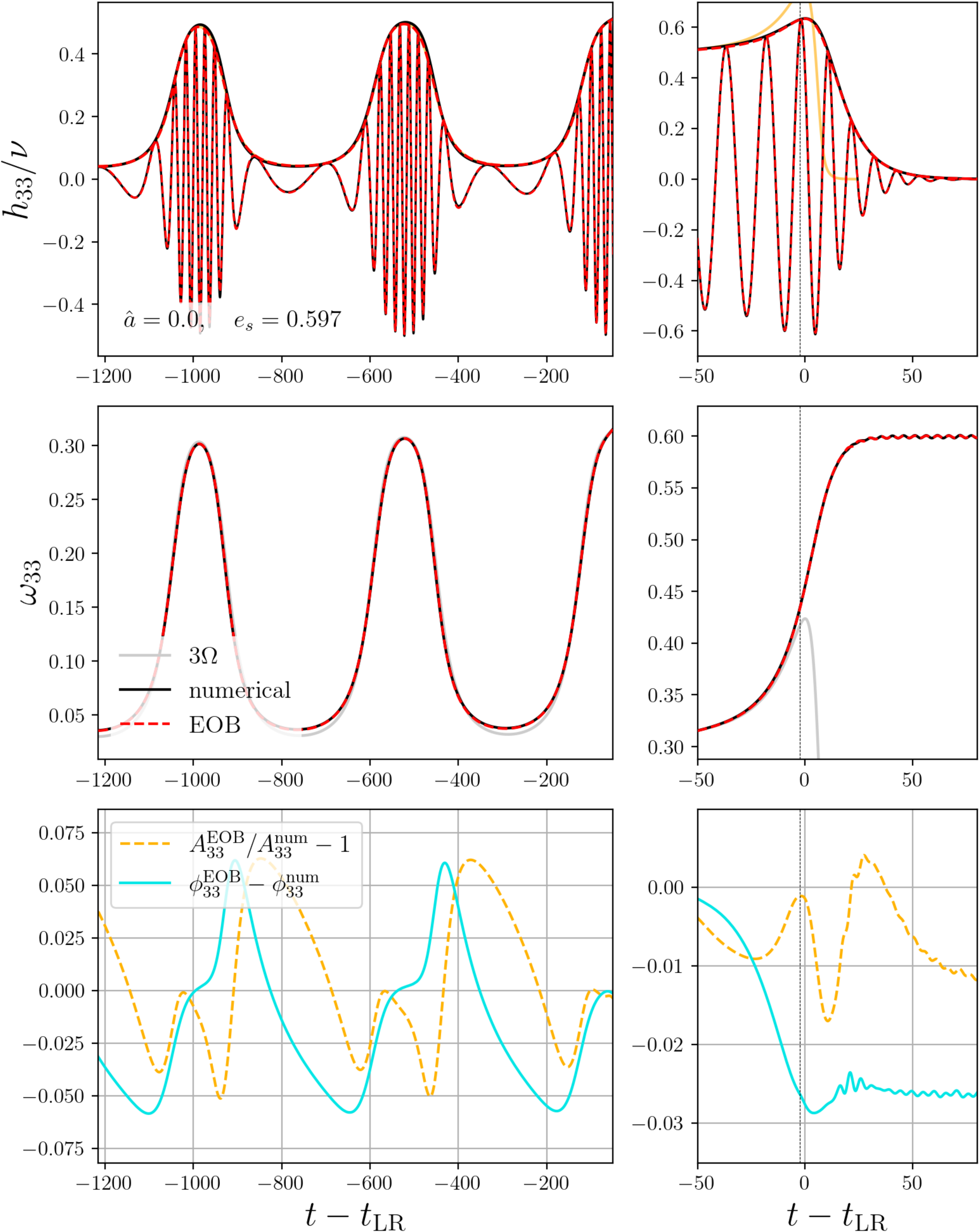}
	\includegraphics[width=0.32\textwidth]{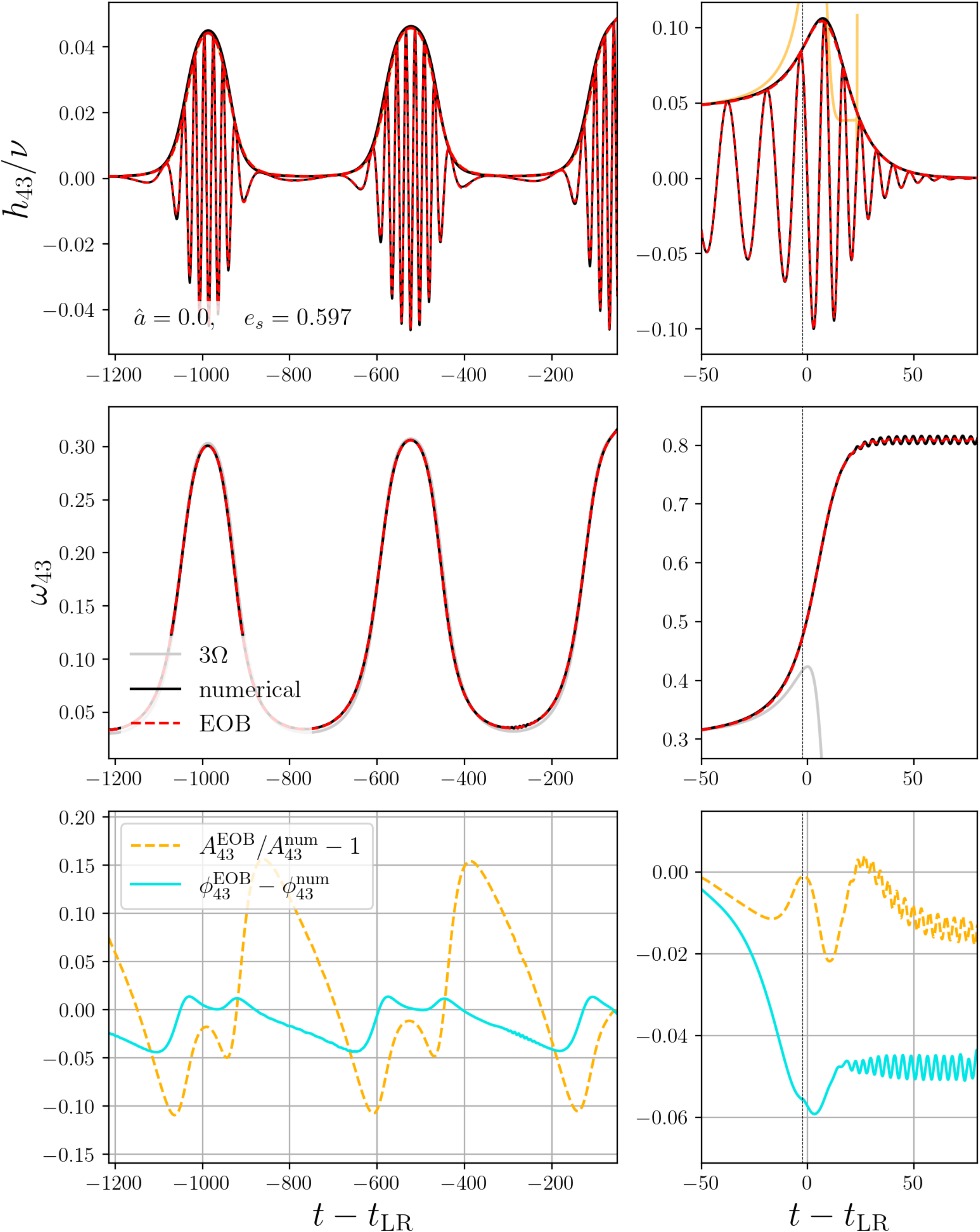}
	\includegraphics[width=0.32\textwidth]{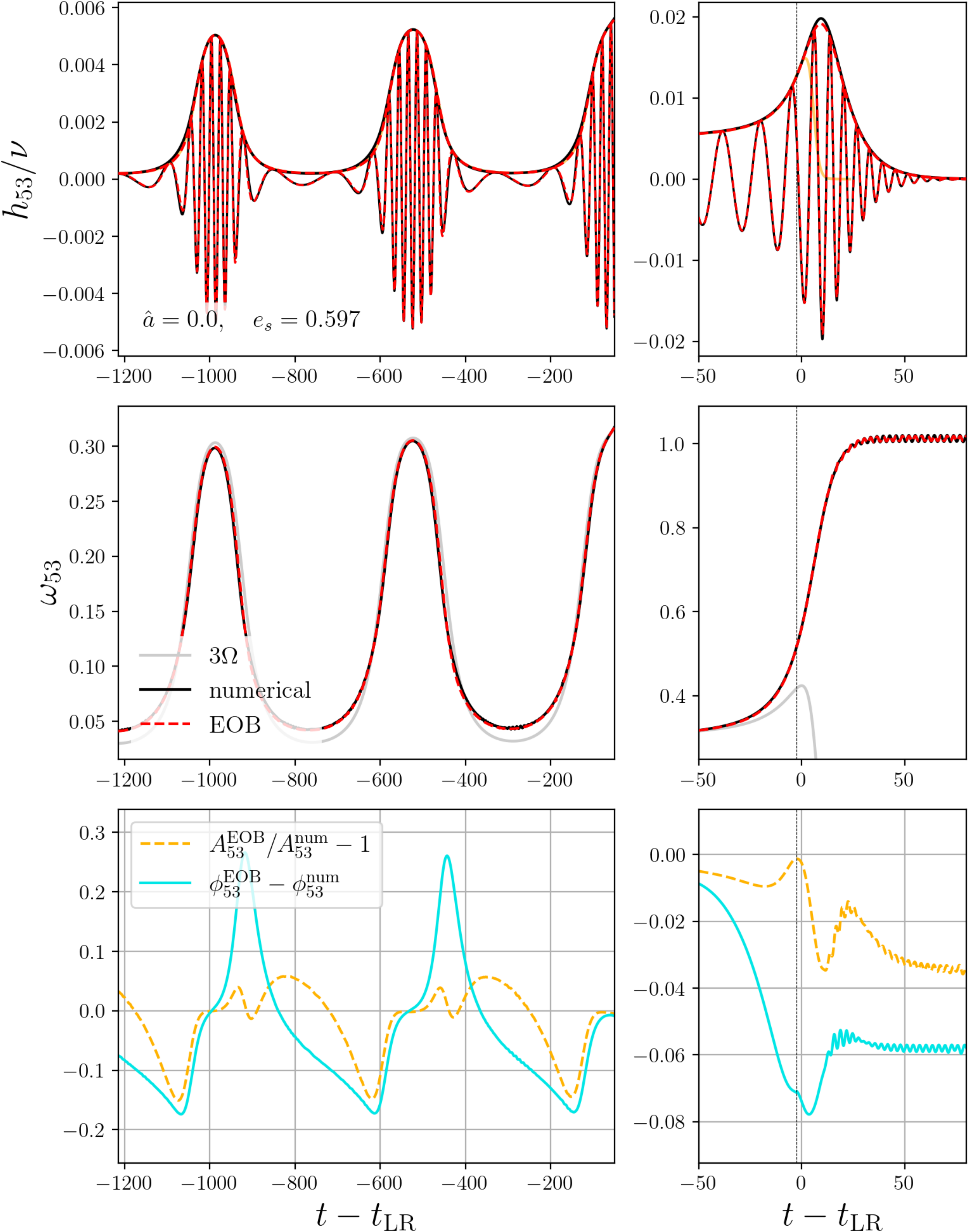}\\
	\hline
	\vspace{0.3cm}
	\includegraphics[width=0.32\textwidth]{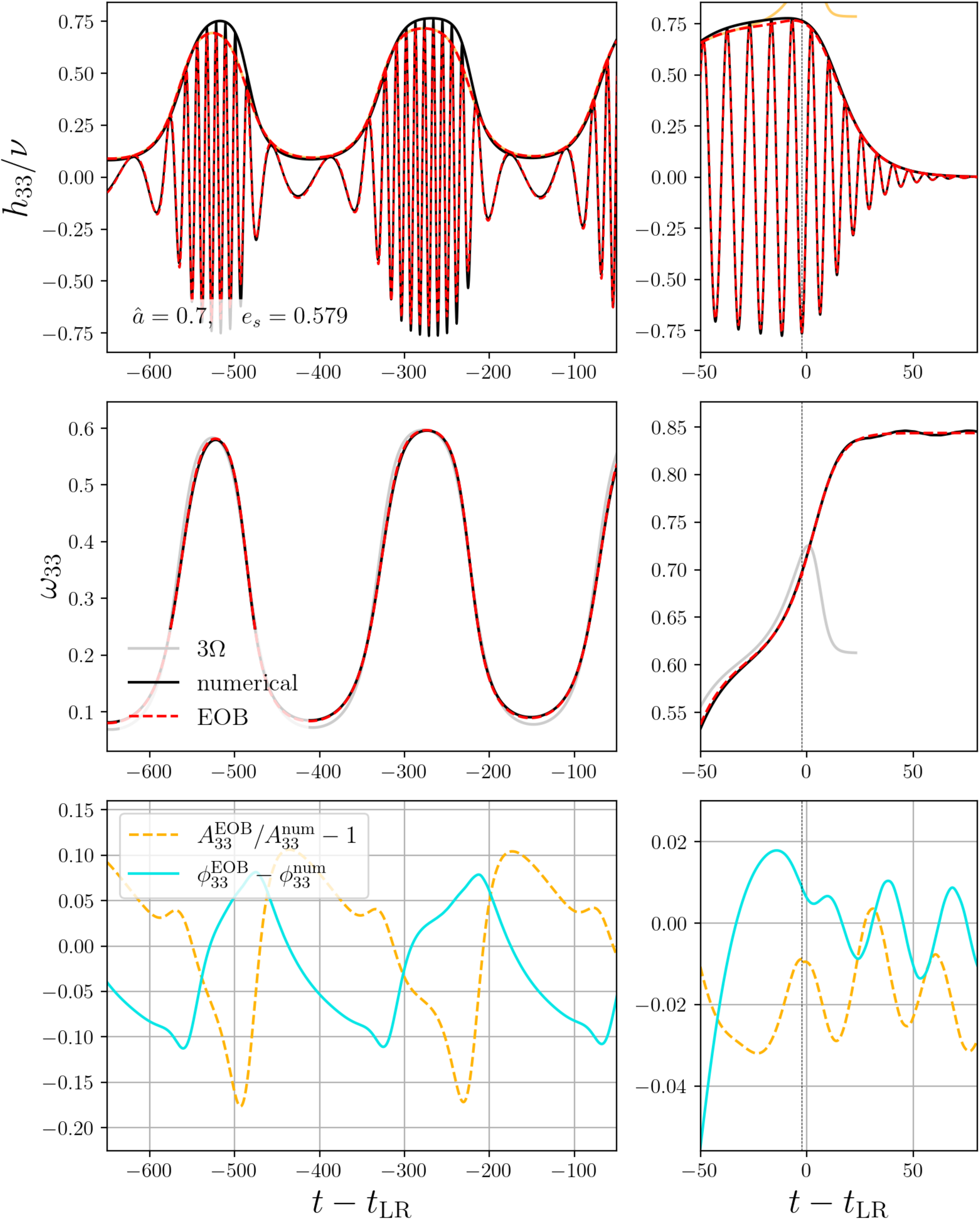}
	\includegraphics[width=0.32\textwidth]{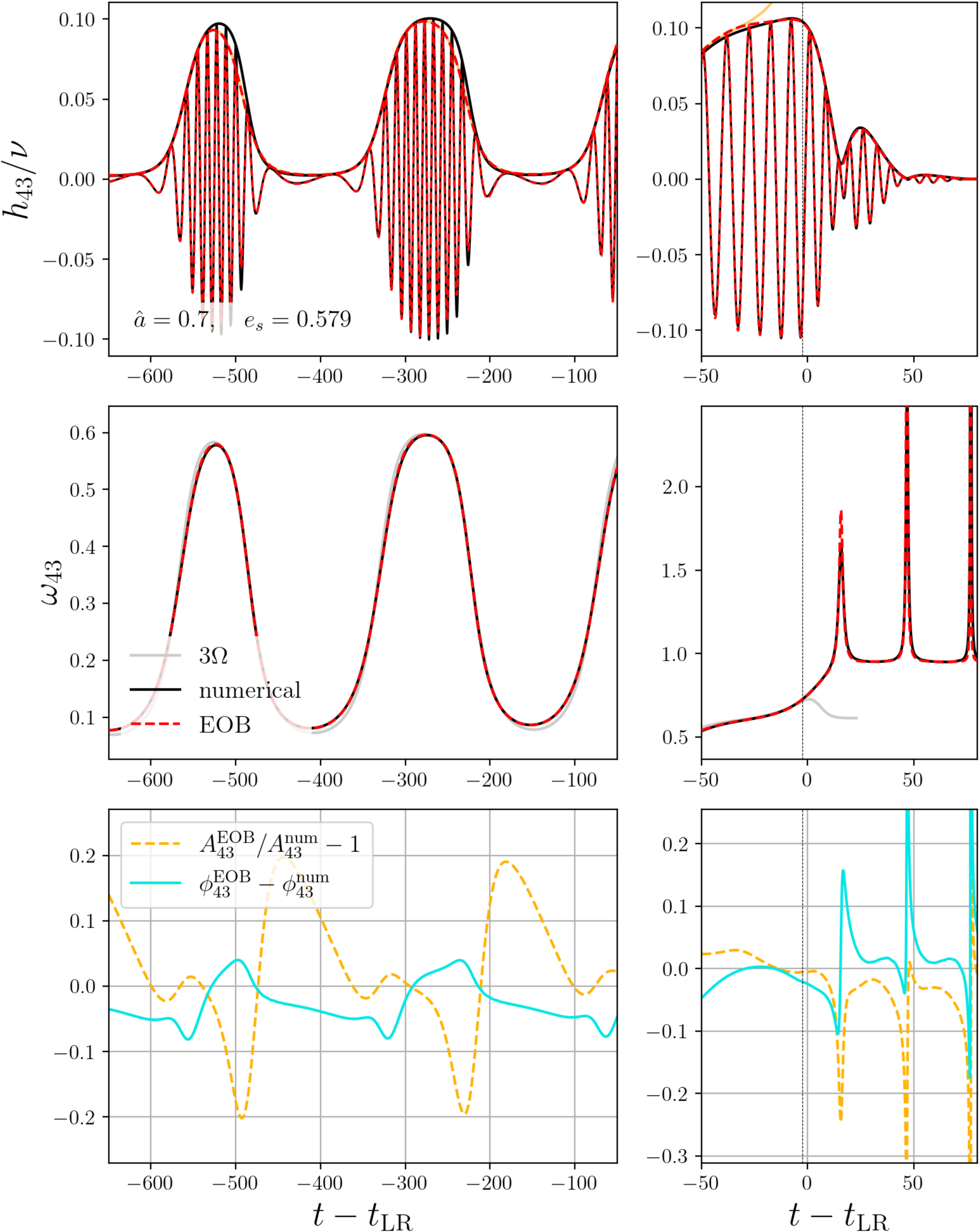}
	\includegraphics[width=0.32\textwidth]{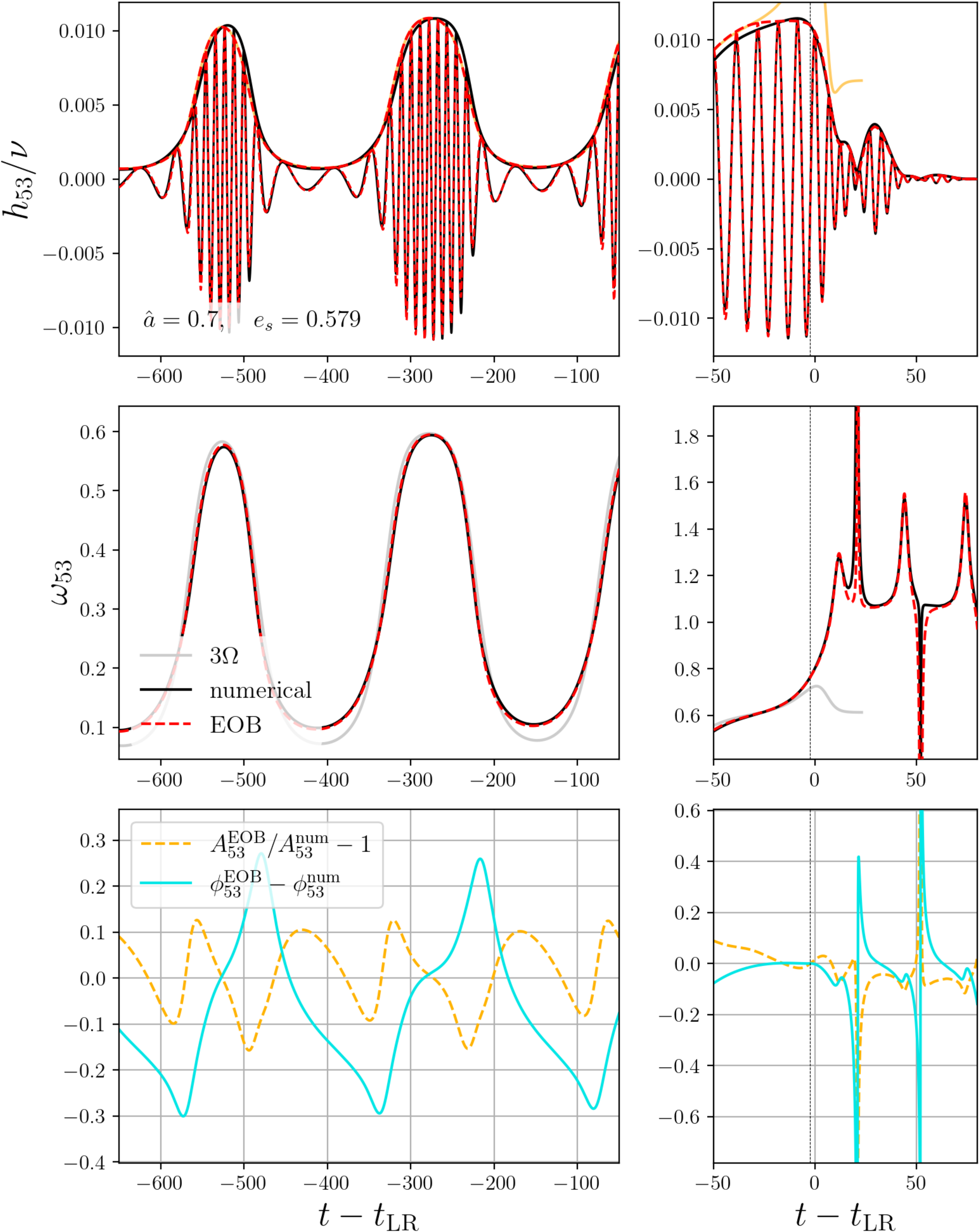}\\
	\hline
	\vspace{0.3cm}
	\includegraphics[width=0.32\textwidth]{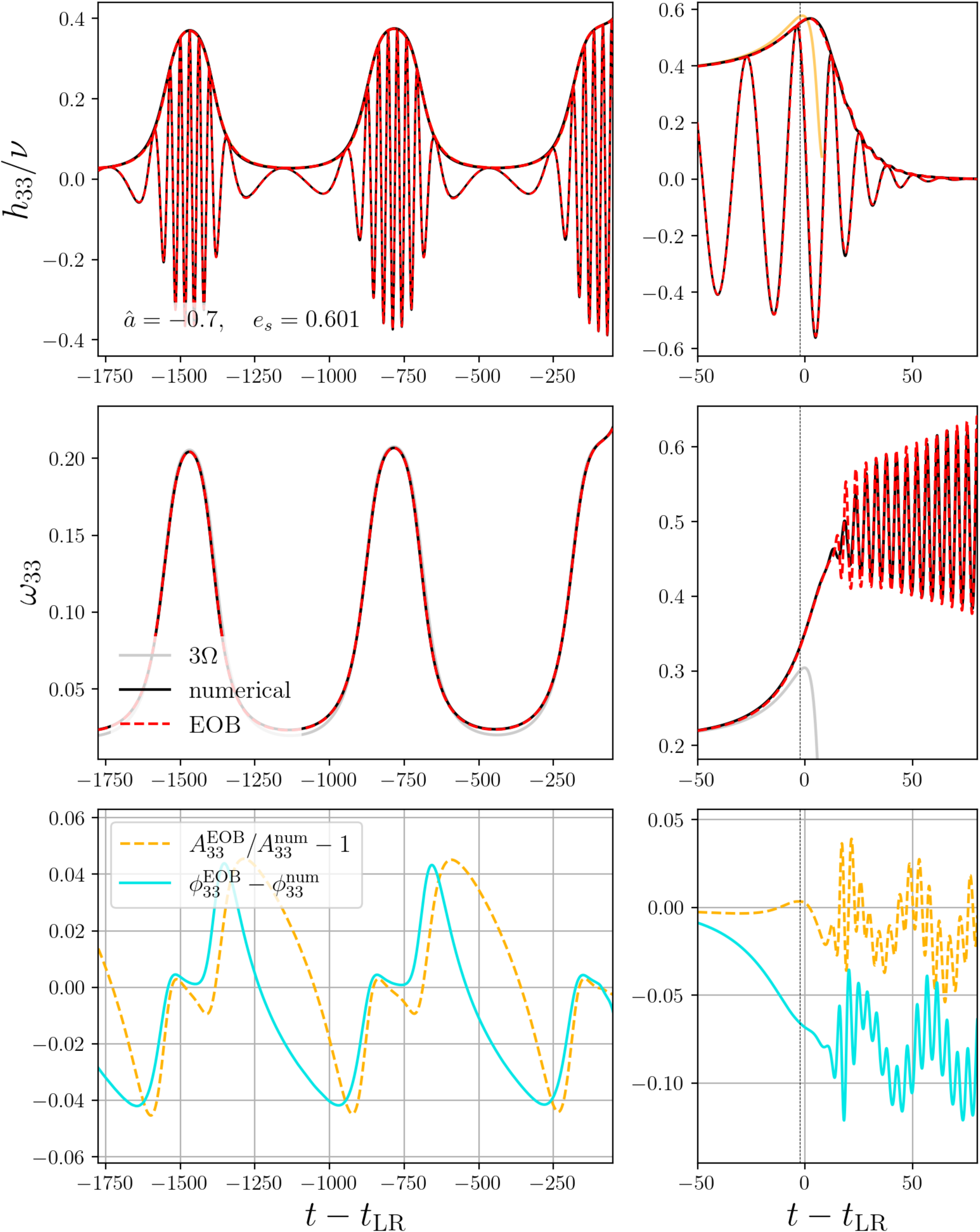}
	\includegraphics[width=0.32\textwidth]{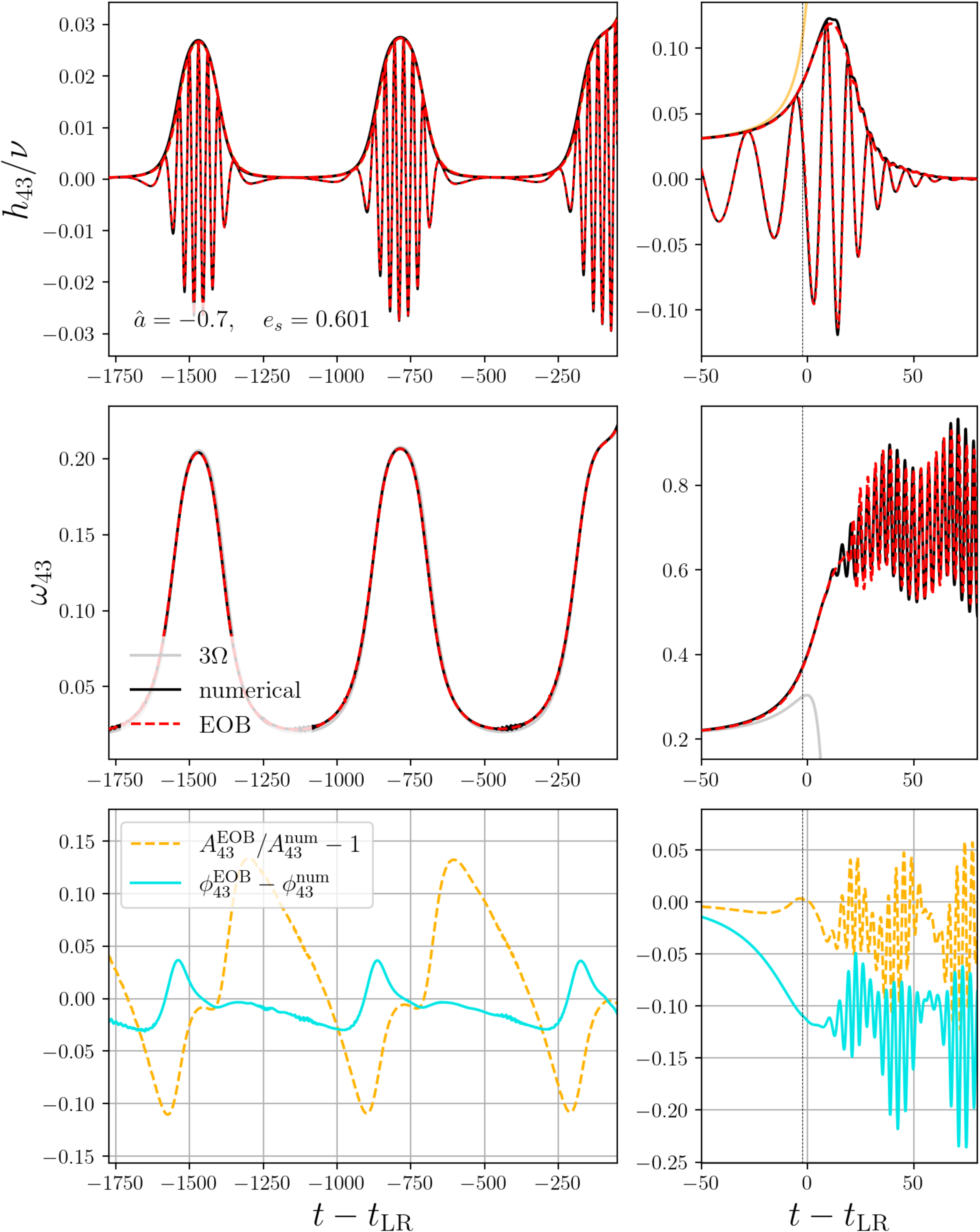}
	\includegraphics[width=0.32\textwidth]{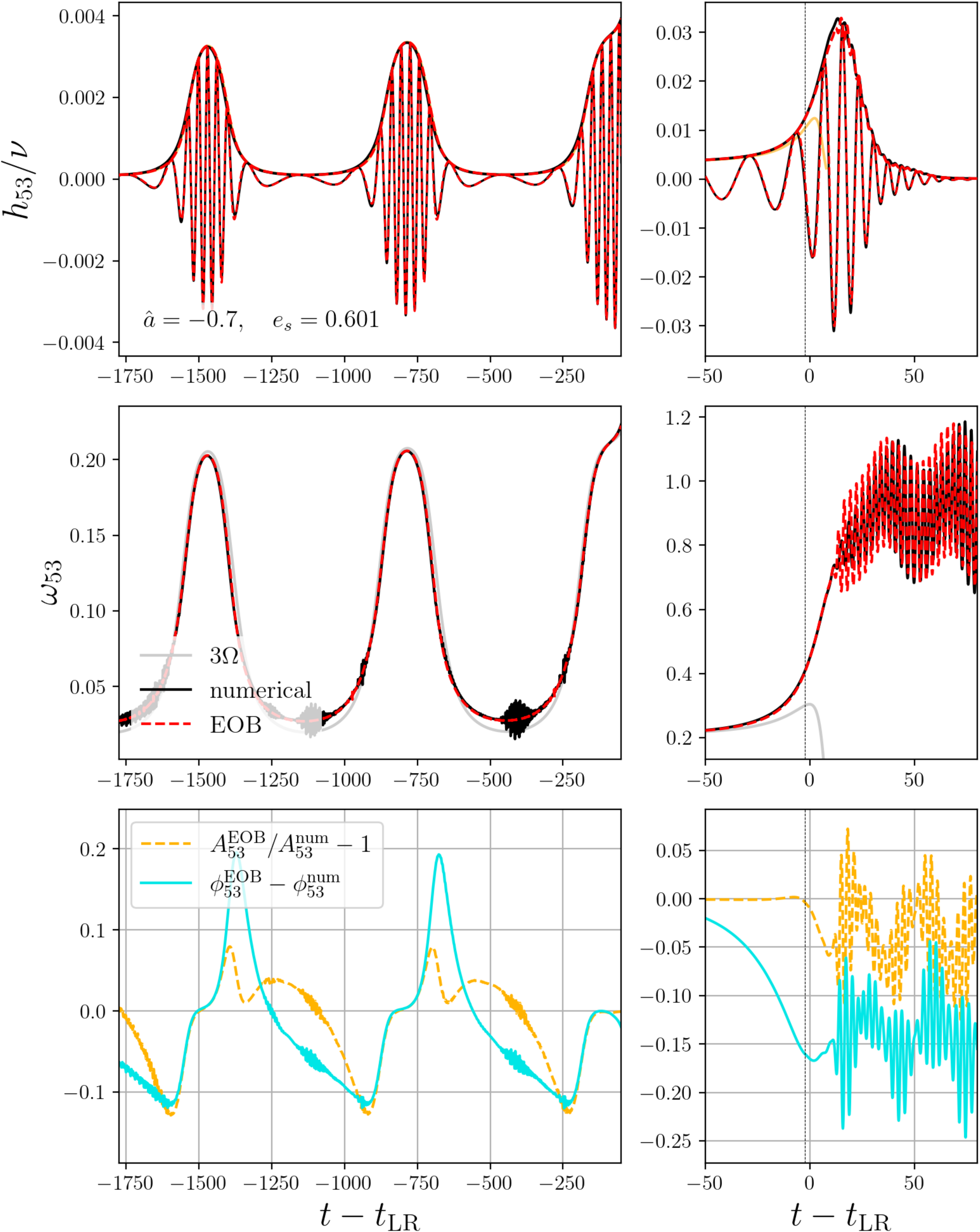}
	 \caption{\label{fig:comparisons_m3_e06} Analytical/numerical comparisons for quasi-circular systems with 
	 $\ha=0,\pm 0.7$ and $\esep\simeq 0.6$ for the $m=3$ modes considered in this work. 
	 The EOB complete waveforms are highlighted 
	 in dashed red, while numerical results in black. We also report the inspiral-plunge
	 amplitude without NQC corrections in the upper panels (orange) and $3 \Omega$ in the middle
	 panels. For each case, we show phase (azul) and relative amplitude differences (dashed orange).}
	\end{center} 
\end{figure*}

\clearpage
\bibliography{refs20260319.bib,refs_loc20260319.bib}

\end{document}